\documentclass[floatfix,aps,prx,twocolumn,letterpaper,lengthcheck,superscriptaddress,showpacs,amssymb,amsmath,amsfonts,nofootinbib,altaffilletter,nopreprintnumbers,showpacs,longbibliography]{revtex4-2}
\pdfoutput=1
\usepackage[utf8]{inputenc}
\usepackage{xspace}
\usepackage{acronym}
\usepackage{booktabs}
\usepackage{amsmath,amssymb,graphicx}
\usepackage[english]{babel}
\usepackage{upgreek}
\usepackage{color}
\usepackage[colorlinks, linkcolor=blue, pdfborder={0 0 0}, breaklinks=true]{hyperref}
\usepackage[table]{xcolor}
\definecolor{lightgray}{gray}{0.9} % table alternating line colors
\usepackage{tabularx}
\usepackage{multirow}
\usepackage{longtable}

\newcolumntype{x}[1]{>{\centering\arraybackslash\hspace{0pt}}p{#1}}
\newcolumntype{Y}{>{\centering\arraybackslash}X}
\newcolumntype{Z}{>{\raggedright\arraybackslash}X}

\newenvironment{PE_table}{\setlength{\tabcolsep}{0pt}}{}
\newenvironment{event_table}{\setlength{\tabcolsep}{0pt}}{}
\usepackage{scrextend}
\usepackage{textgreek}
\usepackage{graphicx}
\usepackage[caption=false]{subfig}

\usepackage{makerobust}

\def\mystrut{\vrule height 9.3pt depth 3.1pt width 0pt}

\renewcommand{\today}{\number\day\space\ifcase\month\or
  January\or February\or March\or April\or May\or June\or
  July\or August\or September\or October\or November\or December\fi
  \space\number\year}

\definecolor{danger-red}{rgb}{0.8, 0.4, 0.0}
\definecolor{warning-amber}{rgb}{0.9, 0.6, 0.0}
\definecolor{ok-green}{rgb}{0.0, 0.6, 0.5}
\newcommand{\reviewed}[1]{{#1}}

\DeclareOldFontCommand{\it}{\normalfont\itshape}{\mathit}

\begin{document}

% Content macros
\input{event_macros}
%%%%%%%%%%%%%%%%%%%%%%%%%%%%%%%%%%%%%%%%%%%%%%%%%%%%%%%%%%%%%%%%%%%%%%%%%%%%%%%
% A place to keep maths symbol and unit definitions                           %
%                                                                             %
% If you want to add numbers, please do so in result_numbers.tex.             %
% If you want to add acronyms or abbreviations, use acronyms.tex.             %
%                                                                             %
% Symbols should follow https://dcc.ligo.org/LIGO-T2000185                    %    
%                                                                             %
%%%%%%%%%%%%%%%%%%%%%%%%%%%%%%%%%%%%%%%%%%%%%%%%%%%%%%%%%%%%%%%%%%%%%%%%%%%%%%%

%%%%%%%%%%%%%%%%%
%%%%% units %%%%%
%%%%%%%%%%%%%%%%%

\newcommand{\Msun}{\ensuremath{{M}_\odot}}
\newcommand\Mpcyr{\ensuremath{\mathrm{Mpc}^{3}\,\mathrm{yr}}}
\newcommand\Gpcyr{\ensuremath{\mathrm{Gpc}^{3}\,\mathrm{yr}}}
\newcommand\perMpcyr{\ensuremath{\mathrm{Mpc}^{-3}\,\mathrm{yr}^{-1}}}
\newcommand\perGpcyr{\ensuremath{\mathrm{Gpc}^{-3}\,\mathrm{yr}^{-1}}}

%%%%%%%%%%%%%%%%%%%
%%%%% symbols %%%%%
%%%%%%%%%%%%%%%%%%%

% mass parameters
\newcommand{\massone}{\ensuremath{m_1}}
\newcommand{\masstwo}{\ensuremath{m_2}}
\newcommand{\Mc}{\ensuremath{\mathcal{M}}}
\newcommand{\Mtot}{\ensuremath{M}}
\newcommand{\Mf}{\ensuremath{M_\mathrm{f}}}
\newcommand{\massratio}{\ensuremath{q}}
\newcommand{\Erad}{\ensuremath{E_\mathrm{rad}}}

% spin parameters
\newcommand{\chieff}{\ensuremath{\chi_\mathrm{eff}}}
\newcommand{\chip}{\ensuremath{\chi_\mathrm{p}}}
\newcommand{\chif}{\ensuremath{\chi_\mathrm{f}}}
\newcommand{\spintilt}[1]{\ensuremath{\theta_{{LS}_{#1}}}}
\newcommand{\spinone}{\ensuremath{\chi_1}}
\newcommand{\vecspinone}{\ensuremath{\vec\chi_1}}
\newcommand{\spintwo}{\ensuremath{\chi_2}}
\newcommand{\vecspintwo}{\ensuremath{\vec\chi_2}}
\newcommand{\LNewton}{\ensuremath{\hat{L}_\mathrm{N}}}
\newcommand{\thetaJN}{\ensuremath{\theta_{JN}}}

% distance
\newcommand{\DL}{\ensuremath{D_\mathrm{L}}}
\newcommand{\DC}{\ensuremath{D_\mathrm{c}}}
\newcommand{\redshift}{\ensuremath{z}}

% PE quantities
\newcommand{\PEprob}{\ensuremath{p}}%
\newcommand{\PEparameter}{\ensuremath{\vec{\theta}}}%
\newcommand{\PEdata}{\ensuremath{{d}}}%
\newcommand{\PEprior}{\ensuremath{\PEprob(\PEparameter)}}%
\newcommand{\PEposterior}{\ensuremath{\PEprob(\PEparameter | \PEdata)}}%
\newcommand{\PElikelihood}{\ensuremath{\PEprob(\PEdata | \PEparameter)}}%
\newcommand{\PEmodelh}{\ensuremath{h_{\mathrm{M}}}}% 

% PE setting
\newcommand{\flow}{\ensuremath{f_\mathrm{low}}}
\newcommand{\fhi}{\ensuremath{f_\mathrm{high}}}
\newcommand{\fsamp}{\ensuremath{f_\mathrm{s}}}
\newcommand{\fNyq}{\ensuremath{f_\mathrm{Nyquist}}}
\newcommand{\alphaRoll}{\ensuremath{\alpha^\mathrm{roll\text{-}off}}}

% Cosmology symbols
\newcommand{\HzeroSymbol}{\ensuremath{H_{0}}}
\newcommand{\WmSymbol}{\ensuremath{\Omega_{\mathrm{m}}}}

% KL divergences
\newcommand{\DKLchip}{\ensuremath{D_\mathrm{KL}^{\chi_\mathrm{p}}}}
\newcommand{\DKLchieff}{\ensuremath{D_\mathrm{KL}^{\chi_\mathrm{eff}}}}

% search pipeline parameters
\newcommand{\rankstat}{\ensuremath{x}}

% others 
\newcommand{\VT}{\ensuremath{\langle VT \rangle}}
\newcommand{\pastro}{\ensuremath{p_{\mathrm{astro}}}}
\newcommand{\pterr}{\ensuremath{p_{\mathrm{terr}}}}
\newcommand{\pbbh}{\ensuremath{p_{\mathrm{BBH}}}}
\newcommand{\pbns}{\ensuremath{p_{\mathrm{BNS}}}}
\newcommand{\pnsbh}{\ensuremath{p_{\mathrm{NSBH}}}}
\newcommand{\comovingv}{\ensuremath{V_\mathrm{c}}}
\newcommand{\comovingvt}{\ensuremath{\langle VT_\mathrm{c} \rangle}}
\newcommand{\injspinmax}{\ensuremath{\chi_\mathrm{max}}}

\input{PE_macros}
\input{PE_macros_with_cuts}
%%%%%%%%%%%%%%%%%%%%%%%%%%%%%%%%%%%%%%%%%%%%%%%%%%%%%%%%%%%%%%%%%%%%%%%%%%%%%%%
% All numbers in the paper should come from a macro                           %
%                                                                             %
% If you are producing a large number of macros, please consider making a     %
% dedicted file as the output of your script.                                 %
% If you only have a small number, please add them below.                     %
% If you want to add acronyms or abbreviations, use acronyms.tex.             %
%                                                                             %
% Definitions of units and and symbols should do in symbol_macros.tex         %    
%                                                                             %
%%%%%%%%%%%%%%%%%%%%%%%%%%%%%%%%%%%%%%%%%%%%%%%%%%%%%%%%%%%%%%%%%%%%%%%%%%%%%%%

% O3b start and stop date
% From https://logbook.virgo-gw.eu/virgo/?r=47511
\newcommand{\RUNSTART}{\reviewed{1 November 2019, 15:00}}  % UTC https://git.ligo.org/publications/O3/o3b-cbc-catalog/-/wikis/Instruments-Section-Review
% From https://logbook.virgo-gw.eu/virgo/?r=48833
\newcommand{\RUNEND}{\reviewed{27 March 2020, 17:00}}   % UTC https://git.ligo.org/publications/O3/o3b-cbc-catalog/-/wikis/Instruments-Section-Review

%% Analysis related macros
\newcommand{\PASTROTHRESHOLD}{\reviewed{\ensuremath{0.5}}}
\newcommand{\PASTROTHRESHOLDPOINTNINE}{\reviewed{\ensuremath{0.9}}}
\newcommand{\CONTAMINATION}{\reviewed{\ensuremath{4\text{--}6}}}
\newcommand{\CONTAMINATIONFRACTION}{\reviewed{\ensuremath{10\text{--}15\%}}}

% Catalog FAR threshold
\newcommand{\FARTHRESHOLD}{\ensuremath{\reviewed{10^{-2}}}}

%number of p-values from BayesWave
\newcommand{\BWPVALNUM}{\ensuremath{\reviewed{12}}}

%number of p-values from BayesWave that are outside the 90% band
\newcommand{\BWPOUTNUM}{\ensuremath{\reviewed{4}}}

%probability of random p-value plot like the one found by BW 
\newcommand{\BWPVALPROB}{\ensuremath{\reviewed{4.1\%}}}

%number of p-values from cWB
\newcommand{\CWBPVALNUM}{\ensuremath{\reviewed{15}}}

%number of p-values from cWB that are outside the 90% band
\newcommand{\CWBPOUTNUM}{\ensuremath{\reviewed{0}}}

%number of cWB events that are reconstructed but not detected 
\newcommand{\CWBRECONLY}{\ensuremath{\reviewed{5}}}

% CBC counterpart threshold
\newcommand{\CBCTHRESHOLD}{\ensuremath{\reviewed{0.1}}}

% Number of public alerts (not including retracted) 
%% From low_latency_numbers.ipynb
\newcommand{\OTHREEBALERTSNOTRETRACTEDCBC}{\reviewed{\ensuremath{22}}} 
\newcommand{\OTHREEBALERTSNOTRETRACTEDBURST}{\reviewed{\ensuremath{1}}} 
\newcommand{\ALERTMEDIANTIME}{\reviewed{\ensuremath{5.8}}} 
\newcommand{\ALERTNINETYFIVEPCTTIME}{\reviewed{\ensuremath{377}}} 
\newcommand{\ALERTFIVEPCTTIME}{\reviewed{\ensuremath{3}}} 
\newcommand{\NUMOTHREEAOPA}{\reviewed{\ensuremath{40}}}   
\newcommand{\NUMOTHREEAOPAUNRETRACTED}{\reviewed{\ensuremath{33}}}
\newcommand{\ALERTMEDIANTIMEOTHREEA}{\reviewed{\ensuremath{7.3}}} 
\newcommand{\ALERTNINETYFIVEPCTTIMEOTHREEA}{\reviewed{\ensuremath{56}}} 
\newcommand{\ALERTFIVEPCTTIMEOTHREEA}{\reviewed{\ensuremath{2}}}  
\newcommand{\MANYALERTS}{\reviewed{\ensuremath{53}}} 
% FAR thresholds
\newcommand{\OPAFARTHRESHCBC}{\reviewed{\ensuremath{\mathrm{1~per~2~months}}}}
\newcommand{\OPAFARTHRESHBURST}{\reviewed{\ensuremath{\mathrm{1~per~year}}}}

% GCN traffic statistics
\newcommand{\GCNTRAFFICALL}{\reviewed{\ensuremath{3463}}} 
\newcommand{\GCNTRAFFICCOUNT}{\reviewed{\ensuremath{1513}}} 
\newcommand{\GCNTRAFFICTHREEACOUNT}{\reviewed{\ensuremath{967}}}  
\newcommand{\GCNTRAFFICTHREEBCOUNT}{\reviewed{\ensuremath{546}}}  
\newcommand{\GCNTRAFFIC}{\reviewed{\ensuremath{44}}} 
\newcommand{\GCNTRAFFICTHREEA}{\reviewed{\ensuremath{64}}}  
\newcommand{\GCNTRAFFICTHREEB}{\reviewed{\ensuremath{36}}}  

%% Vetting latency
\newcommand{\HUMANVETLAT}{\reviewed{\ensuremath{30}}} % Latency of the human vetting by the RRT

%% Window length for grouping events
\newcommand{\WINDOWSEVENTCBC}{\reviewed{\ensuremath{1}}} % Time window length for grouping events in a superevent for CBC 

% Approximate number of different instruments used for observing
\newcommand{\APPROXOBSERVATORIES}{\reviewed{\ensuremath{100}}}

% Number of events followed by different observatories
\newcommand{\DISTKN}{\reviewed{\ensuremath{200~\mathrm{Mpc}}}}
\newcommand{\MAGKN}{\reviewed{\ensuremath{21}}}

% Details of follow-up searches
% Neutrino
\newcommand{\NUMINENERGY}{\reviewed{\ensuremath{1~\mathrm{MeV}}}}
\newcommand{\NUMAXENERGY}{\reviewed{\ensuremath{1~\mathrm{PeV}}}}

% Gamma/X-ray
\newcommand{\GAMMAMAXENERGY}{\reviewed{\ensuremath{1~\mathrm{TeV}}}}

% FRB with possible association to GW190425
\newcommand{\FRB}{\reviewed{FRB~20190425A}}
\newcommand{\FRBDELAY}{\ensuremath{\reviewed{2.5~\mathrm{hr}}}}
\newcommand{\FRBLATE}{\ensuremath{\reviewed{2.5~\mathrm{yr}}}}

% Range macros
% BNS VT single-detector factor
\newcommand{\BNSRANGESNR}{\ensuremath{\reviewed{8}}}
\newcommand{\BNSRANGEMASS}{\ensuremath{\reviewed{1.4\Msun+1.4\Msun}}}
\newcommand{\BNSVTSINGLEFACTOR}{\ensuremath{\reviewed{1.5}}}
\newcommand{\BNSRANGESNRSINGLE}{\ensuremath{\reviewed{12}}}

% O3b numbers
\newcommand{\VIRGORANGE}{\reviewed{\ensuremath{51~\mathrm{Mpc}}}}
\newcommand{\HANFORDRANGE}{\reviewed{\ensuremath{115~\mathrm{Mpc}}}}
\newcommand{\LIVINGSTONRANGE}{\reviewed{\ensuremath{133~\mathrm{Mpc}}}}

% O3a numbers
\newcommand{\VIRGORANGETHREEA}{\reviewed{\ensuremath{45~\mathrm{Mpc}}}}   % https://git.ligo.org/publications/O3/o3b-cbc-catalog/-/wikis/Instruments-Section-Review
\newcommand{\HANFORDRANGETHREEA}{\reviewed{\ensuremath{108~\mathrm{Mpc}}}}   % https://git.ligo.org/publications/O3/o3b-cbc-catalog/-/wikis/Instruments-Section-Review
\newcommand{\LIVINGSTONRANGETHREEA}{\reviewed{\ensuremath{135~\mathrm{Mpc}}}}   % https://git.ligo.org/publications/O3/o3b-cbc-catalog/-/wikis/Instruments-Section-Review

% Order-of-magnitude laswer wavelength for LIGO and Virgo
\newcommand{\APPROXLASERWAVELENGTH}{\reviewed{\ensuremath{1~\mathrm{\upmu{}m}}}}

% Increased noise frequency in LLO
%\newcommand{\LIVINGSTONFREQNOISE}{\reviewed{\ensuremath{20~\mathrm{Hz}}}}

% LIGO sensitivity improvement frequency
\newcommand{\LIGOFREQSENS}{\reviewed{\ensuremath{55~\mathrm{Hz}}}}

% Virgo sensitivity improvement frequency
\newcommand{\VIRGOFREQSENS}{\reviewed{\ensuremath{300~\mathrm{Hz}}}}

% Earthquake noise band (Hz)
\newcommand{\EARTHQUAKELOW}{\ensuremath{\reviewed{0.03}}}
\newcommand{\EARTHQUAKEHIGH}{\ensuremath{\reviewed{0.1}}}
% Microseism noise band (Hz)
\newcommand{\MICROSEISMLOW}{\ensuremath{\reviewed{0.1}}}
\newcommand{\MICROSEISMHIGH}{\ensuremath{\reviewed{0.5}}}
% Anthropogenic noise band (Hz)
\newcommand{\ANTHROPLOW}{\ensuremath{\reviewed{1}}}
\newcommand{\ANTHROPHIGH}{\ensuremath{\reviewed{6}}}

% LLO range changes
\newcommand{\LLORANGEPOWER}{\reviewed{\ensuremath{5~\mathrm{Mpc}}}} % 4.8 rounded
\newcommand{\LLORANGESQZ}{\reviewed{\ensuremath{3~\mathrm{Mpc}}}}   % 3.3 rounded

% LHO range improvement
\newcommand{\LHORANGESQUEEZER}{\reviewed{\ensuremath{7~\mathrm{Mpc}}}}

% Virgo range improvements
\newcommand{\VIRGORANGEFLICKER}{\reviewed{\ensuremath{2~\mathrm{Mpc}}}}
\newcommand{\VIRGORANGEFEEDBACK}{\reviewed{\ensuremath{2\text{--}3~\mathrm{Mpc}}}}
\newcommand{\VIRGORANGEQUANTIZATION}{\reviewed{\ensuremath{2~\mathrm{Mpc}}}}
\newcommand{\VIRGORANGEALIGNMENT}{\reviewed{\ensuremath{1\text{--}2~\mathrm{Mpc}}}}

% Virgo hardware improvements
\newcommand{\VIRGOFEEDBACKACCURACY}{\reviewed{\ensuremath{5}}}
\newcommand{\VIRGOTEMPERATUREACCURACY}{\reviewed{\ensuremath{6~\mathrm{mK}}}}
\newcommand{\VIRGOTEMPERATUREACCURACYPERCENT}{\reviewed{\ensuremath{2\%}}}
\newcommand{\VIRGOETALONFRINGE}{\reviewed{\ensuremath{532~\mathrm{nm}}}}

% Virgo O3b record range
\newcommand{\VIRGORANGERECORD}{\reviewed{\ensuremath{60~\mathrm{Mpc}}}}   % https://git.ligo.org/publications/O3/o3b-cbc-catalog/-/wikis/Instruments-Section-Review
% Virgo O3a record range from https://arxiv.org/abs/2010.14527
\newcommand{\VIRGORANGERECORDTHREEA}{\reviewed{\ensuremath{50~\mathrm{Mpc}}}}   % https://git.ligo.org/publications/O3/o3b-cbc-catalog/-/wikis/Instruments-Section-Review
% Virgo median range before and after 28/01/2020
\newcommand{\VIRGORANGEPRE}{\reviewed{\ensuremath{49~\mathrm{Mpc}}}}
\newcommand{\VIRGORANGEPOST}{\reviewed{\ensuremath{56~\mathrm{Mpc}}}}

% All duty cycles and live times produced using:
% data/bns_vt/calculate_live_times.ipynb

% O3b duty cycles
\newcommand{\VIRGODUTYCYCLE}{\reviewed{\ensuremath{76}}} % 75.64 rounded
\newcommand{\HANFORDDUTYCYCLE}{\reviewed{\ensuremath{79}}} % 78.67 rounded
\newcommand{\LIVINGSTONDUTYCYCLE}{\reviewed{\ensuremath{79}}} % 78.52 rounded 
\newcommand{\ONEDETECTORDUTYCYCLE}{\reviewed{\ensuremath{96.6}}} % 96.57 rounded
\newcommand{\TWODETECTORSDUTYCYCLE}{\reviewed{\ensuremath{85.3}}} % 85.30 rounded
\newcommand{\THREEDETECTORSDUTYCYCLE}{\reviewed{\ensuremath{51.0}}} % 50.96 rounded

% O3a duty cycles
\newcommand{\VIRGODUTYCYCLETHREEA}{\reviewed{\ensuremath{76}}} % 76.26 rounded
\newcommand{\HANFORDDUTYCYCLETHREEA}{\reviewed{\ensuremath{71}}} % 71.14 rounded
\newcommand{\LIVINGSTONDUTYCYCLETHREEA}{\reviewed{\ensuremath{76}}} % 75.68 rounded
\newcommand{\ONEDETECTORDUTYCYCLETHREEA}{\reviewed{\ensuremath{96.8}}} % 96.83 rounded
\newcommand{\TWODETECTORSDUTYCYCLETHREEA}{\reviewed{\ensuremath{81.8}}} % 81.84 rounded
\newcommand{\THREEDETECTORSDUTYCYCLETHREEA}{\reviewed{\ensuremath{44.4}}} % 44.41 rounded

% From https://wiki.ligo.org/Operations/O3OfficialNumbers
% Converted from H1: 10013361s, L1: 9984232s and V1: 9609039s
% Updated to use numbers from data/bns_vt/calculate_live_times.ipynb
% Includes conversion to days
% O3b live times
\newcommand{\VIRGODAYS}{\reviewed{\ensuremath{111.3}}} % 111.25 rounded
\newcommand{\HANFORDDAYS}{\reviewed{\ensuremath{115.7}}} % 115.71 rounded
\newcommand{\LIVINGSTONDAYS}{\reviewed{\ensuremath{115.5}}} % 115.49 rounded
\newcommand{\ONEDETECTORDAYS}{\reviewed{\ensuremath{142.0}}} % 142.04 rounded
\newcommand{\TWODETECTORSDAYS}{\reviewed{\ensuremath{125.5}}} % 125.46 rounded
\newcommand{\THREEDETECTORSDAYS}{\reviewed{\ensuremath{75.0}}} % 74.95 rounded

\newcommand{\VIRGOYEARS}{\reviewed{\ensuremath{0.305}}}
\newcommand{\HANFORDYEARS}{\reviewed{\ensuremath{0.317}}}
\newcommand{\LIVINGSTONYEARS}{\reviewed{\ensuremath{0.316}}}

\newcommand{\VIRGOYEARSINV}{\reviewed{\ensuremath{3.28}}}
\newcommand{\HANFORDYEARSINV}{\reviewed{\ensuremath{3.16}}}
\newcommand{\LIVINGSTONYEARSINV}{\reviewed{\ensuremath{3.16}}}

% O1 live times in days
\newcommand{\HANFORDDAYSONE}{\reviewed{\ensuremath{75.4}}} % 75.35 rounded
\newcommand{\LIVINGSTONDAYSONE}{\reviewed{\ensuremath{65.0}}} % 65.0 rounded
\newcommand{\ONEDETECTORDAYSONE}{\reviewed{\ensuremath{91.0}}} % 91.0 rounded
\newcommand{\TWODETECTORSDAYSONE}{\reviewed{\ensuremath{49.4}}} % 49.36 rounded

% O2 live times in days
\newcommand{\VIRGODAYSTWO}{\reviewed{\ensuremath{23.82}}} % 23.82 rounded
\newcommand{\HANFORDDAYSTWO}{\reviewed{\ensuremath{160.4}}} % 160.35 rounded
\newcommand{\LIVINGSTONDAYSTWO}{\reviewed{\ensuremath{147.0}}} % 146.99 rounded
\newcommand{\ONEDETECTORDAYSTWO}{\reviewed{\ensuremath{189.5}}} % 189.48 rounded
\newcommand{\TWODETECTORSDAYSTWO}{\reviewed{\ensuremath{124.4}}} % 124.37 rounded
\newcommand{\THREEDETECTORSDAYSTWO}{\reviewed{\ensuremath{17.3}}} % 17.31 rounded

% O3a live times in days
\newcommand{\VIRGODAYSTHREEA}{\reviewed{\ensuremath{139.6}}} % 139.56 rounded
\newcommand{\HANFORDDAYSTHREEA}{\reviewed{\ensuremath{130.2}}} % 130.19 rounded
\newcommand{\LIVINGSTONDAYSTHREEA}{\reviewed{\ensuremath{138.5}}} % 138.49 rounded
\newcommand{\ONEDETECTORDAYSTHREEA}{\reviewed{\ensuremath{177.2}}} % 177.19 rounded
\newcommand{\TWODETECTORSDAYSTHREEA}{\reviewed{\ensuremath{149.8}}} % 149.77 rounded
\newcommand{\THREEDETECTORSDAYSTHREEA}{\reviewed{\ensuremath{81.3}}} % 81.28 rounded

% Dates of O3b sensitivity plots
\newcommand{\LIVINGSTONSENS}{\reviewed{4 January 2020 02:53:42}} % UTC
\newcommand{\HANFORDSENS}{\reviewed{4 January 2020 18:20:42}} % UTC
\newcommand{\VIRGOSENS}{\reviewed{9 February 2020 01:16:00}} % UTC
% H1 Jan 04 2020 sensitivity, representative best of O3b - C01_CLEAN_SUB60HZ, from https://dcc.ligo.org/LIGO-G2100674
\newcommand{\HANFORDRANGEPLOT}{\reviewed{\ensuremath{114~\mathrm{Mpc}}}} % 113.706 rounded
% L1 Jan 04 2020 sensitivity, representative best of O3b - C01_CLEAN_SUB60HZ, from https://dcc.ligo.org/LIGO-G2100675
\newcommand{\LIVINGSTONRANGEPLOT}{\reviewed{\ensuremath{133~\mathrm{Mpc}}}} % 133.489 rounded
% I. Nardecchia email
\newcommand{\VIRGORANGEPLOT}{\reviewed{\ensuremath{59~\mathrm{Mpc}}}}
% Date of Virgo upgrade with impact on BNS range
\newcommand{\VIRGORANGESPLIT}{\reviewed{28 January 2020}}
% Date of LHO upgrade with impact on BNS range
\newcommand{\LHORANGESPLIT}{\reviewed{2 January 2020}}
% Reaction train tracking dates
\newcommand{\LLORTRACK}{\reviewed{7 January 2020}} %7 Jan 2020 at Livingston llo_alog51594, +67 days
\newcommand{\LHORTRACK}{\reviewed{14 January 2020}} %14 Jan 2020 Hanford lho_alog54506, +74 days

% Interferometer macros
\newcommand{\HANFORDOPTICSDIFFFREQ}{\ensuremath{\reviewed{430}}} % Hz
\newcommand{\HANFORDUNEXPLAINEDFREQLOW}{\ensuremath{\reviewed{30}}} % Hz
\newcommand{\HANFORDUNEXPLAINEDFREQHIGH}{\ensuremath{\reviewed{100}}} % Hz
\newcommand{\HANFORDANGULARFREQ}{\ensuremath{\reviewed{30}}} % Hz

% Line subtraction
\newcommand{\USPOWERGRIDFREQ}{\ensuremath{\reviewed{60}}}% US 60 Hz mains power
\newcommand{\POWERGRIDHARMONIC}{\ensuremath{\reviewed{300}}}% Frequency (Hz) of max power gird harmonic subracted
\newcommand{\EUPOWERGRIDFREQ}{\ensuremath{\reviewed{50}}}% European 50 Hz mains power

% Other search specifications
% IMBH component mass criterion
\newcommand{\IMBHSEARCHMASS}{\ensuremath{\reviewed{65} \Msun}}
\newcommand{\BURSTDURATIONBOUND}{\ensuremath{\mathcal{O}(\reviewed{1})~\mathrm{s}}}

% Virgo systematic error macros
\newcommand{\VIRGOSYSTEMATICLOWERFREQ}{\ensuremath{\reviewed{46}}}
\newcommand{\VIRGOSYSTEMATICHIGHERFREQ}{\ensuremath{\reviewed{51}}}
\newcommand{\VIRGOADDITIONALSYSTEMATICLOWERFREQ}{\ensuremath{\reviewed{49.5}}}
\newcommand{\VIRGOADDITIONALSYSTEMATICHIGHERFREQ}{\ensuremath{\reviewed{50.5}}}
\newcommand{\VIRGOSUPPRESSLOWERFREQ}{\ensuremath{\reviewed{46}}}
\newcommand{\VIRGOSUPPRESSHERFREQ}{\ensuremath{\reviewed{51}}}
\newcommand{\VIRGOMECHANICALRESFREQ}{\ensuremath{\reviewed{49}}}
\newcommand{\VIRGOSYSTEMATICERRORPERCENTNOMINAL}{\ensuremath{\reviewed{5}}}
\newcommand{\VIRGOSYSTEMATICERRORPHASENOMINAL}{\ensuremath{\reviewed{35}}}
\newcommand{\VIRGOSYSTEMATICERRORPERCENTINCREASED}{\ensuremath{\reviewed{40}}}
\newcommand{\VIRGOSYSTEMATICERRORPHASEINCREASED}{\ensuremath{\reviewed{600}}}

% PCAL system macros
\newcommand{\PCALUNCERTAINTYLIGO}{\reviewed{\ensuremath{1}}}
\newcommand{\PCALUNCERTAINTYVIRGO}{\reviewed{\ensuremath{1.8}}}

% Virgo laser power
% Virgo power
\newcommand{\VIRGOPOWERTHREEA}{\reviewed{\ensuremath{18~\mathrm{W}}}}
\newcommand{\VIRGOPOWER}{\reviewed{\ensuremath{26~\mathrm{W}}}}

% Data Details
% Glitch and data quality information
\newcommand{\MITIGATIONEVENTS}{\reviewed{\ensuremath{7}}}

\newcommand{\LHOCATONE}{\reviewed{\ensuremath{0.30}}}
\newcommand{\LLOCATONE}{\reviewed{\ensuremath{1.68}}}
\newcommand{\VIRGOCATONE}{\reviewed{\ensuremath{0.21}}}

\newcommand{\LHOCATTWOCBC}{\reviewed{\ensuremath{0.02}}}
\newcommand{\LLOCATTWOCBC}{\reviewed{\ensuremath{0.28}}}
\newcommand{\VIRGOCATTWOCBC}{\reviewed{--}}

\newcommand{\LHOCATTWOBURST}{\reviewed{\ensuremath{0.52}}}
\newcommand{\LLOCATTWOBURST}{\reviewed{\ensuremath{0.50}}}
\newcommand{\VIRGOCATTWOBURST}{\reviewed{--}}

\newcommand{\LHOCATTHREEBURST}{\reviewed{\ensuremath{0.41}}}
\newcommand{\LLOCATTHREEBURST}{\reviewed{\ensuremath{0.17}}}
\newcommand{\VIRGOCATTHREEBURST}{\reviewed{--}}

\newcommand{\LHOGATING}{\reviewed{\ensuremath{0.01}}}
\newcommand{\LLOGATING}{\reviewed{\ensuremath{0.01}}}
\newcommand{\VIRGOGATING}{\reviewed{--}}

\newcommand{\LHOIDQ}{\reviewed{--}}
\newcommand{\LLOIDQ}{\reviewed{--}}
\newcommand{\VIRGOIDQ}{\reviewed{--}}

% Data mitigation
\newcommand{\GAUSSIANNOISETHRESH}{\ensuremath{\reviewed{0.01}}} % p-value hreshold for deglitching
\newcommand{\OLDNSBHFLOW}{\ensuremath{\reviewed{25}}} % Original lower frequency cut for GW200115 (Hz)

% Glitch info
\newcommand{\GLITCHSNRMIN}{\ensuremath{\reviewed{6.5}}} % Threshold SNR for glitch rates
\newcommand{\GLITCHBANDLOW}{\ensuremath{\reviewed{20}}} % Lower frequency of band for Figure (Hz)
\newcommand{\GLITCHBANDHIGH}{\ensuremath{\reviewed{2048}}} % Upper frequency of band for Figure (Hz)
\newcommand{\GLITCHTIMESTRIDE}{\ensuremath{\reviewed{2048}}} % Time interval over which glitch rate is calculated for Figure (s)

% Glitch rates
% H1
\newcommand{\LHORATEA}{\ensuremath{\reviewed{0.29}}}
\newcommand{\LHORATEB}{\ensuremath{\reviewed{0.32}}}

% RC tracking at LHO
\newcommand{\LHOPRERCRATE}{\ensuremath{\reviewed{0.82}}}
\newcommand{\LHOPOSTRCRATE}{\ensuremath{\reviewed{0.18}}}

% L1
\newcommand{\LLORATEA}{\ensuremath{\reviewed{1.10}}}
\newcommand{\LLORATEB}{\ensuremath{\reviewed{1.17}}}

% V1
\newcommand{\VIRGORATEA}{\ensuremath{\reviewed{0.47}}}
\newcommand{\VIRGORATEB}{\ensuremath{\reviewed{1.11}}}

% Definition of low frequency glitches
\newcommand{\LOWFREQGLITCH}{\ensuremath{\reviewed{50}}}

% Light scattering gltiches
\newcommand{\SCATTERINGSNRMIN}{\ensuremath{\reviewed{10}}} % Threshold SNR for ligh scattering stats
\newcommand{\SCATTERINGRATELLO}{\ensuremath{\reviewed{44}}} % Percentage of glitches at LLO 
\newcommand{\SCATTERINGRATELHO}{\ensuremath{\reviewed{45}}} % Percentage of glitches at LHO 
\newcommand{\SLOWARCHDURATION}{\ensuremath{\reviewed{2.0\text{--}2.5}}} % Slow scattering duration (s)
\newcommand{\FASTARCHDURATION}{\ensuremath{\reviewed{0.2\text{--}0.3}}} % Fast scattering duration (s)
\newcommand{\FASTSCATTERLOWFREQ}{\ensuremath{\reviewed{20}}} % Fast scattering impact lower frequency (Hz)
\newcommand{\FASTSCATTERHIGHFREQ}{\ensuremath{\reviewed{60}}} % Fast scattering impact upper frequency (Hz)
\newcommand{\FASTSCATTERMAXFREQ}{\ensuremath{\reviewed{120}}} % Fast scattering impact maximum frequency (Hz)
\newcommand{\FASTSCATTERQFREQ}{\ensuremath{\reviewed{4}}} % Fast scattering high-Q resonance frequency

% Virgo glitches
\newcommand{\VIRGOGLITCHFHIGH}{\ensuremath{\reviewed{40}}} % Most Virgo glitches have central frequency below this (Hz)
\newcommand{\VIRGOLOWFRATE}{\ensuremath{\reviewed{80}}} % Percent glitches with central frequencies below \VIRGOGLITCHFHIGH Hz
\newcommand{\VIRGOTOSEA}{\ensuremath{\reviewed{15}}} % Distance to the sea (km)

% Vetos
\newcommand{\NUMCAT}{\ensuremath{\reviewed{3}}} % Number of DQ flag categories
\newcommand{\FLAGPERCENT}{\ensuremath{\reviewed{1}}} % Typical percentage of time flagged per detector

% 200219 glitch time window
\newcommand{\DQWINDOW}{\ensuremath{\reviewed{1~\mathrm{s}}}} % Approximate time range for glitches around trigger

% Searches
\newcommand{\PyCBCOfflineLSNRNSBH}{\ensuremath{\reviewed{13.1}}\xspace} % PyCBC SNR can be found on results page https://ldas-jobs.ligo.caltech.edu/~gareth.davies/o3/runs/hlv/c01/a31_singles_rerun/3._single_triggers/3.17_L1_loudest_all_time_duration_gt_4.0_sec/#A1
\newcommand{\MBTAOfflineLSNRNSBH}{\ensuremath{\reviewed{13.2}}\xspace}
\newcommand{\GWTCTWOFINALBNSPASTRO}{\ensuremath{\reviewed{0.78}}\xspace}

% PE Numbers
\newcommand{\PEMinimumFreq}{\ensuremath{\reviewed{20~\mathrm{Hz}}}}%
\newcommand{\PEPowerLoss}{\ensuremath{\reviewed{1\%}}}%
\newcommand{\PENyquistFactor}{\ensuremath{\reviewed{0.875}}}%
\newcommand{\PEMaximumSampling}{\ensuremath{\reviewed{4096~\mathrm{Hz}}}}%
\newcommand{\PEUberSampling}{\ensuremath{\reviewed{8192~\mathrm{Hz}}}}%
\newcommand{\PEMassRatioMin}{\ensuremath{\reviewed{0.05}}}%
\newcommand{\PEMassRatioStretch}{\ensuremath{\reviewed{0.02}}}%
\newcommand{\PEBHMassThreshold}{\ensuremath{\reviewed{3}{\Msun}}}%
\newcommand{\PEBHMassThresholdProb}{\ensuremath{\reviewed{\PEpercentBBH{GW200322G}}{\%}}}%
\newcommand{\PECCMassGapUpper}{\ensuremath{\reviewed{5}{\Msun}}}%
\newcommand{\PEPISNLowerThreshold}{\ensuremath{\reviewed{65}{\Msun}}}%
\newcommand{\PEPISNUpperThreshold}{\ensuremath{\reviewed{120}{\Msun}}}%
\newcommand{\IMBHThreshold}{\ensuremath{\reviewed{100}{\Msun}}}%
\newcommand{\PEReferenceSpinMag}{\ensuremath{\reviewed{0.8}}}%
\newcommand{\PEReferenceMassRatio}{\ensuremath{\reviewed{0.1}}}%
\newcommand{\HzeroValue}{\ensuremath{\reviewed{67.9~\mathrm{km\,s^{-1}\,Mpc^{-1}}}}}%{\kilo\meter\,\second^{-1}\,\mega\parsec^{-1}}}
\newcommand{\WmValue}{\ensuremath{\reviewed{0.3065}}}%
\newcommand{\PEPostMergerTime}{\reviewed{\ensuremath{2~\mathrm{s}}}}% Standard duration post-merger for seglen
\newcommand{\PSDLargeValue}{\reviewed{\ensuremath{1}}}% PSD notch value
\newcommand{\MaxEllEm}{\reviewed{\ensuremath{(3,3)}}}% Max (l,m) for setting fhi
\newcommand{\SEOBNREllEm}{\reviewed{\ensuremath{(2,2),(2,1),(3,3),(4,4),(5,5)}}}% SEOBNRv4HM (l,m) 
\newcommand{\IMRPhenomEllEm}{\reviewed{\ensuremath{(2,2),(2,1),(3,3),(3,2),(4,4)}}}% IMRPhenomXHM (l,m) 
\newcommand{\NRSurOrbits}{\reviewed{\ensuremath{20}}}% Approximate orbits before merger

\newcommand{\TypicalFinalSpin}{\ensuremath{0.7}}% Equal mass BBH merger spin

\newcommand{\LITERATURELOWSPIN}{\ensuremath{\reviewed{0.1}}}% BH spin with efficient angular moment transport
\newcommand{\LITERATURECHESPIN}{\ensuremath{\reviewed{0.3\text{--}0.5}}}% BH spin from chemically homogeneous evolution
\newcommand{\LITERATUREPBHSPIN}{\ensuremath{\reviewed{0.01}}}% Primordial BH spin 
\newcommand{\LITERATUREMMAX}{\ensuremath{\reviewed{2.1\text{--}2.7\Msun}}}% NS max mass

\newcommand{\EdgeOnThetaJN}{\ensuremath{\reviewed{\pi/2}}} % Edge-on theta_JN ~ pi/2 rad (90 deg)

% GW200208_222617 modes
\newcommand{\GWTwentyZeroTwoZeroEightTwentyTwoPeakTotalMass}{\ensuremath{\reviewed{175 \Msun}}} % Additional peak of source total mass ~ 150--200 Msun

% GW190814 GWTC-2.1 results https://arxiv.org/abs/2108.01045
\newcommand{\GWNineteenZeroEightFourteenMassRatio}{\ensuremath{\reviewed{0.11^{+0.01}_{-0.01}}}} % Median and 90% interval
\newcommand{\GWNineteenZeroEightFourteenMassOne}{\ensuremath{\reviewed{23.3^{+1.4}_{-1.4}\Msun}}}  % Median and 90% interval
\newcommand{\GWNineteenZeroEightFourteenMassTwo}{\ensuremath{\reviewed{2.6^{+0.1}_{-0.1}\Msun}}}  % Median and 90% interval
\newcommand{\GWNineteenZeroEightFourteenSpinOneUpper}{\ensuremath{\reviewed{0.08}}} % 90% upper limit
\newcommand{\GWNineteenZeroEightFourteenChiPUpper}{\ensuremath{\reviewed{0.07}}} % 90% upper limit

% GWTC-2.1 https://arxiv.org/abs/2108.01045 events
\newcommand{\GWNineteenZeroFourZeroThreeMassRatio}{\ensuremath{\reviewed{0.23^{+0.57}_{-0.12}}}} % GW190403_051519 median and 90% interval
\newcommand{\GWNineteenZeroFourZeroThreeChiEff}{\ensuremath{\reviewed{0.68^{+0.16}_{-0.43}}}} % GW190403_051519 median and 90% interval
\newcommand{\GWNineteenZeroFourZeroThreeDistance}{\ensuremath{\reviewed{8.28^{+6.72}_{-4.29}~\mathrm{Gpc}}}} % GW190403_051519 median and 90% interval

% From https://arxiv.org/abs/1910.09528
\newcommand{\IASSeventeenZeroEightSeventeenAMassOne}{\ensuremath{\reviewed{56^{+16}_{-10}\Msun}}} %170817A
\newcommand{\IASSeventeenZeroEightSeventeenAMassTwo}{\ensuremath{\reviewed{40^{+10}_{-11}\Msun}}} %170817A

% From https://arxiv.org/abs/1806.02751
\newcommand{\ThompsonEtAlMass}{\ensuremath{\reviewed{3.3^{+2.8}_{-0.7}\Msun}}} % 2MASS J05215658+4359220 2-sigma
\newcommand{\ThompsonEtAlConfidence}{\ensuremath{\reviewed{95\%}}} % 2MASS J05215658+4359220 2-sigma

% From https://arxiv.org/abs/2101.02212
\newcommand{\JayasingheEtAlMass}{\ensuremath{\reviewed{3.04 \pm 0.06\Msun}}} % V723 Mon 1-sigma
\newcommand{\JayasingheEtAlConfidence}{\ensuremath{\reviewed{68\%}}} % V723 Mon 1-sigma

% From https://arxiv.org/abs/0711.0925
\newcommand{\FreireEtAlMass}{\ensuremath{\reviewed{2.74 \pm 0.21\Msun}}} % J1748-2021B 1-sigma
\newcommand{\FreireEtAlConfidence}{\ensuremath{\reviewed{68\%}}} % J1748-2021B 1-sigma

% From https://arxiv.org/abs/1509.08805
\newcommand{\MartinezEtAlMass}{\ensuremath{\reviewed{1.174 \pm 0.004\Msun}}} % J0453+1559 companion 1-sigma
\newcommand{\MartinezEtAlConfidence}{\ensuremath{\reviewed{68\%}}} % J0453+1559 companion 1-sigma

% From https://arxiv.org/abs/1002.0514
\newcommand{\FerdmanEtAlMass}{\ensuremath{\reviewed{1.24 \pm 0.11\Msun}}} % J1802-2124 companion 1-sigma
\newcommand{\FerdmanEtAlConfidence}{\ensuremath{\reviewed{68\%}}} % J1802-2124 companion 1-sigma

% From https://arxiv.org/abs/1502.07126
\newcommand{\FalangaEtAlSMCXOneMass}{\ensuremath{\reviewed{1.21 \pm 0.12\Msun}}} % SMC X-1 1-sigma
\newcommand{\FalangaEtAlFourUFifteenThirtyEightMass}{\ensuremath{\reviewed{1.02 \pm 0.17\Msun}}} % 4U 1538-522 1-sigma
\newcommand{\FalangaEtAlConfidence}{\ensuremath{\reviewed{68\%}}} % 1-sigma

% O3b retraction rates, before and after R0 tracking, in percentage
\newcommand{\PREOPARETRATE}{\reviewed{\ensuremath{0.55}}}
\newcommand{\POSTOPARETRATE}{\reviewed{\ensuremath{0.21}}}

% VT parameters
\newcommand{\VTLOGNORMALWIDTH}{\reviewed{\ensuremath{0.1}}}
\newcommand{\VTFIRSTDETECTIONLIKEMASS}{\reviewed{\ensuremath{35 \Msun}}}
\newcommand{\VTNSLIKEMASS}{\reviewed{\ensuremath{1.5 \Msun}}}
\newcommand{\VTMASSTWENTY}{\reviewed{\ensuremath{20 \Msun}}}
\newcommand{\VTMASSTEN}{\reviewed{\ensuremath{10 \Msun}}}
\newcommand{\VTMASSFIVE}{\reviewed{\ensuremath{5 \Msun}}}
\newcommand{\VTDETECTEDRANGEMASSES}{\reviewed{\VTMASSTWENTY{}, \VTMASSTEN{} and \VTMASSFIVE{}}}

% S200114f
\newcommand{\CCThreshold}{\ensuremath{\reviewed{0.8}}} % network correlation coefficient threshold

% Input laser power at power recycling mirror (W)
\newcommand{\HANFORDPOWERONE}{\reviewed{\ensuremath{21~\mathrm{W}}}}	
\newcommand{\HANFORDPOWERTWO}{\reviewed{\ensuremath{26~\mathrm{W}}}}	
\newcommand{\HANFORDPOWERTHREEA}{\reviewed{\ensuremath{34~\mathrm{W}}}} 
\newcommand{\HANFORDPOWER}{\reviewed{\ensuremath{34~\mathrm{W}}}}	 
\newcommand{\LIVINGSTONPOWERONE}{\reviewed{\ensuremath{22~\mathrm{W}}}} 
\newcommand{\LIVINGSTONPOWERTWO}{\reviewed{\ensuremath{25~\mathrm{W}}}}
\newcommand{\LIVINGSTONPOWERTHREEA}{\reviewed{\ensuremath{44~\mathrm{W}}}} 
\newcommand{\LIVINGSTONPOWER}{\reviewed{\ensuremath{40~\mathrm{W}}}}    
\newcommand{\VIRGOPOWERTWO}{\reviewed{\ensuremath{10~\mathrm{W}}}}      

% Power recycling gain (PGR)
\newcommand{\HANFORDPRGONE}{\reviewed{\ensuremath{38}}}	
\newcommand{\HANFORDPRGTWO}{\reviewed{\ensuremath{40}}}	
\newcommand{\HANFORDPRGTHREEA}{\reviewed{\ensuremath{44}}} 
\newcommand{\HANFORDPRG}{\reviewed{\ensuremath{44}}}	 
\newcommand{\LIVINGSTONPRGONE}{\reviewed{\ensuremath{38}}} 
\newcommand{\LIVINGSTONPRGTWO}{\reviewed{\ensuremath{36}}}
\newcommand{\LIVINGSTONPRGTHREEA}{\reviewed{\ensuremath{47}}} 
\newcommand{\LIVINGSTONPRG}{\reviewed{\ensuremath{42}}}    
\newcommand{\VIRGOPRGTWO}{\reviewed{\ensuremath{38}}}      
\newcommand{\VIRGOPRGTHREEA}{\reviewed{\ensuremath{36}}}
\newcommand{\VIRGOPRG}{\reviewed{\ensuremath{34}}}

% Non pipeline-specific macros
\newcommand{\PASTROBBHBOUNDARY}{\reviewed{\ensuremath{3 \Msun}}}
\newcommand{\OPAFARTHRESH}{\reviewed{\ensuremath{1.2~\mathrm{yr}^{-1}}}}
\newcommand{\PASTROTHRESH}{\PASTROTHRESHOLD}
\newcommand{\PASTROUNCERT}{\reviewed{\ensuremath{0.1}}}
\newcommand{\OPAFARTHRESHMONTH}{one per two months}
\newcommand{\OPAFARTHRESHYR}{\reviewed{\ensuremath{6~\mathrm{yr}^{-1}}}}
\newcommand{\LOWSNR}{\reviewed{\ensuremath{10}}}
\newcommand{\RETRACTIONLOWSNR}{\reviewed{\ensuremath{5}}}
\newcommand{\SUBTHRESHOLDFAR}{\reviewed{\ensuremath{2.0~\mathrm{day}^{-1}}}}

\newcommand{\COMPONENTBANKMINMASS}{\reviewed{\ensuremath{1 \Msun}}}

\newcommand{\EXPREALINSUBTHRESH}{\reviewed{\ensuremath{7}}}
\newcommand{\SUBTHRESHOLDPURITY}{\reviewed{\ensuremath{0.01}}}
\newcommand{\VTRATIOTWOTHRESH}{\reviewed{\ensuremath{1.2\text{--}1.3}}}

% Order of magnitude number of high-significance NSBH detections
\newcommand{\NUMNSBH}{\reviewed{\ensuremath{1}}} 

% PyCBC macros

% Define the name of the hyperbank and focused-bbh analyses
\newcommand{\PYCBCHYPERBANK}{\PYCBC{}-broad}
\newcommand{\PYCBCBBH}{\PYCBC{}-BBH}
\newcommand{\PYCBCLIVE}{\PYCBC{}~Live}

% matched-filter snr thresholds
\newcommand{\PYCBCSNRTHRESH}{\reviewed{\ensuremath{4.0}}}

% Bank properties
\newcommand{\PYCBCBANKTMIN}{\reviewed{\ensuremath{0.15~\mathrm{s}}}} % Minimum template duration
\newcommand{\PYCBCBANKWFMASS}{\reviewed{\ensuremath{4 \Msun}}} % Switch from TF2 to SEOBNR

% Hyperbank search template bank:
\newcommand{\PYCBCHYPERBANKTOTALMASSMAX}{\reviewed{\ensuremath{500 \Msun}}}

% BBH search template bank:
\newcommand{\PYCBCBBHMASSONEMIN}{\reviewed{\ensuremath{5 \Msun{}}}}
\newcommand{\PYCBCBBHMASSONEMAX}{\reviewed{\ensuremath{350 \Msun{}}}}
\newcommand{\PYCBCBBHMASSTWOMIN}{\reviewed{\ensuremath{5 \Msun{}}}}
\newcommand{\PYCBCBBHTOTALMASSMIN}{\reviewed{\ensuremath{10 \Msun{}}}}
\newcommand{\PYCBCBBHTOTALMASSMAX}{\reviewed{\ensuremath{500 \Msun{}}}}
\newcommand{\PYCBCBBHQMIN}{\reviewed{\ensuremath{1/3}}}
\newcommand{\PYCBCBBHQMAX}{\reviewed{\ensuremath{1}}}
\newcommand{\PYCBCBBHSPINMIN}{\reviewed{\ensuremath{-0.998}}}
\newcommand{\PYCBCBBHSPINMAX}{\reviewed{\ensuremath{0.998}}}

\newcommand{\PYCBCBBHMAXCHIRPMASSWEIGHTING}{\reviewed{\ensuremath{40 \Msun{}}}}

% Reviewed
% See https://git.ligo.org/RatesAndPopulations/pycbc-multi-ifo-p_astro/-/tree/master/config
\newcommand{\PYCBCPASTROMINMCHIRPNSBH}{\reviewed{\ensuremath{2.176}}}
\newcommand{\PYCBCPASTROMINMCHIRPBBH}{\reviewed{\ensuremath{4.353}}}
\newcommand{\PYCBCPASTROBBHMASSMIN}{\reviewed{\ensuremath{5 \Msun}}}

% GW191219 is close to the boundary for NSBH vs BBH;
% what would pastro be if the bounary mchirp was increased?
% result from https://ldas-jobs.ligo.caltech.edu/~thomas.dent/o3/rates/allo3_pastro/nsbh/test_mc_6plus6/H1L1V1-PYCBC_ALLO3_HYPERBANK_MCHIRP_2P176_5P223-1260808298-1.json
\newcommand{\PYCBCPASTROALTERNATIVEMCHIRPBOUND}{\reviewed{\ensuremath{0.085}}}

% Macros to define which events were manually removed from pycbc background data
\newcommand{\PYCBCSINGLESREMOVALSIGNIFICANCE}{\reviewed{\ensuremath{10^{-2}~\mathrm{yr}^{-1}}}}
\newcommand{\PYCBCSINGLESREMOVALWINDOW}{\reviewed{\ensuremath{1~\mathrm{s}} either side of each event}}

% See the executive summary for events which meet the removal criteria
% https://wiki.ligo.org/CBC/Searches/PycbcHyperbankO3HlvExecutiveSummary
% Reviewed by Greg Mendell
\newcommand{\PYCBCHYPERBANKREMOVEDSINGLES}{\FULLNAME{GW200112H}{} and \FULLNAME{GW200202F}{}}

% See the executive summary for events which meet the removal criteria
% https://wiki.ligo.org/CBC/Searches/PyCBCFocussedBbhO3HlvExecutiveSummary
\newcommand{\PYCBCBBHREMOVEDSINGLES}{\FULLNAME{GW200112H}}

% pipeline livetime
\newcommand{\PYCBCLIVETIME}{\reviewed{\ensuremath{124.2~\mathrm{days}}}}

% GstLAL macros
% gstlal template bank lowest minimum match 
\newcommand{\GSTLALTEMPLATEMINMATCH}{\reviewed{\ensuremath{0.97}}}
\newcommand{\GSTLALBANKMINMASS}{\reviewed{\ensuremath{2 \Msun}}}
\newcommand{\GSTLALBANKWFCHIRPMASS}{\reviewed{\ensuremath{1.73 \Msun}}} % Switch from TF2 to SEOBNR
\newcommand{\GSTLALBANKMAXMASS}{\reviewed{\ensuremath{758 \Msun}}}
\newcommand{\GSTLALBANKNSSPINLOW}{\reviewed{\ensuremath{-0.05}}}
\newcommand{\GSTLALBANKNSSPINHIGH}{\reviewed{\ensuremath{0.05}}}
\newcommand{\GSTLALBANKBHSPINLOW}{\reviewed{\ensuremath{-0.999}}}
\newcommand{\GSTLALBANKBHSPINHIGH}{\reviewed{\ensuremath{0.999}}}
\newcommand{\GSTLALWINDOW}{\ensuremath{\reviewed{1~\mathrm{s}}}} 

% Reviewed
% p-astro distribution parameters
\newcommand{\GSTLALPASTROBBHMASSMIN}{\reviewed{\ensuremath{3}}}
\newcommand{\GSTLALPASTROBBHMASSMAX}{\reviewed{\ensuremath{300}}}
\newcommand{\GSTLALPASTROBBHSPINMAX}{\reviewed{\ensuremath{0.99}}}
\newcommand{\GSTLALPASTROBBHMAXREDSHIFT}{\reviewed{\ensuremath{3.76}}}
\newcommand{\GSTLALPASTROBBHKAPPAREDSHIFT}{\reviewed{\ensuremath{0}}}
\newcommand{\GSTLALPASTROBNSMASSMIN}{\reviewed{\ensuremath{1}}}
\newcommand{\GSTLALPASTROBNSMASSMAX}{\reviewed{\ensuremath{3}}}
\newcommand{\GSTLALPASTROBNSSPINMAX}{\reviewed{\ensuremath{0.05}}}
\newcommand{\GSTLALPASTROBNSMAXREDSHIFT}{\reviewed{\ensuremath{0.16}}}
\newcommand{\GSTLALPASTROBNSKAPPAREDSHIFT}{\reviewed{\ensuremath{0}}}
\newcommand{\GSTLALPASTRONSBHMINBBHMASS}{\reviewed{\ensuremath{3}}}
\newcommand{\GSTLALPASTRONSBHMAXBBHMASS}{\reviewed{\ensuremath{300}}}
\newcommand{\GSTLALPASTRONSBHMINBNSMASS}{\reviewed{\ensuremath{1}}}
\newcommand{\GSTLALPASTRONSBHMAXBNSMASS}{\reviewed{\ensuremath{3}}}
\newcommand{\GSTLALPASTRONSBHMAXBBHSPIN}{\reviewed{\ensuremath{0.99}}}
\newcommand{\GSTLALPASTRONSBHMAXBNSSPIN}{\reviewed{\ensuremath{0.4}}}
\newcommand{\GSTLALPASTRONSBHMAXREDSHIFT}{\reviewed{\ensuremath{0.80}}}
\newcommand{\GSTLALPASTRONSBHKAPPAREDSHIFT}{\reviewed{\ensuremath{0}}}

% Not reviewed
% Numbers from GW200105 discovery paper
\newcommand{\NSBHPREVREPORTEDFAR}{\reviewed{\ensuremath{0.36~\mathrm{yr}^{-1}}}}
\newcommand{\NSBHPAPEROBSERVINGEND}{\reviewed{22 January 2020}}

% Absolute pastro difference for GW200105 with different priors
\newcommand{\GSTLALALLSKYPASTRODIFFPRIORS}{\reviewed{\ensuremath{0.045}}}

% MBTA macros
% matched-filter snr thresholds
\newcommand{\MBTASNRTHRESHMIN}{\reviewed{\ensuremath{4.5}}}
\newcommand{\MBTASNRTHRESHMAX}{\reviewed{\ensuremath{4.8}}}
\newcommand{\MBTASNRTHRESHREGIONFOURH}{\reviewed{\ensuremath{9.5}}}
\newcommand{\MBTASNRTHRESHREGIONFOURL}{\reviewed{\ensuremath{11.3}}}
\newcommand{\MBTASNRTHRESHREGIONFOURV}{\reviewed{\ensuremath{12}}}
\newcommand{\MBTACRSTHRESHOLDTOREMOVESINGLES}{\reviewed{\ensuremath{10}}}
\newcommand{\MBTATYPICALFARATCRSTEN}{\reviewed{\ensuremath{0.2~\mathrm{yr}^{-1}}}}

% template bank
\newcommand{\MBTABANKMINMASS}{\reviewed{\ensuremath{2 \Msun}}}
\newcommand{\MBTACOMPONENTBANKMAXMASS}{\reviewed{\ensuremath{195 \Msun}}}
\newcommand{\MBTABANKMAXMASS}{\reviewed{\ensuremath{200 \Msun}}}
\newcommand{\MBTABANKNSMASS}{\reviewed{\ensuremath{2 \Msun}}}
\newcommand{\MBTABANKMAXNSBHMASS}{\reviewed{\ensuremath{100 \Msun}}}
\newcommand{\MBTABANKMAXNSSPIN}{\reviewed{\ensuremath{0.05}}}
\newcommand{\MBTABANKMAXBHSPIN}{\reviewed{\ensuremath{0.997}}}

% p-astro
\newcommand{\MBTAPASTROBINS}{\reviewed{\ensuremath{165}}}
\newcommand{\MBTAPASTROBNSMASSMAX}{\reviewed{\ensuremath{2.5 \Msun}}}
\newcommand{\MBTAPASTROBBHMASSMIN}{\reviewed{\ensuremath{5 \Msun}}}
\newcommand{\MBTAPASTROBBHKAPPAREDSHIFT}{\reviewed{\ensuremath{0}}}
\newcommand{\MBTAPASTROPLPALPHA}{\reviewed{\ensuremath{2.5}}}
\newcommand{\MBTAPASTROPLPBETAQ}{\reviewed{\ensuremath{1.5}}}
\newcommand{\MBTAPASTROPLPMMIN}{\reviewed{\ensuremath{5}}}
\newcommand{\MBTAPASTROPLPMMAX}{\reviewed{\ensuremath{80}}}
\newcommand{\MBTAPASTROPLPLAMBDA}{\reviewed{\ensuremath{0.1}}}
\newcommand{\MBTAPASTROPLPMUM}{\reviewed{\ensuremath{34}}}
\newcommand{\MBTAPASTROPLPSIGMAM}{\reviewed{\ensuremath{5}}}
\newcommand{\MBTAPASTROPLPDELTAM}{\reviewed{\ensuremath{3.5}}}
\newcommand{\MBTAEXAMPLEFOREGROUNDRATEDENSITY}{\reviewed{\ensuremath{0.109~\mathrm{yr}^{-1}}}}
\newcommand{\MBTAEXAMPLEBACKGROUNDRATEDENSITY}{\reviewed{\ensuremath{0.023~\mathrm{yr}^{-1}}}}
\newcommand{\MBTAEXAMPLEBINSIGNALFRACTION}{\reviewed{\ensuremath{11\%}}}
\newcommand{\MBTAEXAMPLEBINBBHTEMPLATESFRACTION}{\reviewed{\ensuremath{0.008\%}}}

% livetime
\newcommand{\MBTALIVETIME}{\reviewed{\ensuremath{124.5~\mathrm{days}}}}

% SPIIR macros
\newcommand{\SPIIRLATENCY}{\reviewed{\ensuremath{7\text{--}10~\mathrm{s}}}}
\newcommand{\SPIIRTIMESHIFTS}{\reviewed{\ensuremath{100}}}
\newcommand{\SPIIRSNRTHRESH}{\reviewed{\ensuremath{4}}}

% template bank
\newcommand{\SPIIRPRIMARYBANKMINMASS}{\reviewed{\ensuremath{1.1 \Msun}}}
\newcommand{\SPIIRPRIMARYBANKMAXMASS}{\reviewed{\ensuremath{100 \Msun}}}

% cWB macros
\newcommand{\CWBMINFREQ}{\reviewed{\ensuremath{16~\mathrm{Hz}}}}
\newcommand{\CWBMAXFREQ}{\reviewed{\ensuremath{512~\mathrm{Hz}}}}
\newcommand{\CWBCENTRALFREQ}{\reviewed{\ensuremath{80~\mathrm{Hz}}}}
\newcommand{\CWBMCHIRP}{\reviewed{\ensuremath{150 \Msun}}}
\newcommand{\CWBTSLIDE}{\reviewed{\ensuremath{1~\mathrm{s}}}}
\newcommand{\CWBTBKG}{\reviewed{\ensuremath{10^3~\mathrm{yr}}}}

% livetime
\newcommand{\CWBLIVETIME}{\reviewed{\ensuremath{94.9~\mathrm{days}}}}

% Reviewed
% Injection distribution parameters
\newcommand{\INJBBHMASSMIN}{\reviewed{\ensuremath{2}}}
\newcommand{\INJBBHMASSMAX}{\reviewed{\ensuremath{100}}}
\newcommand{\INJBBHMASSONEPOWER}{\reviewed{\ensuremath{-2.35}}}
\newcommand{\INJBBHSPINMAX}{\reviewed{\ensuremath{0.998}}}
\newcommand{\INJBBHMAXREDSHIFT}{\reviewed{\ensuremath{1.9}}}
\newcommand{\INJBBHKAPPAREDSHIFT}{\reviewed{\ensuremath{1}}}
\newcommand{\INJBNSMASSMIN}{\reviewed{\ensuremath{1}}}
\newcommand{\INJBNSMASSMAX}{\reviewed{\ensuremath{2.5}}}
\newcommand{\INJBNSSPINMAX}{\reviewed{\ensuremath{0.4}}}
\newcommand{\INJBNSMAXREDSHIFT}{\reviewed{\ensuremath{0.15}}}
\newcommand{\INJBNSKAPPAREDSHIFT}{\reviewed{\ensuremath{0}}}
\newcommand{\INJNSBHMINBBHMASS}{\reviewed{\ensuremath{2.5}}}
\newcommand{\INJNSBHMAXBBHMASS}{\reviewed{\ensuremath{60}}}
\newcommand{\INJNSBHMINBNSMASS}{\reviewed{\ensuremath{1}}}
\newcommand{\INJNSBHMAXBNSMASS}{\reviewed{\ensuremath{2.5}}}
\newcommand{\INJNSBHPOWERBBHMASS}{\reviewed{\ensuremath{-2.35}}}
\newcommand{\INJNSBHMAXBBHSPIN}{\reviewed{\ensuremath{0.998}}}
\newcommand{\INJNSBHMAXBNSSPIN}{\reviewed{\ensuremath{0.4}}}
\newcommand{\INJNSBHMAXREDSHIFT}{\reviewed{\ensuremath{0.25}}}
\newcommand{\INJNSBHKAPPAREDSHIFT}{\reviewed{\ensuremath{0}}}
\newcommand{\INJMODELSWITCHMASS}{\reviewed{\ensuremath{9}}}

\newcommand{\DETECTIONMASSLIMIT}{\reviewed{\ensuremath{100}}} 

% Min FAR in IMBH search for GW191109A
\newcommand{\IMBHMINFAR}{\reviewed{\ensuremath{10^{-3}~\mathrm{yr}^{-1}}}}
 
%cWB-only appendix
%S190804q 190804_083543
\newcommand{\OFFSETCOA}{\reviewed{\ensuremath{1248942961~\mathrm{s}}}}

%S190930ak 190930_234652
\newcommand{\OFFSETCOB}{\reviewed{\ensuremath{1253922430~\mathrm{s}}}}

%S200214br 200214_224526
\newcommand{\TMINCOC}{\reviewed{\ensuremath{\sim 0.5~\mathrm{s}}}}
\newcommand{\TLENCOC}{\reviewed{\ensuremath{\sim 2~\mathrm{s}}}}
\newcommand{\OFFSETCOC}{\reviewed{\ensuremath{1265755544~\mathrm{s}}}}
\newcommand{\FRANGECOC}{\reviewed{\ensuremath{20\text{--}30~\mathrm{Hz}}}}

\input{imrp_macros}
\input{imrp_macros_with_cuts}
\DeclareRobustCommand{\phitwoSEOBminus}[1]{\IfEqCase{#1}{{GW200208K}{2.8}{GW200208G}{2.8}{GW200129D}{2.8}{GW191219E}{2.8}{GW191109A}{2.8}}}
\DeclareRobustCommand{\phitwoSEOBmed}[1]{\IfEqCase{#1}{{GW200208K}{3.2}{GW200208G}{3.2}{GW200129D}{3.2}{GW191219E}{3.1}{GW191109A}{3.1}}}
\DeclareRobustCommand{\phitwoSEOBplus}[1]{\IfEqCase{#1}{{GW200208K}{2.8}{GW200208G}{2.8}{GW200129D}{2.8}{GW191219E}{2.8}{GW191109A}{2.9}}}
\DeclareRobustCommand{\phitwoSEOBtenthpercentile}[1]{\IfEqCase{#1}{{GW200208K}{0.6}{GW200208G}{0.6}{GW200129D}{0.6}{GW191219E}{0.7}{GW191109A}{0.6}}}
\DeclareRobustCommand{\phitwoSEOBnintiethpercentile}[1]{\IfEqCase{#1}{{GW200208K}{5.7}{GW200208G}{5.7}{GW200129D}{5.7}{GW191219E}{5.6}{GW191109A}{5.6}}}
\DeclareRobustCommand{\phioneSEOBminus}[1]{\IfEqCase{#1}{{GW200208K}{2.8}{GW200208G}{2.9}{GW200129D}{2.9}{GW191219E}{2.9}{GW191109A}{2.9}}}
\DeclareRobustCommand{\phioneSEOBmed}[1]{\IfEqCase{#1}{{GW200208K}{3.1}{GW200208G}{3.2}{GW200129D}{3.2}{GW191219E}{3.2}{GW191109A}{3.2}}}
\DeclareRobustCommand{\phioneSEOBplus}[1]{\IfEqCase{#1}{{GW200208K}{2.9}{GW200208G}{2.8}{GW200129D}{2.7}{GW191219E}{2.8}{GW191109A}{2.8}}}
\DeclareRobustCommand{\phioneSEOBtenthpercentile}[1]{\IfEqCase{#1}{{GW200208K}{0.7}{GW200208G}{0.6}{GW200129D}{0.6}{GW191219E}{0.6}{GW191109A}{0.7}}}
\DeclareRobustCommand{\phioneSEOBnintiethpercentile}[1]{\IfEqCase{#1}{{GW200208K}{5.7}{GW200208G}{5.6}{GW200129D}{5.6}{GW191219E}{5.6}{GW191109A}{5.7}}}
\DeclareRobustCommand{\geocenttimeSEOBminus}[1]{\IfEqCase{#1}{{GW200208K}{0.0}{GW200208G}{0.0}{GW200129D}{0.0}{GW191219E}{0.0}{GW191109A}{0.0}}}
\DeclareRobustCommand{\geocenttimeSEOBmed}[1]{\IfEqCase{#1}{{GW200208K}{1265235996.0}{GW200208G}{1265202095.9}{GW200129D}{1264316116.4}{GW191219E}{1260808298.5}{GW191109A}{1257296855.2}}}
\DeclareRobustCommand{\geocenttimeSEOBplus}[1]{\IfEqCase{#1}{{GW200208K}{0.0}{GW200208G}{0.0}{GW200129D}{0.0}{GW191219E}{0.0}{GW191109A}{0.0}}}
\DeclareRobustCommand{\geocenttimeSEOBtenthpercentile}[1]{\IfEqCase{#1}{{GW200208K}{1265235996.0}{GW200208G}{1265202095.9}{GW200129D}{1264316116.4}{GW191219E}{1260808298.5}{GW191109A}{1257296855.2}}}
\DeclareRobustCommand{\geocenttimeSEOBnintiethpercentile}[1]{\IfEqCase{#1}{{GW200208K}{1265235996.0}{GW200208G}{1265202095.9}{GW200129D}{1264316116.4}{GW191219E}{1260808298.5}{GW191109A}{1257296855.2}}}
\DeclareRobustCommand{\chieffSEOBminus}[1]{\IfEqCase{#1}{{GW200208K}{0.38}{GW200208G}{0.26}{GW200129D}{0.12}{GW191219E}{0.10}{GW191109A}{0.26}}}
\DeclareRobustCommand{\chieffSEOBmed}[1]{\IfEqCase{#1}{{GW200208K}{0.34}{GW200208G}{-0.05}{GW200129D}{0.12}{GW191219E}{0.00}{GW191109A}{-0.28}}}
\DeclareRobustCommand{\chieffSEOBplus}[1]{\IfEqCase{#1}{{GW200208K}{0.45}{GW200208G}{0.21}{GW200129D}{0.09}{GW191219E}{0.08}{GW191109A}{0.26}}}
\DeclareRobustCommand{\chieffSEOBtenthpercentile}[1]{\IfEqCase{#1}{{GW200208K}{0.04}{GW200208G}{-0.26}{GW200129D}{0.03}{GW191219E}{-0.08}{GW191109A}{-0.48}}}
\DeclareRobustCommand{\chieffSEOBnintiethpercentile}[1]{\IfEqCase{#1}{{GW200208K}{0.70}{GW200208G}{0.11}{GW200129D}{0.18}{GW191219E}{0.06}{GW191109A}{-0.07}}}
\DeclareRobustCommand{\decSEOBminus}[1]{\IfEqCase{#1}{{GW200208K}{0.45}{GW200208G}{0.070}{GW200129D}{0.094}{GW191219E}{0.69}{GW191109A}{0.15}}}
\DeclareRobustCommand{\decSEOBmed}[1]{\IfEqCase{#1}{{GW200208K}{0.66}{GW200208G}{-0.595}{GW200129D}{0.102}{GW191219E}{-0.24}{GW191109A}{-0.66}}}
\DeclareRobustCommand{\decSEOBplus}[1]{\IfEqCase{#1}{{GW200208K}{0.34}{GW200208G}{0.090}{GW200129D}{0.323}{GW191219E}{1.24}{GW191109A}{0.63}}}
\DeclareRobustCommand{\decSEOBtenthpercentile}[1]{\IfEqCase{#1}{{GW200208K}{0.40}{GW200208G}{-0.649}{GW200129D}{0.024}{GW191219E}{-0.84}{GW191109A}{-0.79}}}
\DeclareRobustCommand{\decSEOBnintiethpercentile}[1]{\IfEqCase{#1}{{GW200208K}{0.94}{GW200208G}{-0.534}{GW200129D}{0.307}{GW191219E}{0.88}{GW191109A}{-0.26}}}
\DeclareRobustCommand{\chipSEOBminus}[1]{\IfEqCase{#1}{{GW200208K}{0.29}{GW200208G}{0.28}{GW200129D}{0.24}{GW191219E}{0.06}{GW191109A}{0.37}}}
\DeclareRobustCommand{\chipSEOBmed}[1]{\IfEqCase{#1}{{GW200208K}{0.38}{GW200208G}{0.37}{GW200129D}{0.35}{GW191219E}{0.07}{GW191109A}{0.70}}}
\DeclareRobustCommand{\chipSEOBplus}[1]{\IfEqCase{#1}{{GW200208K}{0.37}{GW200208G}{0.41}{GW200129D}{0.31}{GW191219E}{0.08}{GW191109A}{0.23}}}
\DeclareRobustCommand{\chipSEOBtenthpercentile}[1]{\IfEqCase{#1}{{GW200208K}{0.14}{GW200208G}{0.13}{GW200129D}{0.15}{GW191219E}{0.02}{GW191109A}{0.42}}}
\DeclareRobustCommand{\chipSEOBnintiethpercentile}[1]{\IfEqCase{#1}{{GW200208K}{0.69}{GW200208G}{0.69}{GW200129D}{0.59}{GW191219E}{0.13}{GW191109A}{0.89}}}
\DeclareRobustCommand{\finalspinSEOBminus}[1]{\IfEqCase{#1}{{GW200208K}{0.24}{GW200208G}{0.11}{GW200129D}{0.04}{GW191219E}{0.06}{GW191109A}{0.16}}}
\DeclareRobustCommand{\finalspinSEOBmed}[1]{\IfEqCase{#1}{{GW200208K}{0.79}{GW200208G}{0.66}{GW200129D}{0.72}{GW191219E}{0.13}{GW191109A}{0.63}}}
\DeclareRobustCommand{\finalspinSEOBplus}[1]{\IfEqCase{#1}{{GW200208K}{0.16}{GW200208G}{0.09}{GW200129D}{0.04}{GW191219E}{0.08}{GW191109A}{0.13}}}
\DeclareRobustCommand{\finalspinSEOBtenthpercentile}[1]{\IfEqCase{#1}{{GW200208K}{0.62}{GW200208G}{0.58}{GW200129D}{0.69}{GW191219E}{0.09}{GW191109A}{0.51}}}
\DeclareRobustCommand{\finalspinSEOBnintiethpercentile}[1]{\IfEqCase{#1}{{GW200208K}{0.91}{GW200208G}{0.73}{GW200129D}{0.75}{GW191219E}{0.19}{GW191109A}{0.73}}}
\DeclareRobustCommand{\massonesourceSEOBminus}[1]{\IfEqCase{#1}{{GW200208K}{13}{GW200208G}{6.2}{GW200129D}{2.4}{GW191219E}{2.6}{GW191109A}{10.0}}}
\DeclareRobustCommand{\massonesourceSEOBmed}[1]{\IfEqCase{#1}{{GW200208K}{33}{GW200208G}{37.6}{GW200129D}{33.5}{GW191219E}{30.6}{GW191109A}{64.6}}}
\DeclareRobustCommand{\massonesourceSEOBplus}[1]{\IfEqCase{#1}{{GW200208K}{50}{GW200208G}{9.4}{GW200129D}{3.6}{GW191219E}{2.8}{GW191109A}{12.1}}}
\DeclareRobustCommand{\massonesourceSEOBtenthpercentile}[1]{\IfEqCase{#1}{{GW200208K}{22}{GW200208G}{32.5}{GW200129D}{31.6}{GW191219E}{28.7}{GW191109A}{56.7}}}
\DeclareRobustCommand{\massonesourceSEOBnintiethpercentile}[1]{\IfEqCase{#1}{{GW200208K}{75}{GW200208G}{44.7}{GW200129D}{36.1}{GW191219E}{32.7}{GW191109A}{73.8}}}
\DeclareRobustCommand{\cosiotaSEOBminus}[1]{\IfEqCase{#1}{{GW200208K}{0.93}{GW200208G}{0.16}{GW200129D}{0.34}{GW191219E}{0.99}{GW191109A}{0.47}}}
\DeclareRobustCommand{\cosiotaSEOBmed}[1]{\IfEqCase{#1}{{GW200208K}{-0.03}{GW200208G}{-0.82}{GW200129D}{0.68}{GW191219E}{0.02}{GW191109A}{-0.50}}}
\DeclareRobustCommand{\cosiotaSEOBplus}[1]{\IfEqCase{#1}{{GW200208K}{0.99}{GW200208G}{0.53}{GW200129D}{0.28}{GW191219E}{0.96}{GW191109A}{1.42}}}
\DeclareRobustCommand{\cosiotaSEOBtenthpercentile}[1]{\IfEqCase{#1}{{GW200208K}{-0.92}{GW200208G}{-0.97}{GW200129D}{0.40}{GW191219E}{-0.95}{GW191109A}{-0.95}}}
\DeclareRobustCommand{\cosiotaSEOBnintiethpercentile}[1]{\IfEqCase{#1}{{GW200208K}{0.91}{GW200208G}{-0.46}{GW200129D}{0.92}{GW191219E}{0.95}{GW191109A}{0.83}}}
\DeclareRobustCommand{\psiSEOBminus}[1]{\IfEqCase{#1}{{GW200208K}{1.3}{GW200208G}{2.8}{GW200129D}{0.87}{GW191219E}{2.9}{GW191109A}{1.5}}}
\DeclareRobustCommand{\psiSEOBmed}[1]{\IfEqCase{#1}{{GW200208K}{1.5}{GW200208G}{3.2}{GW200129D}{1.15}{GW191219E}{3.1}{GW191109A}{1.7}}}
\DeclareRobustCommand{\psiSEOBplus}[1]{\IfEqCase{#1}{{GW200208K}{1.5}{GW200208G}{2.7}{GW200129D}{1.48}{GW191219E}{2.9}{GW191109A}{1.3}}}
\DeclareRobustCommand{\psiSEOBtenthpercentile}[1]{\IfEqCase{#1}{{GW200208K}{0.3}{GW200208G}{0.7}{GW200129D}{0.47}{GW191219E}{0.6}{GW191109A}{0.3}}}
\DeclareRobustCommand{\psiSEOBnintiethpercentile}[1]{\IfEqCase{#1}{{GW200208K}{2.8}{GW200208G}{5.5}{GW200129D}{2.06}{GW191219E}{5.7}{GW191109A}{2.8}}}
\DeclareRobustCommand{\raSEOBminus}[1]{\IfEqCase{#1}{{GW200208K}{0.21}{GW200208G}{0.042}{GW200129D}{0.115}{GW191219E}{0.74}{GW191109A}{0.99}}}
\DeclareRobustCommand{\raSEOBmed}[1]{\IfEqCase{#1}{{GW200208K}{0.27}{GW200208G}{2.437}{GW200129D}{5.554}{GW191219E}{1.39}{GW191109A}{3.43}}}
\DeclareRobustCommand{\raSEOBplus}[1]{\IfEqCase{#1}{{GW200208K}{5.85}{GW200208G}{0.046}{GW200129D}{0.030}{GW191219E}{2.93}{GW191109A}{1.08}}}
\DeclareRobustCommand{\raSEOBtenthpercentile}[1]{\IfEqCase{#1}{{GW200208K}{0.11}{GW200208G}{2.405}{GW200129D}{5.489}{GW191219E}{0.78}{GW191109A}{2.56}}}
\DeclareRobustCommand{\raSEOBnintiethpercentile}[1]{\IfEqCase{#1}{{GW200208K}{1.35}{GW200208G}{2.471}{GW200129D}{5.577}{GW191219E}{4.12}{GW191109A}{4.32}}}
\DeclareRobustCommand{\massonedetSEOBminus}[1]{\IfEqCase{#1}{{GW200208K}{19}{GW200208G}{8.5}{GW200129D}{2.6}{GW191219E}{2.7}{GW191109A}{9.5}}}
\DeclareRobustCommand{\massonedetSEOBmed}[1]{\IfEqCase{#1}{{GW200208K}{53}{GW200208G}{52.8}{GW200129D}{39.0}{GW191219E}{34.4}{GW191109A}{82.8}}}
\DeclareRobustCommand{\massonedetSEOBplus}[1]{\IfEqCase{#1}{{GW200208K}{109}{GW200208G}{12.5}{GW200129D}{4.4}{GW191219E}{2.7}{GW191109A}{13.7}}}
\DeclareRobustCommand{\massonedetSEOBtenthpercentile}[1]{\IfEqCase{#1}{{GW200208K}{35}{GW200208G}{45.9}{GW200129D}{36.9}{GW191219E}{32.4}{GW191109A}{75.1}}}
\DeclareRobustCommand{\massonedetSEOBnintiethpercentile}[1]{\IfEqCase{#1}{{GW200208K}{146}{GW200208G}{62.3}{GW200129D}{42.0}{GW191219E}{36.4}{GW191109A}{93.2}}}
\DeclareRobustCommand{\phijlSEOBminus}[1]{\IfEqCase{#1}{{GW200208K}{2.9}{GW200208G}{3.7}{GW200129D}{4.1}{GW191219E}{2.7}{GW191109A}{2.9}}}
\DeclareRobustCommand{\phijlSEOBmed}[1]{\IfEqCase{#1}{{GW200208K}{3.2}{GW200208G}{3.9}{GW200129D}{4.7}{GW191219E}{3.1}{GW191109A}{3.3}}}
\DeclareRobustCommand{\phijlSEOBplus}[1]{\IfEqCase{#1}{{GW200208K}{2.7}{GW200208G}{2.1}{GW200129D}{1.2}{GW191219E}{2.8}{GW191109A}{2.6}}}
\DeclareRobustCommand{\phijlSEOBtenthpercentile}[1]{\IfEqCase{#1}{{GW200208K}{0.6}{GW200208G}{0.6}{GW200129D}{1.9}{GW191219E}{0.9}{GW191109A}{0.7}}}
\DeclareRobustCommand{\phijlSEOBnintiethpercentile}[1]{\IfEqCase{#1}{{GW200208K}{5.7}{GW200208G}{5.8}{GW200129D}{5.7}{GW191219E}{5.6}{GW191109A}{5.5}}}
\DeclareRobustCommand{\chipinfinityonlyprecavgSEOBminus}[1]{\IfEqCase{#1}{{GW200208K}{0.28}{GW200208G}{0.28}{GW200129D}{0.25}{GW191219E}{0.06}{GW191109A}{0.37}}}
\DeclareRobustCommand{\chipinfinityonlyprecavgSEOBmed}[1]{\IfEqCase{#1}{{GW200208K}{0.38}{GW200208G}{0.37}{GW200129D}{0.36}{GW191219E}{0.07}{GW191109A}{0.69}}}
\DeclareRobustCommand{\chipinfinityonlyprecavgSEOBplus}[1]{\IfEqCase{#1}{{GW200208K}{0.37}{GW200208G}{0.40}{GW200129D}{0.31}{GW191219E}{0.08}{GW191109A}{0.24}}}
\DeclareRobustCommand{\chipinfinityonlyprecavgSEOBtenthpercentile}[1]{\IfEqCase{#1}{{GW200208K}{0.13}{GW200208G}{0.13}{GW200129D}{0.15}{GW191219E}{0.02}{GW191109A}{0.41}}}
\DeclareRobustCommand{\chipinfinityonlyprecavgSEOBnintiethpercentile}[1]{\IfEqCase{#1}{{GW200208K}{0.69}{GW200208G}{0.69}{GW200129D}{0.60}{GW191219E}{0.13}{GW191109A}{0.90}}}
\DeclareRobustCommand{\spintwoxSEOBminus}[1]{\IfEqCase{#1}{{GW200208K}{0.53}{GW200208G}{0.55}{GW200129D}{0.44}{GW191219E}{0.53}{GW191109A}{0.64}}}
\DeclareRobustCommand{\spintwoxSEOBmed}[1]{\IfEqCase{#1}{{GW200208K}{0.00}{GW200208G}{0.00}{GW200129D}{0.00}{GW191219E}{0.00}{GW191109A}{0.00}}}
\DeclareRobustCommand{\spintwoxSEOBplus}[1]{\IfEqCase{#1}{{GW200208K}{0.54}{GW200208G}{0.55}{GW200129D}{0.45}{GW191219E}{0.53}{GW191109A}{0.64}}}
\DeclareRobustCommand{\spintwoxSEOBtenthpercentile}[1]{\IfEqCase{#1}{{GW200208K}{-0.40}{GW200208G}{-0.40}{GW200129D}{-0.32}{GW191219E}{-0.40}{GW191109A}{-0.49}}}
\DeclareRobustCommand{\spintwoxSEOBnintiethpercentile}[1]{\IfEqCase{#1}{{GW200208K}{0.40}{GW200208G}{0.40}{GW200129D}{0.33}{GW191219E}{0.37}{GW191109A}{0.49}}}
\DeclareRobustCommand{\chirpmassdetSEOBminus}[1]{\IfEqCase{#1}{{GW200208K}{6.1}{GW200208G}{4.6}{GW200129D}{1.7}{GW191219E}{0.04}{GW191109A}{9.1}}}
\DeclareRobustCommand{\chirpmassdetSEOBmed}[1]{\IfEqCase{#1}{{GW200208K}{29.2}{GW200208G}{39.1}{GW200129D}{32.1}{GW191219E}{4.81}{GW191109A}{61.5}}}
\DeclareRobustCommand{\chirpmassdetSEOBplus}[1]{\IfEqCase{#1}{{GW200208K}{28.8}{GW200208G}{5.3}{GW200129D}{1.4}{GW191219E}{0.06}{GW191109A}{7.5}}}
\DeclareRobustCommand{\chirpmassdetSEOBtenthpercentile}[1]{\IfEqCase{#1}{{GW200208K}{24.4}{GW200208G}{35.4}{GW200129D}{30.9}{GW191219E}{4.78}{GW191109A}{54.5}}}
\DeclareRobustCommand{\chirpmassdetSEOBnintiethpercentile}[1]{\IfEqCase{#1}{{GW200208K}{49.7}{GW200208G}{43.0}{GW200129D}{33.2}{GW191219E}{4.86}{GW191109A}{67.6}}}
\DeclareRobustCommand{\tiltoneSEOBminus}[1]{\IfEqCase{#1}{{GW200208K}{0.62}{GW200208G}{1.13}{GW200129D}{0.99}{GW191219E}{1.0}{GW191109A}{0.63}}}
\DeclareRobustCommand{\tiltoneSEOBmed}[1]{\IfEqCase{#1}{{GW200208K}{0.77}{GW200208G}{1.75}{GW200129D}{1.38}{GW191219E}{1.7}{GW191109A}{2.06}}}
\DeclareRobustCommand{\tiltoneSEOBplus}[1]{\IfEqCase{#1}{{GW200208K}{1.09}{GW200208G}{1.00}{GW200129D}{1.25}{GW191219E}{1.1}{GW191109A}{0.69}}}
\DeclareRobustCommand{\tiltoneSEOBtenthpercentile}[1]{\IfEqCase{#1}{{GW200208K}{0.24}{GW200208G}{0.87}{GW200129D}{0.57}{GW191219E}{0.8}{GW191109A}{1.56}}}
\DeclareRobustCommand{\tiltoneSEOBnintiethpercentile}[1]{\IfEqCase{#1}{{GW200208K}{1.55}{GW200208G}{2.59}{GW200129D}{2.41}{GW191219E}{2.5}{GW191109A}{2.60}}}
\DeclareRobustCommand{\phionetwoSEOBminus}[1]{\IfEqCase{#1}{{GW200208K}{2.8}{GW200208G}{2.8}{GW200129D}{2.5}{GW191219E}{2.9}{GW191109A}{4.0}}}
\DeclareRobustCommand{\phionetwoSEOBmed}[1]{\IfEqCase{#1}{{GW200208K}{3.1}{GW200208G}{3.1}{GW200129D}{3.0}{GW191219E}{3.2}{GW191109A}{4.2}}}
\DeclareRobustCommand{\phionetwoSEOBplus}[1]{\IfEqCase{#1}{{GW200208K}{2.8}{GW200208G}{2.8}{GW200129D}{2.8}{GW191219E}{2.8}{GW191109A}{1.9}}}
\DeclareRobustCommand{\phionetwoSEOBtenthpercentile}[1]{\IfEqCase{#1}{{GW200208K}{0.7}{GW200208G}{0.7}{GW200129D}{1.0}{GW191219E}{0.7}{GW191109A}{0.4}}}
\DeclareRobustCommand{\phionetwoSEOBnintiethpercentile}[1]{\IfEqCase{#1}{{GW200208K}{5.6}{GW200208G}{5.6}{GW200129D}{5.3}{GW191219E}{5.7}{GW191109A}{5.9}}}
\DeclareRobustCommand{\spinoneySEOBminus}[1]{\IfEqCase{#1}{{GW200208K}{0.52}{GW200208G}{0.46}{GW200129D}{0.39}{GW191219E}{0.10}{GW191109A}{0.75}}}
\DeclareRobustCommand{\spinoneySEOBmed}[1]{\IfEqCase{#1}{{GW200208K}{0.00}{GW200208G}{0.00}{GW200129D}{0.00}{GW191219E}{0.00}{GW191109A}{-0.01}}}
\DeclareRobustCommand{\spinoneySEOBplus}[1]{\IfEqCase{#1}{{GW200208K}{0.52}{GW200208G}{0.47}{GW200129D}{0.40}{GW191219E}{0.10}{GW191109A}{0.77}}}
\DeclareRobustCommand{\spinoneySEOBtenthpercentile}[1]{\IfEqCase{#1}{{GW200208K}{-0.40}{GW200208G}{-0.33}{GW200129D}{-0.29}{GW191219E}{-0.07}{GW191109A}{-0.66}}}
\DeclareRobustCommand{\spinoneySEOBnintiethpercentile}[1]{\IfEqCase{#1}{{GW200208K}{0.40}{GW200208G}{0.33}{GW200129D}{0.30}{GW191219E}{0.07}{GW191109A}{0.66}}}
\DeclareRobustCommand{\costiltoneSEOBminus}[1]{\IfEqCase{#1}{{GW200208K}{1.01}{GW200208G}{0.74}{GW200129D}{1.06}{GW191219E}{0.83}{GW191109A}{0.45}}}
\DeclareRobustCommand{\costiltoneSEOBmed}[1]{\IfEqCase{#1}{{GW200208K}{0.72}{GW200208G}{-0.18}{GW200129D}{0.19}{GW191219E}{-0.09}{GW191109A}{-0.47}}}
\DeclareRobustCommand{\costiltoneSEOBplus}[1]{\IfEqCase{#1}{{GW200208K}{0.27}{GW200208G}{0.99}{GW200129D}{0.73}{GW191219E}{0.89}{GW191109A}{0.61}}}
\DeclareRobustCommand{\costiltoneSEOBtenthpercentile}[1]{\IfEqCase{#1}{{GW200208K}{0.02}{GW200208G}{-0.85}{GW200129D}{-0.74}{GW191219E}{-0.82}{GW191109A}{-0.86}}}
\DeclareRobustCommand{\costiltoneSEOBnintiethpercentile}[1]{\IfEqCase{#1}{{GW200208K}{0.97}{GW200208G}{0.65}{GW200129D}{0.84}{GW191219E}{0.67}{GW191109A}{0.01}}}
\DeclareRobustCommand{\finalmasssourceSEOBminus}[1]{\IfEqCase{#1}{{GW200208K}{11}{GW200208G}{6.8}{GW200129D}{3.0}{GW191219E}{2.6}{GW191109A}{14}}}
\DeclareRobustCommand{\finalmasssourceSEOBmed}[1]{\IfEqCase{#1}{{GW200208K}{46}{GW200208G}{62.6}{GW200129D}{60.1}{GW191219E}{31.7}{GW191109A}{107}}}
\DeclareRobustCommand{\finalmasssourceSEOBplus}[1]{\IfEqCase{#1}{{GW200208K}{47}{GW200208G}{7.5}{GW200129D}{3.7}{GW191219E}{2.7}{GW191109A}{14}}}
\DeclareRobustCommand{\finalmasssourceSEOBtenthpercentile}[1]{\IfEqCase{#1}{{GW200208K}{36}{GW200208G}{57.3}{GW200129D}{57.7}{GW191219E}{29.8}{GW191109A}{96}}}
\DeclareRobustCommand{\finalmasssourceSEOBnintiethpercentile}[1]{\IfEqCase{#1}{{GW200208K}{84}{GW200208G}{68.4}{GW200129D}{63.0}{GW191219E}{33.8}{GW191109A}{118}}}
\DeclareRobustCommand{\totalmassdetSEOBminus}[1]{\IfEqCase{#1}{{GW200208K}{14}{GW200208G}{10}{GW200129D}{3.7}{GW191219E}{2.6}{GW191109A}{17}}}
\DeclareRobustCommand{\totalmassdetSEOBmed}[1]{\IfEqCase{#1}{{GW200208K}{75}{GW200208G}{92}{GW200129D}{74.0}{GW191219E}{35.7}{GW191109A}{144}}}
\DeclareRobustCommand{\totalmassdetSEOBplus}[1]{\IfEqCase{#1}{{GW200208K}{115}{GW200208G}{12}{GW200129D}{3.4}{GW191219E}{2.7}{GW191109A}{16}}}
\DeclareRobustCommand{\totalmassdetSEOBtenthpercentile}[1]{\IfEqCase{#1}{{GW200208K}{63}{GW200208G}{84}{GW200129D}{71.2}{GW191219E}{33.7}{GW191109A}{131}}}
\DeclareRobustCommand{\totalmassdetSEOBnintiethpercentile}[1]{\IfEqCase{#1}{{GW200208K}{170}{GW200208G}{101}{GW200129D}{76.6}{GW191219E}{37.7}{GW191109A}{157}}}
\DeclareRobustCommand{\spintwoySEOBminus}[1]{\IfEqCase{#1}{{GW200208K}{0.51}{GW200208G}{0.54}{GW200129D}{0.45}{GW191219E}{0.57}{GW191109A}{0.64}}}
\DeclareRobustCommand{\spintwoySEOBmed}[1]{\IfEqCase{#1}{{GW200208K}{0.00}{GW200208G}{0.00}{GW200129D}{0.00}{GW191219E}{0.00}{GW191109A}{0.00}}}
\DeclareRobustCommand{\spintwoySEOBplus}[1]{\IfEqCase{#1}{{GW200208K}{0.53}{GW200208G}{0.56}{GW200129D}{0.44}{GW191219E}{0.53}{GW191109A}{0.64}}}
\DeclareRobustCommand{\spintwoySEOBtenthpercentile}[1]{\IfEqCase{#1}{{GW200208K}{-0.37}{GW200208G}{-0.39}{GW200129D}{-0.34}{GW191219E}{-0.40}{GW191109A}{-0.48}}}
\DeclareRobustCommand{\spintwoySEOBnintiethpercentile}[1]{\IfEqCase{#1}{{GW200208K}{0.39}{GW200208G}{0.40}{GW200129D}{0.34}{GW191219E}{0.40}{GW191109A}{0.50}}}
\DeclareRobustCommand{\masstwosourceSEOBminus}[1]{\IfEqCase{#1}{{GW200208K}{6.6}{GW200208G}{7.3}{GW200129D}{3.5}{GW191219E}{0.07}{GW191109A}{13}}}
\DeclareRobustCommand{\masstwosourceSEOBmed}[1]{\IfEqCase{#1}{{GW200208K}{13.8}{GW200208G}{27.7}{GW200129D}{29.9}{GW191219E}{1.17}{GW191109A}{47}}}
\DeclareRobustCommand{\masstwosourceSEOBplus}[1]{\IfEqCase{#1}{{GW200208K}{8.1}{GW200208G}{6.0}{GW200129D}{2.5}{GW191219E}{0.08}{GW191109A}{11}}}
\DeclareRobustCommand{\masstwosourceSEOBtenthpercentile}[1]{\IfEqCase{#1}{{GW200208K}{8.2}{GW200208G}{22.0}{GW200129D}{27.4}{GW191219E}{1.12}{GW191109A}{37}}}
\DeclareRobustCommand{\masstwosourceSEOBnintiethpercentile}[1]{\IfEqCase{#1}{{GW200208K}{19.9}{GW200208G}{32.4}{GW200129D}{31.9}{GW191219E}{1.23}{GW191109A}{56}}}
\DeclareRobustCommand{\chirpmasssourceSEOBminus}[1]{\IfEqCase{#1}{{GW200208K}{3.5}{GW200208G}{3.2}{GW200129D}{1.5}{GW191219E}{0.17}{GW191109A}{7.2}}}
\DeclareRobustCommand{\chirpmasssourceSEOBmed}[1]{\IfEqCase{#1}{{GW200208K}{17.9}{GW200208G}{27.8}{GW200129D}{27.5}{GW191219E}{4.29}{GW191109A}{47.6}}}
\DeclareRobustCommand{\chirpmasssourceSEOBplus}[1]{\IfEqCase{#1}{{GW200208K}{9.8}{GW200208G}{3.6}{GW200129D}{1.8}{GW191219E}{0.14}{GW191109A}{6.8}}}
\DeclareRobustCommand{\chirpmasssourceSEOBtenthpercentile}[1]{\IfEqCase{#1}{{GW200208K}{15.1}{GW200208G}{25.3}{GW200129D}{26.3}{GW191219E}{4.16}{GW191109A}{41.9}}}
\DeclareRobustCommand{\chirpmasssourceSEOBnintiethpercentile}[1]{\IfEqCase{#1}{{GW200208K}{24.7}{GW200208G}{30.5}{GW200129D}{28.9}{GW191219E}{4.41}{GW191109A}{53.0}}}
\DeclareRobustCommand{\costilttwoSEOBminus}[1]{\IfEqCase{#1}{{GW200208K}{1.12}{GW200208G}{0.77}{GW200129D}{1.24}{GW191219E}{0.87}{GW191109A}{0.61}}}
\DeclareRobustCommand{\costilttwoSEOBmed}[1]{\IfEqCase{#1}{{GW200208K}{0.28}{GW200208G}{-0.13}{GW200129D}{0.52}{GW191219E}{-0.04}{GW191109A}{-0.31}}}
\DeclareRobustCommand{\costilttwoSEOBplus}[1]{\IfEqCase{#1}{{GW200208K}{0.67}{GW200208G}{0.97}{GW200129D}{0.44}{GW191219E}{0.94}{GW191109A}{1.00}}}
\DeclareRobustCommand{\costilttwoSEOBtenthpercentile}[1]{\IfEqCase{#1}{{GW200208K}{-0.68}{GW200208G}{-0.82}{GW200129D}{-0.48}{GW191219E}{-0.82}{GW191109A}{-0.85}}}
\DeclareRobustCommand{\costilttwoSEOBnintiethpercentile}[1]{\IfEqCase{#1}{{GW200208K}{0.89}{GW200208G}{0.69}{GW200129D}{0.91}{GW191219E}{0.77}{GW191109A}{0.47}}}
\DeclareRobustCommand{\tilttwoSEOBminus}[1]{\IfEqCase{#1}{{GW200208K}{0.96}{GW200208G}{1.12}{GW200129D}{0.73}{GW191219E}{1.1}{GW191109A}{1.07}}}
\DeclareRobustCommand{\tilttwoSEOBmed}[1]{\IfEqCase{#1}{{GW200208K}{1.29}{GW200208G}{1.70}{GW200129D}{1.03}{GW191219E}{1.6}{GW191109A}{1.89}}}
\DeclareRobustCommand{\tilttwoSEOBplus}[1]{\IfEqCase{#1}{{GW200208K}{1.29}{GW200208G}{1.00}{GW200129D}{1.35}{GW191219E}{1.1}{GW191109A}{0.86}}}
\DeclareRobustCommand{\tilttwoSEOBtenthpercentile}[1]{\IfEqCase{#1}{{GW200208K}{0.47}{GW200208G}{0.81}{GW200129D}{0.43}{GW191219E}{0.7}{GW191109A}{1.08}}}
\DeclareRobustCommand{\tilttwoSEOBnintiethpercentile}[1]{\IfEqCase{#1}{{GW200208K}{2.31}{GW200208G}{2.53}{GW200129D}{2.07}{GW191219E}{2.5}{GW191109A}{2.58}}}
\DeclareRobustCommand{\spintwoSEOBminus}[1]{\IfEqCase{#1}{{GW200208K}{0.40}{GW200208G}{0.38}{GW200129D}{0.36}{GW191219E}{0.39}{GW191109A}{0.50}}}
\DeclareRobustCommand{\spintwoSEOBmed}[1]{\IfEqCase{#1}{{GW200208K}{0.45}{GW200208G}{0.42}{GW200129D}{0.42}{GW191219E}{0.43}{GW191109A}{0.55}}}
\DeclareRobustCommand{\spintwoSEOBplus}[1]{\IfEqCase{#1}{{GW200208K}{0.49}{GW200208G}{0.49}{GW200129D}{0.38}{GW191219E}{0.50}{GW191109A}{0.40}}}
\DeclareRobustCommand{\spintwoSEOBtenthpercentile}[1]{\IfEqCase{#1}{{GW200208K}{0.09}{GW200208G}{0.08}{GW200129D}{0.10}{GW191219E}{0.08}{GW191109A}{0.12}}}
\DeclareRobustCommand{\spintwoSEOBnintiethpercentile}[1]{\IfEqCase{#1}{{GW200208K}{0.87}{GW200208G}{0.84}{GW200129D}{0.73}{GW191219E}{0.85}{GW191109A}{0.91}}}
\DeclareRobustCommand{\massratioSEOBminus}[1]{\IfEqCase{#1}{{GW200208K}{0.31}{GW200208G}{0.29}{GW200129D}{0.169}{GW191219E}{0.005}{GW191109A}{0.24}}}
\DeclareRobustCommand{\massratioSEOBmed}[1]{\IfEqCase{#1}{{GW200208K}{0.42}{GW200208G}{0.75}{GW200129D}{0.901}{GW191219E}{0.038}{GW191109A}{0.73}}}
\DeclareRobustCommand{\massratioSEOBplus}[1]{\IfEqCase{#1}{{GW200208K}{0.49}{GW200208G}{0.22}{GW200129D}{0.083}{GW191219E}{0.006}{GW191109A}{0.22}}}
\DeclareRobustCommand{\massratioSEOBtenthpercentile}[1]{\IfEqCase{#1}{{GW200208K}{0.14}{GW200208G}{0.51}{GW200129D}{0.779}{GW191219E}{0.035}{GW191109A}{0.53}}}
\DeclareRobustCommand{\massratioSEOBnintiethpercentile}[1]{\IfEqCase{#1}{{GW200208K}{0.85}{GW200208G}{0.94}{GW200129D}{0.974}{GW191219E}{0.042}{GW191109A}{0.92}}}
\DeclareRobustCommand{\spintwozSEOBminus}[1]{\IfEqCase{#1}{{GW200208K}{0.44}{GW200208G}{0.52}{GW200129D}{0.39}{GW191219E}{0.57}{GW191109A}{0.53}}}
\DeclareRobustCommand{\spintwozSEOBmed}[1]{\IfEqCase{#1}{{GW200208K}{0.07}{GW200208G}{-0.03}{GW200129D}{0.18}{GW191219E}{-0.01}{GW191109A}{-0.12}}}
\DeclareRobustCommand{\spintwozSEOBplus}[1]{\IfEqCase{#1}{{GW200208K}{0.62}{GW200208G}{0.40}{GW200129D}{0.39}{GW191219E}{0.53}{GW191109A}{0.39}}}
\DeclareRobustCommand{\spintwozSEOBtenthpercentile}[1]{\IfEqCase{#1}{{GW200208K}{-0.25}{GW200208G}{-0.42}{GW200129D}{-0.11}{GW191219E}{-0.44}{GW191109A}{-0.54}}}
\DeclareRobustCommand{\spintwozSEOBnintiethpercentile}[1]{\IfEqCase{#1}{{GW200208K}{0.55}{GW200208G}{0.26}{GW200129D}{0.49}{GW191219E}{0.37}{GW191109A}{0.16}}}
\DeclareRobustCommand{\comovingdistSEOBminus}[1]{\IfEqCase{#1}{{GW200208K}{930}{GW200208G}{490}{GW200129D}{260}{GW191219E}{150}{GW191109A}{460}}}
\DeclareRobustCommand{\comovingdistSEOBmed}[1]{\IfEqCase{#1}{{GW200208K}{2420}{GW200208G}{1610}{GW200129D}{710}{GW191219E}{520}{GW191109A}{1170}}}
\DeclareRobustCommand{\comovingdistSEOBplus}[1]{\IfEqCase{#1}{{GW200208K}{1500}{GW200208G}{510}{GW200129D}{210}{GW191219E}{200}{GW191109A}{600}}}
\DeclareRobustCommand{\comovingdistSEOBtenthpercentile}[1]{\IfEqCase{#1}{{GW200208K}{1680}{GW200208G}{1220}{GW200129D}{500}{GW191219E}{400}{GW191109A}{790}}}
\DeclareRobustCommand{\comovingdistSEOBnintiethpercentile}[1]{\IfEqCase{#1}{{GW200208K}{3480}{GW200208G}{2000}{GW200129D}{890}{GW191219E}{670}{GW191109A}{1620}}}
\DeclareRobustCommand{\iotaSEOBminus}[1]{\IfEqCase{#1}{{GW200208K}{1.3}{GW200208G}{0.67}{GW200129D}{0.52}{GW191219E}{1.3}{GW191109A}{1.68}}}
\DeclareRobustCommand{\iotaSEOBmed}[1]{\IfEqCase{#1}{{GW200208K}{1.6}{GW200208G}{2.54}{GW200129D}{0.83}{GW191219E}{1.6}{GW191109A}{2.10}}}
\DeclareRobustCommand{\iotaSEOBplus}[1]{\IfEqCase{#1}{{GW200208K}{1.3}{GW200208G}{0.44}{GW200129D}{0.40}{GW191219E}{1.4}{GW191109A}{0.82}}}
\DeclareRobustCommand{\iotaSEOBtenthpercentile}[1]{\IfEqCase{#1}{{GW200208K}{0.4}{GW200208G}{2.05}{GW200129D}{0.41}{GW191219E}{0.3}{GW191109A}{0.60}}}
\DeclareRobustCommand{\iotaSEOBnintiethpercentile}[1]{\IfEqCase{#1}{{GW200208K}{2.7}{GW200208G}{2.90}{GW200129D}{1.16}{GW191219E}{2.8}{GW191109A}{2.82}}}
\DeclareRobustCommand{\chieffinfinityonlyprecavgSEOBminus}[1]{\IfEqCase{#1}{{GW200208K}{0.38}{GW200208G}{0.26}{GW200129D}{0.12}{GW191219E}{0.10}{GW191109A}{0.26}}}
\DeclareRobustCommand{\chieffinfinityonlyprecavgSEOBmed}[1]{\IfEqCase{#1}{{GW200208K}{0.34}{GW200208G}{-0.05}{GW200129D}{0.12}{GW191219E}{0.00}{GW191109A}{-0.28}}}
\DeclareRobustCommand{\chieffinfinityonlyprecavgSEOBplus}[1]{\IfEqCase{#1}{{GW200208K}{0.45}{GW200208G}{0.21}{GW200129D}{0.09}{GW191219E}{0.08}{GW191109A}{0.26}}}
\DeclareRobustCommand{\chieffinfinityonlyprecavgSEOBtenthpercentile}[1]{\IfEqCase{#1}{{GW200208K}{0.04}{GW200208G}{-0.26}{GW200129D}{0.03}{GW191219E}{-0.08}{GW191109A}{-0.48}}}
\DeclareRobustCommand{\chieffinfinityonlyprecavgSEOBnintiethpercentile}[1]{\IfEqCase{#1}{{GW200208K}{0.70}{GW200208G}{0.11}{GW200129D}{0.18}{GW191219E}{0.06}{GW191109A}{-0.07}}}
\DeclareRobustCommand{\spinoneSEOBminus}[1]{\IfEqCase{#1}{{GW200208K}{0.55}{GW200208G}{0.31}{GW200129D}{0.30}{GW191219E}{0.07}{GW191109A}{0.38}}}
\DeclareRobustCommand{\spinoneSEOBmed}[1]{\IfEqCase{#1}{{GW200208K}{0.67}{GW200208G}{0.34}{GW200129D}{0.34}{GW191219E}{0.08}{GW191109A}{0.84}}}
\DeclareRobustCommand{\spinoneSEOBplus}[1]{\IfEqCase{#1}{{GW200208K}{0.30}{GW200208G}{0.50}{GW200129D}{0.35}{GW191219E}{0.09}{GW191109A}{0.15}}}
\DeclareRobustCommand{\spinoneSEOBtenthpercentile}[1]{\IfEqCase{#1}{{GW200208K}{0.20}{GW200208G}{0.06}{GW200129D}{0.07}{GW191219E}{0.02}{GW191109A}{0.57}}}
\DeclareRobustCommand{\spinoneSEOBnintiethpercentile}[1]{\IfEqCase{#1}{{GW200208K}{0.93}{GW200208G}{0.74}{GW200129D}{0.62}{GW191219E}{0.15}{GW191109A}{0.97}}}
\DeclareRobustCommand{\totalmasssourceSEOBminus}[1]{\IfEqCase{#1}{{GW200208K}{12}{GW200208G}{7.3}{GW200129D}{3.4}{GW191219E}{2.6}{GW191109A}{15}}}
\DeclareRobustCommand{\totalmasssourceSEOBmed}[1]{\IfEqCase{#1}{{GW200208K}{48}{GW200208G}{65.5}{GW200129D}{63.4}{GW191219E}{31.8}{GW191109A}{112}}}
\DeclareRobustCommand{\totalmasssourceSEOBplus}[1]{\IfEqCase{#1}{{GW200208K}{47}{GW200208G}{8.0}{GW200129D}{4.0}{GW191219E}{2.7}{GW191109A}{15}}}
\DeclareRobustCommand{\totalmasssourceSEOBtenthpercentile}[1]{\IfEqCase{#1}{{GW200208K}{38}{GW200208G}{59.8}{GW200129D}{60.7}{GW191219E}{29.9}{GW191109A}{100}}}
\DeclareRobustCommand{\totalmasssourceSEOBnintiethpercentile}[1]{\IfEqCase{#1}{{GW200208K}{87}{GW200208G}{71.6}{GW200129D}{66.5}{GW191219E}{33.8}{GW191109A}{123}}}
\DeclareRobustCommand{\phaseSEOBminus}[1]{\IfEqCase{#1}{{GW200208K}{2.9}{GW200208G}{2.8}{GW200129D}{2.8}{GW191219E}{2.8}{GW191109A}{2.5}}}
\DeclareRobustCommand{\phaseSEOBmed}[1]{\IfEqCase{#1}{{GW200208K}{3.2}{GW200208G}{3.1}{GW200129D}{3.1}{GW191219E}{3.2}{GW191109A}{3.1}}}
\DeclareRobustCommand{\phaseSEOBplus}[1]{\IfEqCase{#1}{{GW200208K}{2.8}{GW200208G}{2.8}{GW200129D}{2.8}{GW191219E}{2.8}{GW191109A}{2.6}}}
\DeclareRobustCommand{\phaseSEOBtenthpercentile}[1]{\IfEqCase{#1}{{GW200208K}{0.6}{GW200208G}{0.6}{GW200129D}{0.6}{GW191219E}{0.7}{GW191109A}{1.1}}}
\DeclareRobustCommand{\phaseSEOBnintiethpercentile}[1]{\IfEqCase{#1}{{GW200208K}{5.7}{GW200208G}{5.6}{GW200129D}{5.6}{GW191219E}{5.6}{GW191109A}{5.1}}}
\DeclareRobustCommand{\redshiftSEOBminus}[1]{\IfEqCase{#1}{{GW200208K}{0.28}{GW200208G}{0.13}{GW200129D}{0.06}{GW191219E}{0.04}{GW191109A}{0.12}}}
\DeclareRobustCommand{\redshiftSEOBmed}[1]{\IfEqCase{#1}{{GW200208K}{0.65}{GW200208G}{0.40}{GW200129D}{0.17}{GW191219E}{0.12}{GW191109A}{0.28}}}
\DeclareRobustCommand{\redshiftSEOBplus}[1]{\IfEqCase{#1}{{GW200208K}{0.58}{GW200208G}{0.15}{GW200129D}{0.05}{GW191219E}{0.05}{GW191109A}{0.16}}}
\DeclareRobustCommand{\redshiftSEOBtenthpercentile}[1]{\IfEqCase{#1}{{GW200208K}{0.42}{GW200208G}{0.30}{GW200129D}{0.12}{GW191219E}{0.09}{GW191109A}{0.19}}}
\DeclareRobustCommand{\redshiftSEOBnintiethpercentile}[1]{\IfEqCase{#1}{{GW200208K}{1.04}{GW200208G}{0.52}{GW200129D}{0.21}{GW191219E}{0.16}{GW191109A}{0.41}}}
\DeclareRobustCommand{\spinonexSEOBminus}[1]{\IfEqCase{#1}{{GW200208K}{0.51}{GW200208G}{0.46}{GW200129D}{0.40}{GW191219E}{0.10}{GW191109A}{0.74}}}
\DeclareRobustCommand{\spinonexSEOBmed}[1]{\IfEqCase{#1}{{GW200208K}{-0.01}{GW200208G}{0.00}{GW200129D}{0.00}{GW191219E}{0.00}{GW191109A}{-0.02}}}
\DeclareRobustCommand{\spinonexSEOBplus}[1]{\IfEqCase{#1}{{GW200208K}{0.55}{GW200208G}{0.45}{GW200129D}{0.40}{GW191219E}{0.10}{GW191109A}{0.76}}}
\DeclareRobustCommand{\spinonexSEOBtenthpercentile}[1]{\IfEqCase{#1}{{GW200208K}{-0.39}{GW200208G}{-0.33}{GW200129D}{-0.30}{GW191219E}{-0.08}{GW191109A}{-0.66}}}
\DeclareRobustCommand{\spinonexSEOBnintiethpercentile}[1]{\IfEqCase{#1}{{GW200208K}{0.41}{GW200208G}{0.32}{GW200129D}{0.29}{GW191219E}{0.08}{GW191109A}{0.64}}}
\DeclareRobustCommand{\symmetricmassratioSEOBminus}[1]{\IfEqCase{#1}{{GW200208K}{0.118}{GW200208G}{0.030}{GW200129D}{0.005}{GW191219E}{0.004}{GW191109A}{0.023}}}
\DeclareRobustCommand{\symmetricmassratioSEOBmed}[1]{\IfEqCase{#1}{{GW200208K}{0.208}{GW200208G}{0.245}{GW200129D}{0.249}{GW191219E}{0.035}{GW191109A}{0.244}}}
\DeclareRobustCommand{\symmetricmassratioSEOBplus}[1]{\IfEqCase{#1}{{GW200208K}{0.042}{GW200208G}{0.005}{GW200129D}{0.001}{GW191219E}{0.005}{GW191109A}{0.006}}}
\DeclareRobustCommand{\symmetricmassratioSEOBtenthpercentile}[1]{\IfEqCase{#1}{{GW200208K}{0.107}{GW200208G}{0.224}{GW200129D}{0.246}{GW191219E}{0.032}{GW191109A}{0.227}}}
\DeclareRobustCommand{\symmetricmassratioSEOBnintiethpercentile}[1]{\IfEqCase{#1}{{GW200208K}{0.248}{GW200208G}{0.250}{GW200129D}{0.250}{GW191219E}{0.039}{GW191109A}{0.250}}}
\DeclareRobustCommand{\thetajnSEOBminus}[1]{\IfEqCase{#1}{{GW200208K}{1.3}{GW200208G}{0.68}{GW200129D}{0.51}{GW191219E}{1.3}{GW191109A}{1.47}}}
\DeclareRobustCommand{\thetajnSEOBmed}[1]{\IfEqCase{#1}{{GW200208K}{1.6}{GW200208G}{2.52}{GW200129D}{0.76}{GW191219E}{1.6}{GW191109A}{2.05}}}
\DeclareRobustCommand{\thetajnSEOBplus}[1]{\IfEqCase{#1}{{GW200208K}{1.3}{GW200208G}{0.45}{GW200129D}{0.40}{GW191219E}{1.3}{GW191109A}{0.82}}}
\DeclareRobustCommand{\thetajnSEOBtenthpercentile}[1]{\IfEqCase{#1}{{GW200208K}{0.4}{GW200208G}{2.03}{GW200129D}{0.36}{GW191219E}{0.4}{GW191109A}{0.81}}}
\DeclareRobustCommand{\thetajnSEOBnintiethpercentile}[1]{\IfEqCase{#1}{{GW200208K}{2.7}{GW200208G}{2.89}{GW200129D}{1.10}{GW191219E}{2.8}{GW191109A}{2.75}}}
\DeclareRobustCommand{\luminositydistanceSEOBminus}[1]{\IfEqCase{#1}{{GW200208K}{2.0}{GW200208G}{0.83}{GW200129D}{0.33}{GW191219E}{0.19}{GW191109A}{0.67}}}
\DeclareRobustCommand{\luminositydistanceSEOBmed}[1]{\IfEqCase{#1}{{GW200208K}{4.0}{GW200208G}{2.26}{GW200129D}{0.83}{GW191219E}{0.58}{GW191109A}{1.50}}}
\DeclareRobustCommand{\luminositydistanceSEOBplus}[1]{\IfEqCase{#1}{{GW200208K}{4.7}{GW200208G}{1.04}{GW200129D}{0.29}{GW191219E}{0.26}{GW191109A}{1.06}}}
\DeclareRobustCommand{\luminositydistanceSEOBtenthpercentile}[1]{\IfEqCase{#1}{{GW200208K}{2.4}{GW200208G}{1.59}{GW200129D}{0.56}{GW191219E}{0.44}{GW191109A}{0.94}}}
\DeclareRobustCommand{\luminositydistanceSEOBnintiethpercentile}[1]{\IfEqCase{#1}{{GW200208K}{7.1}{GW200208G}{3.03}{GW200129D}{1.08}{GW191219E}{0.78}{GW191109A}{2.28}}}
\DeclareRobustCommand{\costhetajnSEOBminus}[1]{\IfEqCase{#1}{{GW200208K}{0.93}{GW200208G}{0.17}{GW200129D}{0.32}{GW191219E}{0.98}{GW191109A}{0.50}}}
\DeclareRobustCommand{\costhetajnSEOBmed}[1]{\IfEqCase{#1}{{GW200208K}{-0.04}{GW200208G}{-0.81}{GW200129D}{0.72}{GW191219E}{0.02}{GW191109A}{-0.46}}}
\DeclareRobustCommand{\costhetajnSEOBplus}[1]{\IfEqCase{#1}{{GW200208K}{1.00}{GW200208G}{0.55}{GW200129D}{0.25}{GW191219E}{0.95}{GW191109A}{1.30}}}
\DeclareRobustCommand{\costhetajnSEOBtenthpercentile}[1]{\IfEqCase{#1}{{GW200208K}{-0.92}{GW200208G}{-0.97}{GW200129D}{0.46}{GW191219E}{-0.92}{GW191109A}{-0.93}}}
\DeclareRobustCommand{\costhetajnSEOBnintiethpercentile}[1]{\IfEqCase{#1}{{GW200208K}{0.91}{GW200208G}{-0.45}{GW200129D}{0.94}{GW191219E}{0.93}{GW191109A}{0.69}}}
\DeclareRobustCommand{\radiatedenergySEOBminus}[1]{\IfEqCase{#1}{{GW200208K}{1.2}{GW200208G}{0.72}{GW200129D}{0.37}{GW191219E}{0.0043}{GW191109A}{1.3}}}
\DeclareRobustCommand{\radiatedenergySEOBmed}[1]{\IfEqCase{#1}{{GW200208K}{2.0}{GW200208G}{2.87}{GW200129D}{3.24}{GW191219E}{0.0860}{GW191109A}{4.4}}}
\DeclareRobustCommand{\radiatedenergySEOBplus}[1]{\IfEqCase{#1}{{GW200208K}{1.8}{GW200208G}{0.76}{GW200129D}{0.35}{GW191219E}{0.0044}{GW191109A}{1.2}}}
\DeclareRobustCommand{\radiatedenergySEOBtenthpercentile}[1]{\IfEqCase{#1}{{GW200208K}{1.1}{GW200208G}{2.31}{GW200129D}{2.96}{GW191219E}{0.0827}{GW191109A}{3.4}}}
\DeclareRobustCommand{\radiatedenergySEOBnintiethpercentile}[1]{\IfEqCase{#1}{{GW200208K}{3.2}{GW200208G}{3.45}{GW200129D}{3.50}{GW191219E}{0.0894}{GW191109A}{5.4}}}
\DeclareRobustCommand{\spinonezSEOBminus}[1]{\IfEqCase{#1}{{GW200208K}{0.48}{GW200208G}{0.49}{GW200129D}{0.34}{GW191219E}{0.10}{GW191109A}{0.43}}}
\DeclareRobustCommand{\spinonezSEOBmed}[1]{\IfEqCase{#1}{{GW200208K}{0.42}{GW200208G}{-0.03}{GW200129D}{0.03}{GW191219E}{0.00}{GW191109A}{-0.36}}}
\DeclareRobustCommand{\spinonezSEOBplus}[1]{\IfEqCase{#1}{{GW200208K}{0.49}{GW200208G}{0.31}{GW200129D}{0.39}{GW191219E}{0.08}{GW191109A}{0.46}}}
\DeclareRobustCommand{\spinonezSEOBtenthpercentile}[1]{\IfEqCase{#1}{{GW200208K}{0.00}{GW200208G}{-0.39}{GW200129D}{-0.23}{GW191219E}{-0.08}{GW191109A}{-0.72}}}
\DeclareRobustCommand{\spinonezSEOBnintiethpercentile}[1]{\IfEqCase{#1}{{GW200208K}{0.82}{GW200208G}{0.18}{GW200129D}{0.34}{GW191219E}{0.05}{GW191109A}{0.01}}}
\DeclareRobustCommand{\loglikelihoodSEOBminus}[1]{\IfEqCase{#1}{{GW200208K}{7.6}{GW200208G}{6.1}{GW200129D}{5.8}{GW191219E}{6.0}{GW191109A}{7.1}}}
\DeclareRobustCommand{\loglikelihoodSEOBmed}[1]{\IfEqCase{#1}{{GW200208K}{16.8}{GW200208G}{49.7}{GW200129D}{342.5}{GW191219E}{28.5}{GW191109A}{138.4}}}
\DeclareRobustCommand{\loglikelihoodSEOBplus}[1]{\IfEqCase{#1}{{GW200208K}{4.5}{GW200208G}{3.0}{GW200129D}{4.9}{GW191219E}{5.0}{GW191109A}{6.5}}}
\DeclareRobustCommand{\loglikelihoodSEOBtenthpercentile}[1]{\IfEqCase{#1}{{GW200208K}{11.6}{GW200208G}{45.5}{GW200129D}{338.0}{GW191219E}{24.0}{GW191109A}{132.9}}}
\DeclareRobustCommand{\loglikelihoodSEOBnintiethpercentile}[1]{\IfEqCase{#1}{{GW200208K}{20.4}{GW200208G}{52.1}{GW200129D}{346.3}{GW191219E}{32.4}{GW191109A}{143.5}}}
\DeclareRobustCommand{\masstwodetSEOBminus}[1]{\IfEqCase{#1}{{GW200208K}{13}{GW200208G}{10.8}{GW200129D}{3.9}{GW191219E}{0.07}{GW191109A}{17}}}
\DeclareRobustCommand{\masstwodetSEOBmed}[1]{\IfEqCase{#1}{{GW200208K}{24}{GW200208G}{39.0}{GW200129D}{34.9}{GW191219E}{1.31}{GW191109A}{61}}}
\DeclareRobustCommand{\masstwodetSEOBplus}[1]{\IfEqCase{#1}{{GW200208K}{12}{GW200208G}{8.6}{GW200129D}{2.4}{GW191219E}{0.07}{GW191109A}{13}}}
\DeclareRobustCommand{\masstwodetSEOBtenthpercentile}[1]{\IfEqCase{#1}{{GW200208K}{13}{GW200208G}{30.6}{GW200129D}{32.1}{GW191219E}{1.26}{GW191109A}{47}}}
\DeclareRobustCommand{\masstwodetSEOBnintiethpercentile}[1]{\IfEqCase{#1}{{GW200208K}{32}{GW200208G}{45.8}{GW200129D}{36.8}{GW191219E}{1.37}{GW191109A}{72}}}
\DeclareRobustCommand{\finalmassdetSEOBminus}[1]{\IfEqCase{#1}{{GW200208K}{14}{GW200208G}{9.2}{GW200129D}{3.3}{GW191219E}{2.6}{GW191109A}{15}}}
\DeclareRobustCommand{\finalmassdetSEOBmed}[1]{\IfEqCase{#1}{{GW200208K}{72}{GW200208G}{87.9}{GW200129D}{70.3}{GW191219E}{35.6}{GW191109A}{138}}}
\DeclareRobustCommand{\finalmassdetSEOBplus}[1]{\IfEqCase{#1}{{GW200208K}{111}{GW200208G}{10.7}{GW200129D}{3.0}{GW191219E}{2.7}{GW191109A}{15}}}
\DeclareRobustCommand{\finalmassdetSEOBtenthpercentile}[1]{\IfEqCase{#1}{{GW200208K}{60}{GW200208G}{80.6}{GW200129D}{67.8}{GW191219E}{33.6}{GW191109A}{126}}}
\DeclareRobustCommand{\finalmassdetSEOBnintiethpercentile}[1]{\IfEqCase{#1}{{GW200208K}{166}{GW200208G}{95.9}{GW200129D}{72.5}{GW191219E}{37.6}{GW191109A}{150}}}
\DeclareRobustCommand{\PEpercentBNSSEOB}[1]{\IfEqCase{#1}{{GW200208K}{0}{GW200208G}{0}{GW200129D}{0}{GW191219E}{0}{GW191109A}{0}}}
\DeclareRobustCommand{\PEpercentNSBHSEOB}[1]{\IfEqCase{#1}{{GW200208K}{0}{GW200208G}{0}{GW200129D}{0}{GW191219E}{100}{GW191109A}{0}}}
\DeclareRobustCommand{\PEpercentBBHSEOB}[1]{\IfEqCase{#1}{{GW200208K}{100}{GW200208G}{100}{GW200129D}{100}{GW191219E}{0}{GW191109A}{100}}}
\DeclareRobustCommand{\PEpercentMassGapSEOB}[1]{\IfEqCase{#1}{{GW200208K}{0}{GW200208G}{0}{GW200129D}{0}{GW191219E}{0}{GW191109A}{0}}}
\DeclareRobustCommand{\percentmassonelessthanthreeSEOB}[1]{\IfEqCase{#1}{{GW200208K}{0}{GW200208G}{0}{GW200129D}{0}{GW191219E}{0}{GW191109A}{0}}}
\DeclareRobustCommand{\percentmasstwolessthanthreeSEOB}[1]{\IfEqCase{#1}{{GW200208K}{0}{GW200208G}{0}{GW200129D}{0}{GW191219E}{100}{GW191109A}{0}}}
\DeclareRobustCommand{\percentmassonelessthanfiveSEOB}[1]{\IfEqCase{#1}{{GW200208K}{0}{GW200208G}{0}{GW200129D}{0}{GW191219E}{0}{GW191109A}{0}}}
\DeclareRobustCommand{\percentmasstwolessthanfiveSEOB}[1]{\IfEqCase{#1}{{GW200208K}{0}{GW200208G}{0}{GW200129D}{0}{GW191219E}{100}{GW191109A}{0}}}
\DeclareRobustCommand{\percentmassonemorethansixtyfiveSEOB}[1]{\IfEqCase{#1}{{GW200208K}{17}{GW200208G}{0}{GW200129D}{0}{GW191219E}{0}{GW191109A}{48}}}
\DeclareRobustCommand{\percentmasstwomorethansixtyfiveSEOB}[1]{\IfEqCase{#1}{{GW200208K}{0}{GW200208G}{0}{GW200129D}{0}{GW191219E}{0}{GW191109A}{0}}}
\DeclareRobustCommand{\percentmassonemorethanonetwentySEOB}[1]{\IfEqCase{#1}{{GW200208K}{0}{GW200208G}{0}{GW200129D}{0}{GW191219E}{0}{GW191109A}{0}}}
\DeclareRobustCommand{\percentmasstwomorethanonetwentySEOB}[1]{\IfEqCase{#1}{{GW200208K}{0}{GW200208G}{0}{GW200129D}{0}{GW191219E}{0}{GW191109A}{0}}}
\DeclareRobustCommand{\percentmassfinalmorethanonehundredSEOB}[1]{\IfEqCase{#1}{{GW200208K}{2}{GW200208G}{0}{GW200129D}{0}{GW191219E}{0}{GW191109A}{80}}}
\DeclareRobustCommand{\percentchieffmorethanzeroSEOB}[1]{\IfEqCase{#1}{{GW200208K}{93}{GW200208G}{34}{GW200129D}{95}{GW191219E}{45}{GW191109A}{4}}}
\DeclareRobustCommand{\percentchiefflessthanzeroSEOB}[1]{\IfEqCase{#1}{{GW200208K}{7}{GW200208G}{66}{GW200129D}{5}{GW191219E}{55}{GW191109A}{96}}}
\DeclareRobustCommand{\percentchionemorethanpointeightSEOB}[1]{\IfEqCase{#1}{{GW200208K}{31}{GW200208G}{7}{GW200129D}{1}{GW191219E}{0}{GW191109A}{59}}}
\DeclareRobustCommand{\percentchitwomorethanpointeightSEOB}[1]{\IfEqCase{#1}{{GW200208K}{15}{GW200208G}{13}{GW200129D}{5}{GW191219E}{14}{GW191109A}{23}}}
\DeclareRobustCommand{\percentanychimorethanpointeightSEOB}[1]{\IfEqCase{#1}{{GW200208K}{42}{GW200208G}{19}{GW200129D}{6}{GW191219E}{15}{GW191109A}{73}}}
\DeclareRobustCommand{\percentchieffinfinityonlyprecavgmorethanzeroSEOB}[1]{\IfEqCase{#1}{{GW200208K}{93}{GW200208G}{34}{GW200129D}{95}{GW191219E}{45}{GW191109A}{4}}}
\DeclareRobustCommand{\percentchieffinfinityonlyprecavglessthanzeroSEOB}[1]{\IfEqCase{#1}{{GW200208K}{7}{GW200208G}{66}{GW200129D}{5}{GW191219E}{55}{GW191109A}{96}}}
\DeclareRobustCommand{\sinthetajnSEOBminus}[1]{\IfEqCase{#1}{{GW200208K}{0.5}{GW200208K}{0.41}{GW200208K}{0.44}{GW200208K}{0.45}{GW200208K}{0.52}}}
\DeclareRobustCommand{\sinthetajnSEOBmed}[1]{\IfEqCase{#1}{{GW200208K}{0.69}{GW200208K}{0.57}{GW200208K}{0.69}{GW200208K}{0.63}{GW200208K}{0.76}}}
\DeclareRobustCommand{\sinthetajnSEOBplus}[1]{\IfEqCase{#1}{{GW200208K}{0.3}{GW200208K}{0.35}{GW200208K}{0.23}{GW200208K}{0.35}{GW200208K}{0.23}}}
\DeclareRobustCommand{\sinthetajnSEOBtenthpercentile}[1]{\IfEqCase{#1}{{GW200208K}{0.27}{GW200208K}{0.24}{GW200208K}{0.35}{GW200208K}{0.26}{GW200208K}{0.34}}}
\DeclareRobustCommand{\sinthetajnSEOBnintiethpercentile}[1]{\IfEqCase{#1}{{GW200208K}{0.97}{GW200208K}{0.87}{GW200208K}{0.89}{GW200208K}{0.95}{GW200208K}{0.98}}}
\DeclareRobustCommand{\abscosthetajnSEOBminus}[1]{\IfEqCase{#1}{{GW200208K}{0.6}{GW200208K}{0.44}{GW200208K}{0.32}{GW200208K}{0.61}{GW200208K}{0.56}}}
\DeclareRobustCommand{\abscosthetajnSEOBmed}[1]{\IfEqCase{#1}{{GW200208K}{0.72}{GW200208K}{0.82}{GW200208K}{0.72}{GW200208K}{0.77}{GW200208K}{0.65}}}
\DeclareRobustCommand{\abscosthetajnSEOBplus}[1]{\IfEqCase{#1}{{GW200208K}{0.26}{GW200208K}{0.17}{GW200208K}{0.25}{GW200208K}{0.21}{GW200208K}{0.32}}}
\DeclareRobustCommand{\abscosthetajnSEOBtenthpercentile}[1]{\IfEqCase{#1}{{GW200208K}{0.24}{GW200208K}{0.49}{GW200208K}{0.46}{GW200208K}{0.3}{GW200208K}{0.18}}}
\DeclareRobustCommand{\abscosthetajnSEOBnintiethpercentile}[1]{\IfEqCase{#1}{{GW200208K}{0.96}{GW200208K}{0.97}{GW200208K}{0.94}{GW200208K}{0.97}{GW200208K}{0.94}}}
\DeclareRobustCommand{\minthetajnfromhalfpiSEOBminus}[1]{\IfEqCase{#1}{{GW200208K}{1.26}{GW200208K}{0.45}{GW200208K}{0.34}{GW200208K}{1.32}{GW200208K}{0.82}}}
\DeclareRobustCommand{\minthetajnfromhalfpiSEOBmed}[1]{\IfEqCase{#1}{{GW200208K}{-0.04}{GW200208K}{-0.95}{GW200208K}{0.58}{GW200208K}{0.01}{GW200208K}{-0.48}}}
\DeclareRobustCommand{\minthetajnfromhalfpiSEOBplus}[1]{\IfEqCase{#1}{{GW200208K}{0.76}{GW200208K}{0.68}{GW200208K}{0.18}{GW200208K}{0.72}{GW200208K}{1.16}}}
\DeclareRobustCommand{\minthetajnfromhalfpiSEOBtenthpercentile}[1]{\IfEqCase{#1}{{GW200208K}{-1.17}{GW200208K}{-1.32}{GW200208K}{0.33}{GW200208K}{-1.18}{GW200208K}{-1.18}}}
\DeclareRobustCommand{\minthetajnfromhalfpiSEOBnintiethpercentile}[1]{\IfEqCase{#1}{{GW200208K}{0.66}{GW200208K}{-0.46}{GW200208K}{0.75}{GW200208K}{0.68}{GW200208K}{0.58}}}
\DeclareRobustCommand{\phitwoJS}[1]{\IfEqCase{#1}{{GW200208K}{0.001}{GW200208G}{0.000}{GW200129D}{0.002}{GW191219E}{0.001}{GW191109A}{0.001}}}
\DeclareRobustCommand{\spinonezinfinityonlyprecavgJS}[1]{\IfEqCase{#1}{{GW200208K}{0.104}{GW200208G}{0.007}{GW200129D}{0.023}{GW191219E}{0.024}{GW191109A}{0.056}}}
\DeclareRobustCommand{\phioneJS}[1]{\IfEqCase{#1}{{GW200208K}{0.001}{GW200208G}{0.002}{GW200129D}{0.012}{GW191219E}{0.001}{GW191109A}{0.005}}}
\DeclareRobustCommand{\chieffJS}[1]{\IfEqCase{#1}{{GW200208K}{0.130}{GW200208G}{0.011}{GW200129D}{0.061}{GW191219E}{0.024}{GW191109A}{0.086}}}
\DeclareRobustCommand{\geocenttimeJS}[1]{\IfEqCase{#1}{{GW200208K}{nan}{GW200208G}{nan}{GW200129D}{inf}{GW191219E}{nan}{GW191109A}{nan}}}
\DeclareRobustCommand{\cosiotaJS}[1]{\IfEqCase{#1}{{GW200208K}{0.005}{GW200208G}{0.007}{GW200129D}{0.204}{GW191219E}{0.024}{GW191109A}{0.127}}}
\DeclareRobustCommand{\decJS}[1]{\IfEqCase{#1}{{GW200208K}{0.122}{GW200208G}{0.034}{GW200129D}{0.052}{GW191219E}{0.002}{GW191109A}{0.078}}}
\DeclareRobustCommand{\chipJS}[1]{\IfEqCase{#1}{{GW200208K}{0.017}{GW200208G}{0.004}{GW200129D}{0.431}{GW191219E}{0.129}{GW191109A}{0.064}}}
\DeclareRobustCommand{\finalspinJS}[1]{\IfEqCase{#1}{{GW200208K}{0.173}{GW200208G}{0.009}{GW200129D}{0.146}{GW191219E}{0.090}{GW191109A}{0.076}}}
\DeclareRobustCommand{\massonesourceJS}[1]{\IfEqCase{#1}{{GW200208K}{0.245}{GW200208G}{0.002}{GW200129D}{0.245}{GW191219E}{0.078}{GW191109A}{0.007}}}
\DeclareRobustCommand{\psiJS}[1]{\IfEqCase{#1}{{GW200208K}{0.001}{GW200208G}{0.300}{GW200129D}{0.022}{GW191219E}{0.293}{GW191109A}{0.130}}}
\DeclareRobustCommand{\raJS}[1]{\IfEqCase{#1}{{GW200208K}{0.128}{GW200208G}{0.070}{GW200129D}{0.028}{GW191219E}{0.002}{GW191109A}{0.087}}}
\DeclareRobustCommand{\massonedetJS}[1]{\IfEqCase{#1}{{GW200208K}{0.228}{GW200208G}{0.001}{GW200129D}{0.277}{GW191219E}{0.050}{GW191109A}{0.048}}}
\DeclareRobustCommand{\phijlJS}[1]{\IfEqCase{#1}{{GW200208K}{0.003}{GW200208G}{0.042}{GW200129D}{0.634}{GW191219E}{0.015}{GW191109A}{0.013}}}
\DeclareRobustCommand{\chipinfinityonlyprecavgJS}[1]{\IfEqCase{#1}{{GW200208K}{0.018}{GW200208G}{0.004}{GW200129D}{0.430}{GW191219E}{0.129}{GW191109A}{0.049}}}
\DeclareRobustCommand{\spintwoxJS}[1]{\IfEqCase{#1}{{GW200208K}{0.003}{GW200208G}{0.000}{GW200129D}{0.039}{GW191219E}{0.005}{GW191109A}{0.004}}}
\DeclareRobustCommand{\tiltoneJS}[1]{\IfEqCase{#1}{{GW200208K}{0.030}{GW200208G}{0.006}{GW200129D}{0.129}{GW191219E}{0.073}{GW191109A}{0.091}}}
\DeclareRobustCommand{\phionetwoJS}[1]{\IfEqCase{#1}{{GW200208K}{0.001}{GW200208G}{0.001}{GW200129D}{0.060}{GW191219E}{0.000}{GW191109A}{0.034}}}
\DeclareRobustCommand{\chirpmassdetJS}[1]{\IfEqCase{#1}{{GW200208K}{0.166}{GW200208G}{0.009}{GW200129D}{0.075}{GW191219E}{0.019}{GW191109A}{0.060}}}
\DeclareRobustCommand{\spinoneyJS}[1]{\IfEqCase{#1}{{GW200208K}{0.012}{GW200208G}{0.004}{GW200129D}{0.170}{GW191219E}{0.047}{GW191109A}{0.036}}}
\DeclareRobustCommand{\costiltoneJS}[1]{\IfEqCase{#1}{{GW200208K}{0.034}{GW200208G}{0.006}{GW200129D}{0.118}{GW191219E}{0.067}{GW191109A}{0.087}}}
\DeclareRobustCommand{\peakluminosityJS}[1]{\IfEqCase{#1}{{GW200208K}{0.153}{GW200208G}{0.009}{GW200129D}{0.212}{GW191219E}{0.008}{GW191109A}{0.061}}}
\DeclareRobustCommand{\betaJS}[1]{\IfEqCase{#1}{{GW200208K}{0.097}{GW200208G}{0.012}{GW200129D}{0.432}{GW191219E}{0.195}{GW191109A}{0.111}}}
\DeclareRobustCommand{\finalmasssourceJS}[1]{\IfEqCase{#1}{{GW200208K}{0.243}{GW200208G}{0.003}{GW200129D}{0.009}{GW191219E}{0.079}{GW191109A}{0.023}}}
\DeclareRobustCommand{\totalmassdetJS}[1]{\IfEqCase{#1}{{GW200208K}{0.224}{GW200208G}{0.006}{GW200129D}{0.065}{GW191219E}{0.050}{GW191109A}{0.078}}}
\DeclareRobustCommand{\spintwoyJS}[1]{\IfEqCase{#1}{{GW200208K}{0.005}{GW200208G}{0.001}{GW200129D}{0.037}{GW191219E}{0.005}{GW191109A}{0.009}}}
\DeclareRobustCommand{\tiltoneinfinityonlyprecavgJS}[1]{\IfEqCase{#1}{{GW200208K}{0.030}{GW200208G}{0.005}{GW200129D}{0.110}{GW191219E}{0.073}{GW191109A}{0.094}}}
\DeclareRobustCommand{\masstwosourceJS}[1]{\IfEqCase{#1}{{GW200208K}{0.064}{GW200208G}{0.004}{GW200129D}{0.247}{GW191219E}{0.029}{GW191109A}{0.030}}}
\DeclareRobustCommand{\costiltoneinfinityonlyprecavgJS}[1]{\IfEqCase{#1}{{GW200208K}{0.033}{GW200208G}{0.004}{GW200129D}{0.102}{GW191219E}{0.067}{GW191109A}{0.094}}}
\DeclareRobustCommand{\chiptwospinJS}[1]{\IfEqCase{#1}{{GW200208K}{0.019}{GW200208G}{0.002}{GW200129D}{0.357}{GW191219E}{0.127}{GW191109A}{0.093}}}
\DeclareRobustCommand{\tilttwoinfinityonlyprecavgJS}[1]{\IfEqCase{#1}{{GW200208K}{0.008}{GW200208G}{0.001}{GW200129D}{0.081}{GW191219E}{0.002}{GW191109A}{0.040}}}
\DeclareRobustCommand{\spintwozinfinityonlyprecavgJS}[1]{\IfEqCase{#1}{{GW200208K}{0.010}{GW200208G}{0.001}{GW200129D}{0.063}{GW191219E}{0.002}{GW191109A}{0.045}}}
\DeclareRobustCommand{\chirpmasssourceJS}[1]{\IfEqCase{#1}{{GW200208K}{0.141}{GW200208G}{0.003}{GW200129D}{0.109}{GW191219E}{0.044}{GW191109A}{0.033}}}
\DeclareRobustCommand{\psiJJS}[1]{\IfEqCase{#1}{{GW200208K}{0.013}{GW200208G}{0.241}{GW200129D}{0.120}{GW191219E}{0.136}{GW191109A}{0.057}}}
\DeclareRobustCommand{\tilttwoJS}[1]{\IfEqCase{#1}{{GW200208K}{0.008}{GW200208G}{0.000}{GW200129D}{0.025}{GW191219E}{0.002}{GW191109A}{0.015}}}
\DeclareRobustCommand{\massratioJS}[1]{\IfEqCase{#1}{{GW200208K}{0.169}{GW200208G}{0.003}{GW200129D}{0.265}{GW191219E}{0.026}{GW191109A}{0.010}}}
\DeclareRobustCommand{\spintwozJS}[1]{\IfEqCase{#1}{{GW200208K}{0.011}{GW200208G}{0.001}{GW200129D}{0.033}{GW191219E}{0.002}{GW191109A}{0.050}}}
\DeclareRobustCommand{\comovingdistJS}[1]{\IfEqCase{#1}{{GW200208K}{0.010}{GW200208G}{0.005}{GW200129D}{0.056}{GW191219E}{0.047}{GW191109A}{0.088}}}
\DeclareRobustCommand{\iotaJS}[1]{\IfEqCase{#1}{{GW200208K}{0.003}{GW200208G}{0.008}{GW200129D}{0.198}{GW191219E}{0.026}{GW191109A}{0.153}}}
\DeclareRobustCommand{\chieffinfinityonlyprecavgJS}[1]{\IfEqCase{#1}{{GW200208K}{0.130}{GW200208G}{0.011}{GW200129D}{0.061}{GW191219E}{0.024}{GW191109A}{0.086}}}
\DeclareRobustCommand{\spinoneJS}[1]{\IfEqCase{#1}{{GW200208K}{0.165}{GW200208G}{0.007}{GW200129D}{0.388}{GW191219E}{0.101}{GW191109A}{0.033}}}
\DeclareRobustCommand{\totalmasssourceJS}[1]{\IfEqCase{#1}{{GW200208K}{0.241}{GW200208G}{0.003}{GW200129D}{0.006}{GW191219E}{0.079}{GW191109A}{0.028}}}
\DeclareRobustCommand{\phaseJS}[1]{\IfEqCase{#1}{{GW200208K}{0.013}{GW200208G}{0.034}{GW200129D}{0.035}{GW191219E}{0.015}{GW191109A}{0.378}}}
\DeclareRobustCommand{\redshiftJS}[1]{\IfEqCase{#1}{{GW200208K}{0.010}{GW200208G}{0.005}{GW200129D}{0.056}{GW191219E}{0.048}{GW191109A}{0.088}}}
\DeclareRobustCommand{\spinonexJS}[1]{\IfEqCase{#1}{{GW200208K}{0.009}{GW200208G}{0.005}{GW200129D}{0.235}{GW191219E}{0.061}{GW191109A}{0.055}}}
\DeclareRobustCommand{\symmetricmassratioJS}[1]{\IfEqCase{#1}{{GW200208K}{0.174}{GW200208G}{0.003}{GW200129D}{0.326}{GW191219E}{0.062}{GW191109A}{0.008}}}
\DeclareRobustCommand{\thetajnJS}[1]{\IfEqCase{#1}{{GW200208K}{0.004}{GW200208G}{0.008}{GW200129D}{0.070}{GW191219E}{0.115}{GW191109A}{0.125}}}
\DeclareRobustCommand{\luminositydistanceJS}[1]{\IfEqCase{#1}{{GW200208K}{0.010}{GW200208G}{0.005}{GW200129D}{0.056}{GW191219E}{0.049}{GW191109A}{0.087}}}
\DeclareRobustCommand{\costhetajnJS}[1]{\IfEqCase{#1}{{GW200208K}{0.002}{GW200208G}{0.008}{GW200129D}{0.072}{GW191219E}{0.106}{GW191109A}{0.113}}}
\DeclareRobustCommand{\viewingangleJS}[1]{\IfEqCase{#1}{{GW200208K}{0.003}{GW200208G}{0.003}{GW200129D}{0.070}{GW191219E}{0.163}{GW191109A}{0.098}}}
\DeclareRobustCommand{\invertedmassratioJS}[1]{\IfEqCase{#1}{{GW200208K}{0.192}{GW200208G}{0.003}{GW200129D}{0.271}{GW191219E}{0.058}{GW191109A}{0.009}}}
\DeclareRobustCommand{\costilttwoinfinityonlyprecavgJS}[1]{\IfEqCase{#1}{{GW200208K}{0.007}{GW200208G}{0.001}{GW200129D}{0.072}{GW191219E}{0.002}{GW191109A}{0.037}}}
\DeclareRobustCommand{\loglikelihoodJS}[1]{\IfEqCase{#1}{{GW200208K}{0.071}{GW200208G}{0.040}{GW200129D}{0.123}{GW191219E}{0.067}{GW191109A}{0.010}}}
\DeclareRobustCommand{\radiatedenergyJS}[1]{\IfEqCase{#1}{{GW200208K}{0.007}{GW200208G}{0.009}{GW200129D}{0.154}{GW191219E}{0.038}{GW191109A}{0.069}}}
\DeclareRobustCommand{\masstwodetJS}[1]{\IfEqCase{#1}{{GW200208K}{0.046}{GW200208G}{0.007}{GW200129D}{0.233}{GW191219E}{0.068}{GW191109A}{0.020}}}
\DeclareRobustCommand{\spinonezJS}[1]{\IfEqCase{#1}{{GW200208K}{0.105}{GW200208G}{0.009}{GW200129D}{0.008}{GW191219E}{0.024}{GW191109A}{0.061}}}
\DeclareRobustCommand{\spintwoJS}[1]{\IfEqCase{#1}{{GW200208K}{0.005}{GW200208G}{0.001}{GW200129D}{0.083}{GW191219E}{0.005}{GW191109A}{0.033}}}
\DeclareRobustCommand{\costilttwoJS}[1]{\IfEqCase{#1}{{GW200208K}{0.007}{GW200208G}{0.000}{GW200129D}{0.022}{GW191219E}{0.002}{GW191109A}{0.015}}}
\DeclareRobustCommand{\finalmassdetJS}[1]{\IfEqCase{#1}{{GW200208K}{0.223}{GW200208G}{0.005}{GW200129D}{0.096}{GW191219E}{0.050}{GW191109A}{0.079}}}
\DeclareRobustCommand{\phitwoKL}[1]{\IfEqCase{#1}{{GW200208K}{0.00}{GW200208G}{0.00}{GW200129D}{0.00}{GW191219E}{0.00}{GW191109A}{0.00}}}
\DeclareRobustCommand{\phioneKL}[1]{\IfEqCase{#1}{{GW200208K}{0.01}{GW200208G}{0.00}{GW200129D}{0.00}{GW191219E}{0.00}{GW191109A}{0.00}}}
\DeclareRobustCommand{\cosiotaKL}[1]{\IfEqCase{#1}{{GW200208K}{0.16}{GW200208G}{1.62}{GW200129D}{1.50}{GW191219E}{0.55}{GW191109A}{0.27}}}
\DeclareRobustCommand{\chieffKL}[1]{\IfEqCase{#1}{{GW200208K}{1.19}{GW200208G}{0.41}{GW200129D}{1.63}{GW191219E}{1.27}{GW191109A}{1.18}}}
\DeclareRobustCommand{\decKL}[1]{\IfEqCase{#1}{{GW200208K}{1.53}{GW200208G}{3.19}{GW200129D}{2.41}{GW191219E}{0.23}{GW191109A}{1.64}}}
\DeclareRobustCommand{\chipKL}[1]{\IfEqCase{#1}{{GW200208K}{0.07}{GW200208G}{0.06}{GW200129D}{0.23}{GW191219E}{2.16}{GW191109A}{0.80}}}
\DeclareRobustCommand{\massonesourceKL}[1]{\IfEqCase{#1}{{GW200208K}{1.40}{GW200208G}{1.88}{GW200129D}{2.97}{GW191219E}{3.03}{GW191109A}{2.09}}}
\DeclareRobustCommand{\psiKL}[1]{\IfEqCase{#1}{{GW200208K}{0.01}{GW200208G}{38.22}{GW200129D}{0.42}{GW191219E}{43.92}{GW191109A}{0.01}}}
\DeclareRobustCommand{\raKL}[1]{\IfEqCase{#1}{{GW200208K}{2.21}{GW200208G}{4.75}{GW200129D}{5.17}{GW191219E}{1.04}{GW191109A}{1.14}}}
\DeclareRobustCommand{\massonedetKL}[1]{\IfEqCase{#1}{{GW200208K}{2.77}{GW200208G}{2.50}{GW200129D}{4.23}{GW191219E}{3.17}{GW191109A}{3.27}}}
\DeclareRobustCommand{\phijlKL}[1]{\IfEqCase{#1}{{GW200208K}{0.00}{GW200208G}{0.06}{GW200129D}{0.56}{GW191219E}{0.06}{GW191109A}{0.02}}}
\DeclareRobustCommand{\spintwoxKL}[1]{\IfEqCase{#1}{{GW200208K}{0.01}{GW200208G}{0.01}{GW200129D}{0.06}{GW191219E}{inf}{GW191109A}{0.02}}}
\DeclareRobustCommand{\phionetwoKL}[1]{\IfEqCase{#1}{{GW200208K}{0.00}{GW200208G}{0.00}{GW200129D}{0.06}{GW191219E}{0.00}{GW191109A}{0.14}}}
\DeclareRobustCommand{\tiltoneKL}[1]{\IfEqCase{#1}{{GW200208K}{0.86}{GW200208G}{0.06}{GW200129D}{0.03}{GW191219E}{0.03}{GW191109A}{0.67}}}
\DeclareRobustCommand{\chirpmassdetKL}[1]{\IfEqCase{#1}{{GW200208K}{3.36}{GW200208G}{1.21}{GW200129D}{3.08}{GW191219E}{2.55}{GW191109A}{1.74}}}
\DeclareRobustCommand{\spinoneyKL}[1]{\IfEqCase{#1}{{GW200208K}{0.02}{GW200208G}{0.06}{GW200129D}{0.12}{GW191219E}{1.41}{GW191109A}{0.32}}}
\DeclareRobustCommand{\costiltoneKL}[1]{\IfEqCase{#1}{{GW200208K}{0.84}{GW200208G}{0.05}{GW200129D}{0.03}{GW191219E}{0.02}{GW191109A}{0.66}}}
\DeclareRobustCommand{\totalmassdetKL}[1]{\IfEqCase{#1}{{GW200208K}{3.43}{GW200208G}{2.41}{GW200129D}{4.12}{GW191219E}{3.16}{GW191109A}{2.89}}}
\DeclareRobustCommand{\spintwoyKL}[1]{\IfEqCase{#1}{{GW200208K}{0.02}{GW200208G}{0.01}{GW200129D}{0.05}{GW191219E}{inf}{GW191109A}{0.02}}}
\DeclareRobustCommand{\masstwosourceKL}[1]{\IfEqCase{#1}{{GW200208K}{0.56}{GW200208G}{3.93}{GW200129D}{7.87}{GW191219E}{3.07}{GW191109A}{3.43}}}
\DeclareRobustCommand{\chirpmasssourceKL}[1]{\IfEqCase{#1}{{GW200208K}{2.10}{GW200208G}{2.65}{GW200129D}{4.90}{GW191219E}{1.31}{GW191109A}{2.07}}}
\DeclareRobustCommand{\tilttwoKL}[1]{\IfEqCase{#1}{{GW200208K}{0.08}{GW200208G}{0.03}{GW200129D}{0.31}{GW191219E}{0.00}{GW191109A}{0.17}}}
\DeclareRobustCommand{\massratioKL}[1]{\IfEqCase{#1}{{GW200208K}{0.25}{GW200208G}{1.86}{GW200129D}{3.35}{GW191219E}{4.05}{GW191109A}{1.98}}}
\DeclareRobustCommand{\comovingdistKL}[1]{\IfEqCase{#1}{{GW200208K}{4.45}{GW200208G}{4.71}{GW200129D}{9.08}{GW191219E}{inf}{GW191109A}{6.30}}}
\DeclareRobustCommand{\spintwozKL}[1]{\IfEqCase{#1}{{GW200208K}{0.10}{GW200208G}{0.08}{GW200129D}{0.36}{GW191219E}{inf}{GW191109A}{0.22}}}
\DeclareRobustCommand{\iotaKL}[1]{\IfEqCase{#1}{{GW200208K}{0.20}{GW200208G}{1.55}{GW200129D}{1.46}{GW191219E}{0.63}{GW191109A}{0.30}}}
\DeclareRobustCommand{\spinoneKL}[1]{\IfEqCase{#1}{{GW200208K}{0.12}{GW200208G}{0.15}{GW200129D}{0.34}{GW191219E}{2.36}{GW191109A}{0.89}}}
\DeclareRobustCommand{\totalmasssourceKL}[1]{\IfEqCase{#1}{{GW200208K}{1.86}{GW200208G}{1.94}{GW200129D}{3.11}{GW191219E}{3.03}{GW191109A}{1.83}}}
\DeclareRobustCommand{\phaseKL}[1]{\IfEqCase{#1}{{GW200208K}{0.00}{GW200208G}{0.00}{GW200129D}{0.00}{GW191219E}{0.00}{GW191109A}{0.12}}}
\DeclareRobustCommand{\redshiftKL}[1]{\IfEqCase{#1}{{GW200208K}{4.40}{GW200208G}{4.70}{GW200129D}{9.02}{GW191219E}{inf}{GW191109A}{6.22}}}
\DeclareRobustCommand{\spinonexKL}[1]{\IfEqCase{#1}{{GW200208K}{0.02}{GW200208G}{0.07}{GW200129D}{0.11}{GW191219E}{1.38}{GW191109A}{0.30}}}
\DeclareRobustCommand{\symmetricmassratioKL}[1]{\IfEqCase{#1}{{GW200208K}{0.34}{GW200208G}{2.54}{GW200129D}{5.03}{GW191219E}{3.28}{GW191109A}{2.63}}}
\DeclareRobustCommand{\thetajnKL}[1]{\IfEqCase{#1}{{GW200208K}{0.19}{GW200208G}{1.48}{GW200129D}{1.59}{GW191219E}{0.34}{GW191109A}{0.19}}}
\DeclareRobustCommand{\luminositydistanceKL}[1]{\IfEqCase{#1}{{GW200208K}{4.39}{GW200208G}{4.71}{GW200129D}{8.98}{GW191219E}{inf}{GW191109A}{6.19}}}
\DeclareRobustCommand{\costhetajnKL}[1]{\IfEqCase{#1}{{GW200208K}{0.14}{GW200208G}{1.56}{GW200129D}{1.65}{GW191219E}{0.31}{GW191109A}{0.18}}}
\DeclareRobustCommand{\invertedmassratioKL}[1]{\IfEqCase{#1}{{GW200208K}{0.45}{GW200208G}{2.66}{GW200129D}{4.14}{GW191219E}{3.09}{GW191109A}{2.80}}}
\DeclareRobustCommand{\masstwodetKL}[1]{\IfEqCase{#1}{{GW200208K}{1.74}{GW200208G}{1.76}{GW200129D}{3.43}{GW191219E}{3.31}{GW191109A}{1.69}}}
\DeclareRobustCommand{\spinonezKL}[1]{\IfEqCase{#1}{{GW200208K}{1.09}{GW200208G}{0.16}{GW200129D}{0.20}{GW191219E}{1.46}{GW191109A}{0.89}}}
\DeclareRobustCommand{\spintwoKL}[1]{\IfEqCase{#1}{{GW200208K}{0.02}{GW200208G}{0.03}{GW200129D}{0.14}{GW191219E}{inf}{GW191109A}{0.02}}}
\DeclareRobustCommand{\costilttwoKL}[1]{\IfEqCase{#1}{{GW200208K}{0.07}{GW200208G}{0.03}{GW200129D}{0.31}{GW191219E}{0.00}{GW191109A}{0.16}}}
\DeclareRobustCommand{\equalmassSD}[1]{\IfEqCase{#1}{{GW200208K}{1.849}{GW200208G}{4.083}{GW200129D}{11.568}{GW191219E}{0.000}{GW191109A}{2.680}}}
\DeclareRobustCommand{\nonspinningSD}[1]{\IfEqCase{#1}{{GW200208K}{0.276}{GW200208G}{1.660}{GW200129D}{1.025}{GW191219E}{0.261}{GW191109A}{0.000}}}
\newcommand{\phitwoSEOBuncert}[1]{\ensuremath{ \phitwoSEOBmed{#1}_{-\phitwoSEOBminus{#1}}^{+\phitwoSEOBplus{#1}}  } } 
\newcommand{\phioneSEOBuncert}[1]{\ensuremath{ \phioneSEOBmed{#1}_{-\phioneSEOBminus{#1}}^{+\phioneSEOBplus{#1}}  } } 
\newcommand{\geocenttimeSEOBuncert}[1]{\ensuremath{ \geocenttimeSEOBmed{#1}_{-\geocenttimeSEOBminus{#1}}^{+\geocenttimeSEOBplus{#1}}  } } 
\newcommand{\chieffSEOBuncert}[1]{\ensuremath{ \chieffSEOBmed{#1}_{-\chieffSEOBminus{#1}}^{+\chieffSEOBplus{#1}}  } } 
\newcommand{\decSEOBuncert}[1]{\ensuremath{ \decSEOBmed{#1}_{-\decSEOBminus{#1}}^{+\decSEOBplus{#1}}  } } 
\newcommand{\chipSEOBuncert}[1]{\ensuremath{ \chipSEOBmed{#1}_{-\chipSEOBminus{#1}}^{+\chipSEOBplus{#1}}  } } 
\newcommand{\finalspinSEOBuncert}[1]{\ensuremath{ \finalspinSEOBmed{#1}_{-\finalspinSEOBminus{#1}}^{+\finalspinSEOBplus{#1}}  } } 
\newcommand{\massonesourceSEOBuncert}[1]{\ensuremath{ \massonesourceSEOBmed{#1}_{-\massonesourceSEOBminus{#1}}^{+\massonesourceSEOBplus{#1}}  } } 
\newcommand{\cosiotaSEOBuncert}[1]{\ensuremath{ \cosiotaSEOBmed{#1}_{-\cosiotaSEOBminus{#1}}^{+\cosiotaSEOBplus{#1}}  } } 
\newcommand{\psiSEOBuncert}[1]{\ensuremath{ \psiSEOBmed{#1}_{-\psiSEOBminus{#1}}^{+\psiSEOBplus{#1}}  } } 
\newcommand{\raSEOBuncert}[1]{\ensuremath{ \raSEOBmed{#1}_{-\raSEOBminus{#1}}^{+\raSEOBplus{#1}}  } } 
\newcommand{\massonedetSEOBuncert}[1]{\ensuremath{ \massonedetSEOBmed{#1}_{-\massonedetSEOBminus{#1}}^{+\massonedetSEOBplus{#1}}  } } 
\newcommand{\phijlSEOBuncert}[1]{\ensuremath{ \phijlSEOBmed{#1}_{-\phijlSEOBminus{#1}}^{+\phijlSEOBplus{#1}}  } } 
\newcommand{\chipinfinityonlyprecavgSEOBuncert}[1]{\ensuremath{ \chipinfinityonlyprecavgSEOBmed{#1}_{-\chipinfinityonlyprecavgSEOBminus{#1}}^{+\chipinfinityonlyprecavgSEOBplus{#1}}  } } 
\newcommand{\spintwoxSEOBuncert}[1]{\ensuremath{ \spintwoxSEOBmed{#1}_{-\spintwoxSEOBminus{#1}}^{+\spintwoxSEOBplus{#1}}  } } 
\newcommand{\chirpmassdetSEOBuncert}[1]{\ensuremath{ \chirpmassdetSEOBmed{#1}_{-\chirpmassdetSEOBminus{#1}}^{+\chirpmassdetSEOBplus{#1}}  } } 
\newcommand{\tiltoneSEOBuncert}[1]{\ensuremath{ \tiltoneSEOBmed{#1}_{-\tiltoneSEOBminus{#1}}^{+\tiltoneSEOBplus{#1}}  } } 
\newcommand{\phionetwoSEOBuncert}[1]{\ensuremath{ \phionetwoSEOBmed{#1}_{-\phionetwoSEOBminus{#1}}^{+\phionetwoSEOBplus{#1}}  } } 
\newcommand{\spinoneySEOBuncert}[1]{\ensuremath{ \spinoneySEOBmed{#1}_{-\spinoneySEOBminus{#1}}^{+\spinoneySEOBplus{#1}}  } } 
\newcommand{\costiltoneSEOBuncert}[1]{\ensuremath{ \costiltoneSEOBmed{#1}_{-\costiltoneSEOBminus{#1}}^{+\costiltoneSEOBplus{#1}}  } } 
\newcommand{\finalmasssourceSEOBuncert}[1]{\ensuremath{ \finalmasssourceSEOBmed{#1}_{-\finalmasssourceSEOBminus{#1}}^{+\finalmasssourceSEOBplus{#1}}  } } 
\newcommand{\totalmassdetSEOBuncert}[1]{\ensuremath{ \totalmassdetSEOBmed{#1}_{-\totalmassdetSEOBminus{#1}}^{+\totalmassdetSEOBplus{#1}}  } } 
\newcommand{\spintwoySEOBuncert}[1]{\ensuremath{ \spintwoySEOBmed{#1}_{-\spintwoySEOBminus{#1}}^{+\spintwoySEOBplus{#1}}  } } 
\newcommand{\masstwosourceSEOBuncert}[1]{\ensuremath{ \masstwosourceSEOBmed{#1}_{-\masstwosourceSEOBminus{#1}}^{+\masstwosourceSEOBplus{#1}}  } } 
\newcommand{\chirpmasssourceSEOBuncert}[1]{\ensuremath{ \chirpmasssourceSEOBmed{#1}_{-\chirpmasssourceSEOBminus{#1}}^{+\chirpmasssourceSEOBplus{#1}}  } } 
\newcommand{\tilttwoSEOBuncert}[1]{\ensuremath{ \tilttwoSEOBmed{#1}_{-\tilttwoSEOBminus{#1}}^{+\tilttwoSEOBplus{#1}}  } } 
\newcommand{\massratioSEOBuncert}[1]{\ensuremath{ \massratioSEOBmed{#1}_{-\massratioSEOBminus{#1}}^{+\massratioSEOBplus{#1}}  } } 
\newcommand{\spintwozSEOBuncert}[1]{\ensuremath{ \spintwozSEOBmed{#1}_{-\spintwozSEOBminus{#1}}^{+\spintwozSEOBplus{#1}}  } } 
\newcommand{\comovingdistSEOBuncert}[1]{\ensuremath{ \comovingdistSEOBmed{#1}_{-\comovingdistSEOBminus{#1}}^{+\comovingdistSEOBplus{#1}}  } } 
\newcommand{\iotaSEOBuncert}[1]{\ensuremath{ \iotaSEOBmed{#1}_{-\iotaSEOBminus{#1}}^{+\iotaSEOBplus{#1}}  } } 
\newcommand{\chieffinfinityonlyprecavgSEOBuncert}[1]{\ensuremath{ \chieffinfinityonlyprecavgSEOBmed{#1}_{-\chieffinfinityonlyprecavgSEOBminus{#1}}^{+\chieffinfinityonlyprecavgSEOBplus{#1}}  } } 
\newcommand{\spinoneSEOBuncert}[1]{\ensuremath{ \spinoneSEOBmed{#1}_{-\spinoneSEOBminus{#1}}^{+\spinoneSEOBplus{#1}}  } } 
\newcommand{\totalmasssourceSEOBuncert}[1]{\ensuremath{ \totalmasssourceSEOBmed{#1}_{-\totalmasssourceSEOBminus{#1}}^{+\totalmasssourceSEOBplus{#1}}  } } 
\newcommand{\phaseSEOBuncert}[1]{\ensuremath{ \phaseSEOBmed{#1}_{-\phaseSEOBminus{#1}}^{+\phaseSEOBplus{#1}}  } } 
\newcommand{\redshiftSEOBuncert}[1]{\ensuremath{ \redshiftSEOBmed{#1}_{-\redshiftSEOBminus{#1}}^{+\redshiftSEOBplus{#1}}  } } 
\newcommand{\spinonexSEOBuncert}[1]{\ensuremath{ \spinonexSEOBmed{#1}_{-\spinonexSEOBminus{#1}}^{+\spinonexSEOBplus{#1}}  } } 
\newcommand{\symmetricmassratioSEOBuncert}[1]{\ensuremath{ \symmetricmassratioSEOBmed{#1}_{-\symmetricmassratioSEOBminus{#1}}^{+\symmetricmassratioSEOBplus{#1}}  } } 
\newcommand{\thetajnSEOBuncert}[1]{\ensuremath{ \thetajnSEOBmed{#1}_{-\thetajnSEOBminus{#1}}^{+\thetajnSEOBplus{#1}}  } } 
\newcommand{\luminositydistanceSEOBuncert}[1]{\ensuremath{ \luminositydistanceSEOBmed{#1}_{-\luminositydistanceSEOBminus{#1}}^{+\luminositydistanceSEOBplus{#1}}  } } 
\newcommand{\costhetajnSEOBuncert}[1]{\ensuremath{ \costhetajnSEOBmed{#1}_{-\costhetajnSEOBminus{#1}}^{+\costhetajnSEOBplus{#1}}  } } 
\newcommand{\loglikelihoodSEOBuncert}[1]{\ensuremath{ \loglikelihoodSEOBmed{#1}_{-\loglikelihoodSEOBminus{#1}}^{+\loglikelihoodSEOBplus{#1}}  } } 
\newcommand{\radiatedenergySEOBuncert}[1]{\ensuremath{ \radiatedenergySEOBmed{#1}_{-\radiatedenergySEOBminus{#1}}^{+\radiatedenergySEOBplus{#1}}  } } 
\newcommand{\masstwodetSEOBuncert}[1]{\ensuremath{ \masstwodetSEOBmed{#1}_{-\masstwodetSEOBminus{#1}}^{+\masstwodetSEOBplus{#1}}  } } 
\newcommand{\spinonezSEOBuncert}[1]{\ensuremath{ \spinonezSEOBmed{#1}_{-\spinonezSEOBminus{#1}}^{+\spinonezSEOBplus{#1}}  } } 
\newcommand{\spintwoSEOBuncert}[1]{\ensuremath{ \spintwoSEOBmed{#1}_{-\spintwoSEOBminus{#1}}^{+\spintwoSEOBplus{#1}}  } } 
\newcommand{\costilttwoSEOBuncert}[1]{\ensuremath{ \costilttwoSEOBmed{#1}_{-\costilttwoSEOBminus{#1}}^{+\costilttwoSEOBplus{#1}}  } } 
\newcommand{\finalmassdetSEOBuncert}[1]{\ensuremath{ \finalmassdetSEOBmed{#1}_{-\finalmassdetSEOBminus{#1}}^{+\finalmassdetSEOBplus{#1}}  } }

    \newcommand{\luminositydistanceleast}{200105F}

    \newcommand{\luminositydistanceleastpercent}{50}

    \newcommand{\luminositydistancemost}{GW200308G}

    \newcommand{\luminositydistancemostpercent}{45}

    \newcommand{\luminositydistanceleastsecond}{GW200115A}

    \newcommand{\luminositydistanceleastpercentsecond}{30}

    \newcommand{\luminositydistancemostsecond}{GW200220E}

    \newcommand{\luminositydistancemostpercentsecond}{22}

    \newcommand{\massonesourceleast}{GW200115A}

    \newcommand{\massonesourceleastpercent}{96}

    \newcommand{\massonesourcemost}{GW200220E}

    \newcommand{\massonesourcemostpercent}{44}

    \newcommand{\massonesourceleastsecond}{200105F}

    \newcommand{\massonesourceleastpercentsecond}{ 3}

    \newcommand{\massonesourcemostsecond}{GW200308G}

    \newcommand{\massonesourcemostpercentsecond}{23}

    \newcommand{\masstwosourceleast}{GW191219E}

    \newcommand{\masstwosourceleastpercent}{93}

    \newcommand{\masstwosourcemost}{GW200220E}

    \newcommand{\masstwosourcemostpercent}{74}

    \newcommand{\masstwosourceleastsecond}{GW200115A}

    \newcommand{\masstwosourceleastpercentsecond}{ 7}

    \newcommand{\masstwosourcemostsecond}{GW191109A}

    \newcommand{\masstwosourcemostpercentsecond}{17}

    \newcommand{\totalmasssourceleast}{GW200115A}

    \newcommand{\totalmasssourceleastpercent}{99}

    \newcommand{\totalmasssourcemost}{GW200220E}

    \newcommand{\totalmasssourcemostpercent}{67}

    \newcommand{\totalmasssourceleastsecond}{200105F}

    \newcommand{\totalmasssourceleastpercentsecond}{ 1}

    \newcommand{\totalmasssourcemostsecond}{GW200308G}

    \newcommand{\totalmasssourcemostpercentsecond}{18}

    \newcommand{\chirpmasssourceleast}{GW200115A}

    \newcommand{\chirpmasssourceleastpercent}{100}

    \newcommand{\chirpmasssourcemost}{GW200220E}

    \newcommand{\chirpmasssourcemostpercent}{80}

    \newcommand{\chirpmasssourceleastsecond}{GW200322G}

    \newcommand{\chirpmasssourceleastpercentsecond}{ 0}

    \newcommand{\chirpmasssourcemostsecond}{GW200308G}

    \newcommand{\chirpmasssourcemostpercentsecond}{12}

    \newcommand{\massratioleast}{GW191219E}

    \newcommand{\massratioleastpercent}{97}

    \newcommand{\massratiomost}{GW200129D}

    \newcommand{\massratiomostpercent}{ 6}

    \newcommand{\massratioleastsecond}{GW200322G}

    \newcommand{\massratioleastpercentsecond}{ 3}

    \newcommand{\massratiomostsecond}{GW200311L}

    \newcommand{\massratiomostpercentsecond}{ 6}

    \newcommand{\chieffleast}{GW191109A}

    \newcommand{\chieffleastpercent}{32}

    \newcommand{\chieffmost}{GW200208K}

    \newcommand{\chieffmostpercent}{37}

    \newcommand{\chieffleastsecond}{GW200115A}

    \newcommand{\chieffleastpercentsecond}{17}

    \newcommand{\chieffmostsecond}{GW200322G}

    \newcommand{\chieffmostpercentsecond}{22}

    \newcommand{\chieffinfinityonlyprecavgleast}{GW191109A}

    \newcommand{\chieffinfinityonlyprecavgleastpercent}{32}

    \newcommand{\chieffinfinityonlyprecavgmost}{GW200208K}

    \newcommand{\chieffinfinityonlyprecavgmostpercent}{37}

    \newcommand{\chieffinfinityonlyprecavgleastsecond}{GW200115A}

    \newcommand{\chieffinfinityonlyprecavgleastpercentsecond}{17}

    \newcommand{\chieffinfinityonlyprecavgmostsecond}{GW200322G}

    \newcommand{\chieffinfinityonlyprecavgmostpercentsecond}{22}

    \newcommand{\chipleast}{200105F}

    \newcommand{\chipleastpercent}{23}

    \newcommand{\chipmost}{GW200129D}

    \newcommand{\chipmostpercent}{ 9}

    \newcommand{\chipleastsecond}{GW191219E}

    \newcommand{\chipleastpercentsecond}{22}

    \newcommand{\chipmostsecond}{GW191127B}

    \newcommand{\chipmostpercentsecond}{ 8}

    \newcommand{\chipinfinityonlyprecavgleast}{200105F}

    \newcommand{\chipinfinityonlyprecavgleastpercent}{23}

    \newcommand{\chipinfinityonlyprecavgmost}{GW200129D}

    \newcommand{\chipinfinityonlyprecavgmostpercent}{ 9}

    \newcommand{\chipinfinityonlyprecavgleastsecond}{GW191219E}

    \newcommand{\chipinfinityonlyprecavgleastpercentsecond}{22}

    \newcommand{\chipinfinityonlyprecavgmostsecond}{GW191127B}

    \newcommand{\chipinfinityonlyprecavgmostpercentsecond}{ 8}

    \newcommand{\finalspinleast}{GW191219E}

    \newcommand{\finalspinleastpercent}{98}

    \newcommand{\finalspinmost}{GW200208K}

    \newcommand{\finalspinmostpercent}{36}

    \newcommand{\finalspinleastsecond}{GW200322G}

    \newcommand{\finalspinleastpercentsecond}{ 1}

    \newcommand{\finalspinmostsecond}{GW200322G}

    \newcommand{\finalspinmostpercentsecond}{21}

    \newcommand{\luminositydistanceleastBBH}{GW191216G}

    \newcommand{\luminositydistanceleastpercentBBH}{68}

    \newcommand{\luminositydistancemostBBH}{GW200308G}

    \newcommand{\luminositydistancemostpercentBBH}{45}

    \newcommand{\luminositydistanceleastsecondBBH}{GW200202F}

    \newcommand{\luminositydistanceleastpercentsecondBBH}{26}

    \newcommand{\luminositydistancemostsecondBBH}{GW200220E}

    \newcommand{\luminositydistancemostpercentsecondBBH}{22}

    \newcommand{\massonesourceleastBBH}{GW200202F}

    \newcommand{\massonesourceleastpercentBBH}{35}

    \newcommand{\massonesourcemostBBH}{GW200220E}

    \newcommand{\massonesourcemostpercentBBH}{44}

    \newcommand{\massonesourceleastsecondBBH}{GW191129G}

    \newcommand{\massonesourceleastpercentsecondBBH}{27}

    \newcommand{\massonesourcemostsecondBBH}{GW200308G}

    \newcommand{\massonesourcemostpercentsecondBBH}{23}

    \newcommand{\masstwosourceleastBBH}{GW191113B}

    \newcommand{\masstwosourceleastpercentBBH}{40}

    \newcommand{\masstwosourcemostBBH}{GW200220E}

    \newcommand{\masstwosourcemostpercentBBH}{73}

    \newcommand{\masstwosourceleastsecondBBH}{GW191129G}

    \newcommand{\masstwosourceleastpercentsecondBBH}{16}

    \newcommand{\masstwosourcemostsecondBBH}{GW191109A}

    \newcommand{\masstwosourcemostpercentsecondBBH}{17}

    \newcommand{\totalmasssourceleastBBH}{GW191129G}

    \newcommand{\totalmasssourceleastpercentBBH}{49}

    \newcommand{\totalmasssourcemostBBH}{GW200220E}

    \newcommand{\totalmasssourcemostpercentBBH}{67}

    \newcommand{\totalmasssourceleastsecondBBH}{GW200202F}

    \newcommand{\totalmasssourceleastpercentsecondBBH}{40}

    \newcommand{\totalmasssourcemostsecondBBH}{GW200308G}

    \newcommand{\totalmasssourcemostpercentsecondBBH}{18}

    \newcommand{\chirpmasssourceleastBBH}{GW191129G}

    \newcommand{\chirpmasssourceleastpercentBBH}{70}

    \newcommand{\chirpmasssourcemostBBH}{GW200220E}

    \newcommand{\chirpmasssourcemostpercentBBH}{80}

    \newcommand{\chirpmasssourceleastsecondBBH}{GW200202F}

    \newcommand{\chirpmasssourceleastpercentsecondBBH}{24}

    \newcommand{\chirpmasssourcemostsecondBBH}{GW200308G}

    \newcommand{\chirpmasssourcemostpercentsecondBBH}{12}

    \newcommand{\massratioleastBBH}{GW200208K}

    \newcommand{\massratioleastpercentBBH}{31}

    \newcommand{\massratiomostBBH}{GW200129D}

    \newcommand{\massratiomostpercentBBH}{ 6}

    \newcommand{\massratioleastsecondBBH}{GW191113B}

    \newcommand{\massratioleastpercentsecondBBH}{22}

    \newcommand{\massratiomostsecondBBH}{GW200311L}

    \newcommand{\massratiomostpercentsecondBBH}{ 6}

    \newcommand{\chieffleastBBH}{GW191109A}

    \newcommand{\chieffleastpercentBBH}{38}

    \newcommand{\chieffmostBBH}{GW200208K}

    \newcommand{\chieffmostpercentBBH}{37}

    \newcommand{\chieffleastsecondBBH}{GW200209E}

    \newcommand{\chieffleastpercentsecondBBH}{ 9}

    \newcommand{\chieffmostsecondBBH}{GW200322G}

    \newcommand{\chieffmostpercentsecondBBH}{22}

    \newcommand{\chieffinfinityonlyprecavgleastBBH}{GW191109A}

    \newcommand{\chieffinfinityonlyprecavgleastpercentBBH}{38}

    \newcommand{\chieffinfinityonlyprecavgmostBBH}{GW200208K}

    \newcommand{\chieffinfinityonlyprecavgmostpercentBBH}{37}

    \newcommand{\chieffinfinityonlyprecavgleastsecondBBH}{GW200209E}

    \newcommand{\chieffinfinityonlyprecavgleastpercentsecondBBH}{ 9}

    \newcommand{\chieffinfinityonlyprecavgmostsecondBBH}{GW200322G}

    \newcommand{\chieffinfinityonlyprecavgmostpercentsecondBBH}{22}

    \newcommand{\chipleastBBH}{GW191113B}

    \newcommand{\chipleastpercentBBH}{15}

    \newcommand{\chipmostBBH}{GW200129D}

    \newcommand{\chipmostpercentBBH}{ 9}

    \newcommand{\chipleastsecondBBH}{GW191105C}

    \newcommand{\chipleastpercentsecondBBH}{ 7}

    \newcommand{\chipmostsecondBBH}{GW191127B}

    \newcommand{\chipmostpercentsecondBBH}{ 8}

    \newcommand{\chipinfinityonlyprecavgleastBBH}{GW191113B}

    \newcommand{\chipinfinityonlyprecavgleastpercentBBH}{15}

    \newcommand{\chipinfinityonlyprecavgmostBBH}{GW200129D}

    \newcommand{\chipinfinityonlyprecavgmostpercentBBH}{ 9}

    \newcommand{\chipinfinityonlyprecavgleastsecondBBH}{GW191105C}

    \newcommand{\chipinfinityonlyprecavgleastpercentsecondBBH}{ 7}

    \newcommand{\chipinfinityonlyprecavgmostsecondBBH}{GW191127B}

    \newcommand{\chipinfinityonlyprecavgmostpercentsecondBBH}{ 8}

    \newcommand{\finalspinleastBBH}{GW191113B}

    \newcommand{\finalspinleastpercentBBH}{45}

    \newcommand{\finalspinmostBBH}{GW200208K}

    \newcommand{\finalspinmostpercentBBH}{36}

    \newcommand{\finalspinleastsecondBBH}{GW191109A}

    \newcommand{\finalspinleastpercentsecondBBH}{ 9}

    \newcommand{\finalspinmostsecondBBH}{GW200322G}

    \newcommand{\finalspinmostpercentsecondBBH}{21}

\DeclareRobustCommand{\ANYCBCVT}[1]{\IfEqCase{#1}{{TENFIVE}{\ensuremath{ 0.13^{ +0.02 }_{ -0.01 } }}{ONEPOINTFIVEONEPOINTFIVE}{\ensuremath{ 3.9^{ +0.1 }_{ -0.2 } \times 10^{-3}}}{FIVEFIVE}{\ensuremath{ 7.4^{ +0.5 }_{ -0.5 } \times 10^{-2}}}{TWENTYTEN}{\ensuremath{ 0.77^{ +0.06 }_{ -0.06 } }}{THIRTYFIVETWENTY}{\ensuremath{ 3.1^{ +0.1 }_{ -0.2 } }}{THIRTYFIVEONEPOINTFIVE}{\ensuremath{ 3.3^{ +0.4 }_{ -0.3 } \times 10^{-2}}}{THIRTYFIVETHIRTYFIVE}{\ensuremath{ 5.3^{ +0.1 }_{ -0.2 } }}{FIVEONEPOINTFIVE}{\ensuremath{ 1.43^{ +0.06 }_{ -0.06 } \times 10^{-2}}}{TWENTYONEPOINTFIVE}{\ensuremath{ 2.9^{ +0.3 }_{ -0.2 } \times 10^{-2}}}{TENTEN}{\ensuremath{ 0.32^{ +0.02 }_{ -0.01 } }}{TWENTYTWENTY}{\ensuremath{ 1.71^{ +0.06 }_{ -0.07 } }}{TENONEPOINTFIVE}{\ensuremath{ 2.1^{ +0.1 }_{ -0.1 } \times 10^{-2}}}}}
\DeclareRobustCommand{\PYCBCALLSKYVT}[1]{\IfEqCase{#1}{{TENFIVE}{\ensuremath{ 0.12^{ +0.01 }_{ -0.02 } }}{ONEPOINTFIVEONEPOINTFIVE}{\ensuremath{ 3.5^{ +0.1 }_{ -0.2 } \times 10^{-3}}}{FIVEFIVE}{\ensuremath{ 6.5^{ +0.5 }_{ -0.4 } \times 10^{-2}}}{TWENTYTEN}{\ensuremath{ 0.56^{ +0.05 }_{ -0.05 } }}{THIRTYFIVETWENTY}{\ensuremath{ 1.9^{ +0.1 }_{ -0.1 } }}{THIRTYFIVEONEPOINTFIVE}{\ensuremath{ 3.1^{ +0.3 }_{ -0.3 } \times 10^{-2}}}{THIRTYFIVETHIRTYFIVE}{\ensuremath{ 3.3^{ +0.1 }_{ -0.1 } }}{FIVEONEPOINTFIVE}{\ensuremath{ 1.21^{ +0.06 }_{ -0.06 } \times 10^{-2}}}{TWENTYONEPOINTFIVE}{\ensuremath{ 2.7^{ +0.2 }_{ -0.2 } \times 10^{-2}}}{TENTEN}{\ensuremath{ 0.27^{ +0.01 }_{ -0.02 } }}{TWENTYTWENTY}{\ensuremath{ 1.14^{ +0.05 }_{ -0.05 } }}{TENONEPOINTFIVE}{\ensuremath{ 1.8^{ +0.1 }_{ -0.1 } \times 10^{-2}}}}}
\DeclareRobustCommand{\PYCBCHIGHMASSVT}[1]{\IfEqCase{#1}{{TENFIVE}{\ensuremath{ 0.11^{ +0.02 }_{ -0.01 } }}{ONEPOINTFIVEONEPOINTFIVE}{\ensuremath{\text{--}}}{FIVEFIVE}{\ensuremath{ 5.0^{ +0.5 }_{ -0.4 } \times 10^{-2}}}{TWENTYTEN}{\ensuremath{ 0.65^{ +0.05 }_{ -0.05 } }}{THIRTYFIVETWENTY}{\ensuremath{ 2.5^{ +0.1 }_{ -0.1 } }}{THIRTYFIVEONEPOINTFIVE}{\ensuremath{\text{--}}}{THIRTYFIVETHIRTYFIVE}{\ensuremath{ 4.3^{ +0.2 }_{ -0.1 } }}{FIVEONEPOINTFIVE}{\ensuremath{\text{--}}}{TWENTYONEPOINTFIVE}{\ensuremath{\text{--}}}{TENTEN}{\ensuremath{ 0.28^{ +0.02 }_{ -0.02 } }}{TWENTYTWENTY}{\ensuremath{ 1.42^{ +0.06 }_{ -0.05 } }}{TENONEPOINTFIVE}{\ensuremath{\text{--}}}}}
\DeclareRobustCommand{\GSTLALALLSKYVT}[1]{\IfEqCase{#1}{{TENFIVE}{\ensuremath{ 0.10^{ +0.02 }_{ -0.01 } }}{ONEPOINTFIVEONEPOINTFIVE}{\ensuremath{ 2.7^{ +0.1 }_{ -0.1 } \times 10^{-3}}}{FIVEFIVE}{\ensuremath{ 5.8^{ +0.5 }_{ -0.4 } \times 10^{-2}}}{TWENTYTEN}{\ensuremath{ 0.60^{ +0.05 }_{ -0.05 } }}{THIRTYFIVETWENTY}{\ensuremath{ 2.3^{ +0.2 }_{ -0.1 } }}{THIRTYFIVEONEPOINTFIVE}{\ensuremath{ 1.8^{ +0.2 }_{ -0.3 } \times 10^{-2}}}{THIRTYFIVETHIRTYFIVE}{\ensuremath{ 4.1^{ +0.1 }_{ -0.1 } }}{FIVEONEPOINTFIVE}{\ensuremath{ 1.12^{ +0.05 }_{ -0.06 } \times 10^{-2}}}{TWENTYONEPOINTFIVE}{\ensuremath{ 1.9^{ +0.2 }_{ -0.2 } \times 10^{-2}}}{TENTEN}{\ensuremath{ 0.26^{ +0.01 }_{ -0.02 } }}{TWENTYTWENTY}{\ensuremath{ 1.34^{ +0.06 }_{ -0.05 } }}{TENONEPOINTFIVE}{\ensuremath{ 1.6^{ +0.1 }_{ -0.1 } \times 10^{-2}}}}}
\DeclareRobustCommand{\MBTAALLSKYVT}[1]{\IfEqCase{#1}{{TENFIVE}{\ensuremath{ 0.10^{ +0.02 }_{ -0.01 } }}{ONEPOINTFIVEONEPOINTFIVE}{\ensuremath{ 3.4^{ +0.1 }_{ -0.1 } \times 10^{-3}}}{FIVEFIVE}{\ensuremath{ 4.5^{ +0.4 }_{ -0.4 } \times 10^{-2}}}{TWENTYTEN}{\ensuremath{ 0.51^{ +0.05 }_{ -0.04 } }}{THIRTYFIVETWENTY}{\ensuremath{ 1.8^{ +0.1 }_{ -0.1 } }}{THIRTYFIVEONEPOINTFIVE}{\ensuremath{ 1.9^{ +0.3 }_{ -0.3 } \times 10^{-2}}}{THIRTYFIVETHIRTYFIVE}{\ensuremath{ 3.3^{ +0.2 }_{ -0.1 } }}{FIVEONEPOINTFIVE}{\ensuremath{ 1.19^{ +0.06 }_{ -0.05 } \times 10^{-2}}}{TWENTYONEPOINTFIVE}{\ensuremath{ 1.9^{ +0.2 }_{ -0.2 } \times 10^{-2}}}{TENTEN}{\ensuremath{ 0.26^{ +0.01 }_{ -0.02 } }}{TWENTYTWENTY}{\ensuremath{ 1.10^{ +0.05 }_{ -0.05 } }}{TENONEPOINTFIVE}{\ensuremath{ 1.5^{ +0.2 }_{ -0.1 } \times 10^{-2}}}}}
\DeclareRobustCommand{\CWBALLSKYVT}[1]{\IfEqCase{#1}{{TENFIVE}{\ensuremath{ 1.3^{ +0.5 }_{ -0.4 } \times 10^{-2}}}{ONEPOINTFIVEONEPOINTFIVE}{\ensuremath{\text{--}}}{FIVEFIVE}{\ensuremath{ 5^{ +1 }_{ -2 } \times 10^{-3}}}{TWENTYTEN}{\ensuremath{ 0.24^{ +0.03 }_{ -0.04 } }}{THIRTYFIVETWENTY}{\ensuremath{ 1.35^{ +0.09 }_{ -0.10 } }}{THIRTYFIVEONEPOINTFIVE}{\ensuremath{\text{--}}}{THIRTYFIVETHIRTYFIVE}{\ensuremath{ 2.6^{ +0.1 }_{ -0.1 } }}{FIVEONEPOINTFIVE}{\ensuremath{\text{--}}}{TWENTYONEPOINTFIVE}{\ensuremath{\text{--}}}{TENTEN}{\ensuremath{ 6.8^{ +0.8 }_{ -0.9 } \times 10^{-2}}}{TWENTYTWENTY}{\ensuremath{ 0.56^{ +0.04 }_{ -0.04 } }}{TENONEPOINTFIVE}{\ensuremath{\text{--}}}}}
\DeclareRobustCommand{\CWBASSOCVT}[1]{\IfEqCase{#1}{{TENFIVE}{\ensuremath{ 1.3^{ +0.5 }_{ -0.4 } \times 10^{-2}}}{ONEPOINTFIVEONEPOINTFIVE}{\ensuremath{\text{--}}}{FIVEFIVE}{\ensuremath{ 5^{ +1 }_{ -2 } \times 10^{-3}}}{TWENTYTEN}{\ensuremath{ 0.23^{ +0.03 }_{ -0.03 } }}{THIRTYFIVETWENTY}{\ensuremath{ 1.29^{ +0.10 }_{ -0.09 } }}{THIRTYFIVEONEPOINTFIVE}{\ensuremath{\text{--}}}{THIRTYFIVETHIRTYFIVE}{\ensuremath{ 2.5^{ +0.1 }_{ -0.1 } }}{FIVEONEPOINTFIVE}{\ensuremath{\text{--}}}{TWENTYONEPOINTFIVE}{\ensuremath{\text{--}}}{TENTEN}{\ensuremath{ 6.7^{ +0.9 }_{ -0.8 } \times 10^{-2}}}{TWENTYTWENTY}{\ensuremath{ 0.55^{ +0.03 }_{ -0.04 } }}{TENONEPOINTFIVE}{\ensuremath{\text{--}}}}}
\DeclareRobustCommand{\VTBINS}[1]{\IfEqCase{#1}{{TENFIVE}{\ensuremath{ 10.0 + 5.0}}{ONEPOINTFIVEONEPOINTFIVE}{\ensuremath{ 1.5 + 1.5}}{FIVEFIVE}{\ensuremath{ 5.0 + 5.0}}{TWENTYTEN}{\ensuremath{ 20.0 + 10.0}}{THIRTYFIVETWENTY}{\ensuremath{ 35.0 + 20.0}}{THIRTYFIVEONEPOINTFIVE}{\ensuremath{ 35.0 + 1.5}}{THIRTYFIVETHIRTYFIVE}{\ensuremath{ 35.0 + 35.0}}{FIVEONEPOINTFIVE}{\ensuremath{ 5.0 + 1.5}}{TWENTYONEPOINTFIVE}{\ensuremath{ 20.0 + 1.5}}{TENTEN}{\ensuremath{ 10.0 + 10.0}}{TWENTYTWENTY}{\ensuremath{ 20.0 + 20.0}}{TENONEPOINTFIVE}{\ensuremath{ 10.0 + 1.5}}}}
\DeclareRobustCommand{\ANYCBCALLOTHREEVT}[1]{\IfEqCase{#1}{{TENFIVE}{\ensuremath{ 0.31^{ +0.02 }_{ -0.02 } }}{ONEPOINTFIVEONEPOINTFIVE}{\ensuremath{ 8.2^{ +0.2 }_{ -0.2 } \times 10^{-3}}}{FIVEFIVE}{\ensuremath{ 0.158^{ +0.008 }_{ -0.007 } }}{TWENTYTEN}{\ensuremath{ 1.56^{ +0.08 }_{ -0.08 } }}{THIRTYFIVETWENTY}{\ensuremath{ 6.4^{ +0.2 }_{ -0.2 } }}{THIRTYFIVEONEPOINTFIVE}{\ensuremath{ 6.6^{ +0.5 }_{ -0.5 } \times 10^{-2}}}{THIRTYFIVETHIRTYFIVE}{\ensuremath{ 11.2^{ +0.2 }_{ -0.2 } }}{FIVEONEPOINTFIVE}{\ensuremath{ 2.96^{ +0.09 }_{ -0.09 } \times 10^{-2}}}{TWENTYONEPOINTFIVE}{\ensuremath{ 6.0^{ +0.3 }_{ -0.4 } \times 10^{-2}}}{TENTEN}{\ensuremath{ 0.72^{ +0.02 }_{ -0.03 } }}{TWENTYTWENTY}{\ensuremath{ 3.57^{ +0.09 }_{ -0.09 } }}{TENONEPOINTFIVE}{\ensuremath{ 4.5^{ +0.1 }_{ -0.2 } \times 10^{-2}}}}}
\DeclareRobustCommand{\PYCBCALLSKYALLOTHREEVT}[1]{\IfEqCase{#1}{{TENFIVE}{\ensuremath{ 0.27^{ +0.02 }_{ -0.03 } }}{ONEPOINTFIVEONEPOINTFIVE}{\ensuremath{ 7.3^{ +0.2 }_{ -0.2 } \times 10^{-3}}}{FIVEFIVE}{\ensuremath{ 0.138^{ +0.007 }_{ -0.007 } }}{TWENTYTEN}{\ensuremath{ 1.14^{ +0.07 }_{ -0.07 } }}{THIRTYFIVETWENTY}{\ensuremath{ 3.9^{ +0.2 }_{ -0.1 } }}{THIRTYFIVEONEPOINTFIVE}{\ensuremath{ 6.2^{ +0.4 }_{ -0.5 } \times 10^{-2}}}{THIRTYFIVETHIRTYFIVE}{\ensuremath{ 6.9^{ +0.1 }_{ -0.2 } }}{FIVEONEPOINTFIVE}{\ensuremath{ 2.47^{ +0.08 }_{ -0.08 } \times 10^{-2}}}{TWENTYONEPOINTFIVE}{\ensuremath{ 5.4^{ +0.3 }_{ -0.3 } \times 10^{-2}}}{TENTEN}{\ensuremath{ 0.56^{ +0.03 }_{ -0.02 } }}{TWENTYTWENTY}{\ensuremath{ 2.38^{ +0.07 }_{ -0.08 } }}{TENONEPOINTFIVE}{\ensuremath{ 3.8^{ +0.1 }_{ -0.2 } \times 10^{-2}}}}}
\DeclareRobustCommand{\PYCBCHIGHMASSALLOTHREEVT}[1]{\IfEqCase{#1}{{TENFIVE}{\ensuremath{ 0.26^{ +0.02 }_{ -0.02 } }}{ONEPOINTFIVEONEPOINTFIVE}{\ensuremath{\text{--}}}{FIVEFIVE}{\ensuremath{ 0.108^{ +0.006 }_{ -0.006 } }}{TWENTYTEN}{\ensuremath{ 1.32^{ +0.07 }_{ -0.08 } }}{THIRTYFIVETWENTY}{\ensuremath{ 5.3^{ +0.2 }_{ -0.2 } }}{THIRTYFIVEONEPOINTFIVE}{\ensuremath{\text{--}}}{THIRTYFIVETHIRTYFIVE}{\ensuremath{ 9.2^{ +0.2 }_{ -0.2 } }}{FIVEONEPOINTFIVE}{\ensuremath{\text{--}}}{TWENTYONEPOINTFIVE}{\ensuremath{\text{--}}}{TENTEN}{\ensuremath{ 0.59^{ +0.03 }_{ -0.02 } }}{TWENTYTWENTY}{\ensuremath{ 2.99^{ +0.09 }_{ -0.08 } }}{TENONEPOINTFIVE}{\ensuremath{\text{--}}}}}
\DeclareRobustCommand{\GSTLALALLSKYALLOTHREEVT}[1]{\IfEqCase{#1}{{TENFIVE}{\ensuremath{ 0.25^{ +0.02 }_{ -0.03 } }}{ONEPOINTFIVEONEPOINTFIVE}{\ensuremath{ 5.8^{ +0.2 }_{ -0.1 } \times 10^{-3}}}{FIVEFIVE}{\ensuremath{ 0.129^{ +0.007 }_{ -0.006 } }}{TWENTYTEN}{\ensuremath{ 1.25^{ +0.07 }_{ -0.07 } }}{THIRTYFIVETWENTY}{\ensuremath{ 4.9^{ +0.2 }_{ -0.2 } }}{THIRTYFIVEONEPOINTFIVE}{\ensuremath{ 3.8^{ +0.3 }_{ -0.4 } \times 10^{-2}}}{THIRTYFIVETHIRTYFIVE}{\ensuremath{ 8.8^{ +0.2 }_{ -0.2 } }}{FIVEONEPOINTFIVE}{\ensuremath{ 2.34^{ +0.08 }_{ -0.08 } \times 10^{-2}}}{TWENTYONEPOINTFIVE}{\ensuremath{ 3.9^{ +0.2 }_{ -0.3 } \times 10^{-2}}}{TENTEN}{\ensuremath{ 0.59^{ +0.02 }_{ -0.03 } }}{TWENTYTWENTY}{\ensuremath{ 2.82^{ +0.08 }_{ -0.08 } }}{TENONEPOINTFIVE}{\ensuremath{ 3.5^{ +0.1 }_{ -0.2 } \times 10^{-2}}}}}
\DeclareRobustCommand{\MBTAALLSKYALLOTHREEVT}[1]{\IfEqCase{#1}{{TENFIVE}{\ensuremath{ 0.23^{ +0.02 }_{ -0.02 } }}{ONEPOINTFIVEONEPOINTFIVE}{\ensuremath{ 7.0^{ +0.2 }_{ -0.2 } \times 10^{-3}}}{FIVEFIVE}{\ensuremath{ 9.8^{ +0.6 }_{ -0.6 } \times 10^{-2}}}{TWENTYTEN}{\ensuremath{ 1.10^{ +0.06 }_{ -0.07 } }}{THIRTYFIVETWENTY}{\ensuremath{ 3.9^{ +0.2 }_{ -0.1 } }}{THIRTYFIVEONEPOINTFIVE}{\ensuremath{ 3.7^{ +0.4 }_{ -0.4 } \times 10^{-2}}}{THIRTYFIVETHIRTYFIVE}{\ensuremath{ 7.4^{ +0.2 }_{ -0.2 } }}{FIVEONEPOINTFIVE}{\ensuremath{ 2.41^{ +0.08 }_{ -0.08 } \times 10^{-2}}}{TWENTYONEPOINTFIVE}{\ensuremath{ 3.4^{ +0.3 }_{ -0.2 } \times 10^{-2}}}{TENTEN}{\ensuremath{ 0.53^{ +0.02 }_{ -0.02 } }}{TWENTYTWENTY}{\ensuremath{ 2.41^{ +0.07 }_{ -0.08 } }}{TENONEPOINTFIVE}{\ensuremath{ 3.3^{ +0.2 }_{ -0.1 } \times 10^{-2}}}}}
\DeclareRobustCommand{\CWBALLSKYALLOTHREEVT}[1]{\IfEqCase{#1}{{TENFIVE}{\ensuremath{ 3.6^{ +0.9 }_{ -0.8 } \times 10^{-2}}}{ONEPOINTFIVEONEPOINTFIVE}{\ensuremath{\text{--}}}{FIVEFIVE}{\ensuremath{ 1.1^{ +0.2 }_{ -0.2 } \times 10^{-2}}}{TWENTYTEN}{\ensuremath{ 0.48^{ +0.05 }_{ -0.05 } }}{THIRTYFIVETWENTY}{\ensuremath{ 2.7^{ +0.1 }_{ -0.2 } }}{THIRTYFIVEONEPOINTFIVE}{\ensuremath{\text{--}}}{THIRTYFIVETHIRTYFIVE}{\ensuremath{ 5.5^{ +0.1 }_{ -0.2 } }}{FIVEONEPOINTFIVE}{\ensuremath{\text{--}}}{TWENTYONEPOINTFIVE}{\ensuremath{\text{--}}}{TENTEN}{\ensuremath{ 0.15^{ +0.01 }_{ -0.01 } }}{TWENTYTWENTY}{\ensuremath{ 1.19^{ +0.05 }_{ -0.05 } }}{TENONEPOINTFIVE}{\ensuremath{\text{--}}}}}
\DeclareRobustCommand{\CWBASSOCALLOTHREEVT}[1]{\IfEqCase{#1}{{TENFIVE}{\ensuremath{ 3.6^{ +0.9 }_{ -0.9 } \times 10^{-2}}}{ONEPOINTFIVEONEPOINTFIVE}{\ensuremath{\text{--}}}{FIVEFIVE}{\ensuremath{ 1.1^{ +0.2 }_{ -0.2 } \times 10^{-2}}}{TWENTYTEN}{\ensuremath{ 0.47^{ +0.05 }_{ -0.05 } }}{THIRTYFIVETWENTY}{\ensuremath{ 2.6^{ +0.1 }_{ -0.1 } }}{THIRTYFIVEONEPOINTFIVE}{\ensuremath{\text{--}}}{THIRTYFIVETHIRTYFIVE}{\ensuremath{ 5.3^{ +0.2 }_{ -0.1 } }}{FIVEONEPOINTFIVE}{\ensuremath{\text{--}}}{TWENTYONEPOINTFIVE}{\ensuremath{\text{--}}}{TENTEN}{\ensuremath{ 0.15^{ +0.01 }_{ -0.01 } }}{TWENTYTWENTY}{\ensuremath{ 1.17^{ +0.06 }_{ -0.05 } }}{TENONEPOINTFIVE}{\ensuremath{\text{--}}}}}

\DeclareRobustCommand{\skyarea}[1]{\IfEqCase{#1}{{GW200322G}{18000}{GW200316I}{190}{GW200311L}{35}{GW200308G}{12000}{GW200306A}{4600}{GW200302A}{6000}{GW200225B}{370}{GW200224H}{51}{GW200220H}{3200}{GW200220E}{3000}{GW200219D}{700}{GW200216G}{2900}{GW200210B}{1800}{GW200209E}{730}{GW200208K}{1900}{GW200208G}{48}{GW200202F}{160}{GW200129D}{54}{GW200128C}{2600}{GW200115A}{720}{GW200112H}{4300}{200105F}{9600}{GW191230H}{1100}{GW191222A}{2000}{GW191219E}{1500}{GW191216G}{910}{GW191215G}{530}{GW191204G}{310}{GW191204A}{3400}{GW191129G}{850}{GW191127B}{980}{GW191126C}{1400}{GW191113B}{3600}{GW191109A}{1600}{GW191105C}{640}{GW191103A}{2500}}}
\DeclareRobustCommand{\skyvol}[1]{\IfEqCase{#1}{{GW200322G}{\textcolor{red}{-100.0}}{GW200316I}{0.036}{GW200311L}{0.0058}{GW200308G}{\textcolor{red}{-100.0}}{GW200306A}{5.0}{GW200302A}{2.8}{GW200225B}{0.081}{GW200224H}{0.024}{GW200220H}{11}{GW200220E}{25}{GW200219D}{1.8}{GW200216G}{11}{GW200210B}{0.23}{GW200209E}{2.2}{GW200208K}{12}{GW200208G}{0.048}{GW200202F}{0.0023}{GW200129D}{0.0044}{GW200128C}{6.9}{GW200115A}{0.0063}{GW200112H}{0.8}{200105F}{0.042}{GW191230H}{4.4}{GW191222A}{3.4}{GW191219E}{0.074}{GW191216G}{0.0074}{GW191215G}{0.43}{GW191204G}{0.0094}{GW191204A}{3.4}{GW191129G}{0.059}{GW191127B}{3.6}{GW191126C}{0.6}{GW191113B}{2.3}{GW191109A}{0.49}{GW191105C}{0.13}{GW191103A}{0.32}}}
\newcommand{\minareaevent}{GW200311L}
\newcommand{\maxareaevent}{200105F}
\newcommand{\minvolevent}{GW200202F}
\newcommand{\maxvolevent}{GW200220E}

\DeclareRobustCommand{\skyareawithcut}[1]{\IfEqCase{#1}{{GW200322G}{6500}{GW200316I}{190}{GW200311L}{35}{GW200308G}{2000}{GW200306A}{4600}{GW200302A}{6000}{GW200225B}{370}{GW200224H}{50}{GW200220H}{3200}{GW200220E}{3000}{GW200219D}{700}{GW200216G}{2900}{GW200210B}{1800}{GW200209E}{730}{GW200208K}{2000}{GW200208G}{30}{GW200202F}{170}{GW200129D}{130}{GW200128C}{2600}{GW200115A}{370}{GW200112H}{4300}{200105F}{7900}{GW191230H}{1100}{GW191222A}{2000}{GW191219E}{1500}{GW191216G}{490}{GW191215G}{530}{GW191204G}{350}{GW191204A}{3700}{GW191129G}{850}{GW191127B}{980}{GW191126C}{1400}{GW191113B}{3600}{GW191109A}{1600}{GW191105C}{640}{GW191103A}{2500}}}
\DeclareRobustCommand{\skyvolwithcut}[1]{\IfEqCase{#1}{{GW200322G}{\textcolor{red}{-100.0}}{GW200316I}{0.036}{GW200311L}{0.0058}{GW200308G}{11}{GW200306A}{5.0}{GW200302A}{2.8}{GW200225B}{0.081}{GW200224H}{0.023}{GW200220H}{11}{GW200220E}{25}{GW200219D}{1.8}{GW200216G}{11}{GW200210B}{0.23}{GW200209E}{2.2}{GW200208K}{12}{GW200208G}{0.031}{GW200202F}{0.0024}{GW200129D}{0.011}{GW200128C}{6.9}{GW200115A}{0.0024}{GW200112H}{0.8}{200105F}{0.036}{GW191230H}{4.4}{GW191222A}{3.4}{GW191219E}{0.087}{GW191216G}{0.0037}{GW191215G}{0.43}{GW191204G}{0.012}{GW191204A}{3.8}{GW191129G}{0.059}{GW191127B}{3.6}{GW191126C}{0.6}{GW191113B}{2.3}{GW191109A}{0.49}{GW191105C}{0.13}{GW191103A}{0.32}}}
\newcommand{\minareaeventwithcut}{GW200208G}
\newcommand{\maxareaeventwithcut}{200105F}
\newcommand{\minvoleventwithcut}{GW200202F}
\newcommand{\maxvoleventwithcut}{GW200220E}

\newcommand{\NgalBestLocTwentyZeroOneFifteenbandK}{\ensuremath{5400}} 
\newcommand{\CompMinBestLocTwentyZeroOneFifteenbandK}{\ensuremath{73\%}} 
\newcommand{\CompMaxBestLocTwentyZeroOneFifteenbandK}{\ensuremath{19\%}} 
\newcommand{\NgalBestLocTwentyZeroOneFifteenbandbJ}{\ensuremath{19100}} 
\newcommand{\CompMinBestLocTwentyZeroOneFifteenbandbJ}{\ensuremath{100\%}} 
\newcommand{\CompMaxBestLocTwentyZeroOneFifteenbandbJ}{\ensuremath{85\%}} 

\newcommand{\NgalBestLocTwentyZeroTwoZeroTwobandK}{\ensuremath{1500}} 
\newcommand{\CompMinBestLocTwentyZeroTwoZeroTwobandK}{\ensuremath{59\%}} 
\newcommand{\CompMaxBestLocTwentyZeroTwoZeroTwobandK}{\ensuremath{7\%}} 
\newcommand{\NgalBestLocTwentyZeroTwoZeroTwobandbJ}{\ensuremath{10400}} 
\newcommand{\CompMinBestLocTwentyZeroTwoZeroTwobandbJ}{\ensuremath{66\%}} 
\newcommand{\CompMaxBestLocTwentyZeroTwoZeroTwobandbJ}{\ensuremath{13\%}}

\title{GWTC-3: Compact Binary Coalescences Observed by LIGO and Virgo
during the Second Part of the Third Observing Run}

\author{R.~Abbott}
\affiliation{LIGO Laboratory, California Institute of Technology, Pasadena, CA 91125, USA}
\author{T.~D.~Abbott}
\affiliation{Louisiana State University, Baton Rouge, LA 70803, USA}
\author{F.~Acernese}
\affiliation{Dipartimento di Farmacia, Universit\`a di Salerno, I-84084 Fisciano, Salerno, Italy}
\affiliation{INFN, Sezione di Napoli, Complesso Universitario di Monte S. Angelo, I-80126 Napoli, Italy}
\author{K.~Ackley}
\affiliation{OzGrav, School of Physics \& Astronomy, Monash University, Clayton 3800, Victoria, Australia}
\author{C.~Adams}
\affiliation{LIGO Livingston Observatory, Livingston, LA 70754, USA}
\author{N.~Adhikari}
\affiliation{University of Wisconsin-Milwaukee, Milwaukee, WI 53201, USA}
\author{R.~X.~Adhikari}
\affiliation{LIGO Laboratory, California Institute of Technology, Pasadena, CA 91125, USA}
\author{V.~B.~Adya}
\affiliation{OzGrav, Australian National University, Canberra, Australian Capital Territory 0200, Australia}
\author{C.~Affeldt}
\affiliation{Max Planck Institute for Gravitational Physics (Albert Einstein Institute), D-30167 Hannover, Germany}
\affiliation{Leibniz Universit\"at Hannover, D-30167 Hannover, Germany}
\author{D.~Agarwal}
\affiliation{Inter-University Centre for Astronomy and Astrophysics, Pune 411007, India}
% ...

\collaboration{The LIGO Scientific Collaboration, the Virgo Collaboration, and the KAGRA Collaboration}

%\author{The LIGO Scientific Collaboration}
%\author{The Virgo Collaboration}
%\author{The KAGRA Scientific Collaboration}

%\date[\relax]{Compiled: \today, revision: \input{gitID.txt}}
\date{\today}

\begin{abstract}
The third Gravitational-Wave Transient Catalog (GWTC-3) describes signals detected with Advanced LIGO and Advanced Virgo up to the end of their third observing run. 
Updating the previous GWTC-2.1, we present candidate gravitational waves from compact binary coalescences during the second half of the third observing run (O3b) between \RUNSTART{}~UTC and \RUNEND{}~UTC.  
There are \NUMEVENTS{} compact binary coalescence candidates identified by at least one of our search algorithms with a probability of astrophysical origin $\pastro{} > \PASTROTHRESHOLD$. 
Of these, \PREVIOUSLYREPORTED{} were previously reported as low-latency public alerts, and \NEWEVENTS{} are reported here for the first time. 
Based upon estimates for the component masses, our O3b candidates with $\pastro{} > \PASTROTHRESHOLD$ are consistent with gravitational-wave signals from binary black holes or neutron star--black hole binaries, and we identify none from binary neutron stars. 
However, from the gravitational-wave data alone, we are not able to measure matter effects that distinguish whether the binary components are neutron stars or black holes. 
The range of inferred component masses is similar to that found with previous catalogs, but the O3b candidates include the first confident observations of neutron star--black hole binaries. 
Including the \NUMEVENTS{} candidates from O3b in addition to those from GWTC-2.1, GWTC-3 contains \TOTALEVENTS{} candidates found by our analysis with $\pastro{} > \PASTROTHRESHOLD$ across the first three observing runs. 
These observations of compact binary coalescences present an unprecedented view of the properties of black holes and neutron stars.

\end{abstract}


\maketitle

% Capitalization
\newcommand\hmm[1]{\ifnum\ifhmode\spacefactor\else2000\fi>1500 \uppercase{#1}\else#1\fi}

% Stylized pipeline names
\newcommand{\GSTLAL}{GstLAL\xspace}
\newcommand{\CWB}{cWB\xspace}
\newcommand{\PYCBC}{PyCBC\xspace}
\newcommand{\MBTA}{MBTA\xspace}
\newcommand{\SPIIR}{SPIIR\xspace}
\newcommand{\MBTAONLINE}{MBTAOnline\xspace}
\newcommand{\LALSUITE}{LALSuite\xspace}
\newcommand{\LALINFERENCE}{LALInference\xspace}
\newcommand{\IDQ}{iDQ\xspace}
\newcommand{\BAYESWAVE}{BayesWave\xspace}
\newcommand{\BAYESTAR}{Bayestar\xspace}
\newcommand{\RIFT}{RIFT\xspace}
\newcommand{\PESUMMARY}{PESummary\xspace}
\newcommand{\GWCELERY}{GWCelery\xspace}

% Stylised code names
\newcommand{\BILBY}{{Bilby}\xspace}
\newcommand{\DYNESTY}{{Dynesty}\xspace}
\newcommand{\BILBYPIPE}{{BilbyPipe}\xspace}
\newcommand{\PBILBY}{{Parallel \BILBY{}}\xspace}
\newcommand{\ASIMOV}{{Asimov}\xspace}
\newcommand{\GWSUBTRACT}{{gwsubtract}\xspace}

\newcommand{\DMT}{{DMT}\xspace}
\newcommand{\DQR}{{DQR}\xspace}
\newcommand{\DQSEGDB}{{DQSEGDB}\xspace}
\newcommand{\GWDETCHAR}{{gwdetchar}\xspace}
\newcommand{\HVETO}{{hveto}\xspace}
\newcommand{\PYTHONVIRGOTOOLS}{{PythonVirgoTools}\xspace}
\newcommand{\OMICRONSCAN}{{Omicron}\xspace}

\newcommand{\HDF}{{HDF5}\xspace}
\newcommand{\PYTHON}{{Python}\xspace}
\newcommand{\NUMPY}{{NumPy}\xspace}
\newcommand{\SCIPY}{{SciPy}\xspace}
\newcommand{\PLT}{{Matplotlib}\xspace}
\newcommand{\SEABORN}{{seaborn}\xspace}
\newcommand{\GWPY}{{GWpy}\xspace}

% Waveform families
\newcommand{\STTFOUR}{\texttt{SpinTaylorT4}}
\newcommand{\TFTWO}{\texttt{TaylorF2}}
\newcommand{\Phenom}{\texttt{Phenom}}
\newcommand{\IMRPhenomC}{\texttt{IMRPhenomC}}
\newcommand{\IMRPhenomD}{\texttt{IMRPhenomD}}
\newcommand{\IMRPhenomXHM}{\texttt{IMRPhenomXHM}}
\newcommand{\IMRPhenomXPHM}{\texttt{IMRPhenomXPHM}}
\newcommand{\SEOBNR}{\texttt{SEOBNR}}
\newcommand{\SEOBNRFOUR}{\texttt{SEOBNRv4}}
\newcommand{\SEOBNRP}{\texttt{SEOBNRv4P}}
\newcommand{\SEOBNRHM}{\texttt{SEOBNRv4HM}}
\newcommand{\SEOBNRPHM}{\texttt{SEOBNRv4PHM}}
\newcommand{\IMRPhenomPNRTidal}{\texttt{IMRPhenomPv2\_NRTidalv2}}
\newcommand{\IMRPhenomNSBH}{\texttt{IMRPhenomNSBH}}
\newcommand{\SEOBNRNSBH}{\texttt{SEOBNRv4\_ROM\_NRTidalv2\_NSBH}}
\newcommand{\SEOBNRROM}{\texttt{SEOBNRv4\_ROM}}
\newcommand{\NRSUR}{\texttt{NRSur7dq4}}

% Stylised Category names
\newcommand{\CATONE}{Category~1\xspace}
\newcommand{\CATTWO}{Category~2\xspace}
\newcommand{\CATTHREE}{Category~3\xspace}

% Data-quality channel and flag names
\newcommand{\LINSBUTRACTCHANNEL}{\reviewed{L1:LSC-POP\_A\_RF9\_I\_ERR\_DQ}\xspace}
\newcommand{\FLAGONETWOFOUR}{\reviewed{Flag~1.24 ($45~\mathrm{MHz}$ Sideband Fluctuations)}\xspace}

% Glitch names
\newcommand{\FASTSCATTER}{\hmm{f}ast scattering\xspace}
\newcommand{\SLOWSCATTER}{\hmm{s}low scattering\xspace}
\newcommand{\BLIP}{\hmm{b}lip\xspace}
\newcommand{\TOMTE}{\hmm{t}omte\xspace}

% Catalog names
\newcommand{\GWTCONE}{GWTC-1\xspace}
\newcommand{\GWTCTWO}{GWTC-2\xspace}
\newcommand{\GWTCTWOFINAL}{GWTC-2.1\xspace}
\newcommand{\GWTCTHREE}{GWTC-3\xspace}

% Galaxy catalog
\newcommand{\GLADEplus}{\texttt{GLADE+}}

% Population model names
\newcommand{\PLPEAK}{\textsc{Power Law + Peak}\xspace}

% ======================
%  ACRONYMS
% ======================
\acrodef{LSC}[LSC]{LIGO Scientific Collaboration}
\acrodef{LVC}[LVC]{LIGO Scientific and Virgo Collaboration}
\acrodef{LVK}[LVK]{LIGO Scientific, Virgo and KAGRA}
\acrodef{aLIGO}{Advanced Laser Interferometer Gravitational-Wave Observatory}
\acrodef{aVirgo}{Advanced Virgo}
\acrodef{LIGO}[LIGO]{Laser Interferometer Gravitational-Wave Observatory}
%\acrodef{IFO}[IFO]{interferometer}
%\acrodef{LHO}[LHO]{LIGO-Hanford}
%\acrodef{LLO}[LLO]{LIGO-Livingston}
\acrodef{O4}[O4]{fourth observing run}
\acrodef{O3}[O3]{third observing run}
\acrodef{O3a}[O3a]{first part of \ac{O3}}
\acrodef{O3b}[O3b]{second part of \ac{O3}}
\acrodef{O2}[O2]{second observing run}
\acrodef{O1}[O1]{first observing run}

\acrodef{GPS}[GPS]{Global Positioning System}
\acrodef{UTC}[UTC]{Coordinated Universal Time}

\acrodef{BH}[BH]{black hole}
\acrodef{BBH}[BBH]{binary black hole}
\acrodef{BNS}[BNS]{binary neutron star}
\acrodef{NS}[NS]{neutron star}
\acrodef{NSBH}[NSBH]{neutron star--black hole binary}
\acrodefplural{NSBH}[NSBHs]{neutron star--black hole binaries}
\acrodef{IMBH}{intermediate-mass black hole}
\acrodef{CBC}[CBC]{compact binary coalescence}
\acrodef{GW}[GW]{gravitational wave}

\acrodef{HL}[HL]{Hanford--Livingston}
\acrodef{HV}[HV]{Hanford--Virgo}
\acrodef{LV}[LV]{Livingston--Virgo}
\acrodef{HLV}[HLV]{Hanford--Livingston--Virgo}

\acrodef{CWB}[\CWB{}]{coherent WaveBurst}
\acrodef{MBTA}[\MBTA{}]{Multi-Band Template Analysis}
\acrodef{SPIIR}[\SPIIR{}]{Summed Parallel Infinite Impulse Response}

\acrodef{SNR}[SNR]{signal-to-noise ratio}
\acrodef{FAR}[FAR]{false alarm rate}
\acrodef{PSD}[PSD]{power spectral density}
\acrodefplural{PSD}[PSDs]{power spectral densities}

\acrodef{GR}[GR]{general relativity}
\acrodef{NR}[NR]{numerical relativity}
\acrodef{PN}[PN]{post-Newtonian}
\acrodef{EOB}[EOB]{effective-one-body}
\acrodef{ROM}[ROM]{reduced-order model}
\acrodef{IMR}[IMR]{inspiral--merger--ringdown}

\acrodef{PDF}[PDF]{probability density function}
\acrodef{PE}[PE]{parameter estimation}
\acrodef{CL}[CL]{credible level}

\acrodef{EOS}[EOS]{equation of state}

\acrodef{LAL}[LAL]{LIGO Algorithm Library}
\acrodef{GPU}[GPU]{graphics processing unit}

\acrodef{KLD}[KLD]{Kullback--Leibler divergence}
\acrodef{JSD}[JSD]{Jensen--Shannon divergence}

\acrodef{GWOSC}[GWOSC]{Gravitational Wave Open Science Center}
\acrodef{GCN}[GCN]{Gamma-ray Coordinate Network}
\acrodef{GraceDB}[GraceDB]{Gravitational Candidate Event Database}

\acrodef{GLADEplus}[\GLADEplus{}]{extended version of the Galaxy List for the Advanced Detector Era}

\section{Introduction}
\label{sec:introduction}

The Advanced \ac{LIGO}~\cite{TheLIGOScientific:2014jea} and Advanced Virgo~\cite{TheVirgo:2014hva} detectors have revealed the Universe's abundance of \ac{GW} sources. 
Here, we present the third \ac{LVK} Collaboration Gravitational-Wave Transient Catalog (\GWTCTHREE{}), which records transient \ac{GW} signals discovered up to the end of \ac{LIGO}--Virgo's \ac{O3}. 
This updates the previous \GWTCTWO{}~\cite{Abbott:2020niy} and \GWTCTWOFINAL{}~\cite{LIGOScientific:2021usb} by including signals found in the \ac{O3b}. 
This period comprises data taken between \RUNSTART{}~\ac{UTC} and \RUNEND{}~\ac{UTC}. 
\GWTCTHREE{} adds \NUMEVENTS{} \ac{GW} candidates from \ac{O3b} that have an inferred probability of astrophysical \ac{CBC} origin of $\pastro{} > \PASTROTHRESHOLD$ based upon the results of our search algorithms. 
Additionally, there are $\ALLBELOWPASTROTHRESHOTHREEB{}$ subthreshold \ac{O3b} candidates that do not meet the \ac{CBC} $\pastro{}$ threshold but have a \ac{FAR} $< \SUBTHRESHOLDFAR{}$. 
With the inclusion of \ac{O3b} candidates, \GWTCTHREE{} is the most comprehensive set of \ac{GW} observations presented to date, and it will further advance our understanding of astrophysics~\cite{LIGOScientific:2021psn}, fundamental physics~\cite{LIGOScientific:2021sio} and cosmology~\cite{Virgo:2021bbr}.

\GWTCTHREE{} contains candidate \acp{GW} from \acp{CBC}: merging binaries consisting of \acp{BH} and \acp{NS}. 
We analyze in detail the properties of candidates with $\pastro{} > \PASTROTHRESHOLD$. 
Previously reported from \ac{O3b} are the \ac{GW} candidates \FULLNAME{GW200115A}{} and \FULLNAME{200105F}{}, which are consistent with originating from \acp{NSBH}~\cite{LIGOScientific:2021qlt}. 
The naming of these \ac{GW} candidates follows the format GWYYMMDD\_hhmmss, encoding the date and \ac{UTC} of the signal. 
In the \GWTCTHREE{} analysis, \FULLNAME{200105F}{} is found to have $\pastro{} < \PASTROTHRESHOLD$; however, it remains a candidate of interest, and is discussed in detail in later sections. 
In addition to \FULLNAME{GW200115A}{} and \FULLNAME{200105F}{}, the \ac{O3b} candidates include \FULLNAME{GW191219E}{} which is consistent with originating from a \ac{NSBH}, and \FULLNAME{GW200210B}{} which could either be from a \ac{NSBH} or from a \ac{BBH}, as its less massive component has a mass ($\masstwo = \masstwosourceuncert{GW200210B}\Msun$, quoting the median and symmetric $90\%$ credible interval) that spans the range for possible \acp{NS} and \acp{BH}. 
All the other candidates are consistent with being \ac{GW} signals from \acp{BBH}, as their inferred component masses are above the theoretical upper limit of the \ac{NS} maximum mass~\cite{Rhoades:1974fn,Kalogera:1996ci}. 
Among the \ac{O3b} candidates with $\pastro{} > \PASTROTHRESHOLD$, we expect $\sim \CONTAMINATIONFRACTION$ of candidates to be false alarms caused by instrumental noise fluctuations; a smaller, higher-purity sample of candidates could be obtained by adopting a stricter threshold. 

During \ac{O3}, low-latency public alerts were issued through \ac{GCN} Notices and Circulars for \ac{GW} candidates found by initial searches of the data~\cite{OPA,Abbott:2020niy}. 
These public alerts enable the astronomy community to search for multimessenger counterparts to potential \ac{GW} signals. 
There were \NUMPUBLICEVENTS{} low-latency candidates reported during \ac{O3b}. 
Of these, \PREVIOUSLYREPORTED{} (excluding \FULLNAME{200105F}{}) survive our detailed analyses to be included as potential \ac{CBC} signals in \GWTCTHREE{}.  
Additionally, \GWTCTHREE{} includes \NEWEVENTS{} candidates with $\pastro{} > \PASTROTHRESHOLD$ that have not been previously presented. 
No confident multimessenger counterparts have currently been reported from the \ac{O3b} candidates (as we review in Appendix~\ref{sec:follow-up}).

\begin{figure}
\begin{center}
\includegraphics[width=\columnwidth]{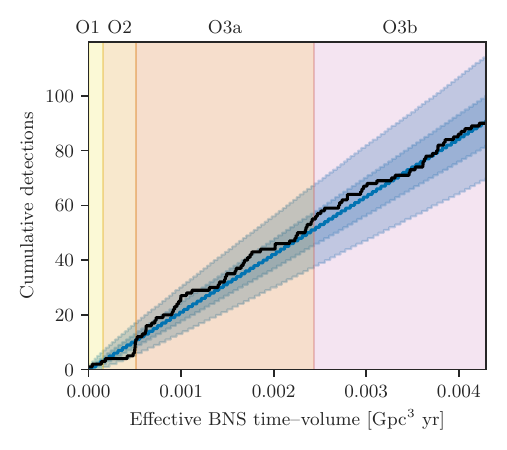}
\end{center}
\caption{\label{fig:VT-events}
The number of \ac{CBC} detection candidates with a probability of astrophysical origin $\pastro{} > \PASTROTHRESHOLD$ versus the detector network's effective surveyed time--volume for \ac{BNS} coalescences~\cite{Abbott:2020niy}.
The colored bands indicate the different observing runs.
The final datasets for \ac{O1}, \ac{O2}, \ac{O3a} and \ac{O3b} consist of \TWODETECTORSDAYSONE~days, \TWODETECTORSDAYSTWO~days, \TWODETECTORSDAYSTHREEA~days (\ONEDETECTORDAYSTHREEA~days) and \TWODETECTORSDAYS~days (\ONEDETECTORDAYS~days) with at least two detectors (one detector) observing, respectively.
The cumulative number of probable candidates is indicated by the solid black line, while the blue line, dark blue band and light blue band are the median, $50\%$ confidence interval and $90\%$ confidence interval for a Poisson distribution fit to the number of candidates at the end of \ac{O3b}. 
}
\end{figure}

The total number of \ac{GW} candidates with $\pastro{} > \PASTROTHRESHOLD$ in \GWTCTHREE{} is \TOTALEVENTS{}, compared with \NUMOBSERVINGONEEVENTS{} candidates found by \ac{LVK} analyses after the end of the \ac{O1}~\cite{Abbott:2016blz,TheLIGOScientific:2016pea}, \NUMGWTCONEEVENTS{} in \GWTCONE{} after the end of the \ac{O2}~\cite{LIGOScientific:2018mvr}, and \NUMGWTCTWOPOINTONEEVENTS{} in \GWTCTWOFINAL{} after the end of the \ac{O3a}~\cite{LIGOScientific:2021usb}. 
Additional candidates have also been reported by other searches of public data~\cite{Zackay:2019tzo,Venumadhav:2019lyq,Nitz:2019hdf,Zackay:2019btq,Nitz:2021uxj,Olsen:2022pin,Davies:2022thw}.
The dramatic increase in the number of \ac{GW} candidates during \ac{O3} was enabled by the improved sensitivity of the detector network. 
A conventional measure of sensitivity is the \ac{BNS} inspiral range, which quantifies the average distance at which a fiducial $\BNSRANGEMASS$ \ac{BNS} could be detected with a \ac{SNR} of $\BNSRANGESNR$~\cite{Finn:1992xs,Allen:2005fk,Chen:2017wpg}. 
During \ac{O3b} observations, the median \ac{BNS} inspiral ranges for \ac{LIGO} Livingston, \ac{LIGO} Hanford and Virgo were \LIVINGSTONRANGE{}, \HANFORDRANGE{} and \VIRGORANGE{}, respectively. 
In Fig.~\ref{fig:VT-events}, we show the growth in the number of candidates in the \ac{LVK} catalog across observing runs. 
Here, the search sensitivity is quantified by the \ac{BNS} time--volume, which should be approximately proportional to the search sensitivity to the overall astrophysical \ac{CBC} population and hence to the number of detections~\cite{Abbott:2020niy}. 
The \ac{BNS} time--volume is defined as the observing time multiplied by the Euclidean sensitive volume for the detector network~\cite{Chen:2017wpg}. 
For \ac{O1} and \ac{O2}, the observing time includes periods when at least two detectors were observing, and the Euclidean sensitive volume is the volume of a sphere with a radius equal to the \ac{BNS} inspiral range of the second most sensitive detector in the network.
For \ac{O3}, to account for the potential of single-detector triggers, the observing time also includes periods when only one detector was observing, and
the radius of the Euclidean sensitive volume is the greater of either (i) the \ac{BNS} inspiral range of the second most sensitive detector, or (ii) the \ac{BNS} inspiral range of the most sensitive detector divided by $\BNSVTSINGLEFACTOR$ (corresponding to a \ac{SNR} threshold of $\BNSRANGESNRSINGLE$)~\cite{Abbott:2020niy}.  
As the sensitivity of the detector network improves~\cite{Abbott:2020qfu}, the rate of discovery increases.

Further searches for \ac{GW} transients in \ac{O3b} data have been conducted focusing on \ac{IMBH} binaries (with a component $\gtrsim \IMBHSEARCHMASS{}$ and a final \ac{BH} $\gtrsim \IMBHThreshold$)~\cite{LIGOScientific:2021tfm}, subsolar-mass binaries~\cite{LIGOScientific:2022hai}, gravitationally lensed signals~\cite{LIGOScientific:2023bwz}, signals coincident with gamma-ray bursts~\cite{LIGOScientific:2021iyk}, cosmic strings~\cite{Abbott:2021ksc}, and both minimally modeled short-duration ($\lesssim\BURSTDURATIONBOUND$, such as from supernova explosions)~\cite{LIGOScientific:2021hoh} and long-duration ($\gtrsim\BURSTDURATIONBOUND$, such as from deformed magnetars or from accretion-disk instabilities)~\cite{LIGOScientific:2021uyj} signals. 
However, no high-significance candidates for types of signals other than the \acp{CBC} reported here have yet been found.

We begin with an overview of the status of the Advanced \ac{LIGO} and Advanced Virgo detectors during \ac{O3b} (Sec.~\ref{sec:instruments}), and then we review the properties and quality of the data used in the analyses (Sec.~\ref{sec:data}). 
We report the significance of the candidates identified by template-based and minimally modeled search analyses, and we compare this set of candidates to the low-latency public \ac{GW} alerts issued during \ac{O3b} (Sec.~\ref{sec:searches}).
We describe the inferred astrophysical parameters for the \ac{O3b} candidates (Sec.~\ref{sec:parameter-estimation}).
Finally, we show the consistency of reconstructed waveforms with those expected for \acp{CBC} (Sec.~\ref{sec:waveform-reconstruction}). 
In the Appendixes, we review public alerts and their multimessenger follow-up (Appendix~\ref{sec:follow-up}); we describe commissioning of the observatories for \ac{O3b} (Appendix~\ref{sec:interferometers}); we detail the data-analysis methods used to assess data quality (Appendix~\ref{sec:data-methods}), search for signals (Appendix~\ref{sec:searches-methods}) and infer source properties (Appendix~\ref{sec:parameter-estimation-methods}), and we discuss the difficulties in assuming a source type when performing a minimally modeled search analysis (Appendix~\ref{sec:cwb-only-events}).
A data release associated with this catalog is available from the \ac{GWOSC}~\cite{gwosc:gwtc3}; this includes calibrated strain time series around significant candidates, detection-pipeline results, parameter-estimation posterior samples, source localizations, and tables of inferred source parameters.

\section{Instruments}
\label{sec:instruments}

The Advanced \ac{LIGO}~\cite{TheLIGOScientific:2014jea} and Advanced Virgo~\cite{TheVirgo:2014hva} instruments are kilometer-scale laser interferometers~\cite{Freise:2009sf,Pitkin:2011yk,Vajente:2019skz}. 
The advanced generation of interferometers began operations in 2015, and observing periods have been alternated with commissioning periods~\cite{Abbott:2020qfu}. 
After \ac{O1}~\cite{Martynov:2016fzi,TheLIGOScientific:2016pea} and \ac{O2}~\cite{LIGOScientific:2018mvr}, the sensitivity of the interferometers has improved significantly~\cite{Buikema:2020dlj,Abbott:2020niy}.  
The main improvements were the adjustment of in-vacuum squeezed-light sources, or \emph{squeezers}, for the \ac{LIGO} Hanford and \ac{LIGO} Livingston interferometers and the increase of the laser power in the Virgo interferometer.
The instrumental changes leading to improved sensitivities during \ac{O3b} are discussed in Appendix~\ref{sec:interferometers}.

Figure~\ref{fig:strain} shows representative sensitivities during \ac{O3b} for \ac{LIGO} Hanford, \ac{LIGO} Livingston and Virgo, as characterized by the amplitude spectral density of the calibrated strain output. 
The sensitivity of the interferometers is primarily limited by the photon shot noise at high frequencies and by a superposition of several noise sources at lower frequencies~\cite{Buikema:2020dlj}.
The narrow-band features include vibrational modes of the suspension fibers, calibration lines, and $\EUPOWERGRIDFREQ~\mathrm{Hz}$ and $\USPOWERGRIDFREQ~\mathrm{Hz}$ electric power harmonics.

\begin{figure} 
\begin{center}
\includegraphics[width=\columnwidth]{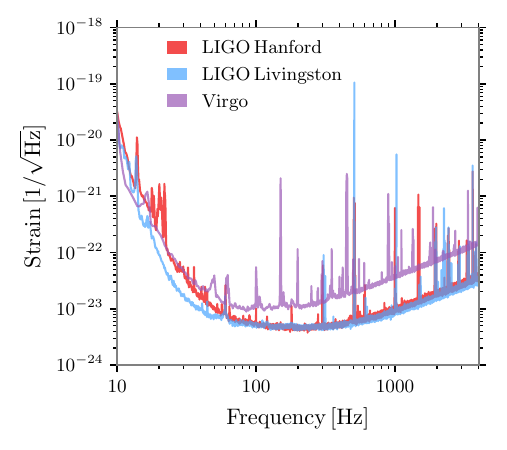}
\end{center}
\caption{\label{fig:strain} Representative amplitude spectral density of the three interferometers' strain sensitivity: \ac{LIGO} Livingston \LIVINGSTONSENS{}~\ac{UTC}, \ac{LIGO} Hanford \HANFORDSENS{}~\ac{UTC}, Virgo \VIRGOSENS{}~\ac{UTC}. 
From the amplitude spectral densities we estimate \ac{BNS} inspiral ranges~\cite{Finn:1992xs,Allen:2005fk,Chen:2017wpg} of \HANFORDRANGEPLOT{}, \LIVINGSTONRANGEPLOT{}, and \VIRGORANGEPLOT{} for \ac{LIGO} Hanford, \ac{LIGO} Livingston and Virgo, respectively.}
\end{figure}

\begin{figure*} 
\begin{center}
\includegraphics[width=\columnwidth]{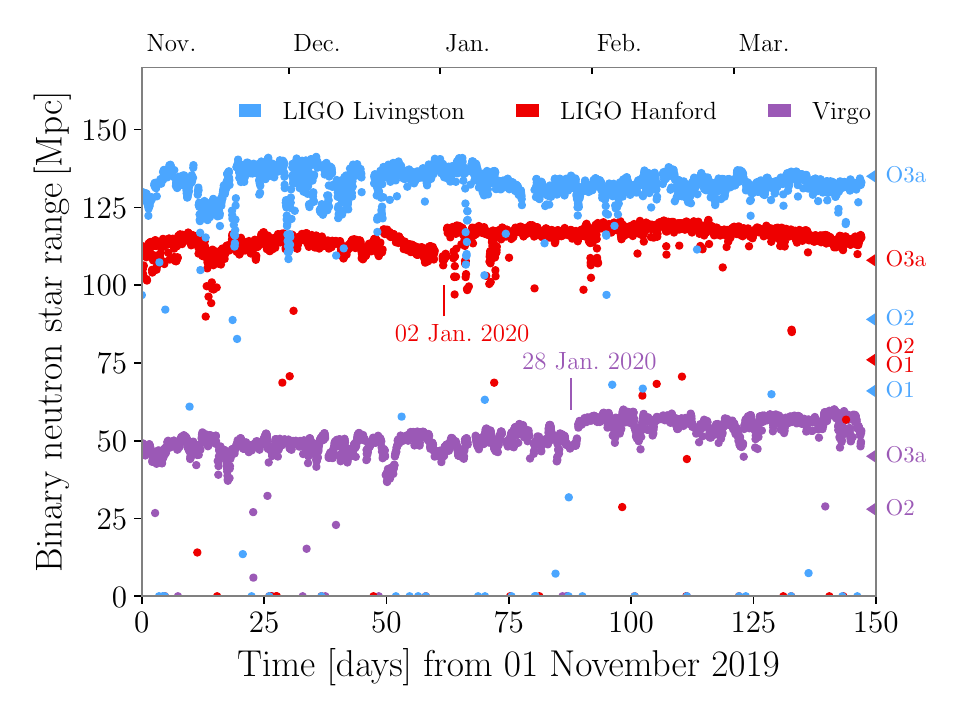}
\includegraphics[width=\columnwidth]{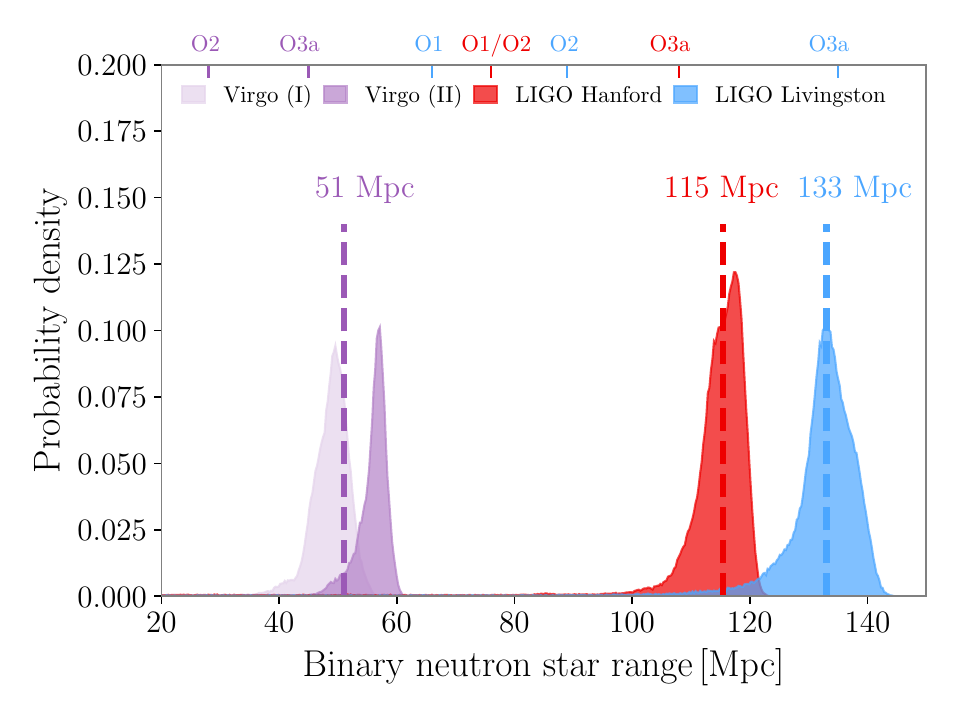} 
\end{center}
\caption{\label{fig:range} The \ac{BNS} inspiral range~\cite{Finn:1992xs,Allen:2005fk,Chen:2017wpg} of the \ac{LIGO} and Virgo detectors. 
\emph{Left}: The range evolution during \ac{O3b}. 
Each data point corresponds to the median value of the range over a one-hour time segment. 
\emph{Right}: Distributions of the range and the median values for the entire duration of \ac{O3b}; the data for Virgo are separately reported for the intervals before (I) and after (II) \VIRGORANGESPLIT{} to illustrate changes in the range following detector improvements.  
An improvement in squeezer performance at \ac{LIGO} Hanford is indicated at \LHORANGESPLIT{}.}
\end{figure*}

The left panel of Fig.~\ref{fig:range} reports the evolution of the detectors' sensitivity over time, as measured by the \ac{BNS} inspiral range~\cite{Finn:1992xs,Allen:2005fk,Chen:2017wpg}. 
Gaps in the range curve are due to maintenance intervals, instrumental failures and earthquakes. 
The epochs marked on the graph correspond to improvements in \ac{LIGO} Hanford (\LHORANGESPLIT{}) and Virgo (\VIRGORANGESPLIT{}) that are discussed in Appendix~\ref{sec:interferometers}.
The median \ac{BNS} inspiral range of Virgo over the whole of \ac{O3b} was \VIRGORANGE{}, while the maximum value reached \VIRGORANGERECORD{}.
For comparison, the median range and the maximum range during \ac{O3a} were \VIRGORANGETHREEA{} and \VIRGORANGERECORDTHREEA{}, respectively.
The \ac{LIGO} Hanford median \ac{BNS} inspiral range improved from \HANFORDRANGETHREEA{} in \ac{O3a} to \HANFORDRANGE{} in \ac{O3b}, primarily due to the squeezed-light~\cite{Caves:1981hw,Barsotti:2018hvm} source adjustments described in Appendix~\ref{sec:interferometers}.
The \ac{LIGO} Livingston median \ac{BNS} inspiral range in \ac{O3b} was \LIVINGSTONRANGE{}, consistent with the \ac{O3a} value of \LIVINGSTONRANGETHREEA{}, with improvements due to squeezing counterbalanced by degradation primarily due to the reduced circulating power.

The duty cycles for the three interferometers, i.e., the fractions of the total \ac{O3b} run duration in which the instruments were observing, were \HANFORDDUTYCYCLE{}\% (\HANFORDDAYS{}~days) for \ac{LIGO} Hanford, \LIVINGSTONDUTYCYCLE{}\% (\LIVINGSTONDAYS{}~days) for \ac{LIGO} Livingston and \VIRGODUTYCYCLE{}\% (\VIRGODAYS{}~days) for Virgo. 
As for previous observing runs, a subset of search analyses rejected additional data based on data-quality metrics, as described in Appendix~\ref{sec:data-methods}.  
The complete three-interferometer network was in observing mode for \THREEDETECTORSDUTYCYCLE{}\% of the time (\THREEDETECTORSDAYS{}~days). 
Moreover, for \ONEDETECTORDUTYCYCLE{}\% of the time (\ONEDETECTORDAYS{}~days) at least one interferometer was observing, while for \TWODETECTORSDUTYCYCLE{}\% (\TWODETECTORSDAYS{}~days) at least two interferometers were observing. 
For comparison, during \ac{O3a} the duty cycles were \HANFORDDUTYCYCLETHREEA{}\%, \LIVINGSTONDUTYCYCLETHREEA{}\% and \VIRGODUTYCYCLETHREEA{}\% for \ac{LIGO} Hanford, \ac{LIGO} Livingston and Virgo, respectively; at least one interferometer was observing \ONEDETECTORDUTYCYCLETHREEA{}\% of the time, and at least two interferometers were observing \TWODETECTORSDUTYCYCLETHREEA{}\% of the time.
The duty cycles for both the Hanford and Livingston interferometers improved from \ac{O3a} to \ac{O3b}. 
This demonstrates a clear improvement in robustness as higher microseism and storm activity were observed during \ac{O3b} compared to \ac{O3a}.
While the fraction of time with at least one detector observing in \ac{O3a} and \ac{O3b} was comparable, the fraction of time with two instruments in observing mode increased, improving the performance of the network for coincident observations.

\section{Data}
\label{sec:data}

Following the approach of previous analyses~\cite{Abbott:2020niy,LIGOScientific:2021usb}, we calibrate the data of each detector to \ac{GW} strain and mitigate known instances of poor data quality before analyzing the \ac{LIGO} and Virgo strain data for astrophysical sources. 
We include segments of data from each detector in our \ac{GW} search analyses only when the detector was operating in a nominal state, and when there were no diagnostic measurements being made that might interfere with \ac{GW} data collection.

Once data are recorded, they are calibrated in near-real time and in higher latency, as described in Sec.~\ref{sec:calibration}. 
We subtract noise from known long-duration, quasistationary instrumental sources~\cite{Vajente:2019ycy,Acernese:2018bfl,Viets:2021aaa}. 
We also exclude time periods containing identified and well-characterized noise likely to interfere with signal extraction from the astrophysical analyses, as described in Sec.~\ref{sec:DQ}.
We thoroughly vet the data surrounding each \ac{GW} candidate for evidence of transient noise, or \emph{glitches}, or other anomalies that could impact accurate assessment of the candidate's significance or accurate source-parameter estimation.
For \ac{GW} candidates found near in time or overlapping with transient noise, we apply additional data-processing steps, including the modeling and subtraction of glitches and linear subtraction of glitches using a witness time series~\cite{Davis:2022ird}, as described in Appendix~\ref{sec:data-methods}. 

\subsection{Calibration and noise subtraction}
\label{sec:calibration}

The dimensionless strain time series measured by the \ac{LIGO} and Virgo detectors are an input to the astrophysical analyses. 
They are reconstructed from different output signals from the detectors and detailed modeling of the response of the detector~\cite{Viets:2017yvy,Acernese:2018bfl}.
The reconstructed strain time series are timestamped following \ac{GPS} time, 
taking into account both the delays introduced in the synchronized distributed-clock timing system and data conditioning along the data-acquisition systems~\cite{Bartos:2010zz}.
The detector responses are described as complex-valued frequency-dependent transfer functions~\cite{Abbott:2016jsd,Acernese:2018bfl}.
Some control-system model parameters, such as the amount of light stored in the interferometer cavities and the gain of the actuators controlling the position of primary optics~\cite{TheLIGOScientific:2014jea}, vary slowly with time throughout operation of the interferometers.
These parameters are monitored and, when possible, aspects of the calibration models are corrected in the strain reconstruction processing~\cite{Tuyenbayev:2016xey,Acernese:2018bfl,Viets:2017yvy}. 
The analysis of the systematic error and uncertainty bounds for calibrated data throughout \ac{O3b} is detailed in previous studies of \ac{LIGO}~\cite{Sun:2020wke,Sun:2021qcg} and Virgo data~\cite{Estevez:2019vna,Rolland:2020vna,VIRGO:2021umk}. 

The three detectors use auxiliary lasers, known as photon calibrators~\cite{Karki:2016pht,Bhattacharjee:2020yxe,Estevez:2020pvj}, to induce fiducial displacement of test masses via photon radiation pressure.
The fiducial displacements are known to better than $\PCALUNCERTAINTYLIGO\%$ in \ac{LIGO} and $\PCALUNCERTAINTYVIRGO\%$ in Virgo and are used to measure interferometer parameters' variation with time, develop accurate models, and establish estimates of systematic error and associated uncertainty.

Calibration models are estimated from a collection of measurements that characterize the full detector response and from other measurements of individual components~\cite{Sun:2020wke,Acernese:2018bfl,Sun:2021qcg}, such as the various electronics and suspension systems, gathered while the detector is offline (roughly once per week).
An initial version of calibrated strain data is produced in low latency throughout an observing period, and the final calibration models are assembled after the completion of an observing period where the detector configuration was stable~\cite{Viets:2017yvy,VIRGO:2021umk}.
As needed, the \ac{GW} strain data stream is then regenerated offline from the optical power variations and the control signals, and the systematic error estimate is updated based on the model used for the offline strain reconstruction.

The best available strain data for each detector have been used for both detection of \ac{GW} signals and estimation of the sources' astrophysical parameters.
For \ac{LIGO}, the offline recalibrated strain data were used~\cite{Sun:2020wke,Sun:2021qcg}.
Initial analysis of Virgo's collection of validation measurements during the run did not motivate offline improvement to the low-latency strain data.
Hence, Virgo's low-latency strain data has been used for all analyses~\cite{Estevez:2019vna,Rolland:2020vna,VIRGO:2021umk}.

After the completion of the run, we identified a narrow-band increased systematic error between $\VIRGOSYSTEMATICLOWERFREQ~\mathrm{Hz}$ and $\VIRGOSYSTEMATICHIGHERFREQ~\mathrm{Hz}$ in Virgo data, mainly related to a control loop designed to damp mechanical resonances of the suspensions at $\VIRGOMECHANICALRESFREQ~\mathrm{Hz}$~\cite{Rolland:2020vna}.
This damping loop was added between \ac{O3a} and \ac{O3b} and ultimately improved the Virgo detector's sensitivity around $\VIRGOMECHANICALRESFREQ~\mathrm{Hz}$.
However, since this damping loop was not included in the calibration models, it resulted in an increased systematic error in the calibrated strain data around $\VIRGOMECHANICALRESFREQ~\mathrm{Hz}$ during \ac{O3b}.
There was also a large increase in the systematic error between $\VIRGOADDITIONALSYSTEMATICLOWERFREQ~\mathrm{Hz}$ and $\VIRGOADDITIONALSYSTEMATICHIGHERFREQ~\mathrm{Hz}$ related to a control loop designed to reduce the electric power-grid line~\cite{VIRGO:2021umk}.
Overall, the Virgo calibration errors in the band $\VIRGOSYSTEMATICLOWERFREQ$--$\VIRGOSYSTEMATICHIGHERFREQ~\mathrm{Hz}$ increased from $\VIRGOSYSTEMATICERRORPERCENTNOMINAL\%$ in amplitude and $ \VIRGOSYSTEMATICERRORPHASENOMINAL~\mathrm{mrad}$ in phase to up to $\VIRGOSYSTEMATICERRORPERCENTINCREASED\%$ in amplitude and $\VIRGOSYSTEMATICERRORPHASEINCREASED~\mathrm{mrad}$ in phase~\cite{VIRGO:2021umk}.
This narrow-band increased systematic error was accounted for in source-parameter estimation by notching out these frequencies (as described in Appendix~\ref{sec:parameter-estimation-methods}).

Known noise sources were subtracted from both the \ac{LIGO} and Virgo strain data.
The sinusoidal excitations used for calibration, known as calibration lines, were subtracted from the \ac{LIGO} strain data.
The $\USPOWERGRIDFREQ~\mathrm{Hz}$ electric power-grid lines were subtracted in the \ac{LIGO} strain data along with the corresponding harmonics up to and including $\POWERGRIDHARMONIC~\mathrm{Hz}$~\cite{Viets:2021aaa}.
Additionally, noise contributions due to nonstationary coupling of the power grid were subtracted from the \ac{LIGO} strain data~\cite{Vajente:2019ycy,Tiwari:2015ofa}.  
Numerous noise sources that limited the Virgo detector's sensitivity were measured and linearly subtracted from the Virgo low-latency strain data using witness auxiliary sensors that measure the source of the noise~\cite{Driggers:2018gii,Davis:2018yrz,Acernese:2018bfl,VIRGO:2021umk}. 
Calibration lines were also subtracted from the Virgo strain data.

All final source-parameter results, waveform reconstructions, and all but one search pipeline used strain data with all noise subtraction applied, as described above. 
The exception is the \ac{CWB} analysis~\cite{Klimenko:2015ypf}, which searches for transient signals without assuming a model template.
Following the \GWTCTWO analysis~\cite{Abbott:2020niy}, \ac{CWB} used \ac{LIGO} strain data with the calibration lines and power-grid lines subtracted, but without the subtraction of the nonstationary coupling of the power grid.
Comparison of analyses using different versions of noise subtraction indicates that the exact noise-subtraction procedure used does not significantly impact the \ac{CWB} search results.

\subsection{Data quality}\label{sec:DQ}

\begin{figure} 
	\begin{center}
		\includegraphics[width=\columnwidth]{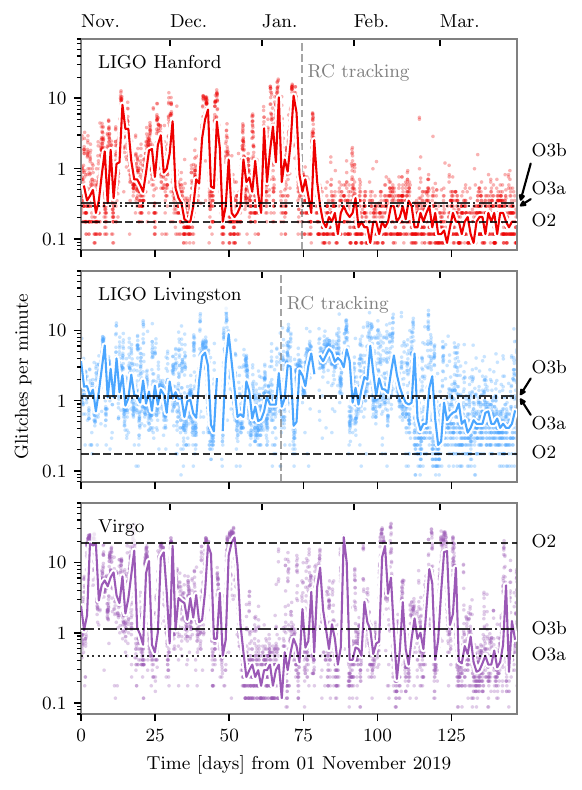}
	\end{center}

	\caption{\label{fig:glitch_rate} The rate of single-interferometer glitches with \ac{SNR} $\rho > \GLITCHSNRMIN$ and peak frequency between $\GLITCHBANDLOW~\mathrm{Hz}$ and $\GLITCHBANDHIGH~\mathrm{Hz}$ identified by Omicron~\cite{Robinet:2020lbf} in each detector during \ac{O3b}. 
Each point represents the average rate per minute, estimated over a $\GLITCHTIMESTRIDE~\mathrm{s}$ interval. 
Continuous curves represent the daily median of the rates. 
Black lines show the median rate over entire runs: dashed for \ac{O2}, dotted for \ac{O3a}, and dash--dotted for \ac{O3b}. 
The vertical dashed lines indicate the implementation of reaction-chain (RC) tracking at the \ac{LIGO} detectors, which reduced the rate of \SLOWSCATTER{} glitches.
	}
\end{figure}

The most limiting source of noise for identification and analysis of transient \ac{GW} sources is frequent, short-duration glitches in \ac{GW} detector data~\cite{TheLIGOScientific:2016zmo,Nuttall:2018xhi,Davis:2021ecd,Virgo:2022ysc}. 
A summary of glitch rates for the three observatories over \ac{O3b} is shown in Fig.~\ref{fig:glitch_rate}. 
Each point corresponds to the average number of glitches per minute with \ac{SNR} $\rho>\GLITCHSNRMIN$ and peak frequency between $\GLITCHBANDLOW~\mathrm{Hz}$ and $\GLITCHBANDHIGH~\mathrm{Hz}$, estimated every $\GLITCHTIMESTRIDE~\mathrm{s}$, as measured with the Omicron algorithm~\cite{Robinet:2020lbf}.
Continuous solid lines indicate the daily median of the corresponding glitch rate. 
In all three detectors, we observed relatively high glitch rates, dominated by glitches below $\sim\LOWFREQGLITCH~\mathrm{Hz}$, corresponding to seasonally bad weather between the beginning of \ac{O3b} and January 2020; some peaks in the glitch rate are also visible in Virgo data during the second half of \ac{O3b} corresponding to persistent unstable weather conditions~\cite{Acernese:2022ozw}.

The horizontal black lines in Fig.~\ref{fig:glitch_rate} indicate the median glitch rates during \ac{O2} (dashed), \ac{O3a} (dotted), and \ac{O3b} (dash--dotted).
With respect to \ac{O3a}, both \ac{LIGO} detectors registered a modest glitch rate increase in \ac{O3b}, with the rate changing from $\LHORATEA~\mathrm{min^{-1}}$ to $\LHORATEB~\mathrm{min^{-1}}$ for Hanford and from $\LLORATEA~\mathrm{min^{-1}}$ to $\LLORATEB~\mathrm{min^{-1}}$ for Livingston; 
this variation was much more pronounced for Virgo, which increased its glitch rate from $\VIRGORATEA~\mathrm{min^{-1}}$ to $\VIRGORATEB~\mathrm{min^{-1}}$.  
As discussed for \GWTCTWO~\cite{Abbott:2020niy}, the increase in glitch rate in the two \ac{LIGO} detectors between \ac{O2} and \ac{O3a} is largely due to scattered-light glitches, and the decrease in Virgo's glitch rate between \ac{O2} and \ac{O3} is due to mitigation of several noise sources.

A large fraction of the \ac{O3b} glitches captured in Fig.~\ref{fig:glitch_rate} are due to light scattering, as described in Appendix~\ref{sec:interferometers}. 
When the relative displacement between a mirror and a nearby moving reflective surface is $\gtrsim\APPROXLASERWAVELENGTH$ (the main laser wavelength) in amplitude, low-frequency ground motion can be up-converted to scattered-light glitches in the sensitive band of \ac{GW} detector data~\cite{Accadia:2010zzb,Ottaway:2012oce}.
During \ac{O3}, approximately \SCATTERINGRATELLO\% and \SCATTERINGRATELHO\% of all the transient noise with \ac{SNR} $\rho > \SCATTERINGSNRMIN$ at \ac{LIGO} Livingston and \ac{LIGO} Hanford, respectively, was due to light scattering.
A high rate of scattered-light glitches is partly a consequence of weather-related high microseismic ground motion at the detector sites during \ac{O3b}~\cite{Soni:2020rbu,Davis:2021ecd,Soni:2021cjy}.

\begin{figure}
	\begin{center}
	\includegraphics[width=\columnwidth]{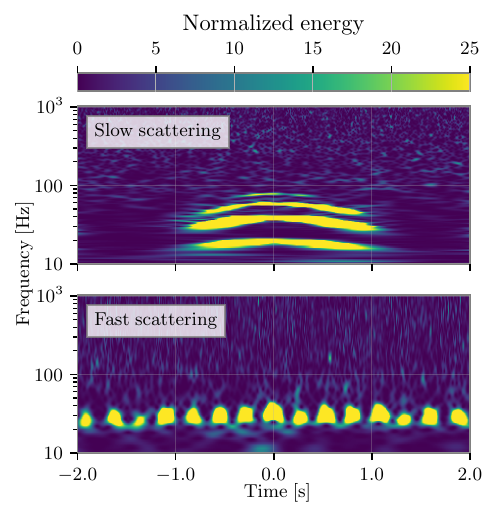}
	\end{center}
	\caption{\label{fig:scattering} Representative spectrograms~\cite{Chatterji:2004qg} of glitches caused by light scattering.
\emph{Top}: \SLOWSCATTER{} appears as long-duration arches in the time--frequency plane.
The multiple arches are due to multiple reflections between the test-mass optics and the scattering surface.
During \ac{O3}, \SLOWSCATTER{} was the most frequent and second most frequent source of transient noise at \ac{LIGO} Hanford and \ac{LIGO} Livingston respectively.
\emph{Bottom}: As compared to \SLOWSCATTER{}, \FASTSCATTER{} transients appear as short-duration, rapidly repeating arches.
	}
\end{figure}

Two separate populations of transient noise due to light scattering known as \emph{\SLOWSCATTER{}} and \emph{\FASTSCATTER{}} polluted \ac{LIGO} data quality in \ac{O3}. 
As illustrated in the spectrograms of Fig.~\ref{fig:scattering}, \SLOWSCATTER{} noise appears as longer-duration ($\sim\SLOWARCHDURATION~\mathrm{s}$) arches in the time--frequency plane, while \FASTSCATTER{} noise appears as short-duration ($\sim\FASTARCHDURATION~\mathrm{s}$) arches~\cite{Soni:2021cjy}.

\SLOWSCATTER{} tends to occur when ground motion is high in the earthquake ($\EARTHQUAKELOW{}$--$\EARTHQUAKEHIGH{}~\mathrm{Hz}$) or microseism ($\MICROSEISMLOW{}$--$\MICROSEISMHIGH{}~\mathrm{Hz}$) frequency bands. 
For the \ac{LIGO} detectors, we found the presence of the \SLOWSCATTER{} arches to be strongly correlated with the relative motion between the end test-mass chain and the reaction-mass chain of the optic suspension system used to control the motion of the test masses.
This led to implementing reaction-chain tracking~\cite{alog:LHO54298,alog:LLO50851} in January 2020 to reduce this relative motion, as discussed in Appendix~\ref{sec:interferometers}.
The rate of glitches associated with \SLOWSCATTER{} significantly decreased after the implementation of the reaction-chain tracking~\cite{Soni:2020rbu}.

Figure~\ref{fig:glitch_rate} shows that the overall \ac{O3b} glitch rate significantly decreased for \ac{LIGO} Hanford after the implementation of the reaction-chain tracking, changing from $\LHOPRERCRATE~\mathrm{min^{-1}}$ to $\LHOPOSTRCRATE~\mathrm{min^{-1}}$. 
Correlated with this drop in glitch rate, the noise background became more stable and the average fraction of \ac{O3b} public alerts that were retracted dropped from $\PREOPARETRATE$ to $\POSTOPARETRATE$.

\FASTSCATTER{} is far more common at \ac{LIGO} Livingston than at \ac{LIGO} Hanford. 
During \ac{O3}, it was the most frequent source of transient noise at Livingston. 
As shown in Fig.~\ref{fig:scattering} and in Appendix~\ref{sec:data-methods}, \FASTSCATTER{} generally affects the \ac{GW} data from $\FASTSCATTERLOWFREQ~\mathrm{Hz}$ to $\FASTSCATTERHIGHFREQ~\mathrm{Hz}$, but occasionally manifests as high as $\FASTSCATTERMAXFREQ~\mathrm{Hz}$. 
Increased ground motion in the anthropogenic ($\ANTHROPLOW{}$--$\ANTHROPHIGH{}~\mathrm{Hz}$) band, usually caused by bad weather conditions and human activity, especially with nearby heavy machinery such as logging trucks, increases the rate of \FASTSCATTER{} glitches.
Physical environment and monitoring tests conducted at \ac{LIGO} Livingston and \ac{LIGO} Hanford found high quality-factor mechanical resonances at frequencies close to $\FASTSCATTERQFREQ~\mathrm{Hz}$~\cite{alog:LLO53025,alog:LLO53364} thought to be related to \FASTSCATTER{}.
The \ac{O4} upgrade plans include damping these resonances and studying the impact on the rate of \FASTSCATTER{} noise.

In Virgo, the initial high glitch rate and the subsequent peaks in Fig.~\ref{fig:glitch_rate} correspond predominantly to high numbers of glitches with central frequencies lower than $\VIRGOGLITCHFHIGH~\mathrm{Hz}$. 
Across \ac{O3b}, $\sim\VIRGOLOWFRATE\%$ of glitches in Virgo with $\rho > \GLITCHSNRMIN$ had central frequencies lower than $\VIRGOGLITCHFHIGH~\mathrm{Hz}$.
These lower-frequency scattered-light glitches are largely the consequence of the activity of the sea, which is $\VIRGOTOSEA~\mathrm{km}$ from the detector site~\cite{Accadia:2010zzb,Acernese:2022ozw}. 

All candidates reported in Table~\ref{tab:events} and Table~\ref{tab:marginal_events} have undergone validation to check for plausible instrumental or environmental causes using the same methods as were applied to \ac{O3a} candidates~\cite{Abbott:2020niy,Davis:2021ecd,Virgo:2022ysc,DQRdocumentation}. 
As discussed in Sec.~\ref{sec:candidates}, none of the \ac{O3b} candidates with \ac{CBC} $\pastro{} > \PASTROTHRESHOLD$ have evidence of instrumental origin, but we identified three marginal candidates (which do not meet the $\pastro{}$ threshold) as likely instrumental in origin.
We also investigated non-Gaussian instrumental artifacts present in the data close to each candidate time that could bias measurements of the source parameters.
In addition to the previously reported \FULLNAME{200105F}{}~\cite{LIGOScientific:2021qlt}, we identified \MITIGATIONEVENTS{} \ac{O3b} candidates in Table~\ref{tab:events} with nearby non-Gaussian artifacts that required mitigation before the data were further analyzed for source-parameter estimation. 
In order to mitigate instrument artifacts present near the time of these candidates, we followed a procedure similar to \ac{O3a}~\cite{Abbott:2020niy}. 
Further details on data-quality mitigation techniques, including data-quality products publicly available via \ac{GWOSC}, are given in Appendix~\ref{sec:data-methods} and in previous \ac{O3} analyses~\cite{Abbott:2020niy,Davis:2021ecd,Virgo:2022kwz}.
The specific mitigation methods applied for each of these candidates are described in Appendix~\ref{sec:parameter-estimation-methods}, with a summary for each candidate reported in Table~\ref{tab:dq_pe_mitigation}.

\section{Candidate identification}
\label{sec:searches}

Identification of candidates and assessment of their significance relative to the background of detector noise is the first step in extracting catalog results. 
This is followed by detailed analyses to estimate source properties (Sec.~\ref{sec:parameter-estimation}) and reconstruct waveforms (Sec.~\ref{sec:waveform-reconstruction}).
We use multiple search algorithms to identify potential \ac{GW} candidates in our data. 
Searches are performed at two different latencies: online searches are run in near-real time as data are collected, and offline searches are completed later, using the final calibrated and cleaned dataset. 
The online analyses allow for the rapid release of public alerts associated with candidates to enable the search for multimessenger counterparts, as described in Appendix~\ref{sec:follow-up}. 
The offline analyses benefit from improved background statistics, extensive data calibration, vetting and conditioning as described in Sec.~\ref{sec:data}, and the ability to perform more computationally expensive calculations to separate signals from background given the relaxation of latency requirements. 
Because of these factors, the offline analyses are more sensitive than the online analyses. 
The increased sensitivity of the offline analyses means that differences in the final candidate list compared to the online results are expected.
While the lowest \ac{FAR} candidates are expected to remain significant, candidates with a higher \ac{FAR} (e.g., near the threshold for public low-latency alerts) are more likely to have changes in significance when reevaluated offline, causing them to move above or below the corresponding threshold for inclusion in this catalog.
The differences between the online and offline search results are further discussed in Sec.~\ref{sec:online-candidates}.
In this catalog, we report on the results of offline analyses performed after the end of \ac{O3b}. 

Our search analyses use different approaches to find candidates, either filtering the data using \ac{CBC} waveform templates to identify matches (described in Sec.~\ref{sec:matched-filter}), or coherently searching data from the detector network for transient signals without assuming a waveform template (described in Sec.~\ref{sec:cwb}). 
We use four pipelines to identify the candidates from \ac{O3b}: three that search using \ac{CBC} waveform templates, \GSTLAL{}~\cite{Messick:2016aqy,Sachdev:2019vvd,Hanna:2019ezx,Cannon:2020qnf}, \ac{MBTA}~\cite{Adams:2015ulm,Aubin:2020goo} and \PYCBC{}~\cite{Allen:2005fk,Allen:2004gu,Canton:2014ena,Usman:2015kfa,Nitz:2017svb,Davies:2020tsx}, and one that searches for transient signals with minimal assumptions about sources, \ac{CWB}~\cite{Klimenko:2004qh,Klimenko:2011hz,Klimenko:2015ypf}. 
The four pipelines used offline were also operated in online configurations, along with the waveform-based \ac{SPIIR} pipeline~\cite{Luan:2011qx,Chu:2017ovg,Chu:2020pjv}, to identify candidate \ac{GW} signals in low latency. 
Of the four pipelines, \ac{CWB}, \GSTLAL{}, and \PYCBC{} were used for offline \ac{LVK} analysis of \ac{O1}~\cite{TheLIGOScientific:2016pea,Abbott:2016ezn}, \ac{O2}~\cite{LIGOScientific:2018mvr} and \ac{O3a}~\cite{Abbott:2020niy,LIGOScientific:2021usb} data, whereas \ac{MBTA} was first used for offline analysis of \ac{O3a}~\cite{LIGOScientific:2021usb}.

There are several technical and configuration differences across the pipelines used in the search analyses. 
While the \ac{CBC} pipelines consider all possible (double or triple) detector combinations to form coincident triggers, \ac{CWB} reports only analysis of pairs of detectors~\cite{LIGOScientific:2021hoh}.
Another significant difference across pipelines is the data baseline used to assign \acp{FAR} to candidates.
The \ac{FAR} is used as a measure of significance, and defines how regularly we would expect to see a noise (nonastrophysical background) trigger with the same, or higher, ranking statistic as the candidate.
\GSTLAL{} compares candidates to a global background from the full \ac{O3b} time span, while \ac{CWB}, \ac{MBTA} and \PYCBC{} use local background from a typical time span of one to a few weeks. 
All pipelines estimate background distributions empirically from the \ac{O3b} data.
Further technical details of the search algorithms are given in Appendix~\ref{sec:searches-methods}. 

\subsection{Modeled search analyses for transient sources}
\label{sec:matched-filter}

The dedicated \ac{CBC} search algorithms use matched filtering~\cite{Wainstein:1962vrq,LIGOScientific:2019hgc}, identifying candidates by correlating the data with templates. 
We use sets of templates, or \emph{banks}, that provide a discrete sampling of the parameter space defined by the binary component masses \massone{} and \masstwo{} (the primary and secondary masses, defining $\massone{} \geq \masstwo{}$), and the corresponding dimensionless spins \vecspinone{} and \vecspintwo{}. 

The signals expected from \acp{CBC} are well characterized by combinations of the binary component parameters. 
To leading order, the phase evolution during inspiral of a binary is determined by the chirp mass~\cite{Peters:1964zz,Blanchet:1995ez},
\begin{equation}
    \Mc{} = \frac{(\massone{} \masstwo{})^{3/5}}{(\massone{} + \masstwo{})^{1/5}}.
    \label{eq:chirp_mass}
\end{equation}
We also use the total mass $\Mtot = \massone{} + \masstwo{}$, and the mass ratio $\massratio{} = \masstwo{} / \massone{} \leq 1$ to describe a binary system. 
The dimensionless component spin $\vec\chi_i = c \vec{S}_i / (Gm_i^2)$, where $\vec{S}_i$ is the spin angular momentum and $i = \{1, 2\}$, can theoretically range in magnitude from $0$ (nonspinning) to $1$ (Kerr limit) for \acp{BH}. 
The two spins are combined to form the effective inspiral spin~\cite{Ajith:2009bn,Santamaria:2010yb} defined as
\begin{equation}
    \chieff{} = \frac{(\massone{} \vecspinone{} + \masstwo{} \vecspintwo{}) \cdot \LNewton}{\Mtot{}},
    \label{eq:eff_inspiral_spin_vec}
\end{equation}
where $\LNewton$ is the unit vector in the direction of the Newtonian orbital angular momentum.
In the modeled search analyses, the spins are assumed to be parallel to $\LNewton$.

The banks cover systems with total masses, redshifted to the detector frame~\cite{Krolak:1987ofj}, ranging from a minimum value $\MBTABANKMINMASS$ for all pipelines to a maximum value of $\MBTABANKMAXMASS$ (\ac{MBTA}), $\PYCBCHYPERBANKTOTALMASSMAX{}$ (\PYCBC{}) or $\GSTLALBANKMAXMASS$ (\GSTLAL{}).
The minimum binary component mass is $\COMPONENTBANKMINMASS$.
Searches for binaries with component masses less than $\COMPONENTBANKMINMASS$ have been completed in complementary analyses~\cite{LIGOScientific:2018glc,LIGOScientific:2019kan,Nitz:2020bdb,Nitz:2021mzz,LIGOScientific:2021job,LIGOScientific:2022hai}.
The \PYCBC{} pipeline performs two search analyses. 
The first is an analysis encompassing a wide parameter space, allowing detection of many different types of \ac{CBC} systems, which we refer to as the \PYCBCHYPERBANK{} analysis.
In addition to this broad analysis, \PYCBC{} is also used in a different configuration, which we refer to as the \PYCBCBBH{} analysis: this analysis is restricted to \ac{BBH} systems with total masses between \PYCBCBBHTOTALMASSMIN{} and \PYCBCBBHTOTALMASSMAX{}, mass ratios in the range $\PYCBCBBHQMIN{} \leq \massratio{} \leq \PYCBCBBHQMAX{}$, and component masses in the range $\PYCBCBBHMASSONEMIN{} \leq \massone{} \leq \PYCBCBBHMASSONEMAX{}$ and $\masstwo{} \geq \PYCBCBBHMASSTWOMIN{}$.
This \PYCBCBBH{} analysis is designed to have higher sensitivity to \ac{BBH} coalescences with component masses that are similar to those of the majority of previously detected systems. 
The range of templates is the same as used for the search of \ac{O3a}~\cite{LIGOScientific:2021usb}.

For each template, the matched-filter correlation produces a time series of \ac{SNR} values for each detector, and peaks in this time series form triggers.
Only triggers with a matched-filter \ac{SNR} exceeding a threshold are considered further in the analysis.
This \ac{SNR} threshold is $\rho > \PYCBCSNRTHRESH$ for \PYCBC{} and \GSTLAL{}, and either $\MBTASNRTHRESHMIN$ or $\MBTASNRTHRESHMAX$, varying across the parameter space, for \ac{MBTA}.
\ac{MBTA} and \PYCBC{} assign a significance to triggers found with consistent binary parameters and times of arrival in at least two detectors, while \GSTLAL{} also does so for single-detector triggers. 
The \ac{SNR} is combined with signal-consistency checks to rank triggers. 
Each pipeline uses a specific ranking statistic and background-estimation method to assess the significance and probability of astrophysical origin of these triggers and coincidences.
Results from the various \ac{CBC} search analyses are expected to differ due to differences in the waveform template banks and in algorithmic choices such as their ranking statistic and assumed signal distributions.
Technical details of the \GSTLAL{}, \ac{MBTA}, \PYCBCHYPERBANK{} and \PYCBCBBH{}, and (online-only) \ac{SPIIR} analyses are given in Appendix~\ref{sec:gstlal-methods}, Appendix~\ref{sec:mbta-methods}, Appendix~\ref{sec:pycbc-methods}, and Appendix~\ref{sec:spiir-methods}, respectively. 

\subsection{Minimally modeled search analyses for transient sources}
\label{sec:cwb}

The \ac{CWB} pipeline searches for generic, short transient signals across a network of \ac{GW} detectors~\cite{Klimenko:2008fu,Necula:2012zz,Klimenko:2015ypf,Salemi:2019uea,Drago:2020kic}.
It provides rapid detection of \ac{GW} transient signals with its online instance, and signal reconstructions and estimates of their significance with the version that runs offline on the final dataset.
Designed to operate without a specific waveform model, \ac{CWB} identifies coherent excess power in multiresolution time--frequency representations of the detector strain data. 
The \ac{SNR} for each detector is estimated from the reconstructed waveforms, and the network \ac{SNR} is calculated by combining the \acp{SNR} from the individual detectors.
The \ac{CWB} search analyses and reconstructions reported in this catalog primarily target \ac{BBH} sources and are limited to the $\CWBMINFREQ$--$\CWBMAXFREQ$ range~\cite{Klimenko:2015ypf} to boost computational efficiency given the expected frequency range of \ac{BBH} signals.
The analysis is further split into two configurations that target high-mass (central frequency $f_\mathrm{c}<\CWBCENTRALFREQ$) and low-mass ($f_\mathrm{c}>\CWBCENTRALFREQ$) \ac{BBH} systems~\cite{Szczepanczyk:2020osv}.
Technical details of the \ac{CWB} analysis are given in Appendix~\ref{sec:cwb-methods}.

\subsection{Probability of astrophysical origin}
\label{sec:pastro}

Our primary criterion for selecting candidates for further study is the probability of astrophysical origin \pastro{}.
In contrast to \ac{FAR}, this measure of significance incorporates our knowledge of the astrophysical rate of signals for different classes of binary systems.
For instance, given the strongly differing rates of detectable signals from \ac{BBH} and \ac{BNS} coalescences, at a given \ac{FAR} the probability of being an astrophysical signal will naturally be different between candidates consistent with \ac{BBH} versus \ac{BNS} origin.
The probability of astrophysical origin is well suited for selecting candidates from a catalog containing results from observing runs of differing sensitivities. 
As the true alarm rate increases with improved sensitivity, the \ac{FAR} needed for a candidate to reach a given $\pastro{}$ will change between observing runs.

In order to estimate \pastro{} and its complement, the probability of terrestrial origin $\pterr{}=1-\pastro{}$, for a candidate, we model foreground and background event rates using a Poisson mixture formalism~\cite{Farr:2013yna}, as in previous \ac{LVK} results~\cite{Abbott:2020niy,TheLIGOScientific:2016pea,Abbott:2016nhf,Abbott:2016drs,LIGOScientific:2021usb}. 
Technical details of the calculation of \pastro{} for each analysis pipeline are given in Appendix~\ref{sec:p-astro-methods}.

For any candidate, \pastro{} depends on the trigger's ranking statistic and where the trigger lies in the parameter space, i.e., the template with which it was found in a matched-filter analysis, or whether it was found in the low- or high-mass configurations of the \ac{CWB} analysis. 
To calculate \pastro{}, we compare the expected number of astrophysical triggers and the expected number of background triggers for the given ranking statistic and measured parameters. 
The number of true astrophysical signals depends on merger rates, which are jointly inferred as part of the \pastro{} estimation method~\cite{Farr:2013yna,Abbott:2020niy,TheLIGOScientific:2016pea,Kapadia:2019uut}, using assumptions about the populations of astrophysical sources, and the detectors' and analysis' sensitivity, which is calculated using simulated signals. 
The number of background triggers is derived from the same background distribution used to estimate \ac{FAR} by the search analyses. 

As we cannot provide full source-parameter estimates for all candidates with \ac{FAR} $<\SUBTHRESHOLDFAR{}$, we instead estimate the probability to originate from different categories of binary source (\ac{BNS}, \ac{NSBH} and \ac{BBH}).
These probabilities are estimated by each pipeline separately and rely primarily on the template masses with which triggers are recovered (see Appendix~\ref{sec:p-astro-methods} for more details).
For calculating \pastro{}, all triggers from the \ac{CWB} analysis are assumed to be from \ac{BBH} sources as \ac{CWB} has a reduced sensitivity to other population types.
The source classes are defined in this calculation via an assumed boundary at $\PEBHMassThreshold{}$: we consider any component with lower mass to be in the \ac{NS} class and any component above as \ac{BH}. 
These classes do not necessarily reflect the true division between \acp{NS} and \acp{BH}. 
The maximum mass of \acp{NS} is not currently known, but $\PEBHMassThreshold{}$ should be a robust upper limit~\cite{Rhoades:1974fn,Kalogera:1996ci}. 
Therefore, the \ac{BBH} category is intended to capture only \acp{BBH}, while the \ac{BNS} and \ac{NSBH} categories should capture all binaries with components that could be \acp{NS} in addition to possibly capturing some \acp{BBH}.

While the same approach is used by all analyses to assess \pastro{} for their candidates, the detailed implementation varies. 
Besides differences in their ranking statistic definition, analyses divide the parameter space in different ways to compute \pastro{}, make slightly different assumptions about the astrophysical populations, have distinct responses to astrophysical sources, and have specific methods to evaluate their background. 
These differences will introduce a variation in results among pipelines. 
Each pipeline is subject to statistical and systematic uncertainties, such as how they respond to the observed noise fluctuations in ranking candidates, and the differences among pipelines mean that these uncertainties are not the same across pipelines. 
The details of these differences among pipelines are given in Appendix~\ref{sec:searches-methods}. 
There is an extra uncertainty for single-detector candidates, where we can assume a conservative upper bound on \ac{FAR} of $1$ per observing time. 
However, we improve upon this estimate by extrapolating the noise background distribution. 
The \pastro{} values given in Sec.~\ref{sec:candidates} represent our current best estimates of the origin of candidates using the information available from search pipelines and detector characterization.

After its calculation, we must set a threshold on \pastro{} for inclusion in the results presented here.
As in previous publications~\cite{LIGOScientific:2018mvr,LIGOScientific:2021usb}, we choose the criterion $\pastro{} > \PASTROTHRESHOLD{}$, such that the selected candidates are all inferred to have a higher probability of astrophysical origin than terrestrial. 
Values of \pastro{} close to $1$ are expected to be robust with respect to uncertainties in the astrophysical populations, whereas cases for which \pastro{} and \pterr{} are comparable are sensitive to such uncertainties. 
Uncertainties are greater for candidates that, if astrophysical, have properties that correspond to a small number of detections in the overall population. 
The mass distributions for \ac{BBH} sources are now sufficiently well constrained~\cite{Abbott:2020gyp} such that we expect related uncertainties on \pastro{} to be small for the bulk of this region; however, at particularly high masses these uncertainties are expected to be larger~\cite{LIGOScientific:2021usb}. 
In contrast to the \ac{BBH} population, the populations of \ac{BNS} and \ac{NSBH} sources remain poorly known~\cite{LIGOScientific:2021psn}. 
Both the shape and the boundaries of the component mass distributions (especially for \acp{NS}) can have a significant impact on the value of \pastro{} inferred for a \ac{BNS} or \ac{NSBH} candidate, and this uncertainty can be greater than $\PASTROUNCERT$ for moderate \pastro{} values near the threshold of \PASTROTHRESHOLD{}~\cite{Andres:2021vew}.  
We therefore expect that inferred values for \pastro{} may change for less significant candidates as our understanding of the population evolves with further observations~\cite{Gaebel:2018poe,Galaudage:2019jdx,Roulet:2020wyq}.

\subsection{Search results}
\label{sec:candidates}

\begin{event_table}
\begin{table*}
\tiny
% Made by ./scripts/generating_tex_macros/make_event_table
% DO NOT EDIT THIS FILE DIRECTLY

\begin{ruledtabular}
\begin{tabular}{l c ccc ccc ccc ccc ccc}
{Name} & {Inst.} & \multicolumn{3}{c}{{\ac{CWB}}} & \multicolumn{3}{c}{{\GSTLAL{}}} & \multicolumn{3}{c}{{\MBTA{}}} & \multicolumn{3}{c}{{\PYCBCHYPERBANK{}}} & \multicolumn{3}{c}{{\PYCBCBBH{}}} \\
\cline{3-5} \cline{6-8} \cline{9-11} \cline{12-14} \cline{15-17}
 &  & \ac{FAR} & \ac{SNR} & $\pastro{}$  & \ac{FAR} & \ac{SNR} & $\pastro{}$  & \ac{FAR} & \ac{SNR} & $\pastro{}$  & \ac{FAR} & \ac{SNR} & $\pastro{}$  & \ac{FAR} & \ac{SNR} & $\pastro{}$  \\
 &  & ($\mathrm{yr}^{-1}$) &  &  & ($\mathrm{yr}^{-1}$) &  &  & ($\mathrm{yr}^{-1}$) &  &  & ($\mathrm{yr}^{-1}$) &  &  & ($\mathrm{yr}^{-1}$) &  &  \\
\hline
\makebox[0pt][l]{\fboxsep0pt\colorbox{lightgray}{\mystrut\hspace*{1.0\linewidth}}}\!\! {\EVENTNAMEBOLD{GW191103A} \FULLNAME{GW191103A}} & \INSTRUMENTS{GW191103A} &
$ \CWBALLSKYMEETSPASTROTHRESH{GW191103A} \CWBALLSKYFAR{GW191103A}$ &
$ \CWBALLSKYMEETSPASTROTHRESH{GW191103A} \CWBALLSKYSNR{GW191103A}$ &
$ \CWBALLSKYMEETSPASTROTHRESH{GW191103A} \CWBALLSKYPASTRO{GW191103A}$ &
$ \GSTLALALLSKYMEETSPASTROTHRESH{GW191103A} \GSTLALALLSKYFAR{GW191103A}$ &
$ \GSTLALALLSKYMEETSPASTROTHRESH{GW191103A} \GSTLALALLSKYSNR{GW191103A}$ &
$ \GSTLALALLSKYMEETSPASTROTHRESH{GW191103A} \GSTLALALLSKYPASTRO{GW191103A}$ &
$ \MBTAALLSKYMEETSPASTROTHRESH{GW191103A} \MBTAALLSKYFAR{GW191103A}$ &
$ \MBTAALLSKYMEETSPASTROTHRESH{GW191103A} \MBTAALLSKYSNR{GW191103A}$ &
$ \MBTAALLSKYMEETSPASTROTHRESH{GW191103A} \MBTAALLSKYPASTRO{GW191103A}$ &
$ \PYCBCALLSKYMEETSPASTROTHRESH{GW191103A} \PYCBCALLSKYFAR{GW191103A}$ &
$ \PYCBCALLSKYMEETSPASTROTHRESH{GW191103A} \PYCBCALLSKYSNR{GW191103A}$ &
$ \PYCBCALLSKYMEETSPASTROTHRESH{GW191103A} \PYCBCALLSKYPASTRO{GW191103A}$ &
$ \PYCBCHIGHMASSMEETSPASTROTHRESH{GW191103A} \PYCBCHIGHMASSFAR{GW191103A}$ &
$ \PYCBCHIGHMASSMEETSPASTROTHRESH{GW191103A} \PYCBCHIGHMASSSNR{GW191103A}$ &
$ \PYCBCHIGHMASSMEETSPASTROTHRESH{GW191103A} \PYCBCHIGHMASSPASTRO{GW191103A} $ \\
\end{tabular}
\end{ruledtabular}

\caption{
\label{tab:events}
Candidate \ac{GW} signals.
The time (\ac{UTC}) of the signal is encoded in the name as GWYYMMDD\_hhmmss (e.g., \FULLNAME{GW200112H}{} occurred on \DATE{GW200112H}{} at \TIME{GW200112H}).
The names of candidates not previously reported are given in \textbf{bold}.
The detectors observing at the merger time of the candidate are indicated using single-letter identifiers (e.g., H for \ac{LIGO} Hanford); these are not necessarily the same detectors that contributed triggers associated with the candidate. 
Where a candidate was found with \pastro{} above the threshold value of $\PASTROTHRESHOLD{}$ by at least one analysis but below the threshold by others, we include in \textit{italics} the results from the other analyses, where available. 
A dash (--) indicates that a candidate was not found by an analysis.
The \NUMSNGLS{} candidates labeled with a dagger ($\dagger$) were found only above threshold in a single detector with the \GSTLAL{} analysis, and the \ac{FAR} estimates were made using significant extrapolation of the background data, meaning that single-detector candidates have higher uncertainty than coincident candidates.
A conservative estimate of the \ac{FAR} for these single-detector candidates is one per live time of the analysis; this is $\sim \HANFORDYEARSINV{}~\mathrm{yr}^{-1}$ for both \ac{LIGO} Hanford and \ac{LIGO} Livingston.
}
\end{table*}
\end{event_table}

There are many potential \ac{GW} sources. 
Hence, in theory, \GWTCTHREE{} could contain a variety of source types. 
However, currently no high-significance (\ac{FAR} $<\FARTHRESHOLD{}~\mathrm{yr^{-1}}$) candidate transients have been reported for sources other than standard, quasicircular \acp{CBC}~\cite{Abbott:2021ksc,LIGOScientific:2021tfm,LIGOScientific:2021hoh,LIGOScientific:2021uyj,LIGOScientific:2021iyk,LIGOScientific:2022jpr}.
Therefore, we limit this \GWTCTHREE{} candidate list to the established source categories of \acp{BNS}, \acp{NSBH} and \acp{BBH}.

Following \GWTCTWOFINAL{}~\cite{LIGOScientific:2021usb}, we select candidates with a probability of an astrophysical \ac{CBC} source $\pastro{} > \PASTROTHRESH{}$ for detailed analysis. 
In applying this criterion, we follow the method used in \GWTCONE{}~\cite{LIGOScientific:2018mvr} and consider only \ac{CWB} candidates that also have a \ac{BBH} counterpart from one of the matched-filter analyses (i.e., a time-coincident candidate with $\pastro{} > \CBCTHRESHOLD{}$).
This is because \ac{CWB} can potentially identify signals from a range of sources, but the calculation of \pastro{} assumes a \ac{CBC} source, and so additional confirmation is needed to verify that the candidate signal is consistent with a \ac{CBC} origin. 
However, all \ac{O3b} \ac{CWB} candidates with $\pastro{} > \PASTROTHRESH{}$ also have $\pastro{} > \PASTROTHRESH{}$ from a matched-filter analysis anyway, except for \FULLNAME{200214K}, which is identified as being of instrumental origin~\cite{LIGOScientific:2021tfm}. 
The requirement that \ac{CWB} candidates have a matched-filter counterpart is discussed further in Appendix~\ref{sec:cwb-only-events}.

We identify \NUMEVENTS{} \ac{CBC} candidates in \ac{O3b} passing our threshold; these include \NEWEVENTS{} new candidates that were not found in low latency and are reported here for the first time. 
Significance estimates for the \ac{CBC} candidates with probability of astrophysical origin $\pastro{} > \PASTROTHRESH{}$ are reported in Table~\ref{tab:events}. 
We report the \ac{FAR}, \ac{SNR} and \pastro{} for each search analysis that finds a trigger when at least one analysis finds the candidate above the threshold for inclusion. 
Additionally, the \acp{SNR} reported from each detector are given in Table~\ref{tab:sngl_ifo_snr} of Appendix~\ref{sec:search_results_appendix}.  
By comparing the sum of \pastro{} values for candidates with $\pastro{} > \PASTROTHRESHOLD{}$ to the number of such candidates for each analysis, we estimate that the expected contamination from triggers of terrestrial origin is $\sim \CONTAMINATIONFRACTION$, or $\sim \CONTAMINATION$ candidates. 
A higher-purity selection of candidates could be obtained by adopting a stricter selection criterion; for example, adopting a threshold of $\pastro{}>\PASTROTHRESHOLDPOINTNINE{}$ would result in a list of $\ALLABOVEPASTROPOINTNINE{}$ \ac{O3b} candidates. 
Probabilities for different source categories (\ac{BNS}, \ac{NSBH} and \ac{BBH}) are included in Table~\ref{tab:multi_pastro} in Appendix~\ref{sec:p-astro-methods}. 
Updated values for $\pastro$ for \ac{O3a} candidates are given in Table~\ref{tab:o3a_pastro} in Appendix~\ref{sec:p-astro-methods}; there is no change to the list of \ac{O3a} candidates with $\pastro{} > \PASTROTHRESH{}$ compared with \GWTCTWOFINAL{}~\cite{LIGOScientific:2021usb}. 
Results from \ac{O1} and \ac{O2} have not been recalculated~\cite{LIGOScientific:2018mvr}. 
The \ac{O3b} candidates bring the total number of \ac{LVK}-reported \ac{CBC} candidates with $\pastro{} > \PASTROTHRESH{}$ to \TOTALEVENTS{}. 

Marginal candidates with $\pastro{} < \PASTROTHRESH{}$ but \ac{FAR} $< \FARTHRESHYR{}~\mathrm{yr^{-1}}$ are discussed further in Sec.~\ref{sec:marginal}.
An extended list of candidates with \ac{FAR} $< \SUBTHRESHOLDFAR$ is available from \ac{GWOSC}~\cite{gwosc:gwtc3}, and discussed in Sec.~\ref{sec:subthreshold}.

\subsubsection{\ac{O3b} online candidates}
\label{sec:online-candidates}

In \ac{O3b}, there were \NUMPUBLICEVENTS{} candidates reported in low latency (see Appendix~\ref{sec:follow-up}).
All candidates identified by the online searches are assigned an internal identifier according to the date on which they occur, for example, \SNAME{200105F}{} for \FULLNAME{200105F}{}.
These online analyses were carried out by the five pipelines: \GSTLAL{}, \MBTAONLINE{}, \PYCBCLIVE{}, \ac{SPIIR} and \ac{CWB}. 
The overall \ac{FAR} threshold for a public alert was set to \OPAFARTHRESHMONTH{} (\OPAFARTHRESHYR{}) for \ac{CBC} sources, meaning that once a trials factor is applied, there was a public-alert threshold of \OPAFARTHRESH{} for each online pipeline.
Candidates found in low latency passing this threshold were disseminated to the public via \ac{GCN} Notices and Circulars.
This allowed for rapid follow-up searching for multimessenger counterparts. 
The online searches are necessarily limited in assessing the noise background as they can only use data collected up to the current time, and hence the \ac{FAR} may be inaccurately calculated if there are sudden changes in the data quality. 
Among the \NUMPUBLICEVENTS{} candidates reported in low latency, \RETRACTIONS{} were later retracted as they were likely due to detector noise. 

None of the \RETRACTIONS{} retracted online candidates were found above our \pastro{} threshold in the offline analyses, and thus are not included in Table~\ref{tab:events}.
There were \NUMPUBLICNOTRECOVERED{} public candidates that did not meet the threshold for inclusion in Table~\ref{tab:events} that were not retracted: 
\begin{itemize}
\item S191205ah was found in low latency by \GSTLAL{} as a low-\ac{SNR} ($\rho < \LOWSNR$) single-detector candidate in \ac{LIGO} Livingston with a \ac{FAR} of $\MISSEDPUBLICEVENTFAR{S191205ah}{}~\mathrm{yr^{-1}}$. 
Such a \ac{FAR} corresponds to modest significance, and thus it is not surprising to find differences in the estimated significance by the initial online analysis and the end-of-run offline analyses.
\GSTLAL{} did not recover an offline trigger at this time with \ac{FAR} $< \SUBTHRESHOLDFAR$. 
\item S191213g was found in low latency by \GSTLAL{} in both \ac{LIGO} Hanford and \ac{LIGO} Livingston, with low network \ac{SNR} and a modest \ac{FAR} of $\MISSEDPUBLICEVENTFAR{S191213g}{}~\mathrm{yr^{-1}}$. 
The offline trigger corresponding to this time was found with \ac{FAR} $> \SUBTHRESHOLDFAR$, so it is not included in this catalog. 
\item The \ac{NSBH} candidate \SNAME{200105F}{} (\FULLNAME{200105F}{}~\cite{LIGOScientific:2021qlt}) is reported as a marginal candidate (see Table~\ref{tab:marginal_events}) and is further discussed below.
\item The \ac{CWB} candidate S200114f was found online in the \ac{HLV} three-detector network with \ac{FAR} of $\MISSEDPUBLICEVENTFAR{S200114f}{}~\mathrm{yr^{-1}}$, meeting the significance threshold for a public alert. 
It was considered for inclusion in the \ac{O3} search for short-duration minimally modeled transients~\cite{LIGOScientific:2021hoh}, but that analysis was uniformly carried out on the \ac{HL} network, where the trigger did not qualify because of its low coherence (\ac{CWB} network correlation coefficient $c_\mathrm{c} < \CCThreshold$). 
This candidate was discussed at length in the context of the search for \ac{IMBH} binaries, where a potential instrumental origin was examined~\cite{LIGOScientific:2021tfm}.
The analysis for the \ac{IMBH} search was carried out using both the \ac{HL} and \ac{HLV} networks, and this candidate came out as marginally significant in the \ac{HLV} network.
In the analysis done for this catalog, this candidate was reported only by the \ac{CWB} pipeline (which performed a two-detector analysis). 
Since the \ac{CWB} \pastro{} is low (\CWBALLSKYPASTRO{200114A}), it does not meet the criteria for inclusion in Table~\ref{tab:events}.
\item S200213t was found in low latency by \GSTLAL{} as a low-\ac{SNR} single-detector candidate in \ac{LIGO} Hanford with a modest \ac{FAR} of $\MISSEDPUBLICEVENTFAR{S200213t}{}~\mathrm{yr^{-1}}$.
Similar to S191205ah and S191213g, there was no offline trigger corresponding to S200213t with a \ac{FAR} $< \SUBTHRESHOLDFAR$, so it does not appear in this catalog. 
Single-detector candidates, such as S191205ah and S200213t, are particularly susceptible to changes in significance due to relatively minor changes in data processing. 
\end{itemize}
The remaining \PREVIOUSLYREPORTED{} candidates reported in low-latency also appear in Table~\ref{tab:events}.

\subsubsection{New \ac{O3b} candidates}
\label{sec:offline-candidates}

The \NEWEVENTS{} new candidates listed in this catalog, not previously shared via \ac{GCN}, are indicated in bold in Table~\ref{tab:events}.
Almost all of these candidates are found with modest significance.
They are all coincident triggers involving at least both of the \ac{LIGO} Hanford and Livingston detectors.
The inferred source properties for all the new candidates (discussed in Sec.~\ref{sec:parameter-estimation}) are consistent with \ac{BBH} masses, with the exceptions of \FULLNAME{GW191219E}{} and \FULLNAME{GW200210B}{} that may be from \acp{NSBH}.

The identification of these new candidates can be attributed to a combination of factors: (i) offline searches benefit from data with better calibration, cleaning and data-quality information, as well as improved algorithms, resulting in better background rejection, and (ii) using a \pastro{} threshold allows us to highlight candidates in source-rich parts of the parameter space, including candidates with an (offline) \ac{FAR} that would not meet the (online) threshold for public alerts.

\subsubsection{Pipeline consistency}
\label{sec:pipeline-comparison}

Not all candidates were found by all pipelines above the \pastro{} threshold of \PASTROTHRESH{}: of the \NUMEVENTS{} candidates, \NUMCWB{} candidates were found by \ac{CWB}, \NUMGSTLAL{} candidates were found by \GSTLAL{} (including the \NUMSNGLS{} candidates found in a single detector), \NUMMBTA{} candidates were found by \ac{MBTA}, and \NUMPYCBC{} candidates were found by one or both of the \PYCBCBBH{} and \PYCBCHYPERBANK{} analyses.
Among the \ac{O3b} candidates, \TWOPIPES{} were found by two or more analysis pipelines, \THREEPIPES{} by three or more pipelines, and \ALLPIPES{} by all pipelines.
We expect the analyses to find different sets of candidates, due to different search methods, tuning, and configuration choices.
The impact of differences among search pipelines will be largest for candidates with low \ac{SNR}; thus, it is expected that such candidates may be identified by only a subset of pipelines.
As methods used by different pipelines will be more or less effective in suppressing specific types of noise artifacts, and the sensitivity of different pipelines will have different dependencies on binary signal parameters, combining information from multiple pipelines should lead to a greater understanding of the population of astrophysical sources~\cite{Banagiri:2023ztt,Sutton:2009up,Biswas:2012ty}. 

Some candidates are unique to a pipeline and not found by other pipelines:
\begin{itemize}
\item The \GSTLAL{} analysis found \NUMUNIQUEGSTLALEVENTS{} unique candidates (see Appendix~\ref{sec:search_results_appendix}); these are both single-detector candidates which had also been reported in low latency. 
As only \GSTLAL{} is configured to identify single-detector signals, we expect a difference among pipelines here.
\item The \MBTA{} analysis found \NUMUNIQUEMBTAEVENTS{} unique candidates, newly reported here, all of which are quiet signals inferred to be from \acp{BBH}. 
These candidates have $\pastro{} > \PASTROTHRESH{}$ even though their \ac{FAR} (integrated over a large parameter space) is high (see Appendix~\ref{sec:p-astro-methods} for further discussion), and their \pterr{} is also significant. 
\FULLNAME{GW191113B}{} may have an unusual mass ratio, and \FULLNAME{GW200322G}{} has significant uncertainties for its inferred source properties (see Sec.~\ref{sec:parameter-estimation}), which may make these signals (if real \acp{GW}) outliers in the astrophysical population; therefore, the \pastro{} for these candidates is more uncertain than for more typical candidates~\cite{Andres:2021vew}.
\item The \PYCBC{} analyses found \NUMUNIQUEPYCBCEVENTS{} unique candidates, all of which are newly reported in this catalog.
Of these, \NUMUNIQUEPYCBCSHAREDEVENTS{} were found by both analyses, \NUMUNIQUEPYCBCBBHEVENTS{} were found in the \PYCBCBBH{} analysis and \NUMUNIQUEPYCBCHYPERBANKEVENTS{} in the \PYCBCHYPERBANK{} analysis.
All the candidates found uniquely by \PYCBC{} are relatively quiet. 
The lowest \ac{FAR}, and therefore most significant, is that of \FULLNAME{GW191103A}: $\PYCBCHIGHMASSFAR{GW191103A}{}~\mathrm{yr}^{-1}$.

The candidate found only by the \PYCBCHYPERBANK{} analysis, \FULLNAME{GW191219E}{}, was found as a potential \ac{NSBH} candidate, with a mass ratio of $\PYCBCALLSKYMASSRATIO{GW191219E}$.
The relatively large asymmetry in the component masses and low mass of the secondary component as identified by the search, $\PYCBCALLSKYMASSTWO{GW191219E} \Msun{}$, meant that the template was not analyzed in the \PYCBCBBH{} analysis.
\FULLNAME{GW191219E}{}, with redshifted chirp mass $\PYCBCALLSKYCHIRPMASS{GW191219E} \Msun$, is included in the same mass bin as the population of significant \ac{BBH} candidates for the estimation of event rates entering \pastro{} (see Appendix~\ref{sec:p-astro-methods}). 
Such a simple binning scheme implies significant modeling uncertainty in \pastro{} for candidates with parameters outside known populations: for instance, with a minor change in bin boundaries that puts the candidate in a different bin from the \ac{BBH} population, its \pastro{} would drop to \PYCBCPASTROALTERNATIVEMCHIRPBOUND{}. 
This example illustrates the sensitivity of $\pastro$ calculations to the assumed astrophysical population. 
For candidates at the edges of (or outside of) the confidently detected populations, like \FULLNAME{GW191219E}{}, there may be large, model-dependent systematic uncertainties in $\pastro{}$.
Future observations will reduce the uncertainty in the rate of similar mergers, and thus enable us to better quantify the origin of \FULLNAME{GW191219E}{}.
\end{itemize}

Despite its high \ac{SNR}, \FULLNAME{GW200129D}{} was identified only by a subset of the search analyses due to a specific set of circumstances.
A data-quality issue in Livingston was reported through active Burst and \ac{CBC} \CATTWO{} flags (and required mitigation, as described in Appendix~\ref{sec:data-methods}).
The \CATTWO{} flags mean that the Livingston data were ignored by the \CWB{}, \MBTA{} and \PYCBC{} analyses.
Moreover, in the \MBTA{} analysis the combination of signal and noise was loud enough to trigger gating in Hanford, but not loud enough in Virgo to create a \ac{HV} coincidence in the high-threshold analysis performed without gating (see Appendix~\ref{sec:mbta-methods} for details about the internal gating procedure used to remove suspected artifacts in the data).
The \PYCBC{} analyses still identified a candidate using only the \ac{HV} data, but the network \ac{SNR} is lower than reported by \GSTLAL{} on account of not including the Livingston data.
In the \CWB{} analysis, the trigger was reconstructed in the \ac{HV} network but was rejected by the postproduction cuts.
The differences in data handling among analyses are expected to lead to such differences in uncommon cases like this. 

\FULLNAME{GW191109A}{}, \FULLNAME{GW200208K}{} and \FULLNAME{GW200220E}{} are candidates with high-mass sources that potentially make them also relevant in the context of the search for \ac{IMBH} binaries~\cite{LIGOScientific:2021tfm}. 
\FULLNAME{GW191109A}{} is a highly significant candidate that was also found in that \ac{IMBH} binary search with a \ac{FAR} as low as \IMBHMINFAR{}, but has a joint posterior distribution for the primary and remnant masses that does not match the strict criteria to be considered as an \ac{IMBH} binary~\cite{LIGOScientific:2021tfm} (see Sec.~\ref{sec:parameter-estimation}). 
\FULLNAME{GW200208K}{} and \FULLNAME{GW200220E}{} are low-\ac{SNR} candidates, which were not identified as significant in the \ac{IMBH} search; this difference is likely due to different choices of ranking statistic between the two searches as well as differences between their noise backgrounds arising from a different parameter space.

\subsubsection{Marginal candidates and \FULLNAME{200105F}{}}
\label{sec:marginal}
\begin{table*}
\tiny
% Made by ./scripts/generating_tex_macros/make_marginal_event_table
% DO NOT EDIT THIS FILE DIRECTLY

\begin{ruledtabular}
\begin{tabular}{l c ccc ccc ccc ccc}
{Name} & {Inst.} & \multicolumn{3}{c}{{\ac{CWB}}} & \multicolumn{3}{c}{{\GSTLAL{}}} & \multicolumn{3}{c}{{\MBTA{}}} & \multicolumn{3}{c}{{\PYCBCHYPERBANK{}}} \\
\cline{3-5} \cline{6-8} \cline{9-11} \cline{12-14}
 &  & \ac{FAR} & \ac{SNR} & $\pastro{}$  & \ac{FAR} & \ac{SNR} & $\pastro{}$  & \ac{FAR} & \ac{SNR} & $\pastro{}$  & \ac{FAR} & \ac{SNR} & $\pastro{}$  \\
 &  & {($\mathrm{yr}^{-1}$)} &  &  & {($\mathrm{yr}^{-1}$)} &  &  & {($\mathrm{yr}^{-1}$)} &  &  & {($\mathrm{yr}^{-1}$)} &  &  \\
\hline
\makebox[0pt][l]{\fboxsep0pt\colorbox{lightgray}{\mystrut\hspace*{1.000000\linewidth}}}\!\! {\EVENTNAMEBOLD{191118N} \FULLNAME{191118N} } & \INSTRUMENTS{191118N} &
$\CWBALLSKYMEETSFARTHRESH{191118N} \CWBALLSKYFAR{191118N}$ &
$\CWBALLSKYMEETSFARTHRESH{191118N} \CWBALLSKYSNR{191118N}$ &
$\CWBALLSKYMEETSFARTHRESH{191118N} \CWBALLSKYPASTRO{191118N}$ &
$\GSTLALALLSKYMEETSFARTHRESH{191118N} \GSTLALALLSKYFAR{191118N}$ &
$\GSTLALALLSKYMEETSFARTHRESH{191118N} \GSTLALALLSKYSNR{191118N}$ &
$\GSTLALALLSKYMEETSFARTHRESH{191118N} \GSTLALALLSKYPASTRO{191118N}$ &
$\MBTAALLSKYMEETSFARTHRESH{191118N} \MBTAALLSKYFAR{191118N}$ &
$\MBTAALLSKYMEETSFARTHRESH{191118N} \MBTAALLSKYSNR{191118N}$ &
$\MBTAALLSKYMEETSFARTHRESH{191118N} \MBTAALLSKYPASTRO{191118N}$ &
$\PYCBCALLSKYMEETSFARTHRESH{191118N} \PYCBCALLSKYFAR{191118N}$ &
$\PYCBCALLSKYMEETSFARTHRESH{191118N} \PYCBCALLSKYSNR{191118N}$ &
$\PYCBCALLSKYMEETSFARTHRESH{191118N} \PYCBCALLSKYPASTRO{191118N}$ \\
\makebox[0pt][l]{\fboxsep0pt{\mystrut\hspace*{1.000000\linewidth}}}\!\! {\EVENTNAMEBOLD{200105F} \FULLNAME{200105F} } & \INSTRUMENTS{200105F} &
$\CWBALLSKYMEETSFARTHRESH{200105F} \CWBALLSKYFAR{200105F}$ &
$\CWBALLSKYMEETSFARTHRESH{200105F} \CWBALLSKYSNR{200105F}$ &
$\CWBALLSKYMEETSFARTHRESH{200105F} \CWBALLSKYPASTRO{200105F}$ &
$\GSTLALALLSKYMEETSFARTHRESH{200105F} \GSTLALALLSKYFAR{200105F}$ &
$\GSTLALALLSKYMEETSFARTHRESH{200105F} \GSTLALALLSKYSNR{200105F}$ &
$\GSTLALALLSKYMEETSFARTHRESH{200105F} \GSTLALALLSKYPASTRO{200105F}$ &
$\MBTAALLSKYMEETSFARTHRESH{200105F} \MBTAALLSKYFAR{200105F}$ &
$\MBTAALLSKYMEETSFARTHRESH{200105F} \MBTAALLSKYSNR{200105F}$ &
$\MBTAALLSKYMEETSFARTHRESH{200105F} \MBTAALLSKYPASTRO{200105F}$ &
$\PYCBCALLSKYMEETSFARTHRESH{200105F} \PYCBCALLSKYFAR{200105F}$ &
$\PYCBCALLSKYMEETSFARTHRESH{200105F} \PYCBCALLSKYSNR{200105F}$ &
$\PYCBCALLSKYMEETSFARTHRESH{200105F} \PYCBCALLSKYPASTRO{200105F}$ \\
\makebox[0pt][l]{\fboxsep0pt\colorbox{lightgray}{\mystrut\hspace*{1.000000\linewidth}}}\!\! {\EVENTNAMEBOLD{200121A} \FULLNAME{200121A} $^{*}$} & \INSTRUMENTS{200121A} &
$\CWBALLSKYMEETSFARTHRESH{200121A} \CWBALLSKYFAR{200121A}$ &
$\CWBALLSKYMEETSFARTHRESH{200121A} \CWBALLSKYSNR{200121A}$ &
$\CWBALLSKYMEETSFARTHRESH{200121A} \CWBALLSKYPASTRO{200121A}$ &
$\GSTLALALLSKYMEETSFARTHRESH{200121A} \GSTLALALLSKYFAR{200121A}$ &
$\GSTLALALLSKYMEETSFARTHRESH{200121A} \GSTLALALLSKYSNR{200121A}$ &
$\GSTLALALLSKYMEETSFARTHRESH{200121A} \GSTLALALLSKYPASTRO{200121A}$ &
$\MBTAALLSKYMEETSFARTHRESH{200121A} \MBTAALLSKYFAR{200121A}$ &
$\MBTAALLSKYMEETSFARTHRESH{200121A} \MBTAALLSKYSNR{200121A}$ &
$\MBTAALLSKYMEETSFARTHRESH{200121A} \MBTAALLSKYPASTRO{200121A}$ &
$\PYCBCALLSKYMEETSFARTHRESH{200121A} \PYCBCALLSKYFAR{200121A}$ &
$\PYCBCALLSKYMEETSFARTHRESH{200121A} \PYCBCALLSKYSNR{200121A}$ &
$\PYCBCALLSKYMEETSFARTHRESH{200121A} \PYCBCALLSKYPASTRO{200121A}$ \\
\makebox[0pt][l]{\fboxsep0pt{\mystrut\hspace*{1.000000\linewidth}}}\!\! {\EVENTNAMEBOLD{200201F} \FULLNAME{200201F} } & \INSTRUMENTS{200201F} &
$\CWBALLSKYMEETSFARTHRESH{200201F} \CWBALLSKYFAR{200201F}$ &
$\CWBALLSKYMEETSFARTHRESH{200201F} \CWBALLSKYSNR{200201F}$ &
$\CWBALLSKYMEETSFARTHRESH{200201F} \CWBALLSKYPASTRO{200201F}$ &
$\GSTLALALLSKYMEETSFARTHRESH{200201F} \GSTLALALLSKYFAR{200201F}$ &
$\GSTLALALLSKYMEETSFARTHRESH{200201F} \GSTLALALLSKYSNR{200201F}$ &
$\GSTLALALLSKYMEETSFARTHRESH{200201F} \GSTLALALLSKYPASTRO{200201F}$ &
$\MBTAALLSKYMEETSFARTHRESH{200201F} \MBTAALLSKYFAR{200201F}$ &
$\MBTAALLSKYMEETSFARTHRESH{200201F} \MBTAALLSKYSNR{200201F}$ &
$\MBTAALLSKYMEETSFARTHRESH{200201F} \MBTAALLSKYPASTRO{200201F}$ &
$\PYCBCALLSKYMEETSFARTHRESH{200201F} \PYCBCALLSKYFAR{200201F}$ &
$\PYCBCALLSKYMEETSFARTHRESH{200201F} \PYCBCALLSKYSNR{200201F}$ &
$\PYCBCALLSKYMEETSFARTHRESH{200201F} \PYCBCALLSKYPASTRO{200201F}$ \\
\makebox[0pt][l]{\fboxsep0pt\colorbox{lightgray}{\mystrut\hspace*{1.000000\linewidth}}}\!\! {\EVENTNAMEBOLD{200214K} \FULLNAME{200214K} $^{*}$} & \INSTRUMENTS{200214K} &
$\CWBALLSKYMEETSFARTHRESH{200214K} \CWBALLSKYFAR{200214K}$ &
$\CWBALLSKYMEETSFARTHRESH{200214K} \CWBALLSKYSNR{200214K}$ &
$\CWBALLSKYMEETSFARTHRESH{200214K} \CWBALLSKYPASTRO{200214K}$ &
$\GSTLALALLSKYMEETSFARTHRESH{200214K} \GSTLALALLSKYFAR{200214K}$ &
$\GSTLALALLSKYMEETSFARTHRESH{200214K} \GSTLALALLSKYSNR{200214K}$ &
$\GSTLALALLSKYMEETSFARTHRESH{200214K} \GSTLALALLSKYPASTRO{200214K}$ &
$\MBTAALLSKYMEETSFARTHRESH{200214K} \MBTAALLSKYFAR{200214K}$ &
$\MBTAALLSKYMEETSFARTHRESH{200214K} \MBTAALLSKYSNR{200214K}$ &
$\MBTAALLSKYMEETSFARTHRESH{200214K} \MBTAALLSKYPASTRO{200214K}$ &
$\PYCBCALLSKYMEETSFARTHRESH{200214K} \PYCBCALLSKYFAR{200214K}$ &
$\PYCBCALLSKYMEETSFARTHRESH{200214K} \PYCBCALLSKYSNR{200214K}$ &
$\PYCBCALLSKYMEETSFARTHRESH{200214K} \PYCBCALLSKYPASTRO{200214K}$ \\
\makebox[0pt][l]{\fboxsep0pt{\mystrut\hspace*{1.000000\linewidth}}}\!\! {\EVENTNAMEBOLD{200219K} \FULLNAME{200219K} $^{*}$} & \INSTRUMENTS{200219K} &
$\CWBALLSKYMEETSFARTHRESH{200219K} \CWBALLSKYFAR{200219K}$ &
$\CWBALLSKYMEETSFARTHRESH{200219K} \CWBALLSKYSNR{200219K}$ &
$\CWBALLSKYMEETSFARTHRESH{200219K} \CWBALLSKYPASTRO{200219K}$ &
$\GSTLALALLSKYMEETSFARTHRESH{200219K} \GSTLALALLSKYFAR{200219K}$ &
$\GSTLALALLSKYMEETSFARTHRESH{200219K} \GSTLALALLSKYSNR{200219K}$ &
$\GSTLALALLSKYMEETSFARTHRESH{200219K} \GSTLALALLSKYPASTRO{200219K}$ &
$\MBTAALLSKYMEETSFARTHRESH{200219K} \MBTAALLSKYFAR{200219K}$ &
$\MBTAALLSKYMEETSFARTHRESH{200219K} \MBTAALLSKYSNR{200219K}$ &
$\MBTAALLSKYMEETSFARTHRESH{200219K} \MBTAALLSKYPASTRO{200219K}$ &
$\PYCBCALLSKYMEETSFARTHRESH{200219K} \PYCBCALLSKYFAR{200219K}$ &
$\PYCBCALLSKYMEETSFARTHRESH{200219K} \PYCBCALLSKYSNR{200219K}$ &
$\PYCBCALLSKYMEETSFARTHRESH{200219K} \PYCBCALLSKYPASTRO{200219K}$ \\
\makebox[0pt][l]{\fboxsep0pt\colorbox{lightgray}{\mystrut\hspace*{1.000000\linewidth}}}\!\! {\EVENTNAMEBOLD{200311H} \FULLNAME{200311H} } & \INSTRUMENTS{200311H} &
$\CWBALLSKYMEETSFARTHRESH{200311H} \CWBALLSKYFAR{200311H}$ &
$\CWBALLSKYMEETSFARTHRESH{200311H} \CWBALLSKYSNR{200311H}$ &
$\CWBALLSKYMEETSFARTHRESH{200311H} \CWBALLSKYPASTRO{200311H}$ &
$\GSTLALALLSKYMEETSFARTHRESH{200311H} \GSTLALALLSKYFAR{200311H}$ &
$\GSTLALALLSKYMEETSFARTHRESH{200311H} \GSTLALALLSKYSNR{200311H}$ &
$\GSTLALALLSKYMEETSFARTHRESH{200311H} \GSTLALALLSKYPASTRO{200311H}$ &
$\MBTAALLSKYMEETSFARTHRESH{200311H} \MBTAALLSKYFAR{200311H}$ &
$\MBTAALLSKYMEETSFARTHRESH{200311H} \MBTAALLSKYSNR{200311H}$ &
$\MBTAALLSKYMEETSFARTHRESH{200311H} \MBTAALLSKYPASTRO{200311H}$ &
$\PYCBCALLSKYMEETSFARTHRESH{200311H} \PYCBCALLSKYFAR{200311H}$ &
$\PYCBCALLSKYMEETSFARTHRESH{200311H} \PYCBCALLSKYSNR{200311H}$ &
$\PYCBCALLSKYMEETSFARTHRESH{200311H} \PYCBCALLSKYPASTRO{200311H}$ \\
\end{tabular}
\end{ruledtabular}

\caption{
\label{tab:marginal_events}
Marginal candidates found by the various analyses.
The candidates in this table have a \ac{FAR} below a threshold of $\FARTHRESHYR{}~\mathrm{yr^{-1}}$ in at least one analysis, but were not found with $\pastro{}$ that meets our threshold for Table~\ref{tab:events} ($\pastro{} > \PASTROTHRESH{}$ from a search analysis, with the additional requirement that \ac{CWB} candidates have a counterpart from a matched-filter analysis). 
The probability of astrophysical origin \pastro{} quoted (i) assumes a \ac{CBC} source, which may not always be applicable for candidates identified by the minimally modeled \ac{CWB} analysis, and (ii) do not factor in data-quality information that was not used by the search algorithms.
Further information on the \ac{CWB}-only candidate \FULLNAME{200214K}{} is available in Appendix~\ref{sec:cwb-only-events}. 
Detector-identifying letters are the same as given in Table~\ref{tab:events}.
The instruments for each candidate are the ones which were operating at the time of the trigger, and are not necessarily the same as those which participated in the detection. 
The candidates are named according to the same convention as in Table~\ref{tab:events} except that here we omit the \ac{GW} prefix for the candidates found to be likely caused by instrumental artifacts, indicated with an asterisk ($\ast$).
Where a candidate was seen below the \ac{FAR} threshold in at least one analysis but above threshold in others, we include in \textit{italics} the information on that trigger from the other analyses as well where available.
As in Table~\ref{tab:events}, the dagger ($\dagger$) indicates a candidate found by a single detector with the \GSTLAL{} analysis.
}
\end{table*}

In Table~\ref{tab:marginal_events} we report the marginal candidates that are found by each analysis below a \ac{FAR} threshold of $\FARTHRESHYR{}~\mathrm{yr^{-1}}$ but do not satisfy the \pastro{} threshold for inclusion in Table~\ref{tab:events}.
The naming of these marginal candidates follows the same YYMMDD\_hhmmss format as that described for the candidates of Table~\ref{tab:events}, except omitting the \ac{GW} prefix for the two candidates found to be caused by instrumental artifacts; for the other marginal candidates, we cannot exclude the possibility that they are quiet \ac{GW} signals.

The marginal candidates \FULLNAME{200121A}{}, \FULLNAME{200214K}{} and \FULLNAME{200219K}{} were found to be likely caused by instrument artifacts.
At the time of \FULLNAME{200121A}{}, \ac{LIGO} Hanford data contain excess power consistent with a \BLIP{} glitch, a common glitch in \ac{LIGO} detector data~\cite{Davis:2021ecd,Cabero:2019orq}.
At the time of \FULLNAME{200214K}{}, \ac{LIGO} Livingston data contained significant excess noise due to \FASTSCATTER{}, while \ac{LIGO} Hanford data showed evidence for a weak scattering arch; this candidate was further examined in the search for \ac{IMBH} binaries~\cite{LIGOScientific:2021tfm}, and is discussed in Appendix~\ref{sec:cwb-only-events}. 
At the time of \FULLNAME{200219K}{}, \ac{LIGO} Hanford data are highly nonstationary, with multiple loud glitches visible within $\DQWINDOW{}$ of the candidate time.

The marginal candidate \FULLNAME{200311H}{} is found by both \ac{MBTA} and \PYCBCHYPERBANK{} with a template consistent with a (redshifted) chirp mass of $\MBTAALLSKYCHIRPMASS{200311H} \Msun$ in both pipelines, and hence, if it were an astrophysical signal, its source would correspond to a \ac{BNS}. 
Its chirp mass is close to that of GW170817~\cite{TheLIGOScientific:2017qsa} and is consistent with Galactic \acp{BNS}~\cite{Farrow:2019xnc}.
Future observations will better constrain the mass distribution of \ac{BNS} mergers and thus enable a more accurate assessment of the origin of this candidate. 

The \ac{NSBH} candidate \FULLNAME{200105F}{}~\cite{LIGOScientific:2021qlt} was found as a single-detector trigger by \GSTLAL{} with a \ac{FAR} of $\MINFAR{200105F}$. 
This is comparable to the previously published value of \NSBHPREVREPORTEDFAR~\cite{LIGOScientific:2021qlt}, which used only data from the beginning of \ac{O3b} until \NSBHPAPEROBSERVINGEND.
\acp{FAR} are not assigned to single-detector triggers by the versions of the \PYCBC{} and \MBTA{} analyses used for these results (more recent developments do allow significance estimates for single-detector triggers in \PYCBC searches~\cite{Nitz:2020naa,Davies:2022thw}); however, \FULLNAME{200105F}{} was also seen by the \PYCBCHYPERBANK{} and \MBTA{} analyses as a Livingston trigger with \acp{SNR} of \PyCBCOfflineLSNRNSBH{} and \MBTAOfflineLSNRNSBH{}, respectively, which were well above the backgrounds for triggers from similar templates. 
Based on $\pastro{}$, \FULLNAME{200105F}{} is listed here as a marginal candidate, despite it being a clear outlier from the background noise~\cite{LIGOScientific:2021qlt}. 
The marginal status of this candidate can at least in part be explained from the underlying assumptions in the candidate's \ac{FAR} estimation and \pastro{} computation.

The empirical background noise distribution available for evaluating the significance of single-detector candidates extends only as far as ranking statistics at which we see one noise trigger per observing time.
In contrast, for multidetector triggers, an extended background estimate can be obtained by constructing unphysical coincidences between triggers in different detectors. 
Consequently, for single-detector candidates like \FULLNAME{200105F}{} that lie outside the background noise distribution, the \ac{FAR} estimation relies on an extrapolation. 
For triggers in the tail of the background distribution, this extrapolation comes with uncertainty that impacts the estimated \ac{FAR}, and this uncertainty also propagates to the noise distribution used in the calculation of \pastro{}~\cite{Abbott:2020niy,LIGOScientific:2021qlt}.

Additionally, the \pastro{} estimation for \ac{NSBH} sources depends on the foreground distribution of ranking statistics as well as their merger rate.
The former is subject to uncertainties coming from a lack of knowledge of the \ac{NSBH} population, while the latter has large error bars due to a paucity of high-significance \ac{NSBH} detections (order $\sim \NUMNSBH$).
Such uncertainties on \pastro{} have a significant impact on marginal candidates whose \pastro{} values hover around \PASTROTHRESH{}.
As a consequence, the moderate \pastro{} value assigned at this time to \FULLNAME{200105F}{} does not allow us to draw a firm conclusion on its origin.
Future observations will likely shed more light on the true provenance of this and similar candidates.

\subsubsection{Subthreshold candidates}
\label{sec:subthreshold}

Following \GWTCTWOFINAL{}~\cite{LIGOScientific:2021usb}, we provide an extended list of \ac{O3b} candidates with \ac{FAR} less than $\SUBTHRESHOLDFAR$ as part of the data products available from \ac{GWOSC}~\cite{gwosc:gwtc3}. 
In addition to the \NUMEVENTS{} \ac{O3b} candidates with $\pastro{} > \PASTROTHRESH{}$ listed in Table~\ref{tab:events}, and the \ALLOTHREEBMARGINAL{} marginal candidates with \ac{FAR} less than $\FARTHRESHYR{}~\mathrm{yr^{-1}}$ listed in Table~\ref{tab:marginal_events}, there are \ALLSUBTHRESHOTHREEB{} further subthreshold \ac{O3b} candidates in the extended list (giving a total of \ALLOTHREEB{} \ac{O3b} candidates in the data release)~\cite{gwosc:gwtc3}.
The subthreshold candidates have not been scrutinized for possible instrumental origin, but the purity of the sample is expected to be low: $\lesssim\SUBTHRESHOLDPURITY$ when considering all subthreshold candidates, as estimated in Sec.~\ref{sec:sub_signal}.

For each subthreshold candidate, we provide estimates of their \pastro{} (assuming a \ac{CBC} source) and localization.
Localization relies on the same tools that were used to provide low-latency localization for public \ac{GW} alerts, namely \BAYESTAR{}~\cite{Singer:2015ema,Singer:2016eax} for \GSTLAL{}, \MBTA{} and \PYCBC{} candidates, and \CWB{} for its own candidates. 

\subsection{Search sensitivity}

\subsubsection{Sensitive hypervolume}
\label{sec:vt}

\begin{table*}
\footnotesize
% Made by ./scripts/generating_tex_macros/make_vt_table
% DO NOT EDIT THIS FILE DIRECTLY

\begin{ruledtabular}
\begin{tabular}{ccc ccccc c}
 \multicolumn{3}{c}{ {Binary masses  (\Msun)} }  & \multicolumn{6}{c}{ {Sensitive hypervolume (\Gpcyr)} } \\
\cline{1-3} \cline{4-9}
  \massone{} & \masstwo{} & \Mc{} & {\CWB{}} & {\GSTLAL{}} & {\MBTA{}} & {\PYCBCHYPERBANK{}} & {\PYCBCBBH{}} & {Any} \\
\hline
\makebox[0pt][l]{\fboxsep0pt\colorbox{lightgray}{\mystrut\hspace*{1.000000\linewidth}}}\!\! $ 35.0 $ & $ 35.0 $ & $30.5 $ &
$ \CWBALLSKYVT{THIRTYFIVETHIRTYFIVE} $ &
$ \GSTLALALLSKYVT{THIRTYFIVETHIRTYFIVE} $ &
$ \MBTAALLSKYVT{THIRTYFIVETHIRTYFIVE} $ &
$ \PYCBCALLSKYVT{THIRTYFIVETHIRTYFIVE} $ &
$ \PYCBCHIGHMASSVT{THIRTYFIVETHIRTYFIVE} $ &
$ \ANYCBCVT{THIRTYFIVETHIRTYFIVE} $ \\
\makebox[0pt][l]{\fboxsep0pt{\mystrut\hspace*{1.000000\linewidth}}}\!\! $ 35.0 $ & $ 20.0 $ & $22.9 $ &
$ \CWBALLSKYVT{THIRTYFIVETWENTY} $ &
$ \GSTLALALLSKYVT{THIRTYFIVETWENTY} $ &
$ \MBTAALLSKYVT{THIRTYFIVETWENTY} $ &
$ \PYCBCALLSKYVT{THIRTYFIVETWENTY} $ &
$ \PYCBCHIGHMASSVT{THIRTYFIVETWENTY} $ &
$ \ANYCBCVT{THIRTYFIVETWENTY} $ \\
\makebox[0pt][l]{\fboxsep0pt\colorbox{lightgray}{\mystrut\hspace*{1.000000\linewidth}}}\!\! $ 35.0 $ & $ 1.5 $ & $5.2 $ &
$ \CWBALLSKYVT{THIRTYFIVEONEPOINTFIVE} $ &
$ \GSTLALALLSKYVT{THIRTYFIVEONEPOINTFIVE} $ &
$ \MBTAALLSKYVT{THIRTYFIVEONEPOINTFIVE} $ &
$ \PYCBCALLSKYVT{THIRTYFIVEONEPOINTFIVE} $ &
$ \PYCBCHIGHMASSVT{THIRTYFIVEONEPOINTFIVE} $ &
$ \ANYCBCVT{THIRTYFIVEONEPOINTFIVE} $ \\
\makebox[0pt][l]{\fboxsep0pt{\mystrut\hspace*{1.000000\linewidth}}}\!\! $ 20.0 $ & $ 20.0 $ & $17.4 $ &
$ \CWBALLSKYVT{TWENTYTWENTY} $ &
$ \GSTLALALLSKYVT{TWENTYTWENTY} $ &
$ \MBTAALLSKYVT{TWENTYTWENTY} $ &
$ \PYCBCALLSKYVT{TWENTYTWENTY} $ &
$ \PYCBCHIGHMASSVT{TWENTYTWENTY} $ &
$ \ANYCBCVT{TWENTYTWENTY} $ \\
\makebox[0pt][l]{\fboxsep0pt\colorbox{lightgray}{\mystrut\hspace*{1.000000\linewidth}}}\!\! $ 20.0 $ & $ 10.0 $ & $12.2 $ &
$ \CWBALLSKYVT{TWENTYTEN} $ &
$ \GSTLALALLSKYVT{TWENTYTEN} $ &
$ \MBTAALLSKYVT{TWENTYTEN} $ &
$ \PYCBCALLSKYVT{TWENTYTEN} $ &
$ \PYCBCHIGHMASSVT{TWENTYTEN} $ &
$ \ANYCBCVT{TWENTYTEN} $ \\
\makebox[0pt][l]{\fboxsep0pt{\mystrut\hspace*{1.000000\linewidth}}}\!\! $ 20.0 $ & $ 1.5 $ & $4.2 $ &
$ \CWBALLSKYVT{TWENTYONEPOINTFIVE} $ &
$ \GSTLALALLSKYVT{TWENTYONEPOINTFIVE} $ &
$ \MBTAALLSKYVT{TWENTYONEPOINTFIVE} $ &
$ \PYCBCALLSKYVT{TWENTYONEPOINTFIVE} $ &
$ \PYCBCHIGHMASSVT{TWENTYONEPOINTFIVE} $ &
$ \ANYCBCVT{TWENTYONEPOINTFIVE} $ \\
\makebox[0pt][l]{\fboxsep0pt\colorbox{lightgray}{\mystrut\hspace*{1.000000\linewidth}}}\!\! $ 10.0 $ & $ 10.0 $ & $8.7 $ &
$ \CWBALLSKYVT{TENTEN} $ &
$ \GSTLALALLSKYVT{TENTEN} $ &
$ \MBTAALLSKYVT{TENTEN} $ &
$ \PYCBCALLSKYVT{TENTEN} $ &
$ \PYCBCHIGHMASSVT{TENTEN} $ &
$ \ANYCBCVT{TENTEN} $ \\
\makebox[0pt][l]{\fboxsep0pt{\mystrut\hspace*{1.000000\linewidth}}}\!\! $ 10.0 $ & $ 5.0 $ & $6.1 $ &
$ \CWBALLSKYVT{TENFIVE} $ &
$ \GSTLALALLSKYVT{TENFIVE} $ &
$ \MBTAALLSKYVT{TENFIVE} $ &
$ \PYCBCALLSKYVT{TENFIVE} $ &
$ \PYCBCHIGHMASSVT{TENFIVE} $ &
$ \ANYCBCVT{TENFIVE} $ \\
\makebox[0pt][l]{\fboxsep0pt\colorbox{lightgray}{\mystrut\hspace*{1.000000\linewidth}}}\!\! $ 10.0 $ & $ 1.5 $ & $3.1 $ &
$ \CWBALLSKYVT{TENONEPOINTFIVE} $ &
$ \GSTLALALLSKYVT{TENONEPOINTFIVE} $ &
$ \MBTAALLSKYVT{TENONEPOINTFIVE} $ &
$ \PYCBCALLSKYVT{TENONEPOINTFIVE} $ &
$ \PYCBCHIGHMASSVT{TENONEPOINTFIVE} $ &
$ \ANYCBCVT{TENONEPOINTFIVE} $ \\
\makebox[0pt][l]{\fboxsep0pt{\mystrut\hspace*{1.000000\linewidth}}}\!\! $ 5.0 $ & $ 5.0 $ & $4.4 $ &
$ \CWBALLSKYVT{FIVEFIVE} $ &
$ \GSTLALALLSKYVT{FIVEFIVE} $ &
$ \MBTAALLSKYVT{FIVEFIVE} $ &
$ \PYCBCALLSKYVT{FIVEFIVE} $ &
$ \PYCBCHIGHMASSVT{FIVEFIVE} $ &
$ \ANYCBCVT{FIVEFIVE} $ \\
\makebox[0pt][l]{\fboxsep0pt\colorbox{lightgray}{\mystrut\hspace*{1.000000\linewidth}}}\!\! $ 5.0 $ & $ 1.5 $ & $2.3 $ &
$ \CWBALLSKYVT{FIVEONEPOINTFIVE} $ &
$ \GSTLALALLSKYVT{FIVEONEPOINTFIVE} $ &
$ \MBTAALLSKYVT{FIVEONEPOINTFIVE} $ &
$ \PYCBCALLSKYVT{FIVEONEPOINTFIVE} $ &
$ \PYCBCHIGHMASSVT{FIVEONEPOINTFIVE} $ &
$ \ANYCBCVT{FIVEONEPOINTFIVE} $ \\
\makebox[0pt][l]{\fboxsep0pt{\mystrut\hspace*{1.000000\linewidth}}}\!\! $ 1.5 $ & $ 1.5 $ & $1.3 $ &
$ \CWBALLSKYVT{ONEPOINTFIVEONEPOINTFIVE} $ &
$ \GSTLALALLSKYVT{ONEPOINTFIVEONEPOINTFIVE} $ &
$ \MBTAALLSKYVT{ONEPOINTFIVEONEPOINTFIVE} $ &
$ \PYCBCALLSKYVT{ONEPOINTFIVEONEPOINTFIVE} $ &
$ \PYCBCHIGHMASSVT{ONEPOINTFIVEONEPOINTFIVE} $ &
$ \ANYCBCVT{ONEPOINTFIVEONEPOINTFIVE} $ \\
\end{tabular}
\end{ruledtabular}

\caption{
\label{tab:vt}
Sensitive hypervolume from \ac{O3b} for the various search analyses with \pastro{} $ > \PASTROTHRESHOLD{}$ at the assessed points in the mass parameter space.
The \emph{Any} results come from calculating the sensitive hypervolume for injections found by at least one search analysis.
For each set of binary masses, the given values are the central points of a log-normal distribution with width \VTLOGNORMALWIDTH{}.
For some regions and analyses, few injections were recovered such that the sensitive hypervolume cannot be accurately estimated; these cases are indicated by a dash (--).
As an example of this, the \PYCBCBBH{} and \ac{CWB} analyses analyzed only injections in the designated \ac{BBH} set, and so no injections were found in the \ac{BNS} or \ac{NSBH} regions.
The injected population is described in Appendix~\ref{sec:p-astro-methods}.
}
\end{table*}

\begin{figure*}
    \centering
    \includegraphics[width=\textwidth]{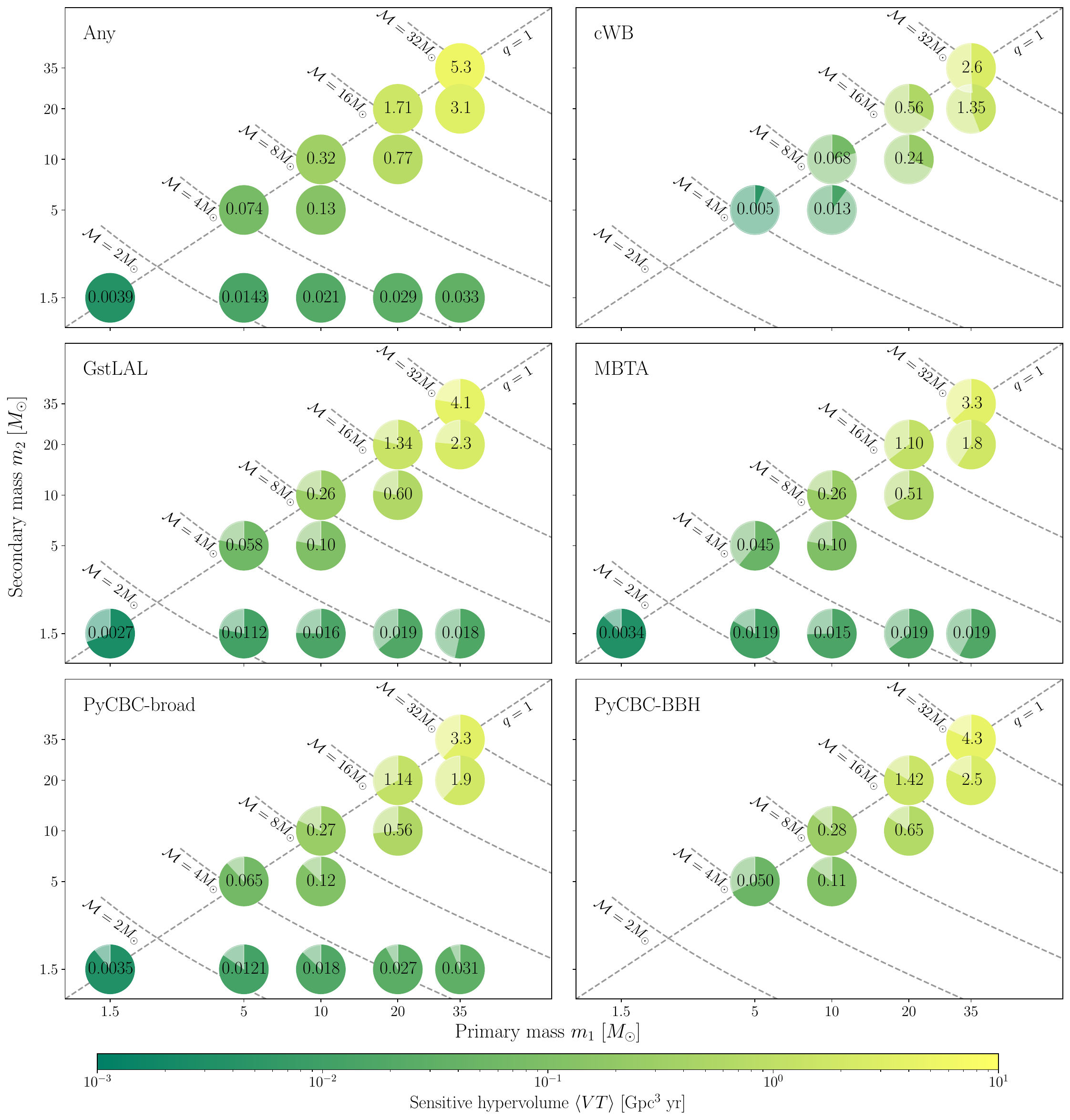}
    \caption{
    \label{fig:vt} 
    Sensitive hypervolume \VT{} from \ac{O3b} for the various searches with $\pastro{} > \PASTROTHRESHOLD{}$ at the assessed points in the mass parameter space.
The \emph{Any} results come from calculating the sensitive hypervolume for injections found by at least one search analysis.
The plotted points correspond to the central points of the log-normal distributions (with widths \VTLOGNORMALWIDTH{}) used for the calculation of \VT{}.
Each point is marked by a pie chart, where the darker portion represents the fraction of the \emph{Any} \VT{} recovered.
The colour of the darker portion corresponds to the value of the sensitive \VT{}, as given by the scale bar.
The values displayed are the same as those given in Table~\ref{tab:vt}.
}
\end{figure*}

To estimate the sensitivity of the search analyses, we calculated a sensitive time--volume hypervolume \VT{} for each analysis during \ac{O3b}. 
This hypervolume represents the sensitivity of each search analysis to a distribution of sources assumed to be uniformly distributed in comoving volume and source-frame time.
The expected number of detections for a search analysis is 
\begin{equation}
\hat{N} = \VT{} R,
\end{equation}
where $R$ is the rate of signals per unit volume and unit observing time.
The different pipeline live times affect their calculated \VT{}. 
The pipeline live times are \CWBLIVETIME{} (\ac{CWB}), $\ONEDETECTORDAYS{}~\mathrm{days}$ (\GSTLAL{}), \MBTALIVETIME{} (\MBTA{}) and \PYCBCLIVETIME{} (both \PYCBC{} analyses).
To estimate \VT{} for each analysis, we add simulated signals (referred to as \emph{injections}) into the data and test how many are recovered. 
The injections we use are designed to cover the detected population of \acp{BBH}, \acp{BNS} and \acp{NSBH}, and are described further in Appendix~\ref{sec:p-astro-methods}.
We use the same sets of simulated signals for each analysis to consistently measure \VT{}, but since the \PYCBCBBH{} and \ac{CWB} analyses are designed to search for \ac{BBH} signals, we use only injections in the designated \ac{BBH} regions for these searches. 
Rather than consider the total rate of signals, we consider signals corresponding to sources with specific masses to parametrize sensitivity to signals across parameter space. 

In Table~\ref{tab:vt} we report the \ac{O3b} \VT{} for simulated signals corresponding to sources with component masses close to the specified values.
In Fig.~\ref{fig:vt}, for each search, we show the variation in the \ac{O3b} \VT{} across the parameter space. 
The injections around the specified points are weighted so that they follow a log-normal distribution about the central mass with a width of \VTLOGNORMALWIDTH{}. 
We also assume component spins are isotropically distributed with uniformly distributed magnitudes up to a maximum spin that depends on the source component mass; if $m_i < \INJBBHMASSMIN{} \Msun$, we assume $\injspinmax{} = \INJBNSSPINMAX{}$ and otherwise assume $\injspinmax{} = \INJBBHSPINMAX{}$.
We consider:
\begin{itemize}
\item \acp{BH} at \VTFIRSTDETECTIONLIKEMASS{}, which corresponds to a GW150914-like system~\cite{TheLIGOScientific:2016wfe,LIGOScientific:2021usb}, and is approximately where we infer a feature (potentially a bump or a break) in the \ac{BH} mass spectrum~\cite{Abbott:2020gyp};
\item \acp{BH} at \VTDETECTEDRANGEMASSES{} to see how sensitivity varies across this range of previously detected \ac{BH} masses;
\item \acp{NS} at \VTNSLIKEMASS{}, close to the canonical \ac{NS} mass.
\end{itemize}
We use several combinations of masses in order to assess our sensitivity to \ac{BNS}, \ac{NSBH}, and (relatively equal-mass) \ac{BBH} systems. 
From the masses considered, the search sensitivity is greatest for $\VTFIRSTDETECTIONLIKEMASS{}+\VTFIRSTDETECTIONLIKEMASS{}$ binaries in all analyses, although our detectors generally survey larger volume for higher-mass populations up to source component masses of $\sim \DETECTIONMASSLIMIT{} \Msun$~\cite{Fishbach:2017zga,Fishbach:2021yvy}. 
Equivalent results for all of \ac{O3} are given in Table~\ref{tab:vt_allo3} in Appendix~\ref{sec:allo3_results}.

The sensitivity results presented in Table~\ref{tab:vt} are obtained considering a detection threshold of $\pastro{} > \PASTROTHRESHOLD{}$, calculated as for our main results.
The \emph{Any} pipeline results come from taking the maximum \pastro{} for an injection from across the analyses, and represent our overall sensitivity to \acp{CBC} in the specified region.

The \ac{CWB} results are obtained using the standard $\pastro{} > \PASTROTHRESHOLD{}$ threshold; however, for candidates reported in Table~\ref{tab:events}, we require that the \ac{CWB} candidates must have an associated trigger from one of the matched-filter analyses, as the \pastro{} calculation performed by \ac{CWB} assumes that the signal is from a \ac{CBC}.
Therefore, we also investigated the \ac{CWB} \VT{} using a cut of $\pastro{} > \PASTROTHRESHOLD{}$ from \ac{CWB} together with the requirement that $\pastro{} > \CBCTHRESHOLD{}$ from at least one matched-filter analysis to match the main results.
We found these values to be comparable; for example the \VT{} for the $\VTMASSFIVE{}+\VTMASSFIVE{}$ bin is unchanged, at $\CWBALLSKYVT{FIVEFIVE}~\Gpcyr$, and the $\VTMASSTEN{}+\VTMASSTEN{}$ bin changes from $\CWBALLSKYVT{TENTEN}~\Gpcyr$ to $\CWBASSOCVT{TENTEN}~\Gpcyr$.
The largest change is in the highest-mass $\VTFIRSTDETECTIONLIKEMASS{}+\VTFIRSTDETECTIONLIKEMASS{}$ bin, where the \VT{} changes from $\CWBALLSKYVT{THIRTYFIVETHIRTYFIVE}~\Gpcyr$ to $\CWBASSOCVT{THIRTYFIVETHIRTYFIVE}~\Gpcyr$. 
Overall, adding the requirement that there be a \ac{CBC} counterpart to \ac{CWB} candidates makes little difference to the search sensitivity calculated from our \ac{CBC} injections.

\subsubsection{Subthreshold signal count via search sensitivities}
\label{sec:sub_signal}

The search sensitivities may also be calculated at the threshold of \ac{FAR} $< \SUBTHRESHOLDFAR$ corresponding to the subthreshold candidate set, enabling us to self-consistently estimate the number of astrophysical signals among these \ALLBELOWPASTROTHRESHOTHREEB{} candidates. 
For an individual search pipeline, if the source population assumed in the \pastro{} calculation is sufficiently close to the (unknown) true population, then the sum of \pastro{} values over a candidate set gives the expectation of the number of true signals in the set~\cite{Creighton:2017aaa}.  
This count of signals is itself a realization of a Poisson process with a mean proportional to the pipeline's \VT{} for the true signal population.  
Hence, if the true population were known, we could scale the sum of \pastro{} values for each pipeline by its \VT{} to obtain an estimate of the signal count at a given threshold for the combined \emph{Any} pipeline analysis. 

In lieu of the true population, we take as reference the \VT{} values for the $\VTFIRSTDETECTIONLIKEMASS{}+\VTFIRSTDETECTIONLIKEMASS{}$ point, as representing the largest proportion of detected signals. 
The resulting estimated signal counts for \emph{Any} pipeline are consistent across pipelines within statistical uncertainties.  
Consistent and similar counts are also obtained for the modeled pipelines if the $\VTMASSTWENTY{}+\VTMASSTWENTY{}$ or $\VTMASSTEN{}+\VTMASSTEN{}$ points are taken as a reference, indicating that the result is not strongly sensitive to an assumed \ac{BBH} mass distribution; the counts for \ac{CWB} do vary, but the \ac{CWB} contribution to the \emph{Any} pipeline sensitivity varies significantly across the parameter space and is subdominant to the modeled pipelines for the lower-mass \ac{BBH} points.
The number of subthreshold signals is then the difference between signal counts (excluding the marginal candidates found to be likely caused by instrumental artifacts) for the thresholds \ac{FAR} $< \SUBTHRESHOLDFAR$  and $\pastro{} > \PASTROTHRESHOLD{}$, which averaged over pipelines yields $\sim \EXPREALINSUBTHRESH{}$, with an expected uncertainty of $\sim \sqrt{\EXPREALINSUBTHRESH{}}$.
This estimate is consistent with the ratio of \VT{} values for \emph{Any} pipeline between the two thresholds, which is \VTRATIOTWOTHRESH{} for the \ac{BBH} mass points.

\section{Source properties}\label{sec:parameter-estimation}

\begin{figure*}
    \centering
    \includegraphics[width=\textwidth]{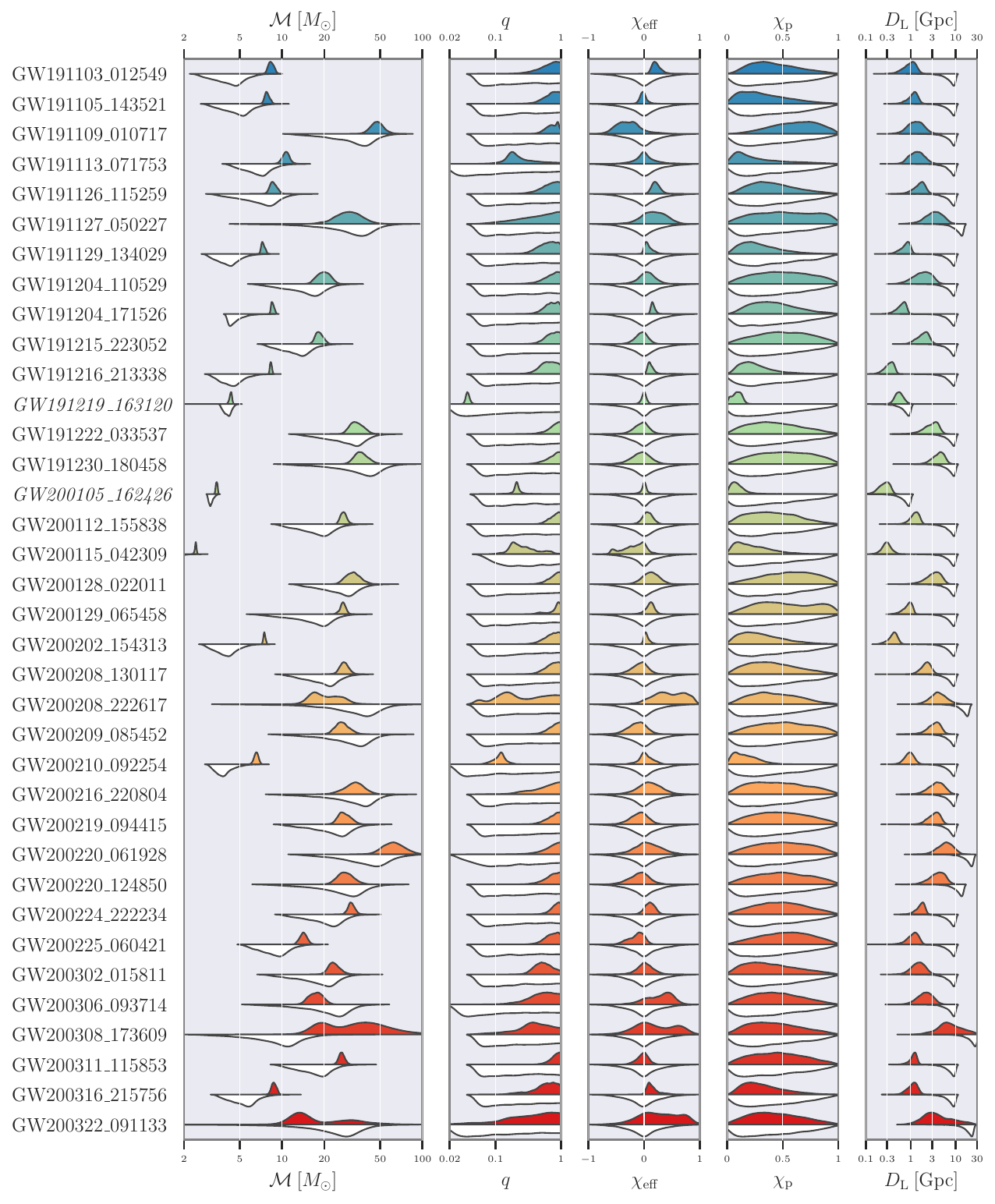}
    \caption{\label{fig:post_1d_multi}Marginal probability distributions for the source chirp mass $\Mc$, mass ratio $\massratio$, effective inspiral spin $\chieff$, effective precession spin $\chip$ and luminosity distance $\DL$ for \ac{O3b} candidates with $\pastro{} > \PASTROTHRESHOLD$ plus \FULLNAME{200105F}{}. 
    The colored upper half of the plot shows the marginal posterior distributions, and the white lower half of the plot shows the marginal prior distributions. 
    The vertical extent of each colored region is proportional to one-dimensional marginal probability distribution at a given parameter value for the corresponding candidate. 
    We highlight with italics \FULLNAME{200105F}{} as it has $\pastro{} < \PASTROTHRESHOLD$, as well as \FULLNAME{GW191219E}{} because of significant uncertainty in its $\pastro{}$ and because it has significant posterior support outside of mass ratios where the waveform models have been calibrated. 
    Results for \FULLNAME{GW200308G}{} and \FULLNAME{GW200322G}{} include a low-likelihood mode at large distances and high masses. 
    Colors correspond to the date of observation.}
\end{figure*}

Having identified candidate signals, we perform a coherent analysis of the data from the \ac{GW} detector network to infer the properties of each source. 
Information about the source parameters is encoded within the amplitude and phase of the \ac{GW} signal recorded by each detector in the network. 
To extract this information, we match model waveform templates to the observed data to calculate the posterior probability of a given set of parameters~\cite{Cutler:1994ys}, assuming that the noise is Gaussian, stationary and uncorrelated between detectors~\cite{LIGOScientific:2019hgc}. 
We use the waveform models \IMRPhenomXPHM{}~\cite{Pratten:2020ceb} and \SEOBNRPHM{}~\cite{Ossokine:2020kjp} to describe \ac{BBH} systems, and \IMRPhenomNSBH{}~\cite{Thompson:2020nei} and \SEOBNRNSBH{}~\cite{Matas:2020wab} to describe matter effects in \ac{NSBH} systems. 
All templates assume quasicircular binaries, with the \ac{BBH} models including the effects of spin precession and higher-order multipole moments~\cite{Garcia-Quiros:2020qpx,Pratten:2020ceb,Cotesta:2018fcv,Ossokine:2020kjp}.
As the higher-order multipole moments and spin precession effects incorporated into the \ac{BBH} waveform templates are more important in describing the signal than the \ac{NSBH} matter effects, we preferentially quote results using the \ac{BBH} waveforms~\cite{LIGOScientific:2021qlt}.
We use an equal combination of \IMRPhenomXPHM{} and \SEOBNRPHM{} samples~\cite{Berry:2015aaa,Ashton:2019leq}.  
Potential systematic uncertainties from differences in waveform modeling are discussed in Sec.~\ref{sec:waveform-systematics}.
Analyses using the \IMRPhenomXPHM{} or \ac{NSBH} waveforms are performed with the \BILBY{} family of codes~\cite{Ashton:2018jfp,Smith:2019ucc,Romero-Shaw:2020owr} and analyses using the \SEOBNRPHM{} waveforms are performed with \RIFT{}~\cite{Pankow:2015cra,Lange:2017wki,Wysocki:2019grj}.
The analysis closely follows the practices from previous studies~\cite{TheLIGOScientific:2016wfe,LIGOScientific:2021usb}, and further details are presented in Appendix~\ref{sec:parameter-estimation-methods}. 

\begin{figure*}
    \centering
    \includegraphics[width=0.8\textwidth]{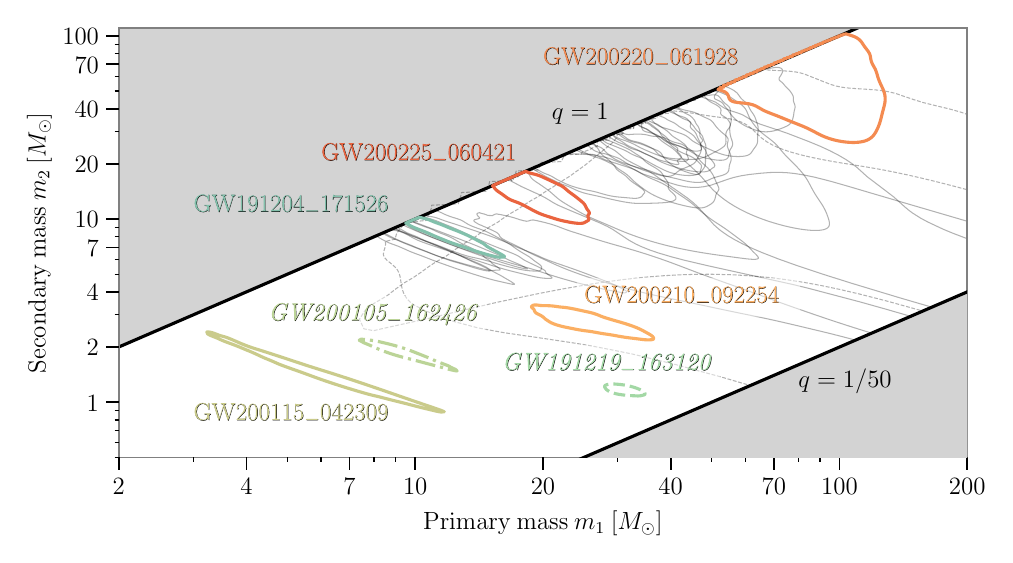} \\    
    \includegraphics[width=0.8\textwidth]{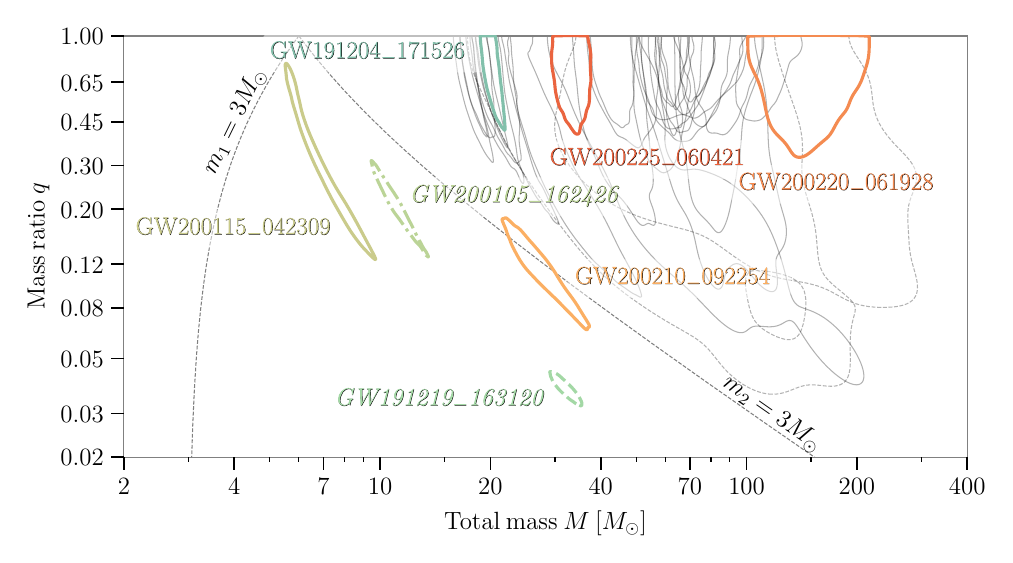}
    \caption{\label{fig:mtotvsqpost} 
    Credible-region contours for the inferred masses of the \ac{O3b} candidates with $\pastro{} > \PASTROTHRESHOLD$ plus \FULLNAME{200105F}{}. 
    \emph{Top}: Results for the primary and secondary component masses $\massone$ and $\masstwo$. 
    The shaded areas indicate regions excluded by the convention $\massone \geq \masstwo$, and by the most extreme mass ratio considered in our analyses (detailed in Appendix~\ref{sec:prior-sampling}).
    \emph{Bottom}: Results for total mass $\Mtot$ and mass ratio $\massratio$.
    Each contour represents the $90\%$ credible region for a different candidate. 
    Highlighted contours are for the \ac{NSBH} candidates \FULLNAME{GW191219E}{}, \FULLNAME{200105F}{} and \FULLNAME{GW200115A}{}; the \ac{NSBH} or low-mass \ac{BBH} candidate \FULLNAME{GW200210B}{}; \FULLNAME{GW191204G}{}, which has inferred $\chieff > 0$; \FULLNAME{GW200225B}{}, which has $\percentchiefflessthanzero{GW200225B}\%$ probability that $\chieff < 0$, and \FULLNAME{GW200220E}{}, which probably has the most massive source of the \ac{O3b} candidates. 
    We highlight with italics \FULLNAME{200105F}{} as it has $\pastro{} < \PASTROTHRESHOLD$, as well as \FULLNAME{GW191219E}{} because of significant uncertainty in its $\pastro{}$ and because it has significant posterior support outside of mass ratios where the waveform models have been calibrated. 
    Results for \FULLNAME{GW200308G}{} and \FULLNAME{GW200322G}{} are indicated with dashed lines to highlight that these include a low-likelihood mode at large distances and high masses, and are particularly prior sensitive.
    The dotted lines delineate regions where the primary and secondary can have a mass below \PEBHMassThreshold. 
    For the region above the $\masstwo{} = \PEBHMassThreshold$ line, both objects in the binary have masses above \PEBHMassThreshold.
    }
\end{figure*}

\begin{figure*}
    \centering
    \includegraphics[width=0.8\textwidth]{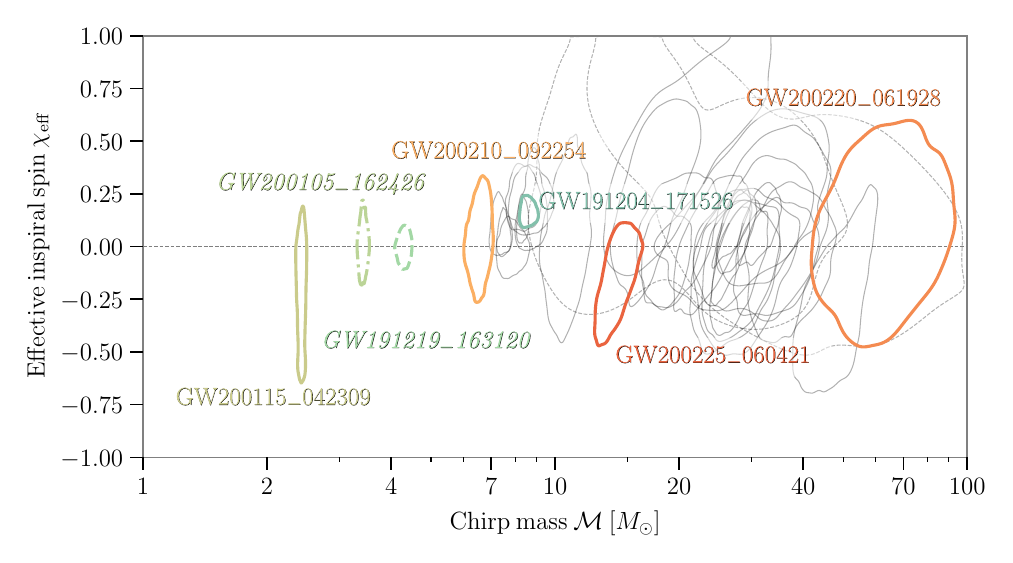}
    \caption{\label{fig:mcchieffpost} Credible-region contours in the plane of chirp mass $\Mc$ and effective inspiral spin $\chieff$ for \ac{O3b} candidates with $\pastro{} > \PASTROTHRESHOLD$ plus \FULLNAME{200105F}{}. 
    Each contour represents the $90\%$ credible region for a different candidate. 
    Highlighted contours are for the \ac{NSBH} candidates \FULLNAME{GW191219E}{}, \FULLNAME{200105F}{} and \FULLNAME{GW200115A}{}; the \ac{NSBH} or low-mass \ac{BBH} candidate \FULLNAME{GW200210B}{}; \FULLNAME{GW191204G}{}, which has inferred $\chieff > 0$; \FULLNAME{GW200225B}{}, which has $\percentchiefflessthanzero{GW200225B}\%$ probability that $\chieff < 0$, and \FULLNAME{GW200220E}{}, which probably has the most massive source of the \ac{O3b} candidates.
    We highlight with italics \FULLNAME{200105F}{} as it has $\pastro{} < \PASTROTHRESHOLD$, as well as \FULLNAME{GW191219E}{} because of significant uncertainty in its $\pastro{}$ and because it has significant posterior support outside of mass ratios where the waveform models have been calibrated. 
     Results for \FULLNAME{GW200308G}{} and \FULLNAME{GW200322G}{} are indicated with dashed lines to highlight that these include a low-likelihood mode at large distances and high masses, and are particularly prior sensitive.}
\end{figure*}

A summary of key results for \ac{O3b} candidates is given in Table~\ref{tab:pe},
and shown in Fig.~\ref{fig:post_1d_multi}, Fig.~\ref{fig:mtotvsqpost} and Fig.~\ref{fig:mcchieffpost}. 
We show results for the \ac{O3b} candidates with $\pastro{} > \PASTROTHRESHOLD$ plus \FULLNAME{200105F}{}, which, despite being a marginal candidate, is a clear outlier from the noise background~\cite{LIGOScientific:2021qlt}. 
On account of its low $\pastro{}$, we highlight \FULLNAME{200105F}{} in figures and tables. 
We similarly highlight \FULLNAME{GW191219E}{} because, as discussed in Sec.~\ref{sec:pipeline-comparison}, the calculated $\pastro$ is especially sensitive to the adopted population model, and, as discussed below, there is significant posterior support for mass ratios outside the range of calibration for the waveform models. 
Following previous analyses~\cite{Abbott:2020niy,LIGOScientific:2021usb}, results are calculated using default priors that are intended to not make strong assumptions about the underlying astrophysical population (e.g., uniform priors are used for redshifted component masses, an isotropic distribution is used for spin orientations, and it is assumed that sources are uniformly distributed in comoving volume and time). 
Posterior samples are available from \ac{GWOSC}~\cite{gwosc:gwtc3}, and the simple form of the prior probability distributions enables the samples to be conveniently reweighted to use alternative prior distributions~\cite{Thrane:2018qnx,Callister:2021gxf}.
Inferences about the underlying population of merging compact binaries are presented in a companion paper~\cite{LIGOScientific:2021psn}.

\begin{PE_table}
    \begin{table*}
    	\begin{ruledtabular}\begin{tabular}{l c c c c c c c c c c c}
Candidate & $\underset{\displaystyle (M_\odot)}{M}$ & $\underset{\displaystyle (M_\odot)}{\mathcal{M}}$ & $\underset{\displaystyle (M_\odot)}{m_1}$ & $\underset{\displaystyle (M_\odot)}{m_2}$ & $\chi_\mathrm{{eff}}$ & $\underset{\displaystyle ({\rm Gpc})}{D_\mathrm{L}}$ & $z$ & $\underset{\displaystyle (M_\odot)}{M_\mathrm{f}}$ & $\chi_\mathrm{f}$ & $\underset{\displaystyle (\mathrm{deg}^2)}{\Delta\Omega}$ & $\mathrm{SNR}$\\ \hline
\FULLNAME{GW191103A} & $\totalmasssourceuncert{GW191103A}$ & $\chirpmasssourceuncert{GW191103A}$ & $\massonesourceuncert{GW191103A}$ & $\masstwosourceuncert{GW191103A}$ & $\chieffinfinityonlyprecavguncert{GW191103A}$ & $\luminositydistanceuncert{GW191103A}$ & $\redshiftuncert{GW191103A}$ & $\finalmasssourceuncert{GW191103A}$ & $\finalspinuncert{GW191103A}$ & $\skyarea{GW191103A}$ & $\networkmatchedfiltersnrIMRPuncert{GW191103A}$\\
\makebox[0pt][l]{\fboxsep0pt\colorbox{lightgray}{\mystrut\hspace*{1.0\linewidth}}}\!\!
\FULLNAME{GW191105C} & $\totalmasssourceuncert{GW191105C}$ & $\chirpmasssourceuncert{GW191105C}$ & $\massonesourceuncert{GW191105C}$ & $\masstwosourceuncert{GW191105C}$ & $\chieffinfinityonlyprecavguncert{GW191105C}$ & $\luminositydistanceuncert{GW191105C}$ & $\redshiftuncert{GW191105C}$ & $\finalmasssourceuncert{GW191105C}$ & $\finalspinuncert{GW191105C}$ & $\skyarea{GW191105C}$ & $\networkmatchedfiltersnrIMRPuncert{GW191105C}$\\
\FULLNAME{GW191109A} & $\totalmasssourceuncert{GW191109A}$ & $\chirpmasssourceuncert{GW191109A}$ & $\massonesourceuncert{GW191109A}$ & $\masstwosourceuncert{GW191109A}$ & $\chieffinfinityonlyprecavguncert{GW191109A}$ & $\luminositydistanceuncert{GW191109A}$ & $\redshiftuncert{GW191109A}$ & $\finalmasssourceuncert{GW191109A}$ & $\finalspinuncert{GW191109A}$ & $\skyarea{GW191109A}$ & $\networkmatchedfiltersnrIMRPuncert{GW191109A}$\\
\makebox[0pt][l]{\fboxsep0pt\colorbox{lightgray}{\mystrut\hspace*{1.0\linewidth}}}\!\!
\FULLNAME{GW191113B} & $\totalmasssourceuncert{GW191113B}$ & $\chirpmasssourceuncert{GW191113B}$ & $\massonesourceuncert{GW191113B}$ & $\masstwosourceuncert{GW191113B}$ & $\chieffinfinityonlyprecavguncert{GW191113B}$ & $\luminositydistanceuncert{GW191113B}$ & $\redshiftuncert{GW191113B}$ & $\finalmasssourceuncert{GW191113B}$ & $\finalspinuncert{GW191113B}$ & $\skyarea{GW191113B}$ & $\networkmatchedfiltersnrIMRPuncert{GW191113B}$\\
\FULLNAME{GW191126C} & $\totalmasssourceuncert{GW191126C}$ & $\chirpmasssourceuncert{GW191126C}$ & $\massonesourceuncert{GW191126C}$ & $\masstwosourceuncert{GW191126C}$ & $\chieffinfinityonlyprecavguncert{GW191126C}$ & $\luminositydistanceuncert{GW191126C}$ & $\redshiftuncert{GW191126C}$ & $\finalmasssourceuncert{GW191126C}$ & $\finalspinuncert{GW191126C}$ & $\skyarea{GW191126C}$ & $\networkmatchedfiltersnrIMRPuncert{GW191126C}$\\
\makebox[0pt][l]{\fboxsep0pt\colorbox{lightgray}{\mystrut\hspace*{1.0\linewidth}}}\!\!
\FULLNAME{GW191127B} & $\totalmasssourceuncert{GW191127B}$ & $\chirpmasssourceuncert{GW191127B}$ & $\massonesourceuncert{GW191127B}$ & $\masstwosourceuncert{GW191127B}$ & $\chieffinfinityonlyprecavguncert{GW191127B}$ & $\luminositydistanceuncert{GW191127B}$ & $\redshiftuncert{GW191127B}$ & $\finalmasssourceuncert{GW191127B}$ & $\finalspinuncert{GW191127B}$ & $\skyarea{GW191127B}$ & $\networkmatchedfiltersnrIMRPuncert{GW191127B}$\\
\FULLNAME{GW191129G} & $\totalmasssourceuncert{GW191129G}$ & $\chirpmasssourceuncert{GW191129G}$ & $\massonesourceuncert{GW191129G}$ & $\masstwosourceuncert{GW191129G}$ & $\chieffinfinityonlyprecavguncert{GW191129G}$ & $\luminositydistanceuncert{GW191129G}$ & $\redshiftuncert{GW191129G}$ & $\finalmasssourceuncert{GW191129G}$ & $\finalspinuncert{GW191129G}$ & $\skyarea{GW191129G}$ & $\networkmatchedfiltersnrIMRPuncert{GW191129G}$\\
\makebox[0pt][l]{\fboxsep0pt\colorbox{lightgray}{\mystrut\hspace*{1.0\linewidth}}}\!\!
\FULLNAME{GW191204A} & $\totalmasssourceuncert{GW191204A}$ & $\chirpmasssourceuncert{GW191204A}$ & $\massonesourceuncert{GW191204A}$ & $\masstwosourceuncert{GW191204A}$ & $\chieffinfinityonlyprecavguncert{GW191204A}$ & $\luminositydistanceuncert{GW191204A}$ & $\redshiftuncert{GW191204A}$ & $\finalmasssourceuncert{GW191204A}$ & $\finalspinuncert{GW191204A}$ & $\skyarea{GW191204A}$ & $\networkmatchedfiltersnrIMRPuncert{GW191204A}$\\
\FULLNAME{GW191204G} & $\totalmasssourceuncert{GW191204G}$ & $\chirpmasssourceuncert{GW191204G}$ & $\massonesourceuncert{GW191204G}$ & $\masstwosourceuncert{GW191204G}$ & $\chieffinfinityonlyprecavguncert{GW191204G}$ & $\luminositydistanceuncert{GW191204G}$ & $\redshiftuncert{GW191204G}$ & $\finalmasssourceuncert{GW191204G}$ & $\finalspinuncert{GW191204G}$ & $\skyarea{GW191204G}$ & $\networkmatchedfiltersnrIMRPuncert{GW191204G}$\\
\makebox[0pt][l]{\fboxsep0pt\colorbox{lightgray}{\mystrut\hspace*{1.0\linewidth}}}\!\!
\FULLNAME{GW191215G} & $\totalmasssourceuncert{GW191215G}$ & $\chirpmasssourceuncert{GW191215G}$ & $\massonesourceuncert{GW191215G}$ & $\masstwosourceuncert{GW191215G}$ & $\chieffinfinityonlyprecavguncert{GW191215G}$ & $\luminositydistanceuncert{GW191215G}$ & $\redshiftuncert{GW191215G}$ & $\finalmasssourceuncert{GW191215G}$ & $\finalspinuncert{GW191215G}$ & $\skyarea{GW191215G}$ & $\networkmatchedfiltersnrIMRPuncert{GW191215G}$\\
\FULLNAME{GW191216G} & $\totalmasssourceuncert{GW191216G}$ & $\chirpmasssourceuncert{GW191216G}$ & $\massonesourceuncert{GW191216G}$ & $\masstwosourceuncert{GW191216G}$ & $\chieffinfinityonlyprecavguncert{GW191216G}$ & $\luminositydistanceuncert{GW191216G}$ & $\redshiftuncert{GW191216G}$ & $\finalmasssourceuncert{GW191216G}$ & $\finalspinuncert{GW191216G}$ & $\skyarea{GW191216G}$ & $\networkmatchedfiltersnrIMRPuncert{GW191216G}$\\
\makebox[0pt][l]{\fboxsep0pt\colorbox{lightgray}{\mystrut\hspace*{1.0\linewidth}}}\!\!
\textit{\FULLNAME{GW191219E}} & $\totalmasssourceuncert{GW191219E}$ & $\chirpmasssourceuncert{GW191219E}$ & $\massonesourceuncert{GW191219E}$ & $\masstwosourceuncert{GW191219E}$ & $\chieffinfinityonlyprecavguncert{GW191219E}$ & $\luminositydistanceuncert{GW191219E}$ & $\redshiftuncert{GW191219E}$ & $\finalmasssourceuncert{GW191219E}$ & $\finalspinuncert{GW191219E}$ & $\skyarea{GW191219E}$ & $\networkmatchedfiltersnrIMRPuncert{GW191219E}$\\
\FULLNAME{GW191222A} & $\totalmasssourceuncert{GW191222A}$ & $\chirpmasssourceuncert{GW191222A}$ & $\massonesourceuncert{GW191222A}$ & $\masstwosourceuncert{GW191222A}$ & $\chieffinfinityonlyprecavguncert{GW191222A}$ & $\luminositydistanceuncert{GW191222A}$ & $\redshiftuncert{GW191222A}$ & $\finalmasssourceuncert{GW191222A}$ & $\finalspinuncert{GW191222A}$ & $\skyarea{GW191222A}$ & $\networkmatchedfiltersnrIMRPuncert{GW191222A}$\\
\makebox[0pt][l]{\fboxsep0pt\colorbox{lightgray}{\mystrut\hspace*{1.0\linewidth}}}\!\!
\FULLNAME{GW191230H} & $\totalmasssourceuncert{GW191230H}$ & $\chirpmasssourceuncert{GW191230H}$ & $\massonesourceuncert{GW191230H}$ & $\masstwosourceuncert{GW191230H}$ & $\chieffinfinityonlyprecavguncert{GW191230H}$ & $\luminositydistanceuncert{GW191230H}$ & $\redshiftuncert{GW191230H}$ & $\finalmasssourceuncert{GW191230H}$ & $\finalspinuncert{GW191230H}$ & $\skyarea{GW191230H}$ & $\networkmatchedfiltersnrIMRPuncert{GW191230H}$\\
\textit{\FULLNAME{200105F}} & $\totalmasssourceuncert{200105F}$ & $\chirpmasssourceuncert{200105F}$ & $\massonesourceuncert{200105F}$ & $\masstwosourceuncert{200105F}$ & $\chieffinfinityonlyprecavguncert{200105F}$ & $\luminositydistanceuncert{200105F}$ & $\redshiftuncert{200105F}$ & $\finalmasssourceuncert{200105F}$ & $\finalspinuncert{200105F}$ & $\skyarea{200105F}$ & $\networkmatchedfiltersnrIMRPuncert{200105F}$\\
\makebox[0pt][l]{\fboxsep0pt\colorbox{lightgray}{\mystrut\hspace*{1.0\linewidth}}}\!\!
\FULLNAME{GW200112H} & $\totalmasssourceuncert{GW200112H}$ & $\chirpmasssourceuncert{GW200112H}$ & $\massonesourceuncert{GW200112H}$ & $\masstwosourceuncert{GW200112H}$ & $\chieffinfinityonlyprecavguncert{GW200112H}$ & $\luminositydistanceuncert{GW200112H}$ & $\redshiftuncert{GW200112H}$ & $\finalmasssourceuncert{GW200112H}$ & $\finalspinuncert{GW200112H}$ & $\skyarea{GW200112H}$ & $\networkmatchedfiltersnrIMRPuncert{GW200112H}$\\
\FULLNAME{GW200115A} & $\totalmasssourceuncert{GW200115A}$ & $\chirpmasssourceuncert{GW200115A}$ & $\massonesourceuncert{GW200115A}$ & $\masstwosourceuncert{GW200115A}$ & $\chieffinfinityonlyprecavguncert{GW200115A}$ & $\luminositydistanceuncert{GW200115A}$ & $\redshiftuncert{GW200115A}$ & $\finalmasssourceuncert{GW200115A}$ & $\finalspinuncert{GW200115A}$ & $\skyarea{GW200115A}$ & $\networkmatchedfiltersnrIMRPuncert{GW200115A}$\\
\makebox[0pt][l]{\fboxsep0pt\colorbox{lightgray}{\mystrut\hspace*{1.0\linewidth}}}\!\!
\FULLNAME{GW200128C} & $\totalmasssourceuncert{GW200128C}$ & $\chirpmasssourceuncert{GW200128C}$ & $\massonesourceuncert{GW200128C}$ & $\masstwosourceuncert{GW200128C}$ & $\chieffinfinityonlyprecavguncert{GW200128C}$ & $\luminositydistanceuncert{GW200128C}$ & $\redshiftuncert{GW200128C}$ & $\finalmasssourceuncert{GW200128C}$ & $\finalspinuncert{GW200128C}$ & $\skyarea{GW200128C}$ & $\networkmatchedfiltersnrIMRPuncert{GW200128C}$\\
\FULLNAME{GW200129D} & $\totalmasssourceuncert{GW200129D}$ & $\chirpmasssourceuncert{GW200129D}$ & $\massonesourceuncert{GW200129D}$ & $\masstwosourceuncert{GW200129D}$ & $\chieffinfinityonlyprecavguncert{GW200129D}$ & $\luminositydistanceuncert{GW200129D}$ & $\redshiftuncert{GW200129D}$ & $\finalmasssourceuncert{GW200129D}$ & $\finalspinuncert{GW200129D}$ & $\skyarea{GW200129D}$ & $\networkmatchedfiltersnrIMRPuncert{GW200129D}$\\
\makebox[0pt][l]{\fboxsep0pt\colorbox{lightgray}{\mystrut\hspace*{1.0\linewidth}}}\!\!
\FULLNAME{GW200202F} & $\totalmasssourceuncert{GW200202F}$ & $\chirpmasssourceuncert{GW200202F}$ & $\massonesourceuncert{GW200202F}$ & $\masstwosourceuncert{GW200202F}$ & $\chieffinfinityonlyprecavguncert{GW200202F}$ & $\luminositydistanceuncert{GW200202F}$ & $\redshiftuncert{GW200202F}$ & $\finalmasssourceuncert{GW200202F}$ & $\finalspinuncert{GW200202F}$ & $\skyarea{GW200202F}$ & $\networkmatchedfiltersnrIMRPuncert{GW200202F}$\\
\FULLNAME{GW200208G} & $\totalmasssourceuncert{GW200208G}$ & $\chirpmasssourceuncert{GW200208G}$ & $\massonesourceuncert{GW200208G}$ & $\masstwosourceuncert{GW200208G}$ & $\chieffinfinityonlyprecavguncert{GW200208G}$ & $\luminositydistanceuncert{GW200208G}$ & $\redshiftuncert{GW200208G}$ & $\finalmasssourceuncert{GW200208G}$ & $\finalspinuncert{GW200208G}$ & $\skyarea{GW200208G}$ & $\networkmatchedfiltersnrIMRPuncert{GW200208G}$\\
\makebox[0pt][l]{\fboxsep0pt\colorbox{lightgray}{\mystrut\hspace*{1.0\linewidth}}}\!\!
\FULLNAME{GW200208K} & $\totalmasssourceuncert{GW200208K}$ & $\chirpmasssourceuncert{GW200208K}$ & $\massonesourceuncert{GW200208K}$ & $\masstwosourceuncert{GW200208K}$ & $\chieffinfinityonlyprecavguncert{GW200208K}$ & $\luminositydistanceuncert{GW200208K}$ & $\redshiftuncert{GW200208K}$ & $\finalmasssourceuncert{GW200208K}$ & $\finalspinuncert{GW200208K}$ & $\skyarea{GW200208K}$ & $\networkmatchedfiltersnrIMRPuncert{GW200208K}$\\
\FULLNAME{GW200209E} & $\totalmasssourceuncert{GW200209E}$ & $\chirpmasssourceuncert{GW200209E}$ & $\massonesourceuncert{GW200209E}$ & $\masstwosourceuncert{GW200209E}$ & $\chieffinfinityonlyprecavguncert{GW200209E}$ & $\luminositydistanceuncert{GW200209E}$ & $\redshiftuncert{GW200209E}$ & $\finalmasssourceuncert{GW200209E}$ & $\finalspinuncert{GW200209E}$ & $\skyarea{GW200209E}$ & $\networkmatchedfiltersnrIMRPuncert{GW200209E}$\\
\makebox[0pt][l]{\fboxsep0pt\colorbox{lightgray}{\mystrut\hspace*{1.0\linewidth}}}\!\!
\FULLNAME{GW200210B} & $\totalmasssourceuncert{GW200210B}$ & $\chirpmasssourceuncert{GW200210B}$ & $\massonesourceuncert{GW200210B}$ & $\masstwosourceuncert{GW200210B}$ & $\chieffinfinityonlyprecavguncert{GW200210B}$ & $\luminositydistanceuncert{GW200210B}$ & $\redshiftuncert{GW200210B}$ & $\finalmasssourceuncert{GW200210B}$ & $\finalspinuncert{GW200210B}$ & $\skyarea{GW200210B}$ & $\networkmatchedfiltersnrIMRPuncert{GW200210B}$\\
\FULLNAME{GW200216G} & $\totalmasssourceuncert{GW200216G}$ & $\chirpmasssourceuncert{GW200216G}$ & $\massonesourceuncert{GW200216G}$ & $\masstwosourceuncert{GW200216G}$ & $\chieffinfinityonlyprecavguncert{GW200216G}$ & $\luminositydistanceuncert{GW200216G}$ & $\redshiftuncert{GW200216G}$ & $\finalmasssourceuncert{GW200216G}$ & $\finalspinuncert{GW200216G}$ & $\skyarea{GW200216G}$ & $\networkmatchedfiltersnrIMRPuncert{GW200216G}$\\
\makebox[0pt][l]{\fboxsep0pt\colorbox{lightgray}{\mystrut\hspace*{1.0\linewidth}}}\!\!
\FULLNAME{GW200219D} & $\totalmasssourceuncert{GW200219D}$ & $\chirpmasssourceuncert{GW200219D}$ & $\massonesourceuncert{GW200219D}$ & $\masstwosourceuncert{GW200219D}$ & $\chieffinfinityonlyprecavguncert{GW200219D}$ & $\luminositydistanceuncert{GW200219D}$ & $\redshiftuncert{GW200219D}$ & $\finalmasssourceuncert{GW200219D}$ & $\finalspinuncert{GW200219D}$ & $\skyarea{GW200219D}$ & $\networkmatchedfiltersnrIMRPuncert{GW200219D}$\\
\FULLNAME{GW200220E} & $\totalmasssourceuncert{GW200220E}$ & $\chirpmasssourceuncert{GW200220E}$ & $\massonesourceuncert{GW200220E}$ & $\masstwosourceuncert{GW200220E}$ & $\chieffinfinityonlyprecavguncert{GW200220E}$ & $\luminositydistanceuncert{GW200220E}$ & $\redshiftuncert{GW200220E}$ & $\finalmasssourceuncert{GW200220E}$ & $\finalspinuncert{GW200220E}$ & $\skyarea{GW200220E}$ & $\networkmatchedfiltersnrIMRPuncert{GW200220E}$\\
\makebox[0pt][l]{\fboxsep0pt\colorbox{lightgray}{\mystrut\hspace*{1.0\linewidth}}}\!\!
\FULLNAME{GW200220H} & $\totalmasssourceuncert{GW200220H}$ & $\chirpmasssourceuncert{GW200220H}$ & $\massonesourceuncert{GW200220H}$ & $\masstwosourceuncert{GW200220H}$ & $\chieffinfinityonlyprecavguncert{GW200220H}$ & $\luminositydistanceuncert{GW200220H}$ & $\redshiftuncert{GW200220H}$ & $\finalmasssourceuncert{GW200220H}$ & $\finalspinuncert{GW200220H}$ & $\skyarea{GW200220H}$ & $\networkmatchedfiltersnrIMRPuncert{GW200220H}$\\
\FULLNAME{GW200224H} & $\totalmasssourceuncert{GW200224H}$ & $\chirpmasssourceuncert{GW200224H}$ & $\massonesourceuncert{GW200224H}$ & $\masstwosourceuncert{GW200224H}$ & $\chieffinfinityonlyprecavguncert{GW200224H}$ & $\luminositydistanceuncert{GW200224H}$ & $\redshiftuncert{GW200224H}$ & $\finalmasssourceuncert{GW200224H}$ & $\finalspinuncert{GW200224H}$ & $\skyarea{GW200224H}$ & $\networkmatchedfiltersnrIMRPuncert{GW200224H}$\\
\makebox[0pt][l]{\fboxsep0pt\colorbox{lightgray}{\mystrut\hspace*{1.0\linewidth}}}\!\!
\FULLNAME{GW200225B} & $\totalmasssourceuncert{GW200225B}$ & $\chirpmasssourceuncert{GW200225B}$ & $\massonesourceuncert{GW200225B}$ & $\masstwosourceuncert{GW200225B}$ & $\chieffinfinityonlyprecavguncert{GW200225B}$ & $\luminositydistanceuncert{GW200225B}$ & $\redshiftuncert{GW200225B}$ & $\finalmasssourceuncert{GW200225B}$ & $\finalspinuncert{GW200225B}$ & $\skyarea{GW200225B}$ & $\networkmatchedfiltersnrIMRPuncert{GW200225B}$\\
\FULLNAME{GW200302A} & $\totalmasssourceuncert{GW200302A}$ & $\chirpmasssourceuncert{GW200302A}$ & $\massonesourceuncert{GW200302A}$ & $\masstwosourceuncert{GW200302A}$ & $\chieffinfinityonlyprecavguncert{GW200302A}$ & $\luminositydistanceuncert{GW200302A}$ & $\redshiftuncert{GW200302A}$ & $\finalmasssourceuncert{GW200302A}$ & $\finalspinuncert{GW200302A}$ & $\skyarea{GW200302A}$ & $\networkmatchedfiltersnrIMRPuncert{GW200302A}$\\
\makebox[0pt][l]{\fboxsep0pt\colorbox{lightgray}{\mystrut\hspace*{1.0\linewidth}}}\!\!
\FULLNAME{GW200306A} & $\totalmasssourceuncert{GW200306A}$ & $\chirpmasssourceuncert{GW200306A}$ & $\massonesourceuncert{GW200306A}$ & $\masstwosourceuncert{GW200306A}$ & $\chieffinfinityonlyprecavguncert{GW200306A}$ & $\luminositydistanceuncert{GW200306A}$ & $\redshiftuncert{GW200306A}$ & $\finalmasssourceuncert{GW200306A}$ & $\finalspinuncert{GW200306A}$ & $\skyarea{GW200306A}$ & $\networkmatchedfiltersnrIMRPuncert{GW200306A}$\\
\FULLNAME{GW200308G}$^{*}$ & $\totalmasssourceuncert{GW200308G}$ & $\chirpmasssourceuncert{GW200308G}$ & $\massonesourceuncert{GW200308G}$ & $\masstwosourceuncert{GW200308G}$ & $\chieffinfinityonlyprecavguncert{GW200308G}$ & $\luminositydistanceuncert{GW200308G}$ & $\redshiftuncert{GW200308G}$ & $\finalmasssourceuncert{GW200308G}$ & $\finalspinuncert{GW200308G}$ & $\skyarea{GW200308G}$ & $\networkmatchedfiltersnrIMRPuncert{GW200308G}$\\
\makebox[0pt][l]{\fboxsep0pt\colorbox{lightgray}{\mystrut\hspace*{1.0\linewidth}}}\!\!
\FULLNAME{GW200311L} & $\totalmasssourceuncert{GW200311L}$ & $\chirpmasssourceuncert{GW200311L}$ & $\massonesourceuncert{GW200311L}$ & $\masstwosourceuncert{GW200311L}$ & $\chieffinfinityonlyprecavguncert{GW200311L}$ & $\luminositydistanceuncert{GW200311L}$ & $\redshiftuncert{GW200311L}$ & $\finalmasssourceuncert{GW200311L}$ & $\finalspinuncert{GW200311L}$ & $\skyarea{GW200311L}$ & $\networkmatchedfiltersnrIMRPuncert{GW200311L}$\\
\FULLNAME{GW200316I} & $\totalmasssourceuncert{GW200316I}$ & $\chirpmasssourceuncert{GW200316I}$ & $\massonesourceuncert{GW200316I}$ & $\masstwosourceuncert{GW200316I}$ & $\chieffinfinityonlyprecavguncert{GW200316I}$ & $\luminositydistanceuncert{GW200316I}$ & $\redshiftuncert{GW200316I}$ & $\finalmasssourceuncert{GW200316I}$ & $\finalspinuncert{GW200316I}$ & $\skyarea{GW200316I}$ & $\networkmatchedfiltersnrIMRPuncert{GW200316I}$\\
\makebox[0pt][l]{\fboxsep0pt\colorbox{lightgray}{\mystrut\hspace*{1.0\linewidth}}}\!\!
\FULLNAME{GW200322G}$^{*}$ & $\totalmasssourceuncert{GW200322G}$ & $\chirpmasssourceuncert{GW200322G}$ & $\massonesourceuncert{GW200322G}$ & $\masstwosourceuncert{GW200322G}$ & $\chieffinfinityonlyprecavguncert{GW200322G}$ & $\luminositydistanceuncert{GW200322G}$ & $\redshiftuncert{GW200322G}$ & $\finalmasssourceuncert{GW200322G}$ & $\finalspinuncert{GW200322G}$ & $\skyarea{GW200322G}$ & $\networkmatchedfiltersnrIMRPuncert{GW200322G}$\\
\end{tabular}\end{ruledtabular}
    	\caption{
    	\label{tab:pe} Median and $90\%$ symmetric credible intervals for selected source parameters, and the $90\%$ credible area for the sky localization for \ac{O3b} candidates with $\pastro{} > \PASTROTHRESHOLD$ plus \FULLNAME{200105F}{}. 
        We highlight with \textit{italics} \FULLNAME{200105F}{} as it has $\pastro{} < \PASTROTHRESHOLD$, as well as \FULLNAME{GW191219E}{} because of significant uncertainty in its $\pastro{}$ and because it has significant posterior support outside of mass ratios where the waveform models have been calibrated. 
        An asterisk ($\ast$) is used to indicate candidates for which the posterior distributions are dominated by potentially unphysical, low-likelihood modes at large distances and high masses, and are particularly prior sensitive. 
    	The columns show source total mass $\Mtot$, chirp mass $\Mc$, component masses $m_i$, effective inspiral spin $\chieff$, luminosity distance $\DL$, redshift $\redshift$, final mass $\Mf$, final spin $\chif$, sky localization $\Delta \Omega$ and the network matched-filter \ac{SNR}. 
	All quoted results are calculated using \ac{BBH} waveform models; values come from averaging \IMRPhenomXPHM{} and \SEOBNRPHM{} results, except for the \ac{SNR} (which is given for \IMRPhenomXPHM{} because the \RIFT{} analysis used for \SEOBNRPHM{} does not output this quantity).  	
    }
    \end{table*}
\end{PE_table}


The \ac{O3b} candidates show a diversity in their source properties. 
Many are similar to previous observations, but some do show unusual features. 
While the mass posterior probability distributions are typically unimodal, some results show multimodal behavior. 
For example, \FULLNAME{GW200129D}{} shows a bimodality in mass ratio that translates to a bimodality in \masstwo{}. 
\FULLNAME{GW200225B}{} and \FULLNAME{GW200306A}{} both show bimodality in their redshifted chirp-mass distributions, although their source-mass distributions (shown in Fig.~\ref{fig:post_1d_multi}) are unimodal, as the additional uncertainty from the inferred redshift is sufficient to broaden the modes such that they merge.
Becaue of the correlations between masses and spins~\cite{Poisson:1995ef,Baird:2012cu,Farr:2015lna}, multimodality in mass distributions may also translate to multiple peaks in the effective inspiral spin distribution. 
Multimodality can arise due to the complexity of the likelihood surface when using waveform models that include higher-order multipole moments~\cite{Estelles:2021jnz,Mehta:2021fgz,Chia:2021mxq,Nitz:2021uxj} and precession~\cite{LIGOScientific:2018hze,Abbott:2020uma}, noise fluctuations for quiet signals~\cite{Huang:2018tqd}, the presence of glitches~\cite{Powell:2018csz,Chatziioannou:2021ezd,Ashton:2021tvz}, or there being multiple overlapping signals in the data (which is unlikely given \ac{O3} sensitivity)~\cite{Relton:2021cax}. 
Therefore, multimodality is expected in a few cases.

Cases with significant multimodality are \FULLNAME{GW200208K}{}, \FULLNAME{GW200308G}{} and \FULLNAME{GW200322G}{}.
These candidates have modest significance with $\pastro{} = \MAXPASTRO{GW200208K}$, $\MAXPASTRO{GW200308G}$ and $\MAXPASTRO{GW200322G}$, respectively, and are each identified with $\pastro > \PASTROTHRESHOLD$ by only one search analysis. 
They have low \acp{SNR}, using \IMRPhenomXPHM{} they are inferred to have $\rho = \networkmatchedfiltersnrIMRPuncert{GW200208K}$, $\networkmatchedfiltersnrIMRPuncert{GW200308G}$ and $\networkmatchedfiltersnrIMRPuncert{GW200322G}$, respectively.
For \FULLNAME{GW200208K}{} the two main modes have comparable likelihoods, indicating comparable fits to the data, while for \FULLNAME{GW200308G}{} and \FULLNAME{GW200322G}{}, there are significant modes with lower likelihoods. 
The posterior probability distributions for \FULLNAME{GW200308G}{} and \FULLNAME{GW200322G}{} both have peaks at lower masses and lower distances, and another broader peak corresponding to higher masses and larger distances; this high-mass, large-distance peak is dominated by the prior. 
The default prior probability distribution (described in Appendix~\ref{sec:prior-sampling}) places significant weight at large distances, and at high masses. 
This means that we can find significant posterior probability at large distances and high masses, even when the likelihood is low. 
Such low-likelihood peaks, corresponding to low \acp{SNR}, may arise due to a random noise fluctuation matching the signal template. 
For \FULLNAME{GW200308G}{} and \FULLNAME{GW200322G}{}, the high-mass and high-distance peak has lower likelihood and posterior support for \acp{SNR} $\rho \sim 0$. 
For such candidates, the multimodality indicates that we cannot separate the possibility of a signal from a lower-mass, closer source from a weaker (potentially vanishing) signal from a higher-mass, more-distant source. 
However, this support for high masses and large distances is driven by our choice of prior, which was not designed to model the astrophysical population of sources. 
Therefore, we consider that the high-likelihood peaks for \FULLNAME{GW200308G}{} and \FULLNAME{GW200322G}{} yield a more plausible estimate of the source parameters, although we cannot exclude the possibility that the low-likelihood peaks describe the sources (assuming that the signals are astrophysical).

All results are given assuming our default priors.
We highlight results for \FULLNAME{GW200308G}{} and \FULLNAME{GW200322G}{} in Table~\ref{tab:pe} and the figures to indicate that these results may be especially sensitive to the choice of prior. 
Using a different prior, such as a population-informed prior~\cite{Mandel:2009pc,Fishbach:2019ckx,Galaudage:2019jdx,Miller:2020zox,Kimball:2020opk,Abbott:2020gyp,Callister:2021fpo}, that has a stronger preference for masses more consistent with other \ac{GW} observations, and a weaker preference for high masses and large distances, would alter results.

\subsection{Masses}\label{sec:mass}

Masses are typically the best constrained binary parameters. 
They are the dominant properties in setting the frequency evolution of the signal, with lower- (higher-) mass systems merging at higher (lower) frequencies. 
While we are typically interested in the source masses, it is the redshifted masses $(1+\redshift)m_i$, where $\redshift$ is the source redshift, that are measured by the detectors~\cite{Krolak:1987ofj}. 
The source masses are calculated by combining the inferred redshifted mass and luminosity distance (see Appendix~\ref{sec:parameter-estimation-methods} for the assumed cosmology). 

Combinations of the two component masses (such as the chirp mass) may be more precisely measured than the individual component masses~\cite{Poisson:1995ef,Baird:2012cu,Farr:2015lna,Vitale:2016avz}. 
However, component masses are most informative about the nature of the source, and indicate whether the compact object is more likely to be a \ac{BH} or a \ac{NS}. 
The maximum \ac{NS} mass is currently uncertain, with estimates ranging over \LITERATUREMMAX{}~\cite{Margalit:2017dij,Ai:2019rre,Shao:2020bzt,Lim:2020zvx,Miller:2021qha,Raaijmakers:2021uju}.
We use $\PEBHMassThreshold$ as a robust upper limit of the maximum \ac{NS} mass~\cite{Rhoades:1974fn,Kalogera:1996ci}, and split the candidates into two categories: \emph{unambiguous \acp{BBH}} where, assuming that the signal is astrophysical, both components of the source were \acp{BH} ($\masstwo > \PEBHMassThreshold$ at $\PEBHMassThresholdProb$ probability), and \emph{potential-\ac{NS} binaries} (in our case, potential-\ac{NSBH} binaries) where at least one component could have been a \ac{NS}. 
Candidates from the two categories are discussed in Sec.~\ref{sec:mass-bh} and Sec.~\ref{sec:mass-tides}, respectively.
As shown in Fig.~\ref{fig:mtotvsqpost}, all of the \NUMEVENTS{} candidates with $\pastro{} > \PASTROTHRESHOLD$ except \FULLNAME{GW191219E}{}, \FULLNAME{GW200115A}{} and \FULLNAME{GW200210B}{} (plus \FULLNAME{200105F}{}) have $\masstwo > \PEBHMassThreshold$, and none of the candidates have posterior support for $\massone < \PEBHMassThreshold$, which would be required for a \ac{BNS} source.
Therefore, we identify the majority of sources as \acp{BBH}. 

\subsubsection{Masses of sources with strictly $\masstwo > 3 \Msun$: Unambiguous \acp{BBH}}\label{sec:mass-bh}

The mass combination with greatest influence on a \ac{CBC} signal's frequency evolution is the chirp mass $\Mc$~\cite{Blanchet:1995ez}. 
The chirp mass's influence on the inspiral means that it is more precisely measured in lower-mass systems, which have more of the inspiral signal in the sensitive frequency band of the detectors~\cite{Ajith:2009fz,Graff:2015bba,Haster:2015cnn,Veitch:2015ela,Islam:2021zee}. 
This is illustrated in Fig.~\ref{fig:mcchieffpost}, which also shows the effective inspiral spin (Sec.~\ref{sec:spin}).  
The modestly significant ($\pastro = \MAXPASTRO{\chirpmasssourcemost}$) \FULLNAME{\chirpmasssourcemost}{} probably has the highest chirp-mass source  of the \ac{O3b} candidates, with $\Mc = \chirpmasssourceuncert{\chirpmasssourcemost}\Msun$. 
Similarly, \FULLNAME{\chirpmasssourceleastBBH}{}'s source probably has the lowest while still being an unambiguous-\ac{BBH} ($\masstwo > \PEBHMassThreshold$) candidate, with $\Mc = \chirpmasssourceuncert{\chirpmasssourceleastBBH}\Msun$. 
The range of chirp masses for the \ac{O3b} candidates is consistent with \GWTCTWOFINAL{}~\cite{Abbott:2020niy,LIGOScientific:2021usb}.

The total mass of the binary $\Mtot$ influences the merger and ringdown of the signal, which constitute a more significant proportion of the observed signal for higher-mass sources~\cite{Echeverria:1989hg,Berti:2005ys,TheLIGOScientific:2016pea}.   
The \ac{O3b} candidates with the highest $\Mtot$ measurements, \FULLNAME{\totalmasssourcemost}{} and (the multimodal) \FULLNAME{\totalmasssourcemostsecond}{}, have lower median $\Mtot$ measurements than \NNAME{GW190521B}{}~\cite{Abbott:2020tfl,LIGOScientific:2021usb}, of $\Mtot = \totalmasssourceuncert{\totalmasssourcemost} \Msun$ and $\totalmasssourceuncert{\totalmasssourcemostsecond} \Msun$, respectively. 
The lowest-mass \ac{O3b} unambiguous-\ac{BBH} candidate is \FULLNAME{\totalmasssourceleastBBH}{}'s source, with $\Mtot = \totalmasssourceuncert{\totalmasssourceleastBBH} \Msun$. 
Posterior probability distributions for the total mass and mass ratio are shown in Fig.~\ref{fig:mtotvsqpost}; the curving degeneracies seen at lower masses are where distributions follow a line of constant chirp mass.

Mass ratios are typically less precisely inferred from \ac{GW} observations than the chirp mass or total mass. 
The mass ratio influences the phase evolution of the inspiral at the \ac{PN} order after the chirp mass~\cite{Cutler:1994ys,Blanchet:1995ez,Poisson:1995ef,Baird:2012cu}.
Most measured mass ratios are consistent with the equal-mass limit $\massratio{} = 1$, as shown in Fig.~\ref{fig:post_1d_multi}.
For example, \FULLNAME{\massratiomostBBH}{} and \FULLNAME{\massratiomostsecondBBH}{} have $\massratio{} \geq \massratiotenthpercentile{\massratiomostBBH}$ and $\geq \massratiotenthpercentile{\massratiomostsecondBBH}$ at $90\%$ probability, respectively. 
However, multiple \ac{BBH} candidates have support for unequal masses. 
\FULLNAME{GW191113B}{}'s source has an inferred $\massratio{} = \massratiouncert{GW191113B}$ ($\massratio{} \leq \massrationintiethpercentile{GW191113B}$ at $90\%$ probability) and \FULLNAME{GW200208K}{}'s source has $\massratio{} = \massratiouncert{GW200208K}$ ($\massratio{} \leq \massrationintiethpercentile{GW200208K}$ at $90\%$ probability). 
Some posterior probability distributions extend outside the calibration range for current waveform models, and hence may be subject to additional systematic uncertainties~\cite{Ossokine:2020kjp,Pratten:2020ceb}. 
Future analysis with waveforms with improved fidelity at more extreme mass ratios should lead to a more complete understanding of these sources. 
\FULLNAME{GW191113B}{} and \FULLNAME{GW200208K}{} have moderate significance ($\pastro =\MAXPASTRO{GW191113B}$ and $\MAXPASTRO{GW200208K}$, respectively), and hence may not be a reflection of the true \ac{BBH} population. 
Using a population-informed prior~\cite{Mandel:2009pc,Fishbach:2019ckx,Galaudage:2019jdx,Miller:2020zox,Kimball:2020opk,Abbott:2020gyp,Callister:2021fpo}, in place of our default uninformative prior, may give greater weight to equal masses~\cite{LIGOScientific:2021psn}. 

Considering individual \ac{BH} masses, the unambiguous-\ac{BBH} candidates have component masses ranging from ${\sim\masstwosourceuncert{\masstwosourceleastBBH} \Msun}$ to ${\sim\massonesourceuncert{\massonesourcemostBBH} \Msun}$. 
Primary masses range from $\massonesourceuncert{\massonesourceleastBBH} \Msun$ for \FULLNAME{\massonesourceleastBBH}{} to $\massonesourceuncert{\massonesourcemostBBH} \Msun$ and $\massonesourceuncert{\massonesourcemostsecondBBH} \Msun$ for \FULLNAME{\massonesourcemostBBH}{} and \FULLNAME{\massonesourcemostsecondBBH}{}, while secondary masses range from $\masstwosourceuncert{\masstwosourceleastBBH} \Msun$ for \FULLNAME{\masstwosourceleastBBH}{} to $\masstwosourceuncert{\masstwosourcemostBBH} \Msun$ for \FULLNAME{\masstwosourcemostBBH}{}. 
The distribution of component masses is analyzed, and its astrophysical implications discussed, in a companion paper~\cite{LIGOScientific:2021psn}.

Given our default prior assumptions, there is a $\percentmassonemorethansixtyfive{GW200220E}\%$ probability that the primary \ac{BH} in \FULLNAME{GW200220E}{} has a mass $\massone > \PEPISNLowerThreshold$; this is approximately the maximum mass of \acp{BH} expected to be formed from stellar collapse before encountering pair-instability supernovae~\cite{Fowler:1964zz,Barkat:1967zz,Fryer:2000my,Belczynski:2016jno,Spera:2017fyx,Stevenson:2019rcw,Mehta:2021fgz}, where the progenitor stars would be disrupted leaving no remnant behind, although there are many physical uncertainties that can impact this maximum mass~\cite{Farmer:2019jed,vanSon:2020zbk,Marchant:2020haw,Costa:2020xbc,Tanikawa:2020abs,Umeda:2020fih,Vink:2020nak,Woosley:2021xba}. 
\FULLNAME{GW191109A}{} has $\percentmassonemorethansixtyfive{GW191109A}\%$ probability that $\massone > \PEPISNLowerThreshold$, while \FULLNAME{GW200208K}{} and \FULLNAME{GW191127B}{} have probabilities $\percentmassonemorethansixtyfive{GW200208K}\%$ and $\percentmassonemorethansixtyfive{GW191127B}\%$, respectively. 
Similarly, \FULLNAME{\masstwosourcemostBBH}{} has a $\percentmasstwomorethansixtyfive{\masstwosourcemostBBH}\%$ probability that its secondary has $\masstwo > \PEPISNLowerThreshold$.
\FULLNAME{GW200220E}{} and \FULLNAME{GW200208K}{}  
have $\percentmassonemorethanonetwenty{GW200220E}\%$ and $\percentmassonemorethanonetwenty{GW200208K}\%$ probabilities that $\massone > \PEPISNUpperThreshold$, respectively, 
which is expected to be approximately the mass where the pair-instability supernova mass gap ends~\cite{Spera:2017fyx,Farmer:2020xne,Renzo:2020lwl,Costa:2020xbc,Mehta:2021fgz}. 

Based upon X-ray binary observations, there is a hypothesized lower \ac{BH} mass gap below $\PECCMassGapUpper$~\cite{Bailyn:1997xt,Ozel:2010su,Farr:2010tu,Kreidberg:2012ud}. 
This may be a signature of the physics of core-collapse supernova explosions~\cite{Fryer:2011cx,Mandel:2020qwb,Zevin:2020gma,Liu:2020uba,Patton:2021gwh}. 
We infer that there are some \acp{BBH} that may have components in this mass gap.
Given our standard prior assumptions, the candidate with the most posterior support for $\masstwo < \PECCMassGapUpper$ is \FULLNAME{GW191113B}{} with $\percentmasstwolessthanfive{GW191113B}\%$ probability. 
None of the unambiguous-\ac{BBH} candidates has a primary mass consistent with being in the lower mass gap. 

The component \ac{BH} masses overlap with those from previous \ac{GW} and electromagnetic observations. 
The range is consistent with observations in \GWTCTWOFINAL{}~\cite{Abbott:2020tfl,LIGOScientific:2021usb}. 
Non-\ac{LVK} analysis of public \ac{GW} data has led to other \ac{BBH} candidates being reported~\cite{Venumadhav:2019tad,Zackay:2019tzo,Venumadhav:2019lyq,Zackay:2019btq,Nitz:2019hdf,Nitz:2021uxj}; these \acp{BBH} have inferred masses and mass ratios that are consistent with the systems found here. 
From these non-\ac{LVK} searches, the marginal candidate GW170817A~\cite{Zackay:2019btq,Roulet:2021hcu} may have the most massive source, with $\massone = \IASSeventeenZeroEightSeventeenAMassOne$ and $\masstwo = \IASSeventeenZeroEightSeventeenAMassTwo$.
While overlapping at lower masses, the \ac{BH} masses inferred from \ac{GW} observations extend above the masses seen in X-ray binaries~\cite{Ozel:2010su,Farr:2010tu,Casares:2013tpa,Casares:2014gma,Corral-Santana:2015fud,Miller-Jones:2021plh}. 
However, these X-ray binaries are largely expected not to form merging \acp{BBH}~\cite{Belczynski:2012yt,TheLIGOScientific:2016htt}: for example, while Cygnus X-1 may form two \acp{BH}, predictions indicate that there is only a small probability that they would merge within a Hubble time~\cite{Neijssel:2021imj}. 
Additionally, X-ray observations are typically drawn from binaries with near solar metallicity. 
Stellar mass loss due to winds increases with metallicity~\cite{Puls:2008mk,Smith:2014txa,Vink:2021dxo}, so stars formed at solar metallicity leave less massive remnants than stars formed at lower metallicity with the same initial mass~\cite{Belczynski:2009xy,Spera:2017fyx,Eldridge:2017cyw,Giacobbo:2017qhh,Neijssel:2019irh,Higgins:2021jux}. 
Studying the masses of \acp{BH} will provide insight into their formation and the lives of their progenitors~\cite{Stevenson:2015bqa,Barrett:2017fcw,Farmer:2020xne,Hall:2020daa,Bavera:2020uch,Bouffanais:2020qds,Zevin:2020gbd,Franciolini:2021tla,Mapelli:2021gyv}.

The remnant \acp{BH} formed from the mergers have masses $\Mf = \Mtot - \Erad/c^2$ where $\Erad$ is the energy radiated as \acp{GW}, which typically corresponds to a few percent of $\Mtot$~\cite{Baker:2008mj,Reisswig:2009vc,Healy:2016lce,Jimenez-Forteza:2016oae}. 
The most massive remnant \ac{BH} among the \ac{O3b} candidates probably corresponds to \FULLNAME{GW200220E}{}, with a final mass of $\finalmasssourceuncert{GW200220E} \Msun$. 
Using our default priors, there is a $\percentmassfinalmorethanonehundred{GW200220E}\%$ probability of its final \ac{BH} mass being above $\IMBHThreshold$ (a conventional threshold for being considered an \ac{IMBH}~\cite{Miller:2003sc,Greene:2019vlv,LIGOScientific:2021tfm}).
Several other systems are consistent with $\Mf > \IMBHThreshold$, including \FULLNAME{GW191109A}{}'s remnant, which has a $\percentmassfinalmorethanonehundred{GW191109A}\%$ probability of exceeding this threshold. 

\subsubsection{Masses of sources with support for $\masstwo < 3 \Msun$: Potential-\ac{NS} binaries}\label{sec:mass-tides}

The candidates \FULLNAME{GW191219E}{}, \FULLNAME{GW200115A}{}, \FULLNAME{GW200210B}{} and \FULLNAME{200105F}{} are all consistent with originating from a source with $\masstwo < \PEBHMassThreshold$.
When a coalescing binary contains a \ac{NS}, matter effects modify the waveform. 
If these effects can be measured, we can identify that the component is a \ac{NS} rather than a \ac{BH}. 
For \ac{O3b} candidates, as discussed in Sec.~\ref{sec:tides}, we find no measurable matter effects.
Without this information, from the \ac{GW} signal we can infer only the component type from their masses. 

As illustrated by Fig.~\ref{fig:post_1d_multi} and Fig.~\ref{fig:mtotvsqpost}, the \ac{O3b} candidates with potential-\ac{NS} binary sources have more extreme mass ratios than the typical \ac{BBH} candidates.  
At $90\%$ probability, the sources of \FULLNAME{GW191219E}{}, \FULLNAME{200105F}{}, \FULLNAME{GW200115A}{} and \FULLNAME{GW200210B}{} have mass ratios $\massratio{} \leq \massrationintiethpercentile{GW191219E}$, $\leq \massrationintiethpercentile{200105F}$, $\leq \massrationintiethpercentile{GW200115A}$ and $\leq \massrationintiethpercentile{GW200210B}$, respectively. 
The mass ratio of \FULLNAME{GW200210B}{}'s source is $\massratio{} = \massratiouncert{GW200210B}$, which is comparable to \NNAME{GW190814H}{}'s $\massratio{} = \GWNineteenZeroEightFourteenMassRatio$~\cite{Abbott:2020khf,LIGOScientific:2021usb}. 
The mass ratio of \FULLNAME{GW191219E}{}'s source is inferred to be $\massratio{} = \massratiouncert{GW191219E}$, which is extremely challenging for waveform modeling, and thus, there may be systematic uncertainties in results for this candidate.

\FULLNAME{\totalmasssourceleast}{}'s source is the lowest total mass \ac{O3b} binary; this potential \ac{NSBH} coalescence has $\Mtot = \totalmasssourceuncert{\totalmasssourceleast} \Msun$. 
Its chirp mass is well measured at $\Mc = \chirpmasssourceuncert{\totalmasssourceleast} \Msun$. 
\FULLNAME{GW200115A}{}'s source has components with masses $\massone = \massonesourceuncert{GW200115A} \Msun$ and $\masstwo = \masstwosourceuncert{GW200115A} \Msun$. 
These results are consistent with previous inferences~\cite{LIGOScientific:2021qlt}, showing that the change in how the \FASTSCATTER{} glitches in Livingston data were mitigated (discussed in Appendix~\ref{sec:data-methods}) does not have a significant impact on this analysis. 
The primary is consistent with being a low-mass \ac{BH}~\cite{LIGOScientific:2021qlt}, and we infer a $\percentmassonelessthanfive{GW200115A}\%$ probability that $\massone < \PECCMassGapUpper$; the secondary is consistent with the masses of known Galactic \acp{NS}~\cite{Alsing:2017bbc,Shao:2020bzt,Fonseca:2021wxt,Reardon:2021gko}. 

\FULLNAME{200105F}{}'s source corresponds to a higher-mass \ac{NSBH} candidate, with $\Mtot = \totalmasssourceuncert{200105F} \Msun$ and $\Mc = \chirpmasssourceuncert{200105F} \Msun$.
The binary components have masses $\massone = \massonesourceuncert{200105F} \Msun$ and $\masstwo = \masstwosourceuncert{200105F} \Msun$, which are consistent with a \ac{BH} and a \ac{NS}, respectively~\cite{LIGOScientific:2021qlt}. 

\FULLNAME{GW200210B}{}'s source has $\Mtot = \totalmasssourceuncert{GW200210B} \Msun$ and $\Mc = \chirpmasssourceuncert{GW200210B} \Msun$, which sit within the range seen for the unambiguous-\acp{BBH} candidates discussed in Sec.~\ref{sec:mass-bh}. 
While the primary is clearly a \ac{BH} with $\massone = \massonesourceuncert{GW200210B} \Msun$, its secondary has $\masstwo = \masstwosourceuncert{GW200210B} \Msun$ with a $\percentmasstwolessthanthree{GW200210B}\%$ probability that $\masstwo < \PEBHMassThreshold$. 
The secondary mass sits within the hypothesized lower mass gap between \acp{NS} and \acp{BH}~\cite{Bailyn:1997xt,Ozel:2010su,Farr:2010tu,Kreidberg:2012ud}. 
The inferred $\masstwo$ is comparable to (i) the $\ThompsonEtAlMass$ ($\ThompsonEtAlConfidence$ confidence) candidate \ac{BH} in the noninteracting binary 2MASS J05215658+4359220~\cite{Thompson:2018ycv}, (ii) the $\JayasingheEtAlMass$ ($\JayasingheEtAlConfidence$ confidence) candidate \ac{BH} binary companion to V723~Mon~\cite{Jayasinghe:2021uqb}, although this binary has alternatively been interpreted as a stripped low-mass giant star with a subgiant companion~\cite{El-Badry:2022xmn}, and (iii) potentially the pulsar J1748$-$2021B's estimated mass of $\FreireEtAlMass$ ($\FreireEtAlConfidence$ confidence) if the assumption of purely relativistic precession (with no contributions from tidal or rotational distortion of the companion) is accurate~\cite{Freire:2007jd}. 
\FULLNAME{GW200210B}{}'s source is similar to \NNAME{GW190814H}{}'s, where the component masses were inferred to be $\massone{} = \GWNineteenZeroEightFourteenMassOne$ and $\masstwo{} = \GWNineteenZeroEightFourteenMassTwo$~\cite{Abbott:2020khf,LIGOScientific:2021usb}.  
\FULLNAME{GW200210B}{}'s source could either be a \ac{BBH} or a \ac{NSBH} system, but given the current understanding of the maximum \ac{NS} mass~\cite{Shao:2020bzt,Essick:2020ghc,Godzieba:2020tjn,Huang:2020cab,Most:2020bba,Tsokaros:2020hli,Tews:2020ylw,Lim:2020zvx}, it is more probable that it is a \ac{BBH}, similar to the case for \NNAME{GW190814H}{}~\cite{Abbott:2020khf}.  

For \FULLNAME{GW191219E}{}, we infer a source with $\Mtot = \totalmasssourceuncert{GW191219E} \Msun$ and $\Mc = \chirpmasssourceuncert{GW191219E} \Msun$. 
It has $\massone = \massonesourceuncert{GW191219E} \Msun$ and $\masstwo = \masstwosourceuncert{GW191219E} \Msun$, which would make the source a clear \ac{NSBH}, assuming that the signal is astrophysical. 
The secondary is probably the least massive compact object among the \ac{O3b} observations, and is comparable to the least massive of known \acp{NS}~\cite{Ozel:2016oaf,Alsing:2017bbc,Shao:2020bzt}: 
for example, the companion to pulsar J0453+1559 that has an estimated mass of $\MartinezEtAlMass$ ($\MartinezEtAlConfidence$ confidence)~\cite{Martinez:2015mya}, although this object has also been suggested to be a white dwarf~\cite{Tauris:2019sho}; the pulsar J1802$-$2124 that has an estimated mass $\FerdmanEtAlMass$ ($\FerdmanEtAlConfidence$ confidence)~\cite{Ferdman:2010rk}, or the \acp{NS} in the high-mass X-ray binaries SMC~X-1 and 4U~1538$-$522 that have inferred masses of $\FalangaEtAlSMCXOneMass$ and $\FalangaEtAlFourUFifteenThirtyEightMass$ ($\FalangaEtAlConfidence$ confidence), respectively~\cite{Falanga:2015mra}.

Measuring the mass distribution of \acp{NS} will illuminate the physical processes that form them. 
Determining the maximum \ac{NS} mass provides a key insight into the properties of \ac{NS} matter~\cite{LIGOScientific:2019eut,Essick:2019ldf,Chatziioannou:2020msi,Godzieba:2020tjn,Tsokaros:2020hli,Tews:2020ylw,Wysocki:2020myz,Golomb:2021tll}, while determining the spectrum of \ac{NS} masses provides an insight into the physics of processes such as supernova explosions~\cite{Tauris:2015xra,Muller:2018utr,Vigna-Gomez:2018dza,Suwa:2018uni,Burrows:2019zce,Tauris:2019sho,Ertl:2019zks,Patton:2021gwh}. 
As the catalog of observations grows, it will be possible to better determine the \ac{NS} mass distribution.

\subsection{Spins}\label{sec:spin}

Spins leave a relatively subtle imprint on the \ac{GW} signal, and so they are more difficult to measure from observations than the masses~\cite{Poisson:1995ef,Baird:2012cu,Vitale:2014mka,Chatziioannou:2014coa,Farr:2015lna,Vitale:2016avz,TheLIGOScientific:2016pea,Pratten:2020igi}. 
Typically, it is not possible to put strong constraints on individual components' spins, as the evolution of the system is primarily determined by mass-weighted combinations of the two component spins~\cite{Damour:2001tu,Blanchet:2013haa,Purrer:2015nkh,Ng:2018neg,Zevin:2020gxf}. 
However, when a binary has unequal masses it may also be possible to constrain the primary spin because $\spinone$ dominates the spin contributions to the signal.
To reflect how the two spins influence the signal, we quote results for two convenient spin parameters: the effective inspiral spin $\chieff$~\cite{Ajith:2009bn,Santamaria:2010yb} and the effective precession spin $\chip$~\cite{Hannam:2013oca,Schmidt:2014iyl}. 

The effective inspiral spin, as defined in Eq.~\eqref{eq:eff_inspiral_spin_vec}, describes the mass-weighted projection of the component spins parallel to the orbital angular momentum, and is approximately conserved throughout the inspiral~\cite{Racine:2008qv} while remaining important in determining evolution through the merger~\cite{Campanelli:2006uy,Reisswig:2009vc,Purrer:2013ojf}. 
The effective inspiral spin influences the length of the inspiral and the transition to merger~\cite{Campanelli:2006uy,Reisswig:2009vc,Blanchet:2013haa,Roulet:2018jbe}. 
A nonzero $\chieff$ indicates the definite presence of spins in the system, with positive values indicating that there is a net spin aligned with the orbital angular momentum, and negative values indicating that there is a net spin antialigned with the orbital angular momentum. 

The effective precession spin, 
\begin{equation}
    \chip =\max \left\{\chi_{1,\perp}, \frac{q (4 q + 3)}{4 + 3 q} \chi_{2,\perp}\right\},
\end{equation}
where $\chi_{i,\perp}$ is the component of spin perpendicular to the direction of the Newtonian orbital angular momentum $\LNewton$, measures the mass-weighted in-plane spin component that contributes to spin precession~\cite{Apostolatos:1994mx,Kidder:1995zr,Hannam:2013oca,Schmidt:2014iyl}. 
With this parametrization, a value of $\chip = 0$ would indicate no spin precession, and a value of $\chip = 1$ indicates maximal precession; typically only weak constraints are placed on \chip{}, so the posterior covers a significant fraction of its prior range~\cite{Abbott:2016izl,Abbott:2020niy,Green:2020ptm}.
Since \chip{} is weakly constrained, the shape of the \chip{} prior often dominates the posterior.
The \chip{} prior tends to zero at $\chip{} = 0$ and peaks at a moderate value of \chip{} that depends on the prior ranges of $\spinone$, $\spintwo$ and $\massratio$, and so an inferred nonzero value does not necessarily imply a measurement of precession. 

As a consequence of orbital precession, $\chip{}$ changes throughout the inspiral. 
However, the tilt angles of a compact binary at a formally infinite separation are well defined~\cite{Gerosa:2015tea}.  
We thus quote the tilt angles and the derived quantities ($\chieff$ and $\chip$) at a fiducial reference point of infinite separation. 
The spins are evolved to infinite separation~\cite{Johnson-McDaniel:2021rvv} using precession-averaged evolution~\cite{Gerosa:2015tea,Chatziioannou:2017tdw} with the orbital angular momentum calculated using higher-order \ac{PN} expressions. 

The spin orientations of a binary can provide clues to its formation channel~\cite{Mandel:2009nx,Vitale:2015tea,Stevenson:2017dlk,Fishbach:2017dwv,Talbot:2017yur,Wysocki:2019grj,Zevin:2020gbd}. 
Dynamically assembled binaries would have no preferred spin orientation, and therefore are expected to have an isotropic distribution of spin orientations (unless embedded in an environment like the disc of an active galactic nucleus where accretion or consecutive mergers can result in an anisotropic spin distribution~\cite{McKernan:2017umu,Yang:2019cbr,McKernan:2019beu,Secunda:2020mhd,Tagawa:2020dxe}); on the other hand, binaries formed through isolated binary evolution are typically expected to have nearly aligned spins, with moderate misalignments arising due to supernova kicks~\cite{Kalogera:1999tq,Rodriguez:2016vmx,Abbott:2017vtc,Callister:2020vyz,Steinle:2020xej,Chan:2020lnd,Fragione:2021qtg}. 
Therefore, negative $\chieff$ or large $\chip$ would be more common in dynamically formed binaries than those formed through isolated evolution. 

Most of the candidates in \ac{O3b} are consistent with $\chieff = 0$. 
However, \FULLNAME{GW191204G}{}'s source has $\chieff = \chieffinfinityonlyprecavguncert{GW191204G}$ with no posterior support at zero, while \FULLNAME{GW191103A}{}, \FULLNAME{GW191126C}{} and \FULLNAME{GW191216G}{} have sources with $\chieff = \chieffinfinityonlyprecavguncert{GW191103A}$, $\chieffinfinityonlyprecavguncert{GW191126C}$ and $\chieffinfinityonlyprecavguncert{GW191216G}$, respectively, and negligible support for $\chieff < 0$. 
Other candidates with significant support for $\chieff > 0$ include \FULLNAME{GW200316I}{}, \FULLNAME{GW200208K}{}, \FULLNAME{GW191129G}{} and \FULLNAME{GW200129D}{} with $\chieff > 0$ at $\percentchieffinfinityonlyprecavgmorethanzero{GW200316I}\%$, $\percentchieffinfinityonlyprecavgmorethanzero{GW200208K}\%$, $\percentchieffinfinityonlyprecavgmorethanzero{GW191129G}\%$ and $\percentchieffinfinityonlyprecavgmorethanzero{GW200129D}\%$ probability, respectively. 
The \ac{O3b} candidates with the most significant support for $\chieff < 0$ are \FULLNAME{GW191109A}{} and \FULLNAME{GW200225B}{} with $\chieff < 0$ at $\percentchieffinfinityonlyprecavglessthanzero{GW191109A}\%$ and $\percentchieffinfinityonlyprecavglessthanzero{GW200225B}\%$ probability, respectively.
As with previous catalogs, there are more systems with $\chieff > 0$ than with $\chieff < 0$~\cite{Abbott:2020niy,Abbott:2020gyp,LIGOScientific:2021usb,Roulet:2021hcu}.

Figure~\ref{fig:post_1d_multi} shows one-dimensional posterior probability distributions for \chieff{} and \chip{}, and Fig.~\ref{fig:mcchieffpost} shows two-dimensional posterior probability distributions for $\Mc$ and $\chieff$. 
\FULLNAME{GW200208K}{} has a high inferred value of $\chieff = \chieffinfinityonlyprecavguncert{GW200208K}$. 
This value is comparable to that inferred for \NNAME{GW190403B}{} ($\pastro = \MAXPASTRO{GW190403B}$, as given in Table~\ref{tab:o3a_pastro} in Appendix~\ref{sec:p-astro-methods}), which has $\chieff = \GWNineteenZeroFourZeroThreeChiEff$~\cite{LIGOScientific:2021usb}. 
Both of these modest-significance candidates correspond to \acp{BBH} that have support for unequal masses. 
For example, \NNAME{GW190403B}{}'s source has $\massratio{} = \GWNineteenZeroFourZeroThreeMassRatio$. 
The \ac{O3b} source with probably the lowest $\chieff$ is \FULLNAME{\chieffinfinityonlyprecavgleast}{}'s, which has $\chieff = \chieffinfinityonlyprecavguncert{\chieffinfinityonlyprecavgleast}$. 
Overall, the range of inferred $\chieff$ values matches the range for previous \ac{LVK} candidates~\cite{LIGOScientific:2021usb} as well as candidates from non-\ac{LVK} analyses (when adopting comparable prior assumptions)~\cite{Nitz:2019hdf,Huang:2020ysn,Pratten:2020ruz,Nitz:2021uxj}.

\begin{figure}
    \centering
    \includegraphics[width=\columnwidth]{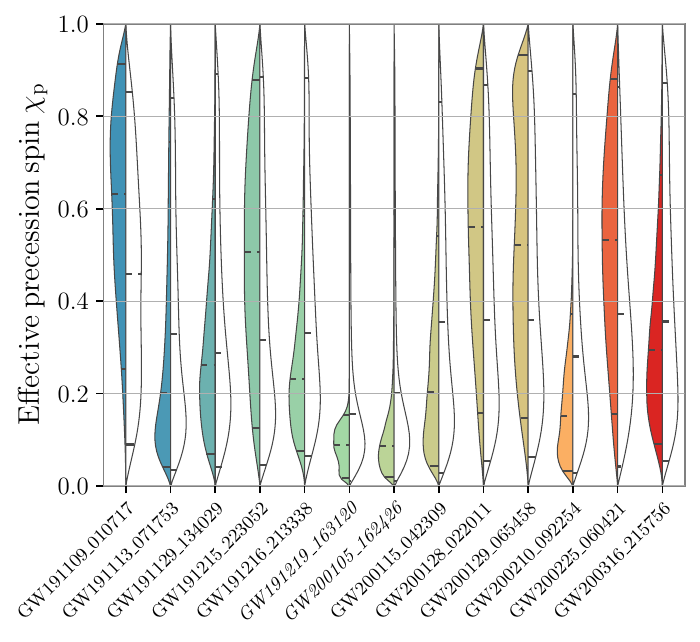}
    \caption{\label{fig:chip} Posterior (left; colored) and effective prior (right; white) probability distributions for the effective precession spin parameter $\chip$ of selected candidates. 
     For each candidate, the prior distribution is conditioned on the posterior probability distribution for the effective inspiral spin $\chieff$ to illustrate how measurement of this quantity is correlated with inference of $\chip$. 
     Horizontal lines mark the median and symmetric $90\%$ interval for the distributions. 
     The candidates selected show the greatest difference between the effective prior and posterior distributions.
     We highlight with italics \FULLNAME{200105F}{} as it has $\pastro{} < \PASTROTHRESHOLD$, as well as \FULLNAME{GW191219E}{} because of significant uncertainty in its $\pastro{}$ and because it has significant posterior support outside of mass ratios where the waveform models have been calibrated. }
\end{figure}

The in-plane spin components are less well constrained than those parallel to the orbital angular momentum. 
Given the constraint that spin magnitudes cannot exceed $1$, a measurement of $\chieff$ influences the permitted values of $\chip$. 
This constraint means that the $\chip$ posterior probability distribution may appear different from its (unrestricted) prior distribution even in cases where the signal contains no measurable information on the in-plane spins~\cite{LIGOScientific:2018mvr,Green:2020ptm}. 
Figure~\ref{fig:chip} shows the $\chip$ posterior probability distribution compared to the prior distribution after conditioning on the $\chieff$ measurement for a selection of candidates~\cite{Abbott:2020niy}. 
These distributions would be the same if no information about the in-plane spin components had been extracted from the signal, and the selected candidates have the greatest difference between the two distributions. 
For many candidates, the $\chip$ posteriors are broad and uninformative. 
\FULLNAME{\chipinfinityonlyprecavgmost}{} (the highest \ac{SNR} \ac{O3b} candidate) has probably the highest inferred $\chip$ of $\chipinfinityonlyprecavguncert{\chipinfinityonlyprecavgmost}$. 
However, this inference is sensitive to the waveform model used, and is discussed in Sec.~\ref{sec:waveform-systematics}. 
\FULLNAME{GW191219E}{} has probably the lowest measurement of the \ac{O3b} candidates, with $\chip \leq \chipinfinityonlyprecavgnintiethpercentile{GW191219E}$ at $90\%$ probability, which is between the measurements for \FULLNAME{200105F}{}~\cite{LIGOScientific:2021qlt} and \NNAME{GW190814H}{}~\cite{Abbott:2020khf,LIGOScientific:2021usb} of $\chip \leq \chipinfinityonlyprecavgnintiethpercentile{200105F}$ and $\leq \GWNineteenZeroEightFourteenChiPUpper$ at $90\%$ probability, respectively. 
Since the mass ratio for this system is beyond the region of calibration for the waveforms, it is not clear how reliable this result is, and further work is needed to characterize the spin.  
For unequal-mass binaries, it is generally easier to observe the effects of precession (or lack thereof), enabling tighter constraints on $\chip$~\cite{Apostolatos:1994mx,Green:2020ptm,Abbott:2020khf,Pratten:2020igi}.

\begin{figure*}
    \centering
    \includegraphics[width=0.3\textwidth]{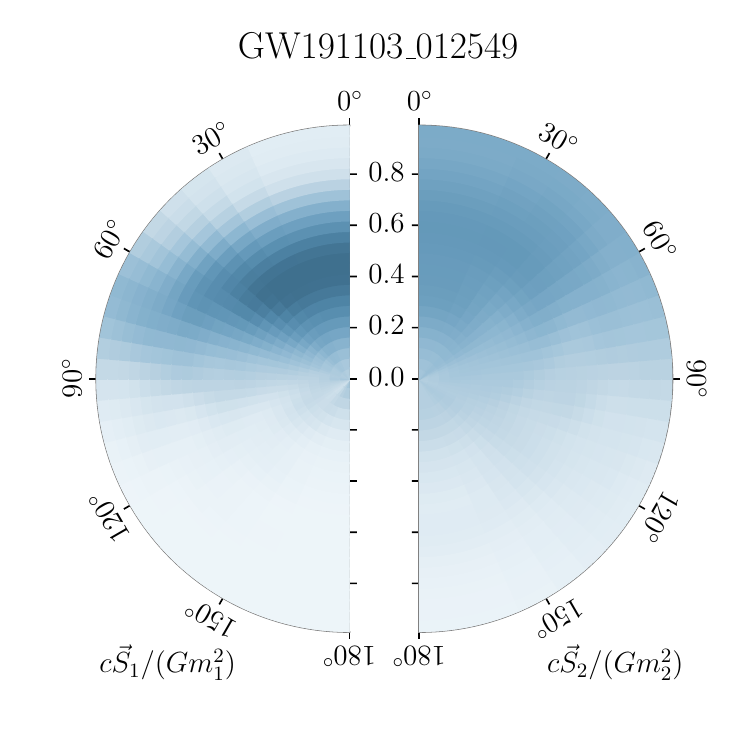}
    \quad
    \includegraphics[width=0.3\textwidth]{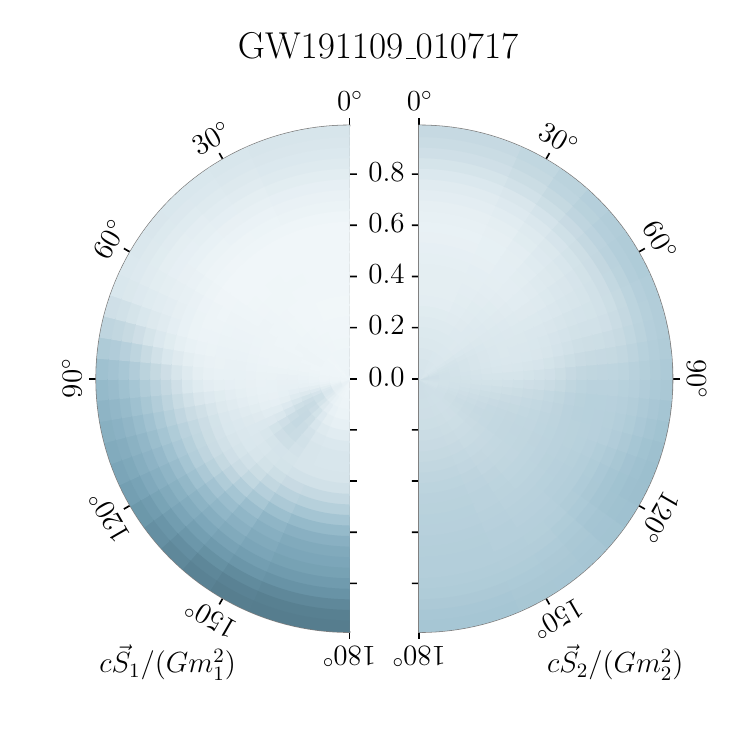}
    \quad
    \includegraphics[width=0.3\textwidth]{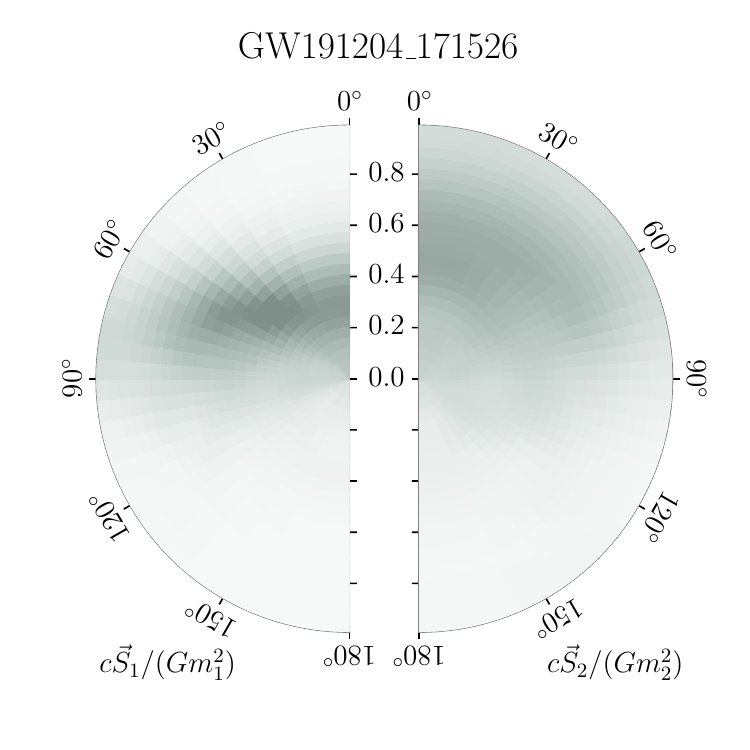}
    \\
    \includegraphics[width=0.3\textwidth]{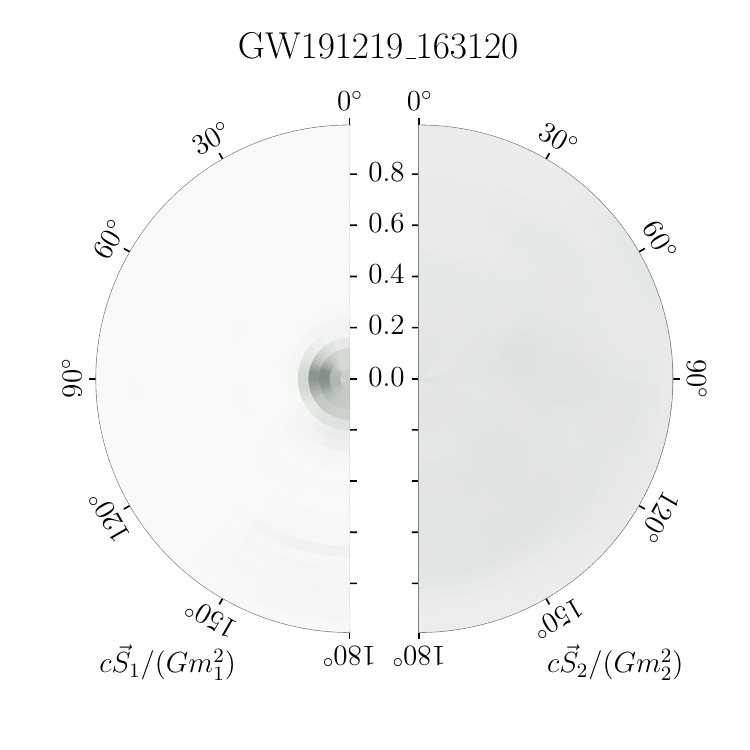}
    \quad
    \includegraphics[width=0.3\textwidth]{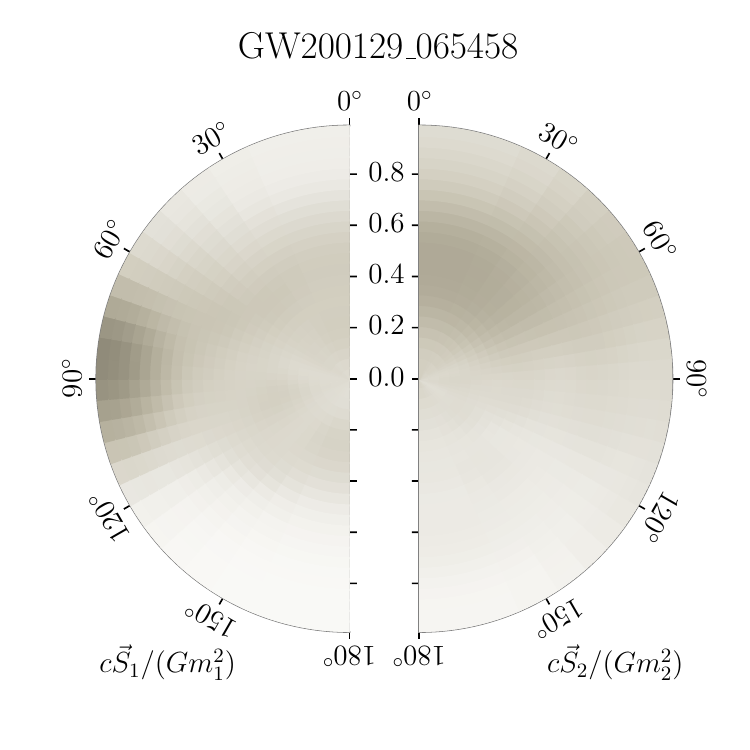}
    \quad
    \includegraphics[width=0.3\textwidth]{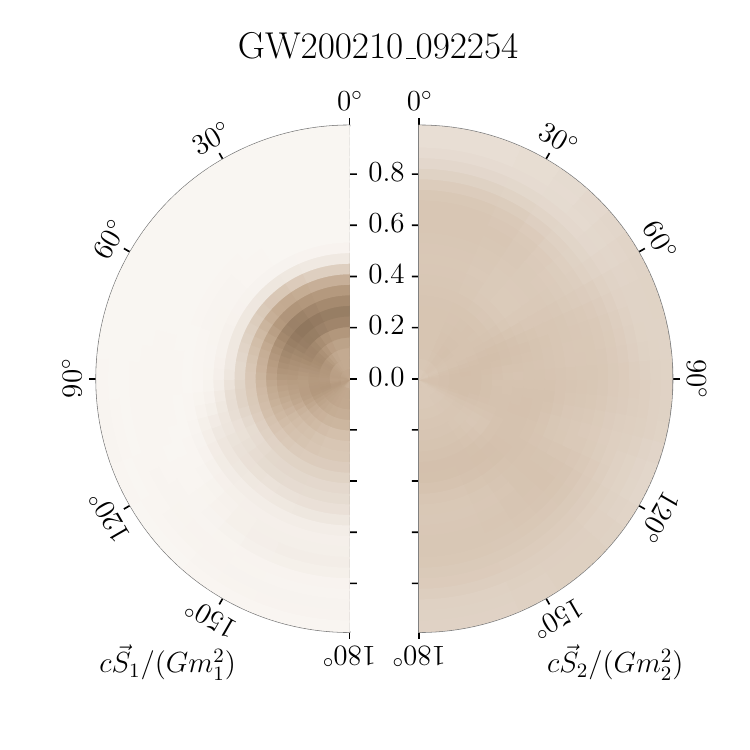}
    \caption{\label{fig:spins} Posterior probability distributions for the dimensionless component spins $\vecspinone{} = c\vec{S}_{1}/(Gm_1^2)$ and $\vecspintwo{} = c\vec{S}_{2}/(Gm_2^2)$ relative to the orbital plane, marginalized over azimuthal angles, for candidates \FULLNAME{GW191103A}{}, \FULLNAME{GW191109A}{}, \FULLNAME{GW191204G}{}, \FULLNAME{GW191219E}{}, \FULLNAME{GW200129D}{} and \FULLNAME{GW200210B}{}, ordered chronologically. 
     \ac{BBH} waveform models are used for all the results shown here.
     \FULLNAME{GW191103A}{} has $\chieff = \chieffinfinityonlyprecavguncert{GW191103A}$ with negligible posterior support at zero. 
     \FULLNAME{GW191109A}{} has $\chieff < 0$ at $\percentchieffinfinityonlyprecavglessthanzero{GW191109A}\%$ probability and $\chip = \chipinfinityonlyprecavguncert{GW191109A}$. 
     \FULLNAME{GW191204G}{} has $\chieff = \chieffinfinityonlyprecavguncert{GW191204G}$ with no posterior support at zero. 
     \FULLNAME{GW191219E}{} is a \ac{NSBH} candidate with $\chip \leq \chipinfinityonlyprecavgnintiethpercentile{GW191219E}$ at $90\%$ probability; this candidate has significant uncertainty in its $\pastro{}$ and has significant posterior support outside of mass ratios where the waveform models have been calibrated.
     \FULLNAME{GW200129D}{} has $\chip = \chipinfinityonlyprecavguncert{GW200129D}$.
     \FULLNAME{GW200210B}{} has $\chip \leq \chipinfinityonlyprecavgnintiethpercentile{GW200210B}$ at $90\%$ probability and mass ratio $\massratio = \massratiouncert{GW200210B}$. 
     In these plots, histogram bins are constructed linearly in spin magnitude and the cosine of the tilt angles such that they contain equal prior probability.}
\end{figure*}

Figure~\ref{fig:spins} shows the posterior probability distributions for the dimensionless spin magnitude $\chi_i$ and tilt angle $\spintilt{i}$ for the binary components of a selection of six \ac{O3b} candidates. 
In most cases, posteriors for the component spin magnitudes are largely uninformative, but for some of the unequal-mass binaries we may constrain $\spinone$~\cite{LIGOScientific:2020stg,Abbott:2020khf,Zevin:2020gxf,Biscoveanu:2020are}.
For \FULLNAME{GW191219E}{}, \FULLNAME{200105F}{} and \FULLNAME{GW200210B}{}, we find $\spinone \leq \spinonenintiethpercentile{GW191219E}$, $\leq \spinonenintiethpercentile{200105F}$ and $\leq \spinonenintiethpercentile{GW200210B}$ at $90\%$ probability, respectively. 
Like \NNAME{GW190814H}{}~\cite{Abbott:2020khf,LIGOScientific:2021usb}, where we inferred $\spinone \leq \GWNineteenZeroEightFourteenSpinOneUpper$, these \acp{NSBH} or \acp{BBH} with low-mass secondaries have negligible support for maximal primary spins. 
Conversely, for the asymmetric \ac{BBH} candidate \FULLNAME{GW200208K}{}, we infer $\spinone \geq \spinonetenthpercentile{GW200208K}$ at $90\%$ probability, with $\percentchionemorethanpointeight{GW200208K}\%$ probability that $\spinone > \PEReferenceSpinMag$. 
These inferred spins are not as extreme as they are for \NNAME{GW190403B}{}'s source~\cite{LIGOScientific:2021usb}. 
With our default prior assumptions, only the \ac{O3a} candidates \NNAME{GW190403B}{}~\cite{LIGOScientific:2021usb}, \NNAME{GW190412B}{}~\cite{LIGOScientific:2020stg,Zevin:2020gxf} and \NNAME{GW190517B}{}~\cite{LIGOScientific:2021usb} lack posterior support for a primary spin of zero. 

The final spin of the merger remnant $\chif$ is determined by conservation of angular momentum, and receives contributions from both the orbital angular momentum at merger and the component spins. 
For equal-mass, nonspinning \acp{BH}, the merger remnant has a spin of $\chif \sim \TypicalFinalSpin$~\cite{Pretorius:2005gq,Gonzalez:2006md,Buonanno:2007sv,GalvezGhersi:2020fvh}. 
As a consequence of the range of mass ratios and spins of the \ac{O3b} candidates, there is a range of final spins from $\chif = \finalspinuncert{GW191219E}$ for \FULLNAME{GW191219E}{} and $\finalspinuncert{GW200210B}$ for \FULLNAME{GW200210B}{} (assuming the \ac{BBH} waveform models are accurate) to $\finalspinuncert{GW200208K}$ for \FULLNAME{GW200208K}{}. 

In comparison to \GWTCTHREE{} observations, spins of \acp{BH} in X-ray binaries span the full range of magnitudes, including near-maximal spins~\cite{Miller:2014aaa,Reynolds:2020jwt,Miller-Jones:2021plh}. 
For low-mass X-ray binaries, it is possible that these spins are grown by accretion from their companion~\cite{Podsiadlowski:2002ww,Fragos:2014cva,Sorensen:2016rnl}; in contrast, for high-mass X-ray binaries there would be insufficient time for accretion to significantly change the spin~\cite{Valsecchi:2010cw,Qin:2018sxk,Miller-Jones:2021plh}. 
The comparison between spins in X-ray binaries and coalescing \ac{BH} binaries may highlight details of their formation and differences in their evolution.

Predictions for \ac{BH} spin magnitudes vary, depending upon the formation channel and assumptions about stellar evolution such as stellar winds or the efficiency of stellar tides~\cite{Kushnir:2016zee,Hotokezaka:2017esv,Belczynski:2017gds,Marchant:2020haw,Bavera:2020uch,Steinle:2020xej}. 
If angular momentum transport is efficient in stars, then \acp{BH} formed from stellar collapse may be born with low ($\lesssim\LITERATURELOWSPIN$) spins~\cite{Qin:2018vaa,Fuller:2019sxi}; for binaries formed via isolated binary evolution, this may mean that the first-born \ac{BH} is expected to have a low spin, although the second-born \ac{BH} may have a larger spin due to tides spinning up its progenitor~\cite{Bavera:2019fkg,Mandel:2020lhv,Olejak:2021iux}. 
The situation may be different if progenitor stars have significant rotation rates, such as for close binary star systems, where tidal locking can lead to chemically homogeneous evolution~\cite{deMink:2009jq,Mandel:2015qlu,Buisson:2020hoq}. 
In this case, predicted \ac{BH} spins are typically $\sim\LITERATURECHESPIN$, and may extend up to the Kerr limit~\cite{Marchant:2016wow,Zevin:2020gbd}. 
Spin could also be imparted by asymmetric supernova explosions~\cite{Chan:2020lnd}. 
For \acp{BBH} embedded in active galactic nuclei discs, accretion can grow spins if they are prograde with respect to the disc, while retrograde spins become smaller before flipping to become prograde, with the rate of evolution depending upon the orientation of the orbit with respect to the disc~\cite{McKernan:2017umu,Secunda:2020mhd,Tagawa:2020dxe}. 
Outside of stellar evolution, primordial \acp{BH} born in the early, radiation-dominated Universe are expected to have small ($\lesssim\LITERATUREPBHSPIN$) spins at formation~\cite{Mirbabayi:2019uph,DeLuca:2019buf,Harada:2020pzb}, but spins could increase through accretion~\cite{Berti:2008af,DeLuca:2020bjf}.
Given the theoretical uncertainties on \ac{BH} spin magnitudes, \ac{GW} (and X-ray) observations may reveal details of \ac{BH} formation; the distribution of spins is analyzed in a companion paper~\cite{LIGOScientific:2021psn}.

\subsection{Tidal effects}\label{sec:tides}

If a binary contains at least one \ac{NS} component, the \ac{GW} signal from the inspiral is influenced by the deformability of \ac{NS} matter. 
Tidal effects are quantified by the dimensionless quadrupole tidal deformability, 
\begin{equation}
    \Lambda_i = \frac{2}{3}k_{2,i}\left[\frac{c^2 R_i}{Gm_i}\right]^5,
\end{equation}
where $k_{2,i}$ is the second Love number and $R_i$ is the component's radius~\cite{Damour:1991yw,Flanagan:2007ix}. 
Quasiuniversal relations~\cite{Yagi:2016bkt} are used to parametrize the effects of \ac{NS} spin-induced deformations in terms of $\Lambda_i$.  
Stiffer \ac{NS} equations of state give larger values of $\Lambda_i$, which accelerates the rate of inspiral. 
\acp{BH} have $\Lambda_i = 0$~\cite{LeTiec:2020spy,Chia:2020yla,Goldberger:2020fot,Charalambous:2021mea}.

On account of their \acp{SNR}, we do not expect to be able to place a lower limit on the tidal deformability for any candidates from \ac{O3b}~\cite{Kumar:2016zlj,Huang:2020pba,Brown:2021seh}. 
Results confirm this, with no analysis showing strong support for matter effects. 
This is consistent with previous observations where it was not possible to determine the nature of the compact objects from the \ac{GW} data alone, such as GW170817~\cite{TheLIGOScientific:2017qsa,LIGOScientific:2019eut} and \NNAME{GW190814H}{}~\cite{Abbott:2020khf}.

\subsection{Localization}\label{sec:localization}

The distance to the source is inferred from the amplitude of the signal as the two are inversely related~\cite{Cutler:1994ys,TheLIGOScientific:2016wfe}. 
Posterior probability distributions for the luminosity distance are shown in Fig.~\ref{fig:post_1d_multi}. 
The closest source found in \ac{O3b} is probably \FULLNAME{\luminositydistanceleast}{}, with an inferred distance of $\DL = \luminositydistanceuncert{\luminositydistanceleast}~\mathrm{Gpc}$ and redshift $\redshift = \redshiftuncert{\luminositydistanceleast}$. 
At $90\%$ probability, \FULLNAME{\luminositydistanceleast}{} has $\DL \leq \luminositydistancenintiethpercentile{\luminositydistanceleast}~\mathrm{Gpc}$.
\FULLNAME{\luminositydistancemost}{} probably has the farthest source (including the high-distance, low-likelihood mode) at $\DL = \luminositydistanceuncert{\luminositydistancemost}~\mathrm{Gpc}$ ($\DL \geq \luminositydistancetenthpercentile{\luminositydistancemost}~\mathrm{Gpc}$ at $90\%$ probability), $\redshift = \redshiftuncert{\luminositydistancemost}$. 
This measurement is comparable to the probably most distant source reported in \GWTCTWOFINAL{}, which is for \NNAME{GW190403B}{} at $\DL = \GWNineteenZeroFourZeroThreeDistance$~\cite{Abbott:2020niy,LIGOScientific:2021usb}. 
As our detectors become more sensitive, it will be possible to observe sources at greater distances. 

The sky localization depends critically upon the number of observatories able to detect a signal~\cite{Wen:2010cr,Singer:2014qca,Abbott:2020qfu}. 
With only a single detector observing, localizations may cover the entire sky. 
The most constrained localizations are achieved when all three observatories record a significant \ac{SNR}.
The \ac{O3b} source with the best sky localization is \FULLNAME{\minareaevent}{}, with a $90\%$ credible area of $\skyarea{\minareaevent}~\mathrm{deg^{2}}$, which was observed with all three detectors. 
As the detector network expands, the typical sky-localization precision will improve~\cite{Pankow:2019oxl,Abbott:2020qfu}. 

The volume localization depends upon both the distance and sky localization. 
The best three-dimensional localizations from \ac{O3b} are for \FULLNAME{\minvolevent}{} and \FULLNAME{GW200115A}{}, which have $90\%$ credible volumes of $\skyvol{\minvolevent}~\mathrm{Gpc^{3}}$ and $\skyvol{GW200115A}~\mathrm{Gpc^{3}}$, respectively. 
These correspond to two of the closest sources, with $\DL = \luminositydistanceuncert{\minvolevent}~\mathrm{Gpc}$ and $\luminositydistanceuncert{GW200115A}~\mathrm{Gpc}$, respectively.
Using the \ac{GLADEplus}~\cite{Dalya:2018cnd,Dalya:2021ewn,Virgo:2021bbr}, the $90\%$ credible volume for \FULLNAME{\minvolevent}{} contains $\sim \NgalBestLocTwentyZeroTwoZeroTwobandK$ galaxies reported in the K band ($\sim \NgalBestLocTwentyZeroTwoZeroTwobandbJ$ in the bJ band), where we estimate the completeness of the galaxy catalog to be $\CompMaxBestLocTwentyZeroTwoZeroTwobandK$--$\CompMinBestLocTwentyZeroTwoZeroTwobandK$ ($\CompMaxBestLocTwentyZeroTwoZeroTwobandbJ$--$\CompMinBestLocTwentyZeroTwoZeroTwobandbJ$). 
Similarly, the $90\%$ credible volume for \FULLNAME{GW200115A}{} contains $\sim \NgalBestLocTwentyZeroOneFifteenbandK$ galaxies in the K band ($\sim \NgalBestLocTwentyZeroOneFifteenbandbJ$ in the bJ band), with estimated completeness of $\CompMaxBestLocTwentyZeroOneFifteenbandK$--$\CompMinBestLocTwentyZeroOneFifteenbandK$ ($\CompMaxBestLocTwentyZeroOneFifteenbandbJ$--$\CompMinBestLocTwentyZeroOneFifteenbandbJ$).
As the typical distance to sources increases, so will the typical localization volume; however, improvements to detector sensitivity will mean that the localization precision for the best localized sources will improve~\cite{DelPozzo:2018dpu,Pankow:2019oxl,Abbott:2020qfu}. 

The localization is crucial to multimessenger follow-up efforts. 
Previously reported candidates have been the target of dedicated follow-up observations. 
The details of currently reported follow-up observations are reviewed in Appendix~\ref{sec:follow-up}. 

\subsection{Waveform systematics}\label{sec:waveform-systematics}

Our inference of the source properties is dependent on being able to accurately calculate the signal waveform given the source parameters~\cite{Littenberg:2012uj,Varma:2014jxa,Kumar:2016dhh,Abbott:2016wiq,LIGOScientific:2018hze,Purrer:2019jcp,Shaik:2019dym,Ramos-Buades:2020noq}. 
The current generation of quasicircular \ac{BBH} waveforms used here (\IMRPhenomXPHM{} and \SEOBNRPHM{}) include higher-order spherical harmonics and model spin precession.    
Since the waveforms include equivalent physical effects, we expect that any differences that exist are attributable to the particular modeling of the relevant physics. 
Additionally, \IMRPhenomXPHM{} uses the stationary phase approximation to trade accuracy for faster waveform evaluation in the frequency domain, which produces less reliable descriptions of massive merger--ringdown dominated signals. 
To assess the effects of waveform uncertainty on our inferences, and to identify discrepancies that require further study, we compare the results obtained with different waveforms.
 
The waveforms are calibrated to nonprecessing \ac{NR} waveforms, and good agreement has been found between the two waveform models for nonprecessing systems~\cite{Colleoni:2020tgc}. 
However, the waveforms are not calibrated to precessing \ac{NR} waveforms and use different approximations to describe precession (discussed in Appendix~\ref{sec:waveforms}). 
The lack of accurate information about precession from \ac{NR} also affects the merger and ringdown portions of the waveform, and the calculation of the quasinormal-mode frequencies. 
Additional issues regarding an accurate description of precessing systems arise for nearly antialigned spins, where approximations used to model spin effects can break down due to a wide opening angle of the precession cone (for more extreme mass ratios), or instabilities in the spin configuration~\cite{Gerosa:2015hba}. 
Generally, waveforms tend to disagree in parts of the parameter space with higher spins and more extreme mass ratios~\cite{Colleoni:2020tgc,Ossokine:2020kjp,Pratten:2020igi,Estelles:2021gvs}, where the number of \ac{NR} waveforms available for calibration are limited.

We find that for almost all the signals analyzed here, the differences between results obtained with the \IMRPhenomXPHM{} and \SEOBNRPHM{} are subdominant compared to the statistical uncertainty.
As for previous observations, differences are typically small, and most noticeable for parameters like the spins \cite{Abbott:2016izl,LIGOScientific:2018hze,LIGOScientific:2018mvr,Abbott:2020niy}. 
In some cases there are differences in the multimodality of the posterior probability distribution. 
Multimodality can be an indication of the complex structure of the waveform and highlight where subtle changes in the modeling may be important.
Examples of candidates where there are differences between \IMRPhenomXPHM{} and \SEOBNRPHM{} are:
\begin{itemize}
\item \FULLNAME{GW191109A}{}, which has significant support for negative $\chieff$ and misaligned spins, where waveform differences may be expected~\cite{Colleoni:2020tgc,Biscoveanu:2021nvg}.
There are differences in the spins and mass ratio inferred with the two waveforms.
Both models show a structured, multimodal joint posterior distribution on $\chieff$, $\massratio$, orbital inclination $\thetaJN$ (the angle between the total angular momentum and the line of sight) and $\chip$, although the modes are overlapping. 
\SEOBNRPHM{} has a posterior probability distribution with two modes separated mostly in $\thetaJN$, one face on and one face off. 
Both modes show similarly high values of $\chip$, and both have $\chieff<0$ with high probability.
\IMRPhenomXPHM{}, however, finds a near-edge-on mode ($\thetaJN \sim \EdgeOnThetaJN$) that prefers more equal component masses, and includes greater support for positive $\chieff$. 
We infer $\chieff = \chieffinfinityonlyprecavgIMRPuncert{GW191109A}$ with \IMRPhenomXPHM{} and $\chieff = \chieffinfinityonlyprecavgSEOBuncert{GW191109A}$ with \SEOBNRPHM{}. 
When a binary is viewed edge on, any precession effects are maximally visible~\cite{Vitale:2014mka,Abbott:2016wiq,Green:2020ptm,Biscoveanu:2021nvg,Krishnendu:2021cyi}.

\item \FULLNAME{GW191219E}{}, which has a large mass asymmetry, with the bulk of the posterior probability distribution outside the range of calibration of the waveforms. 
Despite this, the posteriors obtained with \SEOBNRPHM{} and \IMRPhenomXPHM{} show good agreement overall. 
While the waveforms produce consistent results, there are differences in the inferred inclination, with \IMRPhenomXPHM{} showing less support for near-edge-on orientations; total mass, with \IMRPhenomXPHM{} preferring higher masses, and distance, with \IMRPhenomXPHM{} having less support for larger distances. 
We infer $\massratio = \massratioIMRPuncert{GW191219E}$ with \IMRPhenomXPHM{} and $\massratio = \massratioSEOBuncert{GW191219E}$ with \SEOBNRPHM{}.
Modeling of higher-order multipole moments is particularly important for inferring the properties of systems with unequal masses~\cite{Blanchet:2013haa,Varma:2016dnf,Payne:2019wmy,Mills:2020thr,LIGOScientific:2020stg,Krishnendu:2021cyi}, and it may impact inference of parameters including the mass ratio, inclination and distance~\cite{Graff:2015bba,Kumar:2018hml,Chatziioannou:2019dsz,Kalaghatgi:2019log,Khan:2019kot,Shaik:2019dym,Abbott:2020niy}.

\item \FULLNAME{GW200129D}{}, which has a high \ac{SNR} ($\rho = \networkmatchedfiltersnrIMRPuncert{GW200129D}$ using \IMRPhenomXPHM{}) and was detected in all three detectors. 
While both waveforms show approximately the same $\chieff$, this candidate shows a high $\chip$, as well as stronger support for unequal masses, when analyzed with \IMRPhenomXPHM{}, whereas with \SEOBNRPHM{} it does not exhibit strong evidence for precession and shows more support for equal masses. 
We infer $\chip = \chipinfinityonlyprecavgIMRPuncert{GW200129D}$ and $\massratio = \massratioIMRPuncert{GW200129D}$ with \IMRPhenomXPHM{}, and $\chip = \chipinfinityonlyprecavgSEOBuncert{GW200129D}$ and $\massratio = \massratioSEOBuncert{GW200129D}$ with \SEOBNRPHM{}.
Unlike \FULLNAME{GW191109A}{}, the orbital plane is not viewed edge on to the line of sight, so amplitude modulations from precession of the orbital plane are likely to be less significant. 
However, \FULLNAME{GW200129D}{} has significant support for inclinations up to $\thetaJN \lesssim\thetajnnintiethpercentile{GW200129D}$, where precession and higher-order harmonic content may be important~\cite{Vitale:2014mka,Graff:2015bba,Abbott:2016wiq,Kalaghatgi:2019log,Mills:2020thr,Green:2020ptm,Krishnendu:2021cyi}.
Waveform systematics become more important for higher \ac{SNR} signals, where statistical uncertainties are smaller~\cite{TheLIGOScientific:2016wfe,Purrer:2019jcp}.

\item \FULLNAME{GW200208K}{}, which has a multimodal mass posterior and low \ac{SNR}. 
The preference for the different modes varies between waveforms. 
Of the two main modes, the lower $\massone$ and $\Mtot$ mode is favored by \SEOBNRPHM{}, while the higher $\massone$ and $\Mtot$ mode is favored by \IMRPhenomXPHM{}. 
Additionally, the \IMRPhenomXPHM{} analysis finds an additional minor mode with $\Mtot \sim \GWTwentyZeroTwoZeroEightTwentyTwoPeakTotalMass$ (visible in Fig.~\ref{fig:mtotvsqpost} as a protuberance of the $90\%$ contour). 
The \IMRPhenomXPHM{} analysis also shows a greater preference for higher $\chieff$: we infer $\chieff = \chieffinfinityonlyprecavgIMRPuncert{GW200208K}$ with \IMRPhenomXPHM{}, and $\chieff = \chieffinfinityonlyprecavgSEOBuncert{GW200208K}$ with \SEOBNRPHM{}.

\end{itemize}
Future analyses with enhanced waveforms will update our understanding of the source parameters for these candidates.

\section{Waveform consistency tests}\label{sec:waveform-reconstruction}

Waveforms can be reconstructed from the data using two complementary approaches, either using parameter-estimation methods with templates~\cite{Veitch:2014wba,TheLIGOScientific:2016wfe} or using minimal modeling~\cite{Cornish:2014kda,Klimenko:2015ypf,Salemi:2019owp}.
While the parameter-estimation pipelines directly estimate the match between \ac{CBC} model waveforms and data, \BAYESWAVE{} (Appendix~\ref{sec:data-methods}) and \ac{CWB} (Appendix~\ref{sec:cwb-methods}) reconstruct waveforms making only minimal assumptions on the signal shape~\cite{Cornish:2014kda,Klimenko:2015ypf,Salemi:2019owp}. 
The waveform reconstruction performed by these pipelines uses time--frequency wavelets to identify coherent features in the data, filtering out incoherent noise from the detectors. 
Although there are similarities between the methods used by \ac{CWB}~\cite{Klimenko:2015ypf,Salemi:2019uea} and \BAYESWAVE{}~\cite{Cornish:2014kda,Cornish:2020dwh}, their waveform reconstructions differ in some details. 
In particular, the point estimate returned by \ac{CWB} is the constrained maximum-likelihood reconstruction, while for \BAYESWAVE{} we use the median of the time-domain waveform reconstructions from \BAYESWAVE{}'s posterior probability distribution.
Examples of both types of reconstruction were reported in \GWTCTWO{}~\cite{Abbott:2020niy}. 

Starting from minimally modeled waveform reconstructions we can try to detect unexpected behavior by comparing these reconstructions with the \ac{CBC} waveforms from parameter estimation~\cite{LIGOScientific:2018mvr,Salemi:2019uea,Ghonge:2020suv,Abbott:2020niy,Johnson-McDaniel:2021yge}. 
To test the consistency (or lack thereof) between minimally modeled reconstructions and the \ac{CBC} waveforms, we perform sets of dedicated injections of \ac{CBC} waveform samples from the posterior distributions for the source parameters.
In these simulations the random waveforms are added to background data around the time of the candidates, and the simulated signal is analyzed by the minimally modeled pipelines.  
We call these \emph{off-source} injected waveforms, while the reconstructed waveform of the candidate is our \emph{on-source} result. 

\begin{table*}
% Made by tableV
% DO NOT EDIT THIS FILE DIRECTLY

\begin{ruledtabular}
\begin{tabular}{@{\extracolsep{\fill}}l c c c c}{Candidate} & \multicolumn{2}{c}{\BAYESWAVE{}} & \multicolumn{2}{c}{\ac{CWB}} \\
\cline{2-3} \cline{4-5}
& {On-source match}  & {Off-source match} & {On-source match} & {Off-source match} \\
\hline
\FULLNAME{GW191109A}{} &  $ 0.93$ & $ 0.94_{- 0.10}^{+ 0.04}$ &  $ 0.90$ & $ 0.90_{- 0.05}^{+ 0.04}$ \\
\makebox[0pt][l]{\fboxsep0pt\colorbox{lightgray}{\mystrut\hspace*{1.0\linewidth}}}\!\!
\FULLNAME{GW191127B}{} & -- & -- &  $ 0.86$ & $ 0.83_{- 0.10}^{+ 0.07}$ \\
\FULLNAME{GW191129G}{} &  $ 0.57$ & $ 0.35_{- 0.28}^{+ 0.26}$ & -- & -- \\
\makebox[0pt][l]{\fboxsep0pt\colorbox{lightgray}{\mystrut\hspace*{1.0\linewidth}}}\!\!
\FULLNAME{GW191204G}{} &  $ 0.82$ & $ 0.68_{- 0.30}^{+ 0.14}$ &  $ 0.91$ & $ 0.88_{- 0.07}^{+ 0.04}$ \\
\FULLNAME{GW191215G}{} &  $ 0.79$ & $ 0.65_{- 0.49}^{+ 0.17}$ &  $ 0.86$ & $ 0.80_{- 0.10}^{+ 0.05}$ \\
\makebox[0pt][l]{\fboxsep0pt\colorbox{lightgray}{\mystrut\hspace*{1.0\linewidth}}}\!\!
\FULLNAME{GW191216G}{} &  $ 0.73$ & $ 0.74_{- 0.42}^{+ 0.09}$ & -- & -- \\
\FULLNAME{GW191222A}{} &  $ 0.90$ & $ 0.88_{- 0.16}^{+ 0.06}$ &  $ 0.86$ & $ 0.81_{- 0.13}^{+ 0.07}$ \\
\makebox[0pt][l]{\fboxsep0pt\colorbox{lightgray}{\mystrut\hspace*{1.0\linewidth}}}\!\!
\FULLNAME{GW191230H}{} & -- & -- &  $ 0.85$ & $ 0.78_{- 0.11}^{+ 0.08}$ \\
\FULLNAME{GW200128C}{} & -- & -- &  $ 0.87$ & $ 0.89_{- 0.05}^{+ 0.04}$ \\
\makebox[0pt][l]{\fboxsep0pt\colorbox{lightgray}{\mystrut\hspace*{1.0\linewidth}}}\!\!
\FULLNAME{GW200129D}{} &  $ 0.96$ & $ 0.96_{- 0.06}^{+ 0.02}$ &  $ 0.80$ & $ 0.88_{- 0.09}^{+ 0.05}$ \\
\FULLNAME{GW200208G}{} &  $ 0.73$ & $ 0.74_{- 0.50}^{+ 0.14}$ &  $ 0.78$ & $ 0.79_{- 0.13}^{+ 0.07}$ \\
\makebox[0pt][l]{\fboxsep0pt\colorbox{lightgray}{\mystrut\hspace*{1.0\linewidth}}}\!\!
\FULLNAME{GW200209E}{} & -- & -- &  $ 0.82$ & $ 0.83_{- 0.09}^{+ 0.08}$ \\
\FULLNAME{GW200216G}{} & -- & -- &  $ 0.73$ & $ 0.87_{- 0.13}^{+ 0.05}$ \\
\makebox[0pt][l]{\fboxsep0pt\colorbox{lightgray}{\mystrut\hspace*{1.0\linewidth}}}\!\!
\FULLNAME{GW200219D}{} &  $ 0.81$ & $ 0.74_{- 0.35}^{+ 0.14}$ &  $ 0.81$ & $ 0.85_{- 0.08}^{+ 0.06}$ \\
\FULLNAME{GW200224H}{} &  $ 0.96$ & $ 0.93_{- 0.09}^{+ 0.03}$ &  $ 0.83$ & $ 0.84_{- 0.10}^{+ 0.06}$ \\
\makebox[0pt][l]{\fboxsep0pt\colorbox{lightgray}{\mystrut\hspace*{1.0\linewidth}}}\!\!
\FULLNAME{GW200225B}{} &  $ 0.85$ & $ 0.73_{- 0.38}^{+ 0.12}$ &  $ 0.77$ & $ 0.85_{- 0.13}^{+ 0.07}$ \\
\FULLNAME{GW200311L}{} &  $ 0.94$ & $ 0.90_{- 0.43}^{+ 0.06}$ &  $ 0.93$ & $ 0.92_{- 0.04}^{+ 0.03}$ \\
\end{tabular}
\end{ruledtabular}

\caption{
\label{tab:wfreclist}
List of candidates tested by \BAYESWAVE{} and \ac{CWB} for consistency with the waveform templates used in the inference of source parameters.
We quote the on-source match calculated using the waveform reconstructed for the candidate, and the median and $90\%$ symmetric interval for off-source matches calculated for simulated signals with source parameters consistent with those inferred for the candidate signal.
The values reported in the table correspond to those in Fig.~\ref{fig:match}. 
Dashes (--) correspond to candidates not included in an analysis.
}
\end{table*}

Here, as in \GWTCTWO{}~\cite{Abbott:2020niy}, we measure the waveform \emph{match} (or \emph{overlap}), defined by
\begin{equation}
\mathcal{O}( h_1 , h_2  )= \frac{\langle h_1 | h_2  \rangle}{\sqrt{\langle h_1 | h_1  \rangle \langle h_2 | h_2  \rangle}} , 
\label{eq:overlap}
\end{equation}
where $h_1$ and $h_2$ are two waveforms, $\langle \cdot | \cdot \rangle$ represents the noise-weighted inner product~\cite{Finn:1992wt}, and the match is $-1 \le \mathcal {O}( h_1,h_2  ) \le 1$. 
The theoretical definition of match in Eq.~\eqref{eq:overlap} does not depend on the amplitude of each signal~\cite{Abbott:2020niy}.
However, the addition of noise typically reduces the match value, and the calculated match does depend both on the \ac{SNR} and, in more detail, on the distribution of signal power in time and frequency.
A value of $1$ indicates a perfect coincidence between waveforms, while a value close to $0$ indicates that the correlation between waveforms is nil. 
A theoretically possible value of $-1$ would indicate an improbable perfect anticoincidence. 
The match is larger for signals corresponding to high-mass systems~\cite{Littenberg:2015kpb,Becsy:2016ofp,Pannarale:2018cct,Ghonge:2020suv}. 
The distribution of match values of the off-source injections defines a null distribution for each candidate; this distribution can be used both to estimate the uncertainty of the observed on-source match value and to obtain a p-value from the on-source match.
For each candidate, the match is computed off source between injected waveforms and their reconstructions, while on source it is computed between the point estimate of the actual candidate and the maximum-likelihood estimate provided by source-parameter estimation.

The sets of candidates chosen for the \BAYESWAVE{} and \ac{CWB} consistency tests are different. 
For the \BAYESWAVE{} analysis we consider candidates that are sufficiently loud and short for \BAYESWAVE{} to produce valid signal reconstructions.
The candidates considered by \ac{CWB} are those detected by the search analysis (reported in Table~\ref{tab:events}), plus \CWBRECONLY{} additional candidates that were identified by other search analyses (also reported in Table~\ref{tab:events}). 
These additional \CWBRECONLY{} candidates were reconstructed by the initial stages of the \ac{CWB} search analysis, but did not pass the \ac{CWB} postproduction cuts that are used to identify low-\ac{FAR} candidates (described in Appendix~\ref{sec:cwb-methods}). 
Both lists are reported in Table~\ref{tab:wfreclist}.

\begin{figure*}
\centering
\includegraphics[width=\columnwidth]{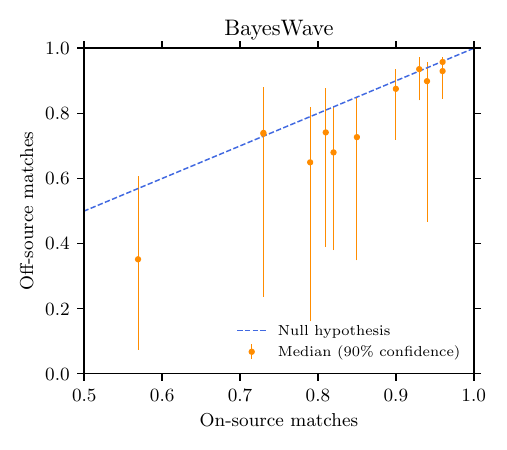}  \quad
\includegraphics[width=\columnwidth]{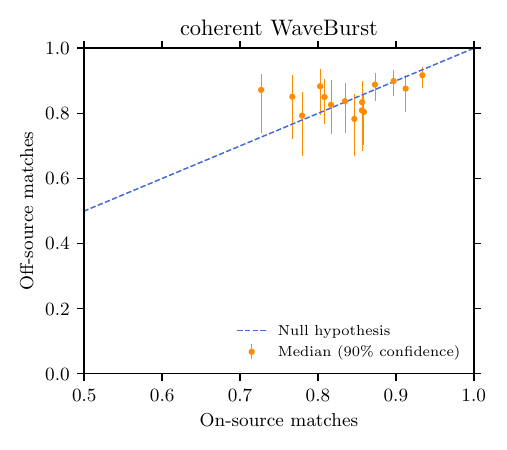} 
\caption{Off-source versus on-source match values for the candidates in \ac{O3b}.
The left and right panels show the results of the \BAYESWAVE{} and \ac{CWB} analyses, respectively. 
The on-source match is estimated comparing the inferred maximum-likelihood \ac{CBC} waveform with point estimates from the minimally modeled waveform reconstructions.
The off-source match is the median value of the match distribution estimated from off-source injection of sample waveforms from the template-based posterior distribution.
The error bars in both panels are given by the symmetric (equal-tailed) $90\%$ confidence interval, and they mark the distance from the null hypothesis (blue dashed line).
The different sizes of the error bars in the two panels is due to the different numbers of off-source injections in the \BAYESWAVE{} and \ac{CWB} analyses. 
}
\label{fig:match}
\end{figure*}

\begin{figure*}
\centering
\includegraphics[width=\columnwidth]{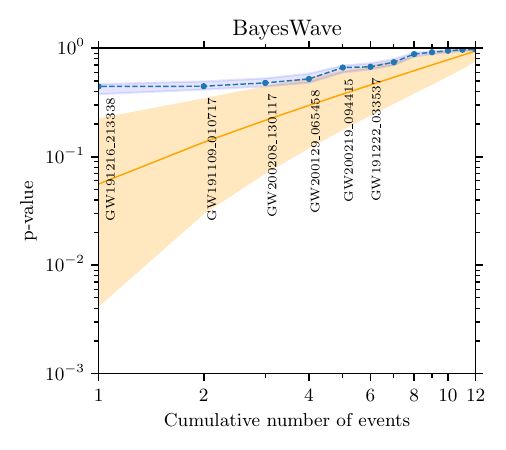} \quad
\includegraphics[width=\columnwidth]{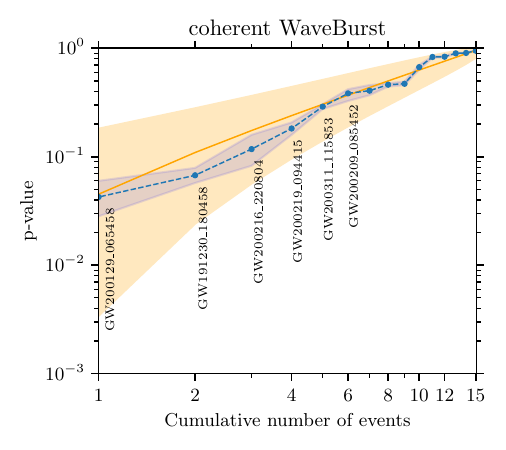} 
\caption{Distribution of p-values for the \ac{O3b} candidates reconstructed by the minimally modeled pipelines.
The left and right panels report the \BAYESWAVE{} and \ac{CWB} results, respectively.
The p-values are sorted in increasing order and graphed against the order number (blue dashed line).
Each p-value is estimated from the observed on-source match value and the related off-source distribution of the match values from off-source injections.
The shadowed orange band is the symmetric $90\%$ interval about the median, represented by the orange solid line. 
The blue band represents the symmetric $90\%$ interval associated with the finite number of off-source injections.
}
\label{fig:pvalue}
\end{figure*}

The waveform consistency tests were carried out with respect to the results calculated by the Bayesian inference library~\BILBY{}~\cite{Ashton:2018jfp,Romero-Shaw:2020owr} using the \IMRPhenomXPHM{} waveform~\cite{Pratten:2020ceb} (details are presented in Appendix~\ref{sec:parameter-estimation-methods}).  
Figure~\ref{fig:match} shows the on-source match values versus the median off-source match values (together with the $90\%$ intervals). 
The match values move to lower values for smaller \ac{SNR}, but the on-source value is still expected to be close to the median of the off-source distribution (blue dashed line in the figure) if the null hypothesis (that the minimally modeled reconstruction does not deviate significantly from the template-based reconstruction) holds. 

Figure~\ref{fig:pvalue} shows the p-values sorted in increasing order~\cite{Salemi:2019uea,Ghonge:2020suv}. 
When the null hypothesis holds, the sorted p-values are expected to remain close to the median value (orange dashed line); the $90\%$ interval that surrounds the median line shows the size of the fluctuations that we expect to observe. 
Any significant deviations \emph{below} the plot diagonal, corresponding to low p-values, point to a set of candidates that show potential disagreement with the waveform templates.
However, the significance of several simultaneous deviations cannot be directly assessed from the $90\%$ interval, which is calculated for \emph{single values}~\cite{LIGOScientific:2012fcp}. 
Since the p-values are sorted in increasing order, the sorting induces a correlation between successive values, and this means that there may be a whole subset of points outside the interval. 
All of the $\CWBPVALNUM{}$ \ac{CWB} p-values are within the $90\%$ interval. 
This is not the case for \BAYESWAVE{}, where it is important to consider the finite-size effect due to the limited number of off-source samples in the match distribution.
The posterior probability distribution for the p-value estimate due to the finite-size effect is $\mathrm{Beta}(n+1,N-n+1)$, where $N$ is the total number of samples, and $n$ is the number of samples with a match value less than the on-source match~\cite{LIGOScientific:2021sio}. 
In addition to the $90\%$ interval associated with the statistical distribution of p-values for the null hypothesis, the p-value plots in Fig.~\ref{fig:pvalue} also display the $90\%$ interval from the finite-size effect: we see that after including this effect there are $\BWPOUTNUM{}$ out of $\BWPVALNUM{}$ incompatible p-values (where the $90\%$ intervals do not overlap). 
Finding $\BWPOUTNUM{}$ incompatible p-values is not an unlikely event, it has a $\sim \BWPVALPROB$ probability of being due to chance.
We conclude that both the match--match and the p-value plots indicate that there is no significant inconsistency between the minimally modeled waveform reconstruction and the results of the parameter-estimation analysis. 
Further checks of the consistency of the signals, focused on the predictions from general relativity, are given in a companion paper~\cite{LIGOScientific:2021sio}.

\section{Conclusion}\label{sec:conclusion}

We have presented the latest \ac{LVK} catalog of \acp{GW}, which contains a total of \TOTALEVENTS{} \ac{CBC} candidate signals with an estimated probability of astrophysical origin $\pastro{} > \PASTROTHRESHOLD$. 
\GWTCTHREE{} builds upon past catalogs of \ac{GW} candidates from \ac{O1}~\cite{TheLIGOScientific:2016pea}, \ac{O2}~\cite{LIGOScientific:2018mvr} and \ac{O3a}~\cite{Abbott:2020niy,LIGOScientific:2021usb}, adding an additional \NUMEVENTS{} candidates from \ac{O3b} with $\pastro{} > \PASTROTHRESHOLD$. 
These include the \ac{NSBH} candidates \FULLNAME{GW191219E}{} and \FULLNAME{GW200115A}, as well as the candidate \FULLNAME{GW200210B}{} that could potentially be either from a \ac{NSBH} or a \ac{BBH}. 
We additionally provide a list of candidates with $\pastro{} < \PASTROTHRESHOLD$ meeting a \ac{FAR} threshold of $< \SUBTHRESHOLDFAR{}$. 
This includes \FULLNAME{200105F}{}, which is estimated to have $\pastro{} = \MAXPASTRO{200105F}$ but is a clear outlier from our background noise distribution, and is inferred to have a \ac{NSBH} source~\cite{LIGOScientific:2021qlt}. 
While we expect $\sim \CONTAMINATION{}$ of the candidates with $\pastro{} > \PASTROTHRESHOLD$ to be false alarms, we also expect $\sim\EXPREALINSUBTHRESH{}$ candidates with $\pastro{} < \PASTROTHRESHOLD$ to be astrophysical \ac{GW} signals. 
\ac{GW} observations of \acp{CBC} provide new insight into diverse areas of physics ranging from binary stellar evolution to gravitation.
Further analysis and interpretation of the \GWTCTHREE{} candidates is conducted in the companion papers~\cite{LIGOScientific:2021psn,LIGOScientific:2021sio,Virgo:2021bbr}.  
As the population of \ac{GW} observations grows, it will be possible to make increasingly detailed measurements of compact-object physics.

The growing catalog of \ac{GW} sources has revealed a diversity of potential \ac{CBC} sources. 
Among the candidates are a few with posterior support for high spins ($\chi_i \gtrsim \PEReferenceSpinMag$) and large mass asymmetries ($\massratio{} \lesssim \PEReferenceMassRatio$). 
Creating waveform models in these regimes is challenging, as the need to maintain accuracy necessitates more complete prescriptions of the underlying physics, including effects such as spin-induced precession~\cite{Williamson:2017evr,Samajdar:2019ulq} plus higher-order multipole moments~\cite{Littenberg:2012uj,Graff:2015bba,Kalaghatgi:2019log,Shaik:2019dym}. 
This task is further complicated by the lack of extensive \ac{NR} waveform catalogs covering these regions of parameter space~\cite{Jani:2016wkt,Boyle:2019kee,Healy:2020vre,Lousto:2020tnb}. 
As sensitivity improves, waveform uncertainty may be a significant source of systematic uncertainty~\cite{Abbott:2016wiq,Purrer:2019jcp}. 
Therefore, to ensure reliable interpretation of \ac{GW} observations in the future, it is imperative to develop improved waveform models that cover a wider range of source properties, and include potentially important additional physics such as orbital eccentricity~\cite{Lower:2018seu,Moore:2019vjj,Romero-Shaw:2019itr,Lenon:2020oza,OShea:2021ugg,Favata:2021vhw}.

Data products associated with \GWTCTHREE{} results are available through \ac{GWOSC}~\cite{gwosc:gwtc3} in addition to the full \ac{O3b} detector strain data~\cite{gwosc:o3b,LIGOScientific:2023vdi}. 
Release of previous observing runs' strain data~\cite{Abbott:2019ebz,LIGOScientific:2023vdi} has enabled multiple independent analyses of \ac{LIGO} and Virgo data, including identification of additional detection candidates~\cite{Nitz:2018imz,Magee:2019vmb,Venumadhav:2019tad,Zackay:2019tzo,Venumadhav:2019lyq,Nitz:2019hdf,Zackay:2019btq,Nitz:2020bdb,Nitz:2021mzz,Nitz:2021uxj,Olsen:2022pin,Chandra:2022ixv}. 
Therefore, we anticipate that further discoveries may come from \ac{O3b} data.%~\cite{Nitz:2021zwj,Mishra:2022ott,Nitz:2022ltl,Davies:2022thw,Szczepanczyk:2022urr,Wang:2023qgw}.

\ac{O3} saw the Advanced \ac{LIGO} and Advanced Virgo detectors reach their greatest sensitivity to date, enabling an unprecedented rate of discovery. 
Coupled to the longer duration of \ac{O3} compared to previous observing runs, this sensitivity has enabled the number of \ac{GW} detections from \ac{O3} to significantly exceed that from \ac{O1} and \ac{O2}. 
The Advanced \ac{LIGO} and Advanced Virgo detectors are currently offline undergoing commissioning to further enhance their performance for \ac{O4}. 
\ac{O4} will also see the joint operation of the KAGRA detector~\cite{Akutsu:2018axf}. 
KAGRA successfully completed a joint observing run with GEO\,600 following the end of \ac{O3} for Advanced \ac{LIGO} and Advanced Virgo~\cite{LIGOScientific:2022myk}, and \ac{O4} will be the first opportunity for observations from the LIGO--Virgo--KAGRA detector network. 
The enhanced \ac{O4} global detector network will further increase the prospects for \ac{GW} and multimessenger discoveries~\cite{Abbott:2020qfu}. 

While the \TOTALEVENTS{} probable \ac{GW} candidates of \GWTCTHREE{} all correspond to \ac{CBC} sources, we anticipate that there are other \ac{GW} signals waiting to be found~\cite{Sathyaprakash:2009xs}. 
These could include new types of transient signal, such as from supernovae~\cite{Abbott:2019pxc}, cosmic strings~\cite{Abbott:2021ksc}, or previously unidentified sources~\cite{LIGOScientific:2021hoh,LIGOScientific:2021uyj}. 
Additionally, we may find long-lived signals such as continuous waves from rotating \acp{NS}~\cite{Abbott:2020mev,Abbott:2021jiu,Abbott:2021oqi,LIGOScientific:2021quq,LIGOScientific:2022pjk,LIGOScientific:2022cgf} or stochastic backgrounds~\cite{Abbott:2021xxi,Abbott:2021jel}. 
As detector sensitivity increases and we observe for longer, we expect more of the \ac{GW} universe to reveal itself.

\acknowledgments
Calibration of the \ac{LIGO} strain data was performed with \GSTLAL{}-based calibration software pipeline~\cite{Viets:2017yvy}.
Calibration of the Virgo strain data is performed with C-based software~\cite{VIRGO:2021umk}. 
Data-quality products and event-validation results were computed using the \DMT{}~\cite{DMTdocumentation}, \DQR{}~\cite{DQRdocumentation}, \DQSEGDB{}~\cite{Fisher:2020pnr}, \GWDETCHAR{}~\cite{gwdetchar-software}, \HVETO{}~\cite{Smith:2011an}, \IDQ{}~\cite{Essick:2020qpo}, \OMICRONSCAN{}~\cite{Robinet:2020lbf} and \PYTHONVIRGOTOOLS{}~\cite{pythonvirgotools} software packages and contributing software tools.  
Analyses in this catalog relied upon the \LALSUITE{} software library~\cite{lalsuite-software}.
The detection of the signals and subsequent significance evaluations in this catalog were performed with the \GSTLAL{}-based inspiral software pipeline~\cite{Messick:2016aqy,Sachdev:2019vvd,Hanna:2019ezx,Cannon:2020qnf}, with the \ac{MBTA} pipeline~\cite{Adams:2015ulm,Aubin:2020goo}, and with the \PYCBC{}~\cite{Usman:2015kfa,Nitz:2017svb,Davies:2020tsx} and the \ac{CWB}~\cite{Klimenko:2004qh,Klimenko:2011hz,Klimenko:2015ypf} packages. 
Estimates of the noise spectra and glitch models were obtained using \BAYESWAVE{}~\cite{Cornish:2014kda,Littenberg:2015kpb,Cornish:2020dwh}. 
Noise subtraction for one candidate was also performed with \GWSUBTRACT{}~\cite{Davis:2022ird}. 
Source-parameter estimation was performed with the \BILBY{} and \PBILBY{} libraries~\cite{Ashton:2018jfp,Smith:2019ucc,Romero-Shaw:2020owr} using the \DYNESTY{} nested sampling package~\cite{Speagle:2019ivv}, and the \RIFT{} library~\cite{Pankow:2015cra,Lange:2017wki,Wysocki:2019grj}, with the \LALINFERENCE{}~\cite{Veitch:2014wba} libraries used for initial analyses.
\PESUMMARY{} was used to postprocess and collate parameter-estimation results~\cite{Hoy:2020vys}.
The various stages of the parameter-estimation analysis were managed with the \ASIMOV{} library~\cite{Williams:2022pgn}. 
Plots were prepared with \PLT{}~\cite{Hunter:2007ouj}, \SEABORN{}~\cite{Waskom:2021psk} and \GWPY{}~\cite{gwpy-software}. 
\NUMPY{}~\cite{Harris:2020xlr} and \SCIPY{}~\cite{Virtanen:2019joe} were used in the preparation of the manuscript.

This material is based upon work supported by NSF’s LIGO Laboratory which is a major facility
fully funded by the National Science Foundation.
The authors also gratefully acknowledge the support of
% the United States National Science Foundation (NSF) for the construction and operation of the
% LIGO Laboratory and Advanced LIGO as well as
the Science and Technology Facilities Council (STFC) of the
United Kingdom, the Max-Planck-Society (MPS), and the State of
Niedersachsen/Germany for support of the construction of Advanced LIGO 
and construction and operation of the GEO\,600 detector. 
Additional support for Advanced LIGO was provided by the Australian Research Council.
The authors gratefully acknowledge the Italian Istituto Nazionale di Fisica Nucleare (INFN),  
the French Centre National de la Recherche Scientifique (CNRS) and
the Netherlands Organization for Scientific Research, 
for the construction and operation of the Virgo detector
and the creation and support  of the EGO consortium. 
The authors also gratefully acknowledge research support from these agencies as well as by 
the Council of Scientific and Industrial Research of India, 
the Department of Science and Technology, India,
the Science \& Engineering Research Board (SERB), India,
the Ministry of Human Resource Development, India,
the Spanish Agencia Estatal de Investigaci\'on,
the Vicepresid\`encia i Conselleria d'Innovaci\'o, Recerca i Turisme and the Conselleria d'Educaci\'o i Universitat del Govern de les Illes Balears,
the Conselleria d'Innovaci\'o, Universitats, Ci\`encia i Societat Digital de la Generalitat Valenciana and
the CERCA Programme Generalitat de Catalunya, Spain,
the National Science Centre of Poland and the Foundation for Polish Science (FNP),
the Swiss National Science Foundation (SNSF),
the Russian Foundation for Basic Research, 
the Russian Science Foundation,
the European Commission,
the European Regional Development Funds (ERDF),
the Royal Society, 
the Scottish Funding Council, 
the Scottish Universities Physics Alliance, 
the Hungarian Scientific Research Fund (OTKA),
the French Lyon Institute of Origins (LIO),
the Belgian Fonds de la Recherche Scientifique (FRS-FNRS), 
Actions de Recherche Concertées (ARC) and
Fonds Wetenschappelijk Onderzoek – Vlaanderen (FWO), Belgium,
the Paris \^{I}le-de-France Region, 
the National Research, Development and Innovation Office Hungary (NKFIH), 
the National Research Foundation of Korea,
the Natural Science and Engineering Research Council Canada,
Canadian Foundation for Innovation (CFI),
the Brazilian Ministry of Science, Technology, and Innovations,
the International Center for Theoretical Physics South American Institute for Fundamental Research (ICTP-SAIFR), 
the Research Grants Council of Hong Kong,
the National Natural Science Foundation of China (NSFC),
the Leverhulme Trust, 
the Research Corporation, 
the Ministry of Science and Technology (MOST), Taiwan,
the United States Department of Energy,
and
the Kavli Foundation.
The authors gratefully acknowledge the support of the NSF, STFC, INFN and CNRS for provision of computational resources.
Computing was performed on the OzSTAR Australian national facility at Swinburne University of Technology, which receives funding in part from the Astronomy National Collaborative Research Infrastructure Strategy (NCRIS) allocation provided by the Australian Government.
We thankfully acknowledge the computer resources at MareNostrum and the technical support provided by Barcelona Supercomputing Center (RES-AECT-2021-2-0021). 
This work was supported by MEXT, JSPS Leading-edge Research Infrastructure Program, JSPS Grant-in-Aid for Specially Promoted Research 26000005, JSPS Grant-in-Aid for Scientific Research on Innovative Areas 2905: JP17H06358, JP17H06361 and JP17H06364, JSPS Core-to-Core Program A.\ Advanced Research Networks, JSPS Grant-in-Aid for Scientific Research (S) 17H06133, the joint research program of the Institute for Cosmic Ray Research, University of Tokyo, National Research Foundation (NRF) and Computing Infrastructure Project of KISTI-GSDC in Korea, Academia Sinica (AS), AS Grid Center (ASGC) and the Ministry of Science and Technology (MoST) in Taiwan under grants including AS-CDA-105-M06, Advanced Technology Center (ATC) of NAOJ, and Mechanical Engineering Center of KEK.

We thank the anonymous journal referees for helpful comments. 
We would like to thank all of the essential workers who put their health at risk during the COVID-19 pandemic, without whom we would not have been able to complete this work.

%\clearpage

\appendix

\section{Low-latency alert system and multimessenger follow-up}\label{sec:follow-up}

Public alerts were issued for \ac{GW} candidates identified by low-latency searches of the data. 
These candidates were cataloged in the \ac{GraceDB}.
Each entry into \ac{GraceDB} is known as an event, and a collection of these within a specific time window is referred to as a \emph{superevent}. 
The time window for \ac{CBC} events was variable based on the spread of events, with a typical value of $\WINDOWSEVENTCBC{}~\mathrm{s}$ symmetric around the merger time. 
The duration of the time window for \ac{CWB} was variable and was reported by the search pipeline for each event. 
One candidate event belonging to the superevent was identified as the preferred event, and its attributes (time, localization, significance, classification and properties)~\cite{Singer:2015ema,Chatterjee:2019avs,Kapadia:2019uut} were inherited by the superevent. 
The \emph{HasRemnant} property indicator was related to the probability of having an electromagnetic counterpart~\cite{Chatterjee:2019avs}, and the \pastro{} classifier assigned a source-category-based astrophysical probability under the assumption that astrophysical and terrestrial triggers occurred as independent Poisson processes~\cite{Farr:2013yna,Kapadia:2019uut}.
The name of a superevent was its uniquely assigned identification in \ac{GraceDB} consisting of three parts: the prefix S (for superevent), the six-digit \ac{UTC} date of the event (YYMMDD), and a lowercase alphabetic suffix.

During \ac{O3}, \ac{CBC} superevents that passed a \ac{FAR} threshold of $\OPAFARTHRESHCBC{}$ and generic transient (\emph{Burst}) superevents that passed a \ac{FAR} threshold of $\OPAFARTHRESHBURST{}$ were distributed as public alerts. 
The individual \ac{FAR} thresholds of each pipeline were corrected by a trials factor to account for the data being analyzed by multiple pipelines.
Generally, multiple pipelines identified the candidate \ac{GW} events distributed as public alerts. 

When a preferred event candidate passed the public-alert threshold, a preliminary alert was queued, while new event candidates were still accepted to be added to the superevent. 
After the preliminary alert reception by the \ac{GCN} broker, the preferred event was revised and a second preliminary Notice was issued, even if the preferred event candidate remained unchanged. 
The alerts were processed by the \GWCELERY{} distributed task queue software~\cite{Magee:2021xdx,OPA}, which organized basic data-quality checks, grouped events from online searches, and initiated localization and inference of source properties. 

As in \ac{O2}~\cite{LIGOScientific:2019gag}, human vetting of the superevents was a critical part of the online program, and was completed once the superevent passed the public-alert threshold. 
The rapid-response team consisted of commissioning, computing and calibration experts from each of the detector sites, search-pipeline experts, detector-characterization experts, and follow-up advocates in charge of the delivery of the initial \ac{GCN} Notice and Circular. 
A data-quality report was also initiated by \GWCELERY{}, and consisted of a semiautomated detector-characterization and data-quality investigation.
It provided a variety of metrics based on auxiliary instrumental and environmental sensors to help the rapid-response team make a decision whether to confirm or retract a candidate.
The preliminary alerts were typically issued within a few minutes of data collection, for which latency due to data transfer between sites and search investigation were largely dominated by the \GWCELERY{} task.
The human vetting and delivery of initial alerts had a median duration of $\sim\HUMANVETLAT{}~\mathrm{min}$.

There were \NUMOTHREEAOPA{} public alerts sent out via \ac{GCN} during \ac{O3a} and \NUMPUBLICEVENTS{} during \ac{O3b}. 
Of these, \NUMOTHREEAOPAUNRETRACTED{} from \ac{O3a} and \NUMPUBLICNOTRETRACTED{} from \ac{O3b} were not retracted; the remaining were retracted on timescales from minutes to days. 
The majority of the retracted public alerts in \ac{O3b} corresponded to candidates with \ac{SNR} $\rho > \RETRACTIONLOWSNR$ in only one detector.
The online search pipelines collect background in real time, leaving them susceptible to new noise sources, and single-detector candidates are especially impacted by uncertainties in the background noise distribution since they cannot rely on coincidence to establish significance. 
Among the remaining \ac{O3b} alerts, \OTHREEBALERTSNOTRETRACTEDCBC{} involved \ac{CBC} candidates, and \OTHREEBALERTSNOTRETRACTEDBURST{} (S200114f) was a Burst candidate, as discussed in Sec.~\ref{sec:online-candidates}.  
The unretracted \ac{O3a} alerts were publicly distributed in $\ALERTMEDIANTIMEOTHREEA{}^{+\ALERTNINETYFIVEPCTTIMEOTHREEA{}}_{-\ALERTFIVEPCTTIMEOTHREEA{}}~\mathrm{min}$, and the \ac{O3b} alerts in $\ALERTMEDIANTIME{}^{+\ALERTNINETYFIVEPCTTIME{}}_{-\ALERTFIVEPCTTIME{}}~\mathrm{min}$ (median and $90\%$ symmetric interval).
One \ac{O3b} candidate, S200303ba, was retracted but never had a preliminary Notice sent out due to problems connecting to the \ac{GCN} broker.
The \ac{GW} candidate alerts generated \GCNTRAFFICCOUNT{} Circulars during \ac{O3} (\GCNTRAFFIC{}\% of \GCNTRAFFICALL{} \ac{GCN} Circulars in the same period), with \GCNTRAFFICTHREEACOUNT{} and \GCNTRAFFICTHREEBCOUNT{} Circulars (\GCNTRAFFICTHREEA{}\% and \GCNTRAFFICTHREEB{}\%) sent during \ac{O3a} and \ac{O3b}, respectively.  

Follow-up observations were made by teams across the astronomical community, culminating in \ac{GCN} Circulars and papers. 
The searches for multimessenger counterparts employed the same variety of observing strategies used for previous observing runs~\cite{LIGOScientific:2019gag}, including archival analysis, prompt searches with all-sky instruments, wide-field tiled searches, targeted searches of potential host galaxies, and deep follow-up of individual sources.
The follow-up effort mobilized a total of about $\APPROXOBSERVATORIES{}$ ground- and space-based instruments such as neutrino observatories, very-high-energy gamma-ray observatories, space-based gamma-ray and X-ray instruments, visible and infrared telescopes, and radio telescopes.
The latency for follow-up observations, analyses, public reporting of results and the process efficiency varied across the collaborations and the multimessenger probe involved.
Additionally, the public alerts enabled amateur astronomers to join professional astronomers in the search for electromagnetic counterparts~\cite{Antier:2020nuy}. 
Summaries of the \ac{O3a} and \ac{O3b} candidates with public alerts and follow-up investigations are reported in Table~\ref{tab:follow-up-o3a} and Table~\ref{tab:follow-up}, respectively.

\begin{table*}
% Made by followup_tables
% DO NOT EDIT THIS FILE DIRECTLY

\begin{ruledtabular}
\begin{tabularx}{\textwidth}{@{\extracolsep{\fill}}l l l l}SID & Event & GCN & Follow-up publications\\ \hline
\makebox[0pt][l]{\fboxsep0pt\colorbox{lightgray}{\mystrut\hspace*{1.0\linewidth}}}\!\!
S190408an & GW190408\_181802 & \cite{GCN24063} & \cite{Abbasi:2021kft,IceCube:2022mma,Abbasi:2023bfo,Abe:2020zpn,Abe:2021jtb,Acero:2020duu,Adriani:2022zli,Antier:2019pzz,Becerra:2021lyq,Cai:2021lmr,Gompertz:2020cur,Graham:2022xxu,Hussain:2019xzb,Kim:2021hhl,Lipunov:2022tno,Lundquist:2019cty,Mo:2023sfz,Paterson:2020mmd,Ridnaia:2020pau,J-GEM:2021aem,Unatlokov:2021erm}\\ 
S190412m  & GW190412 & \cite{GCN24098} & \cite{Abbasi:2021kft,IceCube:2022mma,Abbasi:2023bfo,Abe:2020zpn,Abe:2021jtb,Acero:2020duu,Adriani:2022zli,ANTARES:2023wcj,Antier:2019pzz,Becerra:2021lyq,Cai:2021lmr,Gompertz:2020cur,Graham:2022xxu,Hussain:2019xzb,Kim:2021hhl,Lipunov:2022tno,Mo:2023sfz,Oates:2021eyk,Page:2020tnx,Ridnaia:2020pau,J-GEM:2021aem,Unatlokov:2021erm}\\ 
\makebox[0pt][l]{\fboxsep0pt\colorbox{lightgray}{\mystrut\hspace*{1.0\linewidth}}}\!\!
 & GW190413\_052954 &  & \cite{Abbasi:2021kft,IceCube:2022mma,Abbasi:2023bfo,Abe:2021jtb,ANTARES:2023wcj,Cai:2021lmr,Graham:2022xxu,Mo:2023sfz,Unatlokov:2021erm}\\ 
 & GW190413\_134308 &  & \cite{Abbasi:2021kft,IceCube:2022mma,Abbasi:2023bfo,Abe:2021jtb,ANTARES:2023wcj,Cai:2021lmr,Graham:2022xxu,Mo:2023sfz,Unatlokov:2021erm}\\ 
\makebox[0pt][l]{\fboxsep0pt\colorbox{lightgray}{\mystrut\hspace*{1.0\linewidth}}}\!\!
S190421ar & GW190421\_213856 & \cite{GCN24141} & \cite{Abbasi:2021kft,IceCube:2022mma,Abbasi:2023bfo,Abe:2020zpn,Abe:2021jtb,Acero:2020duu,Adriani:2022zli,ANTARES:2023wcj,Antier:2019pzz,Becerra:2021lyq,Cai:2021lmr,Gompertz:2020cur,Graham:2022xxu,Hussain:2019xzb,Lipunov:2022tno,Mo:2023sfz,Ridnaia:2020pau,J-GEM:2021aem,Unatlokov:2021erm}\\ 
 & GW190424\_180648 &  & \cite{Abbasi:2021kft,Abe:2021jtb,ANTARES:2023wcj,Cai:2021lmr,Graham:2022xxu,Mo:2023sfz,Unatlokov:2021erm}\\ 
\makebox[0pt][l]{\fboxsep0pt\colorbox{lightgray}{\mystrut\hspace*{1.0\linewidth}}}\!\!
S190425z & GW190425 & \cite{GCN24167} & \cite{Abbasi:2021kft,IceCube:2022mma,Abbasi:2023bfo,Abe:2020zpn,Abe:2021jtb,Acero:2020duu,Adriani:2022zli,ANTARES:2023wcj,Antier:2019pzz,Becerra:2021lyq,Boersma:2021gyq,BOREXINO:2023nji,Cai:2021lmr,Chang:2021zdi,Coughlin:2020fwx,deJaeger:2021tcq,Gompertz:2020cur,Graham:2022xxu,Hosseinzadeh:2019ifm,Hussain:2019xzb,Kasliwal:2020wmy,Lipunov:2022tno,Lundquist:2019cty,Mo:2023sfz,Oates:2021eyk,Page:2020tnx,Panther:2022jfg,Pozanenko:2019lwh,Ridnaia:2020pau,J-GEM:2021aem,Unatlokov:2021erm}\\ 
S190426c & GW190426\_152155 & \cite{GCN24237} & \cite{Abbasi:2021kft,Abbasi:2023bfo,Abe:2020zpn,Abe:2021jtb,Acero:2020duu,Adriani:2022zli,ANTARES:2023wcj,Anand:2020mys,Antier:2019pzz,Becerra:2021lyq,BOREXINO:2023nji,Cai:2021lmr,Chang:2021zdi,Coughlin:2020fwx,deJaeger:2021tcq,Goldstein:2019roe,Gompertz:2020cur,Gourdji:2023bbm,Graham:2022xxu,Hosseinzadeh:2019ifm,Hussain:2019xzb,Kasliwal:2020wmy,Kumar:2022hvu,Lipunov:2022tno,Lundquist:2019cty,Mo:2023sfz,Oates:2021eyk,Page:2020tnx,Ridnaia:2020pau,J-GEM:2021aem,Unatlokov:2021erm}\\ 
\makebox[0pt][l]{\fboxsep0pt\colorbox{lightgray}{\mystrut\hspace*{1.0\linewidth}}}\!\!
S190503bf & GW190503\_185404 & \cite{GCN24377} & \cite{Abbasi:2021kft,IceCube:2022mma,Abbasi:2023bfo,Abe:2020zpn,Abe:2021jtb,Acero:2020duu,Adriani:2022zli,ANTARES:2023wcj,Antier:2019pzz,Cai:2021lmr,Graham:2022xxu,Hussain:2019xzb,Kim:2021hhl,Lipunov:2022tno,Mo:2023sfz,Ridnaia:2020pau,Unatlokov:2021erm}\\ 
\textit{S190510g} &  & \cite{GCN24442} & \cite{Abe:2020zpn,Adriani:2022zli,Anand:2020mys,Andreoni:2019kqi,Antier:2019pzz,Chang:2021zdi,Coughlin:2020fwx,Garcia:2020smy,Lipunov:2022tno,Oates:2021eyk,Page:2020tnx,Ridnaia:2020pau,Hussain:2019xzb,Gompertz:2020cur,J-GEM:2021aem}\\ 
\makebox[0pt][l]{\fboxsep0pt\colorbox{lightgray}{\mystrut\hspace*{1.0\linewidth}}}\!\!
S190512at & GW190512\_180714 & \cite{GCN24503} & \cite{Abbasi:2021kft,IceCube:2022mma,Abbasi:2023bfo,HESS:2021gbx,Abe:2020zpn,Abe:2021jtb,Adriani:2022zli,ANTARES:2023wcj,Antier:2019pzz,Ashkar:2020hxe,Cai:2021lmr,Gompertz:2020cur,Graham:2022xxu,Hussain:2019xzb,Lipunov:2022tno,Mo:2023sfz,Ohgami:2021npc,Ridnaia:2020pau,Unatlokov:2021erm}\\ 
S190513bm & GW190513\_205428 & \cite{GCN24522} & \cite{Abbasi:2021kft,IceCube:2022mma,Abbasi:2023bfo,Abe:2020zpn,Abe:2021jtb,Adriani:2022zli,ANTARES:2023wcj,Antier:2019pzz,Becerra:2021lyq,Cai:2021lmr,Gompertz:2020cur,Graham:2022xxu,Hussain:2019xzb,Lipunov:2022tno,Mo:2023sfz,Ridnaia:2020pau,Unatlokov:2021erm}\\ 
\makebox[0pt][l]{\fboxsep0pt\colorbox{lightgray}{\mystrut\hspace*{1.0\linewidth}}}\!\!
 & GW190514\_065416 &  & \cite{Abbasi:2021kft,IceCube:2022mma,Abbasi:2023bfo,Abe:2021jtb,ANTARES:2023wcj,Cai:2021lmr,Graham:2022xxu,Mo:2023sfz,Unatlokov:2021erm}\\ 
S190517h & GW190517\_055101 & \cite{GCN24570} & \cite{Abbasi:2021kft,IceCube:2022mma,Abbasi:2023bfo,Abe:2020zpn,Abe:2021jtb,Acero:2020duu,Adriani:2022zli,ANTARES:2023wcj,Antier:2019pzz,Cai:2021lmr,Gompertz:2020cur,Graham:2022xxu,Hussain:2019xzb,Lipunov:2022tno,Mo:2023sfz,Ridnaia:2020pau,Unatlokov:2021erm}\\ 
\makebox[0pt][l]{\fboxsep0pt\colorbox{lightgray}{\mystrut\hspace*{1.0\linewidth}}}\!\!
\textit{S190518bb} &  & \cite{GCN24591} & \\ 
S190519bj & GW190519\_153544 & \cite{GCN24598} & \cite{Abbasi:2021kft,IceCube:2022mma,Abbasi:2023bfo,Abe:2021jtb,Acero:2020duu,Adriani:2022zli,ANTARES:2023wcj,Cai:2021lmr,Graham:2022xxu,Lipunov:2022tno,Mo:2023sfz,Unatlokov:2021erm}\\ 
\makebox[0pt][l]{\fboxsep0pt\colorbox{lightgray}{\mystrut\hspace*{1.0\linewidth}}}\!\!
S190521g  & GW190521 & \cite{GCN24618} & \cite{Abbasi:2021kft,IceCube:2022mma,Abbasi:2023bfo,Abe:2020zpn,Abe:2021jtb,Acero:2020duu,Adriani:2022zli,ANTARES:2023wcj,Antier:2019pzz,Becerra:2021lyq,Cai:2021lmr,Gompertz:2020cur,Graham:2020gwr,Graham:2022xxu,Hussain:2019xzb,Lipunov:2022tno,Mo:2023sfz,Paterson:2020mmd,Podlesnyi:2020dbm,Ridnaia:2020pau,Unatlokov:2021erm}\\ 
S190521r & GW190521\_074359 & \cite{GCN24629} & \cite{Abbasi:2021kft,IceCube:2022mma,Abbasi:2023bfo,Abe:2020zpn,Abe:2021jtb,Acero:2020duu,Adriani:2022zli,ANTARES:2023wcj,Antier:2019pzz,Becerra:2021lyq,Cai:2021lmr,Gompertz:2020cur,Graham:2022xxu,Hussain:2019xzb,Lipunov:2022tno,Mo:2023sfz,Ridnaia:2020pau,J-GEM:2021aem,Unatlokov:2021erm}\\ 
\makebox[0pt][l]{\fboxsep0pt\colorbox{lightgray}{\mystrut\hspace*{1.0\linewidth}}}\!\!
\textit{S190524q} &  & \cite{GCN24655} & \\ 
 & GW190527\_092055 &  & \cite{Abbasi:2021kft,IceCube:2022mma,Abe:2021jtb,Abbasi:2023bfo,ANTARES:2023wcj,Cai:2021lmr,Graham:2022xxu,Mo:2023sfz,Unatlokov:2021erm}\\ 
\makebox[0pt][l]{\fboxsep0pt\colorbox{lightgray}{\mystrut\hspace*{1.0\linewidth}}}\!\!
S190602aq & GW190602\_175927 & \cite{GCN24716} & \cite{Abbasi:2021kft,IceCube:2022mma,Abbasi:2023bfo,Abe:2020zpn,Abe:2021jtb,Acero:2020duu,Adriani:2022zli,ANTARES:2023wcj,Antier:2019pzz,Cai:2021lmr,Graham:2022xxu,Hussain:2019xzb,Lipunov:2022tno,Mo:2023sfz,Ridnaia:2020pau,Unatlokov:2021erm}\\ 
 & GW190620\_030421 &  & \cite{Abbasi:2021kft,IceCube:2022mma,Abbasi:2023bfo,Abe:2021jtb,ANTARES:2023wcj,Cai:2021lmr,Graham:2022xxu,Mo:2023sfz,Unatlokov:2021erm}\\ 
\makebox[0pt][l]{\fboxsep0pt\colorbox{lightgray}{\mystrut\hspace*{1.0\linewidth}}}\!\!
S190630ag & GW190630\_185205 & \cite{GCN24920} & \cite{Abbasi:2021kft,IceCube:2022mma,Abbasi:2023bfo,Abe:2021jtb,Acero:2020duu,Adriani:2022zli,ANTARES:2023wcj,Cai:2021lmr,Graham:2022xxu,Lipunov:2022tno,Mo:2023sfz,Unatlokov:2021erm}\\ 
S190701ah & GW190701\_203306 & \cite{GCN24948} & \cite{Abbasi:2021kft,IceCube:2022mma,Abbasi:2023bfo,Abe:2020zpn,Acero:2020duu,Adriani:2022zli,ANTARES:2023wcj,Antier:2019pzz,Cai:2021lmr,Graham:2022xxu,Hussain:2019xzb,Lipunov:2022tno,Mo:2023sfz,Ridnaia:2020pau,Unatlokov:2021erm}\\ 
\makebox[0pt][l]{\fboxsep0pt\colorbox{lightgray}{\mystrut\hspace*{1.0\linewidth}}}\!\!
S190706ai & GW190706\_222641 & \cite{GCN24997} & \cite{Abbasi:2021kft,IceCube:2022mma,Abbasi:2023bfo,Abe:2020zpn,Acero:2020duu,Adriani:2022zli,ANTARES:2023wcj,Antier:2019pzz,Becerra:2021lyq,Cai:2021lmr,Gompertz:2020cur,Graham:2022xxu,Hussain:2019xzb,Lipunov:2022tno,Mo:2023sfz,Ridnaia:2020pau,Unatlokov:2021erm}\\ 
S190707q & GW190707\_093326 & \cite{GCN25011} & \cite{Abbasi:2021kft,IceCube:2022mma,Abbasi:2023bfo,Abe:2020zpn,Acero:2020duu,Adriani:2022zli,ANTARES:2023wcj,Antier:2019pzz,Cai:2021lmr,Gompertz:2020cur,Graham:2022xxu,Hussain:2019xzb,Lipunov:2022tno,Mo:2023sfz,Ridnaia:2020pau,Unatlokov:2021erm}\\ 
\makebox[0pt][l]{\fboxsep0pt\colorbox{lightgray}{\mystrut\hspace*{1.0\linewidth}}}\!\!
 & GW190708\_232457 &  & \cite{Abbasi:2021kft,IceCube:2022mma,Abbasi:2023bfo,ANTARES:2023wcj,Cai:2021lmr,Graham:2022xxu,Mo:2023sfz,Unatlokov:2021erm}\\ 
S190718y &  & \cite{GCN25086} & \cite{Abe:2020zpn,Adriani:2022zli,Antier:2019pzz,Gompertz:2020cur,Hussain:2019xzb,Lipunov:2022tno,Oates:2021eyk,Page:2020tnx,Ridnaia:2020pau}\\ 
\makebox[0pt][l]{\fboxsep0pt\colorbox{lightgray}{\mystrut\hspace*{1.0\linewidth}}}\!\!
 & GW190719\_215514 &  & \cite{Abbasi:2021kft,IceCube:2022mma,Abbasi:2023bfo,ANTARES:2023wcj,Cai:2021lmr,Graham:2022xxu,Mo:2023sfz,Unatlokov:2021erm}\\ 
S190720a & GW190720\_000836 & \cite{GCN25112} & \cite{Abbasi:2021kft,IceCube:2022mma,Abbasi:2023bfo,Abe:2020zpn,Adriani:2022zli,ANTARES:2023wcj,Antier:2019pzz,Cai:2021lmr,Gompertz:2020cur,Graham:2022xxu,Hussain:2019xzb,Lipunov:2022tno,Mo:2023sfz,Ridnaia:2020pau,J-GEM:2021aem,Unatlokov:2021erm}\\ 
\makebox[0pt][l]{\fboxsep0pt\colorbox{lightgray}{\mystrut\hspace*{1.0\linewidth}}}\!\!
S190727h & GW190727\_060333 & \cite{GCN25162} & \cite{Abbasi:2021kft,IceCube:2022mma,Abbasi:2023bfo,Abe:2020zpn,Adriani:2022zli,ANTARES:2023wcj,Antier:2019pzz,Cai:2021lmr,Gompertz:2020cur,Graham:2022xxu,Hussain:2019xzb,Lipunov:2022tno,Mo:2023sfz,Ridnaia:2020pau,Unatlokov:2021erm}\\ 
S190728q & GW190728\_064510 & \cite{GCN25183} & \cite{Abbasi:2021kft,IceCube:2022mma,Abbasi:2023bfo,HESS:2021gbx,Abe:2020zpn,Adriani:2022zli,Antier:2019pzz,Ashkar:2020hxe,Cai:2021lmr,Gompertz:2020cur,Graham:2022xxu,Hussain:2019xzb,Keivani:2020utg,Lipunov:2022tno,Mo:2023sfz,Oates:2021eyk,Ridnaia:2020pau,J-GEM:2021aem,Unatlokov:2021erm}\\ 
\makebox[0pt][l]{\fboxsep0pt\colorbox{lightgray}{\mystrut\hspace*{1.0\linewidth}}}\!\!
 & GW190731\_140936 &  & \cite{Abbasi:2021kft,IceCube:2022mma,Abbasi:2023bfo,ANTARES:2023wcj,Cai:2021lmr,Graham:2022xxu,Mo:2023sfz,Unatlokov:2021erm}\\ 
 & GW190803\_022701 &  & \cite{Abbasi:2021kft,IceCube:2022mma,Abbasi:2023bfo,ANTARES:2023wcj,Cai:2021lmr,Graham:2022xxu,Mo:2023sfz,Unatlokov:2021erm}\\ 
\makebox[0pt][l]{\fboxsep0pt\colorbox{lightgray}{\mystrut\hspace*{1.0\linewidth}}}\!\!
\textit{S190808ae} &  & \cite{GCN25295} & \cite{Oates:2021eyk}\\ 
S190814bv & GW190814 & \cite{GCN25320} & \cite{IceCube:2022mma,Abbasi:2023bfo,Abe:2020zpn,Ackley:2020qkz,Adriani:2022zli,ANTARES:2023wcj,Alexander:2021twj,Andreoni:2019qgh,Antier:2019pzz,Becerra:2021lyq,Cai:2021lmr,Coughlin:2019xfb,deWet:2021qdx,Dobie:2019ctw,Gomez:2019tuj,Gompertz:2020cur,Graham:2022xxu,Kasliwal:2020wmy,Kilpatrick:2021aku,Lipunov:2022tno,Mo:2023sfz,Oates:2021eyk,Page:2020tnx,Ridnaia:2020pau,J-GEM:2021aem,Thakur:2020yvu,DES:2021xnt,Vieira:2020lze,Watson:2020iqj}\\ 
\makebox[0pt][l]{\fboxsep0pt\colorbox{lightgray}{\mystrut\hspace*{1.0\linewidth}}}\!\!
\textit{S190816i} &  & \cite{GCN25367} & \\ 
\textit{S190822c} &  & \cite{GCN25439} & \cite{Oates:2021eyk,Page:2020tnx}\\ 
\makebox[0pt][l]{\fboxsep0pt\colorbox{lightgray}{\mystrut\hspace*{1.0\linewidth}}}\!\!
S190828j & GW190828\_063405 & \cite{GCN25497} & \cite{Abbasi:2021kft,IceCube:2022mma,Abbasi:2023bfo,Abe:2020zpn,Adriani:2022zli,ANTARES:2023wcj,Antier:2019pzz,Becerra:2021lyq,Cai:2021lmr,Gompertz:2020cur,Graham:2022xxu,Lipunov:2022tno,Mo:2023sfz,Ridnaia:2020pau,Unatlokov:2021erm}\\ 
S190828l & GW190828\_065509 & \cite{GCN25501} & \cite{Abbasi:2021kft,IceCube:2022mma,Abbasi:2023bfo,Abe:2020zpn,Adriani:2022zli,ANTARES:2023wcj,Antier:2019pzz,Cai:2021lmr,Gompertz:2020cur,Graham:2022xxu,Lipunov:2022tno,Mo:2023sfz,Ridnaia:2020pau,Unatlokov:2021erm}\\ 
\makebox[0pt][l]{\fboxsep0pt\colorbox{lightgray}{\mystrut\hspace*{1.0\linewidth}}}\!\!
\textit{S190829u} &  & \cite{GCN25554} & ,\\ 
S190901ap &  & \cite{GCN25606} & \cite{Abe:2020zpn,Adriani:2022zli,Antier:2019pzz,Chang:2021zdi,Coughlin:2019xfb,deJaeger:2021tcq,Gompertz:2020cur,Kasliwal:2020wmy,Lipunov:2022tno,Paterson:2020mmd,Ridnaia:2020pau,J-GEM:2021aem}\\ 
\makebox[0pt][l]{\fboxsep0pt\colorbox{lightgray}{\mystrut\hspace*{1.0\linewidth}}}\!\!
 & GW190909\_114149 &  & \cite{Abbasi:2021kft,ANTARES:2023wcj,Cai:2021lmr,Graham:2022xxu,Mo:2023sfz}\\ 
S190910d &  & \cite{GCN25693} & \cite{Abe:2020zpn,Adriani:2022zli,Antier:2019pzz,deJaeger:2021tcq,Gompertz:2020cur,Kasliwal:2020wmy,Lipunov:2022tno,Ridnaia:2020pau,Unatlokov:2021erm}\\ 
\makebox[0pt][l]{\fboxsep0pt\colorbox{lightgray}{\mystrut\hspace*{1.0\linewidth}}}\!\!
S190910h &  & \cite{GCN25707} & \cite{Abe:2020zpn,Adriani:2022zli,Antier:2019pzz,Chang:2021zdi,Coughlin:2019xfb,Lipunov:2022tno,Ridnaia:2020pau,Unatlokov:2021erm}\\ 
 & GW190910\_112807 &  & \cite{Abbasi:2021kft,Abbasi:2023bfo,Cai:2021lmr,Mo:2023sfz}\\ 
\makebox[0pt][l]{\fboxsep0pt\colorbox{lightgray}{\mystrut\hspace*{1.0\linewidth}}}\!\!
S190915ak & GW190915\_235702 & \cite{GCN25752} & \cite{Abbasi:2021kft,Abe:2020zpn,Adriani:2022zli,ANTARES:2023wcj,Antier:2019pzz,Becerra:2021lyq,Cai:2021lmr,Gompertz:2020cur,Graham:2022xxu,Lipunov:2022tno,Mo:2023sfz,Ridnaia:2020pau,Unatlokov:2021erm}\\ 
S190923y &  & \cite{GCN25811} & \cite{Abe:2020zpn,Adriani:2022zli,Antier:2019pzz,deJaeger:2021tcq,Gompertz:2020cur,Kasliwal:2020wmy,Lipunov:2022tno,Paterson:2020mmd,Ridnaia:2020pau,J-GEM:2021aem}\\ 
\makebox[0pt][l]{\fboxsep0pt\colorbox{lightgray}{\mystrut\hspace*{1.0\linewidth}}}\!\!
S190924h & GW190924\_021846 & \cite{GCN25828} & \cite{Abbasi:2021kft,IceCube:2022mma,Abbasi:2023bfo,Abe:2020zpn,Adriani:2022zli,ANTARES:2023wcj,Antier:2019pzz,Cai:2021lmr,Gompertz:2020cur,Graham:2022xxu,Lipunov:2022tno,Mo:2023sfz,Ridnaia:2020pau}\\ 
S190928c &  & \cite{GCN25883} & \\ 
\makebox[0pt][l]{\fboxsep0pt\colorbox{lightgray}{\mystrut\hspace*{1.0\linewidth}}}\!\!
 & GW190929\_012149 &  & \cite{Abbasi:2021kft,IceCube:2022mma,Abbasi:2023bfo,ANTARES:2023wcj,Cai:2021lmr,Unatlokov:2021erm}\\ 
S190930s & GW190930\_133541 & \cite{GCN25870} & \cite{Abbasi:2021kft,IceCube:2022mma,Abbasi:2023bfo,Abe:2020zpn,Adriani:2022zli,ANTARES:2023wcj,Antier:2019pzz,Becerra:2021lyq,Cai:2021lmr,Gompertz:2020cur,Graham:2022xxu,Lipunov:2022tno,Mo:2023sfz,Ridnaia:2020pau,J-GEM:2021aem}\\ 
\makebox[0pt][l]{\fboxsep0pt\colorbox{lightgray}{\mystrut\hspace*{1.0\linewidth}}}\!\!
S190930t &  & \cite{GCN25874} & \cite{Abe:2020zpn,Adriani:2022zli,Antier:2019pzz,Gompertz:2020cur,Kasliwal:2020wmy,Lipunov:2022tno,Mo:2023sfz,Oates:2021eyk,Paterson:2020mmd,Ridnaia:2020pau,J-GEM:2021aem}\\ 
\end{tabularx}
\end{ruledtabular}

\caption{\label{tab:follow-up-o3a} Public alerts and follow-up investigations of \ac{O3a} \ac{GW} candidates. 
The columns show the superevent identification (SID), the \ac{GW} candidate name if in offline results~\cite{Abbott:2020niy,LIGOScientific:2021usb}, the \ac{GCN} Circular, and references for follow-up publications. 
Candidates retracted following rapid event-validation checks are marked in \textit{italics}.
Candidates without superevent identifications were found only in the offline searches. 
}
\end{table*}

\begin{table*}
% Made by followup_tables
% DO NOT EDIT THIS FILE DIRECTLY

\begin{ruledtabular}
\begin{tabularx}{\textwidth}{@{\extracolsep{\fill}}l l l l}SID & Event & GCN & Follow-up publications\\ \hline
\makebox[0pt][l]{\fboxsep0pt\colorbox{lightgray}{\mystrut\hspace*{1.0\linewidth}}}\!\!
S191105e & \FULLNAME{GW191105C} & \cite{GCN26182} & \cite{Abbasi:2021kft,IceCube:2022mma,Abbasi:2023bfo,Abe:2020zpn,Adriani:2022zli,ANTARES:2023wcj,Antier:2020nuy,Becerra:2021lyq,Graham:2022xxu,Lipunov:2022tno,Mo:2023sfz,Ridnaia:2020pau,J-GEM:2021aem}\\ 
S191109d & \FULLNAME{GW191109A} & \cite{GCN26202} & \cite{Abbasi:2021kft,IceCube:2022mma,Abbasi:2023bfo,Abe:2020zpn,Adriani:2022zli,ANTARES:2023wcj,Antier:2020nuy,Becerra:2021lyq,Graham:2022xxu,Lipunov:2022tno,Mo:2023sfz,Ridnaia:2020pau}\\ 
\makebox[0pt][l]{\fboxsep0pt\colorbox{lightgray}{\mystrut\hspace*{1.0\linewidth}}}\!\!
\textit{S191110af} &  & \cite{GCN26222} & \cite{Oates:2021eyk,Page:2020tnx}\\ 
\textit{S191110x} &  & \cite{GCN26218} & \\ 
\makebox[0pt][l]{\fboxsep0pt\colorbox{lightgray}{\mystrut\hspace*{1.0\linewidth}}}\!\!
\textit{S191117j} &  & \cite{GCN26254} & \\ 
\textit{S191120aj} &  & \cite{GCN26263} & \\ 
\makebox[0pt][l]{\fboxsep0pt\colorbox{lightgray}{\mystrut\hspace*{1.0\linewidth}}}\!\!
\textit{S191120at} &  & \cite{GCN26265} & \\ 
\textit{S191124be} &  & \cite{GCN26288} & \\ 
\makebox[0pt][l]{\fboxsep0pt\colorbox{lightgray}{\mystrut\hspace*{1.0\linewidth}}}\!\!
S191129u & \FULLNAME{GW191129G} & \cite{GCN26303} & \cite{Abbasi:2021kft,IceCube:2022mma,Abbasi:2023bfo,Abe:2020zpn,Adriani:2022zli,ANTARES:2023wcj,Antier:2020nuy,Graham:2022xxu,Lipunov:2022tno,Mo:2023sfz,Ridnaia:2020pau}\\ 
S191204r & \FULLNAME{GW191204G} & \cite{GCN26334} & \cite{Abbasi:2021kft,IceCube:2022mma,Abbasi:2023bfo,Abe:2020zpn,Adriani:2022zli,ANTARES:2023wcj,Antier:2020nuy,Becerra:2021lyq,Graham:2022xxu,Lipunov:2022tno,Mo:2023sfz,Ridnaia:2020pau}\\ 
\makebox[0pt][l]{\fboxsep0pt\colorbox{lightgray}{\mystrut\hspace*{1.0\linewidth}}}\!\!
S191205ah &  & \cite{GCN26350} & \cite{Abe:2020zpn,Adriani:2022zli,Antier:2020nuy,Becerra:2021lyq,deJaeger:2021tcq,Kasliwal:2020wmy,Lipunov:2022tno,Paterson:2020mmd,J-GEM:2021aem}\\ 
\textit{S191212q} &  & \cite{GCN26394} & \cite{Abbasi:2021kft}\\ 
\makebox[0pt][l]{\fboxsep0pt\colorbox{lightgray}{\mystrut\hspace*{1.0\linewidth}}}\!\!
S191213g &  & \cite{GCN26402} & \cite{Abbasi:2021kft,Abe:2020zpn,Adriani:2022zli,Antier:2020nuy,deJaeger:2021tcq,Gourdji:2023bbm,Kasliwal:2020wmy,Lipunov:2022tno,Oates:2021eyk,Page:2020tnx,Paterson:2020mmd,Ridnaia:2020pau}\\ 
\textit{S191213ai} &  & \cite{GCN26413} & \\ 
\makebox[0pt][l]{\fboxsep0pt\colorbox{lightgray}{\mystrut\hspace*{1.0\linewidth}}}\!\!
S191215w & \FULLNAME{GW191215G} & \cite{GCN26441} & \cite{IceCube:2022mma,Abe:2020zpn,Abbasi:2023bfo,Adriani:2022zli,ANTARES:2023wcj,Antier:2020nuy,Becerra:2021lyq,Graham:2022xxu,Lipunov:2022tno,Mo:2023sfz,Ridnaia:2020pau}\\ 
S191216ap & \FULLNAME{GW191216G} & \cite{GCN26454} & \cite{Abbasi:2021kft,IceCube:2022mma,Abbasi:2023bfo,Abe:2020zpn,Adriani:2022zli,ANTARES:2023wcj,Antier:2020nuy,Becerra:2021lyq,Bhakta:2020vle,Graham:2022xxu,Keivani:2020utg,Lipunov:2022tno,Mo:2023sfz,Oates:2021eyk,Page:2020tnx,Ridnaia:2020pau,J-GEM:2021aem}\\ 
\makebox[0pt][l]{\fboxsep0pt\colorbox{lightgray}{\mystrut\hspace*{1.0\linewidth}}}\!\!
\textit{S191220af} &  & \cite{GCN26513} & \cite{Chang:2021zdi}\\ 
S191222n & \FULLNAME{GW191222A} & \cite{GCN26543} & \cite{Abbasi:2021kft,IceCube:2022mma,Abe:2020zpn,Adriani:2022zli,ANTARES:2023wcj,Antier:2020nuy,Graham:2022xxu,Lipunov:2022tno,Mo:2023sfz,Ridnaia:2020pau}\\ 
\makebox[0pt][l]{\fboxsep0pt\colorbox{lightgray}{\mystrut\hspace*{1.0\linewidth}}}\!\!
\textit{S191225aq} &  & \cite{GCN26585} & \\ 
S200105ae & \FULLNAME{200105F} & \cite{GCN26640} & \cite{Abbasi:2021kft,IceCube:2022mma,Abbasi:2023bfo,Abe:2020zpn,Adriani:2022zli,ANTARES:2023wcj,Anand:2020eyg,Antier:2020nuy,Ashkar:2020hxe,Graham:2022xxu,Kasliwal:2020wmy,Lipunov:2022tno,Paterson:2020mmd,Ridnaia:2020pau,Unatlokov:2021erm}\\ 
\makebox[0pt][l]{\fboxsep0pt\colorbox{lightgray}{\mystrut\hspace*{1.0\linewidth}}}\!\!
\textit{S200106au} &  & \cite{GCN26641} & \\ 
\textit{S200106av} &  & \cite{GCN26641} & \\ 
\makebox[0pt][l]{\fboxsep0pt\colorbox{lightgray}{\mystrut\hspace*{1.0\linewidth}}}\!\!
\textit{S200108v} &  & \cite{GCN26665} & \\ 
S200112r & \FULLNAME{GW200112H} & \cite{GCN26715} & \cite{Abbasi:2021kft,IceCube:2022mma,Abbasi:2023bfo,Abe:2020zpn,Adriani:2022zli,ANTARES:2023wcj,Antier:2020nuy,Graham:2022xxu,Lipunov:2022tno,Mo:2023sfz,Ridnaia:2020pau}\\ 
\makebox[0pt][l]{\fboxsep0pt\colorbox{lightgray}{\mystrut\hspace*{1.0\linewidth}}}\!\!
S200114f &  & \cite{GCN26734} & \cite{Abe:2020zpn,Adriani:2022zli,Becerra:2021lyq,Lipunov:2022tno,Oates:2021eyk,Page:2020tnx,Paterson:2020mmd,Ridnaia:2020pau,J-GEM:2021aem}\\ 
S200115j & \FULLNAME{GW200115A} & \cite{GCN26759} & \cite{Abbasi:2021kft,IceCube:2022mma,Abbasi:2023bfo,Abe:2020zpn,Adriani:2022zli,ANTARES:2023wcj,Anand:2020eyg,Antier:2020nuy,Ashkar:2020hxe,Becerra:2021lyq,deJaeger:2021tcq,Dichiara:2021vjy,Graham:2022xxu,Kasliwal:2020wmy,Lipunov:2022tno,Mo:2023sfz,Oates:2021eyk,Page:2020tnx,Paterson:2020mmd,Ridnaia:2020pau,Unatlokov:2021erm}\\ 
\makebox[0pt][l]{\fboxsep0pt\colorbox{lightgray}{\mystrut\hspace*{1.0\linewidth}}}\!\!
\textit{S200116ah} &  & \cite{GCN26785} & \\ 
S200128d & \FULLNAME{GW200128C} & \cite{GCN26906} & \cite{Abbasi:2021kft,IceCube:2022mma,Abbasi:2023bfo,Abe:2020zpn,Adriani:2022zli,ANTARES:2023wcj,Antier:2020nuy,Graham:2022xxu,Lipunov:2022tno,Mo:2023sfz,Ridnaia:2020pau}\\ 
\makebox[0pt][l]{\fboxsep0pt\colorbox{lightgray}{\mystrut\hspace*{1.0\linewidth}}}\!\!
S200129m & \FULLNAME{GW200129D} & \cite{GCN26926} & \cite{Abbasi:2021kft,IceCube:2022mma,Abbasi:2023bfo,Abe:2020zpn,Adriani:2022zli,ANTARES:2023wcj,Antier:2020nuy,Graham:2022xxu,Lipunov:2022tno,Mo:2023sfz,Ridnaia:2020pau}\\ 
S200208q & \FULLNAME{GW200208G} & \cite{GCN27014} & \cite{Abbasi:2021kft,IceCube:2022mma,Abbasi:2023bfo,Abe:2020zpn,Adriani:2022zli,ANTARES:2023wcj,Antier:2020nuy,Graham:2022xxu,Lipunov:2022tno,Mo:2023sfz,Ridnaia:2020pau}\\ 
\makebox[0pt][l]{\fboxsep0pt\colorbox{lightgray}{\mystrut\hspace*{1.0\linewidth}}}\!\!
S200213t &  & \cite{GCN27042} & \cite{Abbasi:2021kft,Abe:2020zpn,Adriani:2022zli,Antier:2020nuy,Becerra:2021lyq,deJaeger:2021tcq,Gourdji:2023bbm,Kasliwal:2020wmy,Keivani:2020utg,Lipunov:2022tno,Oates:2021eyk,Page:2020tnx,Ridnaia:2020pau,J-GEM:2021aem}\\ 
S200219ac & \FULLNAME{GW200219D} & \cite{GCN27130} & \cite{Abbasi:2021kft,IceCube:2022mma,Abbasi:2023bfo,Abe:2020zpn,Adriani:2022zli,ANTARES:2023wcj,Antier:2020nuy,Becerra:2021lyq,Graham:2022xxu,Lipunov:2020egw,Lipunov:2022tno,Mo:2023sfz,Ridnaia:2020pau,J-GEM:2021aem}\\ 
\makebox[0pt][l]{\fboxsep0pt\colorbox{lightgray}{\mystrut\hspace*{1.0\linewidth}}}\!\!
S200224ca & \FULLNAME{GW200224H} & \cite{GCN27184} & \cite{Abbasi:2021kft,IceCube:2022mma,Abbasi:2023bfo,HESS:2021gbx,Abe:2020zpn,Adriani:2022zli,ANTARES:2023wcj,Antier:2020nuy,Becerra:2021lyq,Graham:2022xxu,Klingler:2020osq,Lipunov:2020vrn,Lipunov:2022tno,Mo:2023sfz,Oates:2021eyk,J-GEM:2023mqd,Page:2020tnx,Paterson:2020mmd,Ridnaia:2020pau,J-GEM:2021aem}\\ 
S200225q & \FULLNAME{GW200225B} & \cite{GCN27193} & \cite{Abbasi:2021kft,IceCube:2022mma,Abbasi:2023bfo,ANTARES:2023wcj,Antier:2020nuy,Abe:2020zpn,Adriani:2022zli,Becerra:2021lyq,Graham:2022xxu,Lipunov:2022tno,Mo:2023sfz,Oates:2021eyk,Page:2020tnx,Ridnaia:2020pau,J-GEM:2021aem}\\ 
\makebox[0pt][l]{\fboxsep0pt\colorbox{lightgray}{\mystrut\hspace*{1.0\linewidth}}}\!\!
S200302c & \FULLNAME{GW200302A} & \cite{GCN27278} & \cite{Abbasi:2021kft,IceCube:2022mma,Abbasi:2023bfo,Abe:2020zpn,Adriani:2022zli,ANTARES:2023wcj,Antier:2020nuy,Graham:2022xxu,Lipunov:2020qyj,Lipunov:2022tno,Mo:2023sfz,Ridnaia:2020pau}\\ 
\textit{S200303ba} &  & \cite{GCN27306} & \\ 
\makebox[0pt][l]{\fboxsep0pt\colorbox{lightgray}{\mystrut\hspace*{1.0\linewidth}}}\!\!
\textit{S200308e} &  & \cite{GCN27347} &  \cite{Abbasi:2023bfo}\\ 
S200311bg & \FULLNAME{GW200311L} & \cite{GCN27358} & \cite{Abbasi:2021kft,IceCube:2022mma,Abbasi:2023bfo,Abe:2020zpn,Adriani:2022zli,ANTARES:2023wcj,Antier:2020nuy,Graham:2022xxu,Lipunov:2022tno,Mo:2023sfz,Ridnaia:2020pau}\\ 
\makebox[0pt][l]{\fboxsep0pt\colorbox{lightgray}{\mystrut\hspace*{1.0\linewidth}}}\!\!
S200316bj & \FULLNAME{GW200316I} & \cite{GCN27387} & \cite{Abbasi:2021kft,Abe:2020zpn,IceCube:2022mma,Abbasi:2023bfo,Adriani:2022zli,Antier:2020nuy,Becerra:2021lyq,Graham:2022xxu,Lipunov:2022tno,Mo:2023sfz,Paterson:2020mmd,Ridnaia:2020pau}\\ 
\end{tabularx}
\end{ruledtabular}

\caption{\label{tab:follow-up} Public alerts and follow-up investigations of \ac{O3b} \ac{GW} candidates. 
The columns show the superevent identification (SID), the \ac{GW} candidate name if in the offline results (including \FULLNAME{200105F}{}), the \ac{GCN} Circular, and references for follow-up publications.
Candidates retracted following rapid event-validation checks are marked in \textit{italics}. 
Candidates without superevent identifications were found only in the offline searches.
}
\end{table*}

The two alerts with the largest number of \ac{GCN} Circulars distributed during \ac{O3a} were \NNAME{GW190814H}{} (\SNAME{GW190814H}{}), whose source was a potential \ac{NSBH} or low-mass \ac{BBH} coalescence~\cite{Abbott:2020khf,Ackley:2020qkz,Alexander:2021twj,Andreoni:2019qgh,deWet:2021qdx,Kilpatrick:2021aku,Thakur:2020yvu,Vieira:2020lze,Watson:2020iqj} and the \ac{BNS} \NNAME{GW190425B}{} (\SNAME{GW190425B}{})~\cite{Abbott:2020uma,Coughlin:2019xfb,Hosseinzadeh:2019ifm}.
A potential association between \NNAME{GW190425B}{} and the fast radio burst \FRB{}~\cite{CHIMEFRB:2021srp}, occurring \FRBDELAY{} after the merger, has been suggested~\cite{Moroianu:2022ccs}, but a late-time optical and radio search \FRBLATE{} postburst was negative~\cite{Panther:2022jfg}. 
S191213g, the first \ac{O3b} \ac{BNS} candidate, had the largest number of \ac{GCN} Circulars during \ac{O3b}, a total of \MANYALERTS{}~\cite{GCN26402} (but it is only in fifth position considering the whole of \ac{O3}).
As discussed in Sec.~\ref{sec:candidates}, S191213g was not identified as a significant candidate in the offline search results. 
The \ac{O3} candidates were predominantly \acp{BBH}, where counterparts are not typically expected unless the system has surrounding gas~\cite{Perna:2016jqh,Stone:2016wzz,Murase:2016etc,Mink:2017npg,McKernan:2019hqs,Bogdanovic:2021aav}. 

The neutrino follow-up involved searches of events with energies ranging from $\sim\NUMINENERGY$ to $\sim\NUMAXENERGY$.
No confirmed neutrino counterpart has been found for any \ac{GW} candidate~\cite{Hussain:2019xzb,Abe:2021jtb,Petkov:2021asd,Acero:2020duu,Abe:2020zpn,Abbasi:2021kft,IceCube:2022mma,Unatlokov:2021erm,ANTARES:2023wcj,Abbasi:2023bfo}.

The gamma and X-ray observations involved energies extending up to $\sim\GAMMAMAXENERGY$.
The majority of high-energy searches reported no candidates~\cite{Ashkar:2020hxe,Podlesnyi:2020dbm,Pozanenko:2019lwh,Abbott:2020yvp,Watson:2020iqj,Page:2020tnx,Klingler:2020osq,Keivani:2020utg,HESS:2021gbx,Adriani:2022zli}.

The optical and near-infrared teams focused mainly on the non-\ac{BBH} systems or well-localized and nearby candidates.
Often multiple optical telescopes worked in synergy for the identification and characterization of counterparts~\cite{Antier:2020nuy,Lipunov:2020qyj,Gompertz:2020cur,Kasliwal:2020wmy,J-GEM:2021aem}.  
Several surveys performed systematic prompt follow-up searches for counterparts for a large number of candidates~\cite{Anand:2020mys,Lundquist:2019cty,Paterson:2020mmd,Becerra:2021lyq,Chang:2021zdi,Kim:2021hhl,Graham:2022xxu,Lipunov:2022tno,Mo:2023sfz}.
No confirmed prompt optical or infrared counterpart has been detected for \ac{O3} candidates.

The follow-up in the radio domain was mostly focused on the characterization of specific candidate counterparts, either neutrino, X-ray or optical candidates~\cite{Bhakta:2020vle,Dobie:2019ctw,Boersma:2021gyq}. 
No confirmed radio counterparts have been reported, with the possible exception of the fast radio burst associated to \NNAME{GW190425B}{}~\cite{Moroianu:2022ccs}. 

Nondetection of electromagnetic counterparts in follow-up searches for candidates where at least one component could be a \ac{NS} can potentially set constraints on the ejected matter; however, current observations cannot provide strong constraints~\cite{Coughlin:2019zqi,Coughlin:2020fwx}. 
It has been suggested that due to their faintness and fast evolution, searches by optical surveys for kilonovae within a distance up to \DISTKN{} require early observations down to magnitude \MAGKN{}~\cite{Carracedo:2020xhd}.
Future counterpart detections as soon as the next observing run are likely to place strong, multimessenger constraints on the equation of state of \acp{NS}, and the Hubble constant~\cite{Coughlin:2018fis,Dietrich:2020efo,Coughlin:2020pbb,Nicholl:2021rcr,Raaijmakers:2021slr}.

Additional specific counterpart searches have been performed after alerts, based on properties of the \ac{GW} candidates and using all-sky, multiwavelength data. 
As an illustration, \NNAME{GW190521B}{}, a signal from a high-mass \ac{BBH}~\cite{Abbott:2020tfl,LIGOScientific:2021usb}, generated interest due to the possible association with an observed flare of the active galactic nucleus AGN~J124942.3+344929~\cite{Graham:2020gwr}. 
This association, while still uncertain~\cite{Ashton:2020kyr,Palmese:2021wcv,DePaolis:2020onl,Nitz:2021uxj}, highlights the potential discoveries that could be made by searching for counterparts to \ac{BBH} coalescences, as well as the scope for detections of counterparts in archival searches.

\section{Observatory evolution}
\label{sec:interferometers}

\begin{figure} 
\begin{center}
\includegraphics[width=\columnwidth]{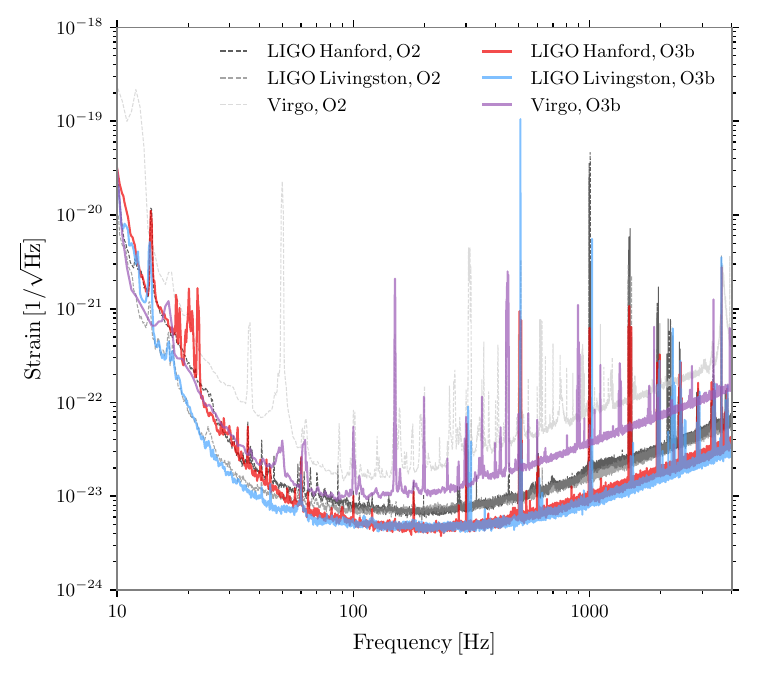}
\end{center}
\caption{
\label{fig:straino3bo2}
Representative noise amplitude spectral densities for \ac{LIGO} Livingston, \ac{LIGO} Hanford, Virgo during \ac{O2} and \ac{O3b}.
}
\end{figure}

From the start of the advanced-detector era in \ac{O1} through to \ac{O3b}, the network of \ac{GW} observatories has undergone a variety of commissioning activities to improve performance~\cite{Abbott:2020qfu}. 
The configurations of the detectors were different in \ac{O1}~\cite{TheLIGOScientific:2016pea}, \ac{O2}~\cite{Abbott:2017vtc,LIGOScientific:2018mvr} and \ac{O3}~\cite{Buikema:2020dlj,Virgo:2022ysc}. 
\ac{O2} marked the first operation of the three-detector \ac{LIGO}--Virgo network, and Fig.~\ref{fig:straino3bo2} shows the \ac{O2} and \ac{O3b} sensitivities for all interferometers to illustrate the evolution in performance. 
Key parameters describing the \ac{LIGO} detectors and Virgo are reported in Table~\ref{tab:inst_table}, specifically: input laser power (estimated at the power recycling mirror, after exiting the input mode cleaner); power recycling gain (the ratio of stored power in the power recycling cavity to the input laser power, which depends on the reflectivities of the test masses and power recycling mirror, as well as the losses in the arms and the power recycling cavity~\cite{Brooks:2021vyi}); presence of signal recycling; adoption of squeezed light, and suspension type. 
The main upgrade from run to run for all interferometers was the increase in the input laser power, which is instrumental in reducing shot noise.
Signal recycling mirrors~\cite{Meers:1988wp} are presently only installed in the \ac{LIGO} interferometers~\cite{TheLIGOScientific:2014jea}.
While in Hanford and Livingston monolithic test-mass suspensions has been operating since \ac{O1}, they were installed in Virgo only for \ac{O3}, replacing the wire suspensions used during \ac{O2}.
Squeezing was implemented during \ac{O3} at all sites~\cite{Tse:2019wcy,Acernese:2019sbr}. 
More details on the hardware and software changes that the \ac{LIGO} Hanford, \ac{LIGO} Livingston and Virgo observatories underwent from \ac{O3a} to \ac{O3b} are given below.

\begin{table*}
\begin{ruledtabular}
\begin{tabularx}{\textwidth}{@{\extracolsep{\fill}}l c c c c c c c c c c c} 
Parameter & \multicolumn{2}{c}{O1} & \multicolumn{3}{c}{O2} & \multicolumn{3}{c}{O3a} & \multicolumn{3}{c}{O3b} \\ 
\cline{2-3} \cline{4-6} \cline{7-9} \cline{10-12}
 & H & L & H & L & V & H & L & V & H & L & V \\ \hline
Input laser power  & \HANFORDPOWERONE{} & \LIVINGSTONPOWERONE{} & \HANFORDPOWERTWO{} & \LIVINGSTONPOWERTWO{} & \VIRGOPOWERTWO{} & \HANFORDPOWERTHREEA{} & \LIVINGSTONPOWERTHREEA{} & \VIRGOPOWERTHREEA{} & \HANFORDPOWER{} & \LIVINGSTONPOWER{} & \VIRGOPOWER{} \\ 
Power recycling gain   & \HANFORDPRGONE{} & \LIVINGSTONPRGONE{} & \HANFORDPRGTWO{} & \LIVINGSTONPRGTWO{} & \VIRGOPRGTWO{} & \HANFORDPRGTHREEA{} & \LIVINGSTONPRGTHREEA{} & \VIRGOPRGTHREEA{} & \HANFORDPRG{} & \LIVINGSTONPRG{} & \VIRGOPRG{} \\ 
Signal recycling       & \checkmark & \checkmark &  \checkmark & \checkmark & $\times$               & \checkmark & \checkmark & $\times$               &  \checkmark & \checkmark & $\times$ \\ 
Squeezing              & $\times$  & $\times$  & $\times$  & $\times$  & $\times$  & \checkmark & \checkmark & \checkmark & \checkmark & \checkmark & \checkmark  \\ 
Suspension type        & Silica & Silica & Silica & Silica & Steel & Silica & Silica & Silica & Silica & Silica & Silica \\ 
\end{tabularx}
\end{ruledtabular}

\caption{
\label{tab:inst_table}
Summary of selected optical and physical parameters of the \ac{LIGO} Hanford (H), \ac{LIGO} Livingston (L), and Virgo (V) interferometers in the advanced-detector era. 
The input laser power is an estimate for the maximum laser power level typically achieved during the observing period, and is the power that would be measured at the power recycling mirror (after the input mode cleaner). 
The suspension type is abbreviated to \emph{Silica} for monolithic fused-silica fibers, and \emph{Steel} for steel wires.
}
\end{table*}

\subsection{LIGO Hanford \& Livingston Observatories}
\label{sec:LIGO-interferometers}

The sensitivities of the Hanford and Livingston interferometers during \ac{O3b} were similar to during \ac{O3a}~\cite{Abbott:2020niy,Buikema:2020dlj}.
The upgrades between \ac{O3a} and \ac{O3b} aimed to address not only noise couplings that affect the range, but also reduce light scattering that degrades data quality, and improve resilience against environmental conditions that affect duty cycle.

High optical power in the interferometer reduces the shot noise.  
The current limit on the maximum circulating power of both \ac{LIGO} interferometers~\cite{Brooks:2021vyi} is from point defects in the test-mass mirror optical coatings which absorb and scatter light. 
Point absorbers appear in both the \ac{LIGO} Hanford and Livingston interferometers, and may be identified using Hartmann wavefront sensors, which can measure distortions created by point defects~\cite{Brooks:2009wiv,aLIGO:2016yis}. 
Prior to \ac{O3b}, both end test masses at \ac{LIGO} Livingston were inspected with a microscope to investigate potential defects.
After this investigation, new point absorbers appeared on both of these end test masses~\cite{alog:LLO49535,alog:LLO49564} for reasons not yet known~\cite{Brooks:2021vyi}. 
These new absorbers resulted in increased optical losses, a reduction in circulating power, and a consequent degradation of the Livingston interferometer's \ac{BNS} inspiral range due to increased shot noise of $\sim\LLORANGEPOWER{}$.

Adjustments to the squeezing subsystem produced the largest range improvements during \ac{O3b} shown in the left panel of Fig.~\ref{fig:range}.
An in-vacuum squeezer was installed for the \ac{O3} run at both \ac{LIGO} sites to improve detector sensitivity above $\sim\LIGOFREQSENS{}$~\cite{Tse:2019wcy}, below which radiation-pressure noise is larger with squeezing than the shot-noise level without squeezing.
The squeezer works by optically pumping a nonlinear crystal to create correlated photons.
The correlations modify the distribution of uncertainty in the quantum state that enters the interferometer~\cite{Caves:1981hw,Barsotti:2018hvm}.
The squeezer crystal has been found to degrade on timescales between a week and a month, reducing the pump light power and diminishing the squeezing below its optimal level.
At \ac{LIGO} Livingston, increased squeezing from moving the spot position on the crystal recovered $\sim\LLORANGESQZ{}$ in \ac{BNS} inspiral range between \ac{O3a} and \ac{O3b}. 
At \ac{LIGO} Hanford, a damaged fiber delivering pump light to the crystal was replaced between \ac{O3a} and \ac{O3b}, allowing a threefold increase in pump power and more squeezing. 
Adjustments done between \ac{O3a} and \ac{O3b}, in conjunction with moving the crystal position and retuning the squeezer on \LHORANGESPLIT{} during \ac{O3b} (shown in Fig.~\ref{fig:range}), produced an improvement of $\sim\LHORANGESQUEEZER{}$ in Hanford's \ac{BNS} inspiral range.

\ac{O3b} included upgrades to the \ac{LIGO} detectors to reduce scattered-light noise.  
Scattered-light noise occurs when a fraction of light gets scattered from its intended path, hits another moving surface, and a part of this light gets reflected back, rejoining the main interferometer beam with a noisy, varying phase~\cite{Accadia:2010zzb,Ottaway:2012oce}. 
This noise can be upconverted to higher harmonics of the surface motion frequencies, causing glitches.
At \ac{LIGO} Livingston, several locations at both end stations were outfitted with improved light baffles to prevent scattered light reflected off the vacuum envelope from recoupling with the main beam.
A particularly important contribution was new baffles installed between \ac{O3a} and \ac{O3b} surrounding a suspended platform that relay a beam transmitted by one end test mass.  
At \ac{LIGO} Hanford, a window in the output optic chain was replaced between \ac{O3a} and \ac{O3b} with one that has a larger incidence angle to ensure the back reflection from the window could not be a source of scattered light. 
Scattered-light noise was found to be correlated to microseismic activity, which is ground motion in the frequency band $\MICROSEISMLOW{}$--$\MICROSEISMHIGH{}~\mathrm{Hz}$ driven primarily by oceanic waves.
During periods of high microseismic activity both Hanford and Livingston interferometers suffer from large relative motion between the end test mass and the reaction mass that is immediately behind the test mass. 
This motion was found to produce a scattered-light noise path contributing to transient noise in the interferometer output~\cite{Soni:2020rbu}. 
This noise was mitigated by implementing reaction-chain tracking, a control loop that makes the reaction mass follow the end test mass, reducing the relative motion. 
Reaction-chain tracking was implemented on \LLORTRACK{} and \LHORTRACK{} at Livingston and Hanford, respectively. 
These efforts to reduce scattered-light noise had a significant effect on data quality by reducing transient noise as discussed in Sec.~\ref{sec:DQ}.

Finally, at \ac{LIGO} Hanford, another source of environmental noise, ground tilt induced by wind on the buildings, was mitigated by installing wind fences that reduce the wind velocity at the end stations~\cite{Nguyen:2021ybi}. 
This has been shown to lower ground tilt. 
The effect on data quality and duty cycle is still being investigated.

While the Hanford and Livingston detectors are nominally the same design~\cite{TheLIGOScientific:2014jea}, differences in environment and implementation result in different sensitivity during \ac{O3b}. 
Hanford has more unexplained noise from $\HANFORDUNEXPLAINEDFREQLOW~\mathrm{Hz}$ to $\HANFORDUNEXPLAINEDFREQHIGH~\mathrm{Hz}$ and more angular control noise below $\HANFORDANGULARFREQ~\mathrm{Hz}$. 
The higher noise above $\HANFORDOPTICSDIFFFREQ~\mathrm{Hz}$ in the Hanford spectrum is due to lower optical power causing increased shot noise as well as higher frequency-dependent losses that degrade the squeezing above the interferometer bandwidth~\cite{McCuller:2021mbn}. 

\subsection{Virgo Observatory}
\label{sec:Virgo-interferometers}

The one month commissioning break between the two observing periods was used to get a better understanding of the Virgo sensitivity and of some of its main limiting noises.
Throughout \ac{O3}, work was continuously carried out to improve the Virgo sensitivity in parallel with the ongoing data taking.
Dedicated tests were made during planned breaks in operation (commissioning, calibration and maintenance), and in-depth data analysis of these tests was performed between breaks to ensure continual improvement. 
This effort culminated during the last three months of \ac{O3b}, as shown by the step in the \ac{BNS} inspiral range evolution in the left panel of Fig.~\ref{fig:range}, and by the bimodal \ac{BNS} inspiral range distribution in the right panel.

The most significant change to the Virgo configuration between \ac{O3a} and \ac{O3b} was the increase of the input power from \VIRGOPOWERTHREEA{} to \VIRGOPOWER{}.
As for the \ac{LIGO} detectors, we found that the optical losses of the arms increased following the increase of the input power.
The presence of absorbing points on the arm-cavity mirrors is suspected~\cite{Brooks:2021vyi}, and mitigation strategies will be implemented before \ac{O4}.

The squeezing system in the Virgo interferometer was implemented before the start of \ac{O3a} and squeezing injection was maintained during the whole of \ac{O3}, with a gain in sensitivity at high frequency~\cite{Acernese:2019sbr,Acernese:2020usj}.
Prior to the start of \ac{O3a}, new high-quantum-efficiency photodiodes were installed at the output (detection) port of the interferometer.
These diodes increased the electronics noise at low frequency, but were improved at the end of January 2020 during a maintenance period, by replacing preamplifiers.
The electronic noise disappeared completely, leading to a \ac{BNS} inspiral range gain of $\sim\VIRGORANGEFLICKER{}$.

Shortly thereafter, an extended period of continuous and stable control of the Virgo detector allowed improvement to the performance of the etalon feedback system designed to reduce the residual asymmetry between the optical linewidths of the interferometer arm cavities~\cite{Brooks:2020txa}. 
To compensate for this asymmetry, the input mirrors of the Virgo Fabry--Perot cavities have parallel faces that create an optical resonator (the etalon) inside the substrate.
To remain close to the optimized working point, it is necessary to reduce the temperature variations of the substrate by using heating belts in the input test-mass towers. 
The implemented feedback requires hours to reach equilibrium, but has a temperature accuracy of \VIRGOTEMPERATUREACCURACY{}, about \VIRGOTEMPERATUREACCURACYPERCENT{} of a full etalon fringe (\VIRGOETALONFRINGE{}).
The \ac{BNS} inspiral range improvement from this etalon feedback control was $\sim\VIRGORANGEFEEDBACK{}$.

During the same period, it was discovered that some channels used as input for the \ac{GW} strain channel reconstruction were numerically limited by quantization errors.
Changing their storage from float to double precision led to an immediate gain of $\sim\VIRGORANGEQUANTIZATION{}$ for the \ac{BNS} inspiral range.

Finally, in the period between the end of January to the beginning of February 2020 the alignment was improved for the injection of the squeezed light into the interferometer~\cite{Acernese:2019sbr,Acernese:2020usj}, a critical parameter of the low-frequency sensitivity.
By mitigating scattered-light noise, the \ac{BNS} inspiral range increased by $\sim\VIRGORANGEALIGNMENT{}$.

All these quasisimultaneous hardware and software improvements led to a significant increase in the \ac{BNS} inspiral range visible in the data after \VIRGORANGESPLIT{} (Fig.~\ref{fig:range}, left panel).
The median range improved from \VIRGORANGEPRE{} (before \VIRGORANGESPLIT{}) to \VIRGORANGEPOST{} (after \VIRGORANGESPLIT{}).
The Virgo sensitivity improved over the whole frequency range, with a larger improvement below about \VIRGOFREQSENS{}, around the minimum of the sensitivity curve and at lower frequencies.

\section{Data-quality methods}
\label{sec:data-methods}

\begin{table*}
\begin{ruledtabular}
%\begin{tabular}{l@{\hskip 0.2in}c@{\hskip 0.2in}c@{\hskip 0.2in}c@{\hskip 0.2in}c@{\hskip 0.2in}c@{\hskip 0.2in}c@{\hskip 0.2in}c}
\begin{tabular}{l c c c c c c c}
{Search pipeline} & \CATONE{} & {\ac{CBC} \CATTWO{}} & {Burst \CATTWO{}} & {Burst \CATTHREE{}} & {Gating} & {\IDQ} \\
\hline
\ac{CWB} & \checkmark  & $\times$ & \checkmark & \checkmark & $\times$  & $\times$  \\
\GSTLAL{} &  \checkmark & $\times$  & $\times$ & $\times$  & $\times$   & \checkmark \\
\ac{MBTA} &  \checkmark & \checkmark  &  $\times$ & $\times$ & $\times$  & $\times$ \\ 
\PYCBC{} & \checkmark & \checkmark & $\times$ & $\times$ & \checkmark  &  $\times$\\
[0.25\normalbaselineskip]
\hline\hline
{Detector} \rule{0pt}{1.0\normalbaselineskip} & \CATONE{} & {\ac{CBC} \CATTWO{}} & {Burst \CATTWO{}} & {Burst \CATTHREE{}} & {Gating} & {\IDQ} \\ 
\hline
\ac{LIGO} Hanford & \LHOCATONE\% & \LHOCATTWOCBC\%  & \LHOCATTWOBURST\% & \LHOCATTHREEBURST\% & \LHOGATING\% & \LHOIDQ{} \\
\ac{LIGO} Livingston & \LLOCATONE\% & \LLOCATTWOCBC\% & \LLOCATTWOBURST\% & \LLOCATTHREEBURST\% & \LLOGATING\% & \LHOIDQ{} \\
Virgo & \VIRGOCATONE \% & \VIRGOCATTWOCBC{} & \VIRGOCATTWOBURST{} & \VIRGOCATTHREEBURST{} & \VIRGOGATING{} & \LHOIDQ{} \\
\end{tabular}
\end{ruledtabular}

\caption{\label{tab:search_dq_veto} 
\emph{Top}: Data-quality products used for noise mitigation by each offline search pipeline. 
Products listed here are publicly available from \ac{GWOSC}~\cite{gwosc:gwtc3}. 
Most analyses employ additional internal noise mitigation methods, including gating~\cite{Sachdev:2019vvd,Aubin:2020goo,Chu:2020pjv,DalCanton:2020vpm,Drago:2020kic}. 
\emph{Bottom}: The percent of single-detector time removed by each of the same veto categories for each detector during \ac{O3b}. 
Veto time values for \ac{LIGO} Hanford and \ac{LIGO} Livingston are reproduced from studies of \ac{O3} detector characterization~\cite{Davis:2021ecd}. 
A dash (--) in a data-quality product's column indicates that it is not produced for the relevant detector, except for \IDQ{} output; \IDQ{} has no associated removed time as it is incorporated directly into the search-pipeline ranking statistic~\cite{Godwin:2020weu}. 
The listed removed time is in addition to the downtime associated with one or more detectors not being in a nominal observing state, as described in Sec.~\ref{sec:instruments}, which is common to all searches. 
}
\end{table*}

Information about the data quality of the detectors is repackaged into products used by astrophysical analyses, including data-quality flags, gating, and \IDQ{} glitch likelihoods, as introduced and discussed below.
Including this information in searches, as summarized in Table~\ref{tab:search_dq_veto} for each offline analysis, increases the total number of detectable signals~\cite{TheLIGOScientific:2017lwt,Godwin:2020weu,Davis:2021ecd}.
The most egregious periods of light-scattering glitches in the \ac{LIGO} detectors are vetoed from the astrophysical analyses through a combination of these veto products, but the rate of scattering glitches was so high in the beginning of \ac{O3b}, especially in \ac{LIGO} Hanford data, that current methods cannot effectively exclude these glitches without losing large stretches of data~\cite{Davis:2021ecd}.  

Data-quality flags are lists of time segments that identify the status of the detectors or the likely presence of a particular instrumental artifact. 
These flags are broken into $\NUMCAT$ categories based on the severity of the data-quality issue and how the flag was designed~\cite{TheLIGOScientific:2017lwt,LIGOScientific:2019hgc,Davis:2021ecd}. 
The amount of time removed by data-quality flags in each detector is typically of order $\FLAGPERCENT\%$.  
Table~\ref{tab:search_dq_veto} shows the cumulative fractional time removed by each category during \ac{O3b}.
The fractional time removed by individual data-quality flags can be found in a summary of flags applied during \ac{O3} for \ac{LIGO} and Virgo~\cite{Davis:2021aaa, Arnaud:2021aaa}.
\CATONE{} flags indicate time periods where data should not be analyzed due to either incorrect configuration of the detector, operator error, or egregious data-quality issues. 
All \ac{GW} searches uniformly use \CATONE{} flag information to exclude these time periods.
\CATTWO{} flags are designed to indicate segments that are predicted to contain non-Gaussian artifacts likely to trigger \ac{GW} searches based on information from auxiliary channels~\cite{Davis:2021ecd}. 
While data during \CATTWO{} flags is still used in analyses to compute estimates of the \ac{PSD}, searches that use \CATTWO{} vetoes do not consider any triggers during these time periods in estimates of significance.
The set of \CATTWO{} flags that are used in analyses is different between the \ac{CBC} analyses that use waveform templates and the Burst analyses that are more waveform agnostic.
Similar to \CATTWO{} flags, \CATTHREE{} flags are used to indicate periods of transient noise, but are constructed using estimates of statistically significant correlations between glitches in auxiliary channels and behavior of \ac{GW} detector data~\cite{Smith:2011an}. 
\CATTHREE{} flags are produced only for use by the Burst analysis \ac{CWB}. 

The gating method removes short-duration artifacts from the data by smoothly rolling the data containing the artifact to zero with an inverse window function, as employed for \ac{LIGO} data during previous observing runs~\cite{Davis:2021ecd}.  
The gating data product referenced in Table~\ref{tab:search_dq_veto} and available from \ac{GWOSC}~\cite{gwosc:gwtc3} was generated using times corresponding to a loud excursion in the data identified with auxiliary channel information.
Most transient search algorithms also employ internal gating methods to exclude noise transients from analysis based only on the amplitude of the glitch.  

The \IDQ{} glitch likelihood uses machine learning to predict the probability that a non-Gaussian transient is present in detector data based only on information from auxiliary channels~\cite{Essick:2020qpo}. 
This likelihood is used by \GSTLAL{} as a part of the search-pipeline ranking statistic to penalize triggers near periods of high \IDQ{} likelihood~\cite{Godwin:2020weu}.
As shown in Table~\ref{tab:search_dq_veto}, \GSTLAL{} incorporates \IDQ{} glitch likelihood information in lieu of applying \CATTWO{} or \CATTHREE{} data-quality flags.

\begin{table*}
\begin{ruledtabular}
%\begin{tabular}{l@{\hskip 0.2in}c@{\hskip 0.2in}c@{\hskip 0.2in}c@{\hskip 0.2in}c@{\hskip 0.2in}c@{\hskip 0.2in}c}
\begin{tabular}{l c c c c c c}
	{Search pipeline} & {State vector} & {LIGO Data Quality vector} & {Virgo online \CATONE{}} & {Virgo \PYCBC{} veto} & {Online gating} \\
\hline
\ac{CWB} & \checkmark  & \checkmark & \checkmark & $\times$ & $\times$ \\
\GSTLAL{} &  \checkmark & \checkmark  & \checkmark & $\times$  & $\times$ \\
\MBTAONLINE{} &  \checkmark & \checkmark  &  \checkmark & $\times$ & $\times$ \\ 
\PYCBCLIVE{} & \checkmark & \checkmark & \checkmark & \checkmark & $\times$ \\
\SPIIR{} & \checkmark & \checkmark & \checkmark & $\times$ & \checkmark \\ 
\end{tabular}
\end{ruledtabular}

\caption{
\label{tab:online_search_dq_veto}
State information and data-quality products used for noise mitigation by each online search pipeline.
For each detector, the state vector defines the times the detector is online and the data is ready for analysis. 
The \ac{LIGO} Data Quality vector and Virgo online \CATONE{} products flag known instrument artifacts during observing mode~\cite{Davis:2021ecd,Acernese:2022jes}. 
Information from online data-quality products is available in a different form for offline searches, e.g., the \CATONE{} and \CATTWO{} data-quality products, as part of the products listed in Table~\ref{tab:search_dq_veto}. 
Most analyses employ additional internal noise mitigation methods, including gating~\cite{Sachdev:2019vvd,Aubin:2020goo,Chu:2020pjv,DalCanton:2020vpm,Drago:2020kic}.
}
\end{table*}

Astrophysical analyses performed online use different data-quality products from those listed in Table~\ref{tab:search_dq_veto}.
Online data-quality products describe the state of the detectors and the data using only products available within the low latency needed for the online search analyses and the sending of public alerts (as detailed in Appendix~\ref{sec:follow-up}). 
For example, the \CATTWO{} and \CATTHREE{} data-quality flags used by offline searches are often informed by follow-up investigations over the scale of weeks or months after data is initially collected. 
Table~\ref{tab:online_search_dq_veto} lists the state vector and data-quality products available online, and which searches use each of these.
Online data-quality products used in \ac{O3}, including online detector state vectors, the \ac{LIGO} Data Quality vector, search-specific online Virgo vetoes, Virgo online \CATONE{} products, and an online version of gating, are described in detail in related papers about \ac{LIGO}~\cite{Davis:2021ecd} and Virgo detector characterization~\cite{Virgo:2022kwz}.

After the event-validation procedures described in Sec.~\ref{sec:DQ}, we assessed whether excess power present within the target analysis time of any candidate was sufficiently nonstationary to require mitigation~\cite{Mozzon:2020gwa,Davis:2022ird}. 
We compared the variance of the noise \ac{PSD} in each identified time--frequency region for consistency with Gaussian noise. 
Time--frequency regions inconsistent with Gaussian noise ($p < \GAUSSIANNOISETHRESH$) were deglitched, as described below, before source-parameter estimation. 
Details of the candidates requiring mitigation are given in Appendix~\ref{sec:datamitigation}

The majority of glitch-subtracted data discussed in Appendix~\ref{sec:datamitigation} were produced with the \BAYESWAVE{} algorithm~\cite{Cornish:2014kda,Cornish:2020dwh}.
\BAYESWAVE{} models localized excess power as a sum of sine--Gaussian wavelets, using a multicomponent model that simultaneously fits signals, glitches and the \ac{PSD} of the Gaussian noise component using a transdimensional Bayesian inference, wherein the number of model components (wavelets, spectral lines and spline control points for the smooth portion of the \ac{PSD}) is allowed to vary, in addition to the parameters that describe each component. 

The signal model reconstructs the plus and cross polarization states of a \ac{GW} signal as a sum of wavelets, which are coherently projected onto the detector network~\cite{Cornish:2020dwh}. 
We use the waveform reconstruction produced by \BAYESWAVE{} in the waveform consistency tests as discussed in Sec.~\ref{sec:waveform-reconstruction}.
The glitch model reconstructs noise transients separately in each detector. 
The spectral model adjusts to take into account the power that gets assigned to the signal and glitch models. 
Central to the \BAYESWAVE{} approach is that the model dimension is not fixed, with both the number of wavelets and their parameters explored using a transdimensional reversible jump Markov-chain Monte Carlo algorithm~\cite{Green:1995mxx}. 
Louder signals generally demand more wavelets. 
In the case of \ac{CBC} signals, high-mass, short-duration signals are generally reconstructed with fewer wavelets than low-mass, longer-duration signals. 

\begin{figure} 
	\begin{center}
	\includegraphics[width=\columnwidth]{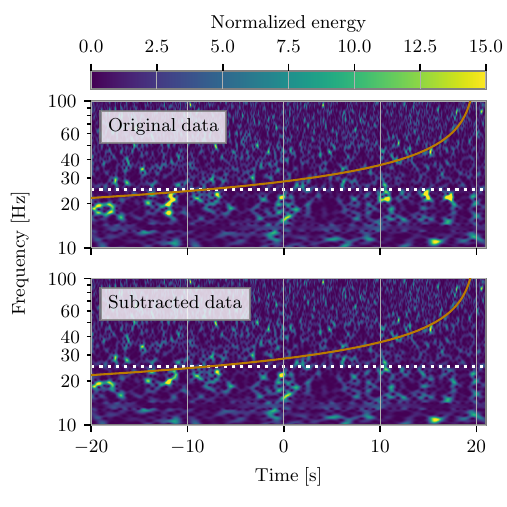}
	\end{center}
   	\caption{\label{fig:GW200115_data_mitigation} A spectrogram~\cite{Chatterji:2004qg} of \ac{LIGO} Livingston data prior to the estimated merger time of \FULLNAME{GW200115A}. 
     The top plot shows the untreated data and the bottom shows the data with some excess power due to \FASTSCATTER{} subtracted~\cite{Cornish:2020dwh}. 
     The estimated signal track is represented as an orange line. 
     A white dashed line shows the lower frequency used for source-parameter estimation for the original \FULLNAME{GW200115A}{} inference ($\flow = \OLDNSBHFLOW~\mathrm{Hz}$)~\cite{LIGOScientific:2021qlt}. 
     	}
\end{figure}

The natural parsimony of Bayesian inference works to ensure that any coherent signal power is assigned to the signal model, while any incoherent noise transients are assigned to the glitch power, since fitting the data with a coherent model requires fewer parameters than fitting the data in each detector independently.  
Thus, this method allows us to remove glitches even if they overlap with a \ac{GW} signal~\cite{TheLIGOScientific:2017qsa,Cornish:2020dwh}. 
Going forward, it may be desirable to perform the glitch fitting and \ac{PSD} estimation in concert with the \ac{CBC} parameter estimation~\cite{Chatziioannou:2021ezd}. 
In the current analysis, the \BAYESWAVE{} algorithm was used to produce cleaned data and point estimates of the \ac{PSD} that were then used in source-parameter estimation (Appendix~\ref{sec:parameter-estimation-methods}). 

An example of glitch subtraction by the \BAYESWAVE{} algorithm is illustrated in Fig.~\ref{fig:GW200115_data_mitigation} for the case of analyzing \FULLNAME{GW200115A}{}. 
The glitches removed here are \FASTSCATTER{}, one of the most common glitches observed in \ac{O3b} \ac{LIGO} Livingston data, as described in Sec.~\ref{sec:DQ}~\cite{Davis:2021ecd,Soni:2021cjy}. 

The mitigated data discussed in Appendix~\ref{sec:datamitigation} for \FULLNAME{GW200129D}{} were produced with the \GWSUBTRACT{} algorithm~\cite{Davis:2022ird}, which employs linear subtraction~\cite{Allen:1999wh,Davis:2018yrz}. 
We used a photodiode monitoring an element of the \ac{LIGO} Livingston detector's input optics (\LINSBUTRACTCHANNEL) identified~\cite{alog:LLO47707} as a linear witness of the glitch. 
The time of the subtracted glitch was also identified as correlated with an auxiliary witness channel by a \ac{CBC} \CATTWO{} data-quality flag~\cite{Davis:2018yrz} defined as \FLAGONETWOFOUR{} in the \ac{O3} \ac{LIGO} data-quality flag summary~\cite{Davis:2021aaa}.

In order to assess the efficacy of glitch subtraction by either method described above for \ac{O3b} candidates, we compared the stationarity of the glitch-subtracted data within the targeted time--frequency window to Gaussian noise.
Glitch-subtracted data consistent with Gaussian noise were deemed sufficiently stationary for parameter estimation. 

Uncertainties related to the glitch modeling are not accounted for in source-parameter estimation. 
To do so would require glitches to be modeled simultaneously with inference of the source properties. 
Analyses estimating both the properties of glitches and signals have found that the glitch modeling uncertainty may impact the inferred source properties in specific parts of the parameter space~\cite{Chatziioannou:2021ezd,Hourihane:2022doe,Payne:2022spz}. 
However, joint glitch and source inference has not yet been incorporated into the inference algorithms used here. 
Both of the glitch subtraction methods used in this work have different associated uncertainties, and may work better in different circumstances; for example, \BAYESWAVE{} works well when its coherent wavelet analysis accurately models the glitch, and \GWSUBTRACT{} works well when there is a reliable witness channel~\cite{Davis:2022ird}.

\section{Candidate-identification methods}
\label{sec:searches-methods}

\subsection{\GSTLAL{}}
\label{sec:gstlal-methods}

The \GSTLAL{} pipeline~\cite{Cannon:2011vi,Privitera:2013xza,Messick:2016aqy,Sachdev:2019vvd,Hanna:2019ezx,Cannon:2020qnf} uses matched filtering in the time domain to detect triggers and coincidences.
We model signals and search for them in the data using the same template bank as for the \GWTCTWO{} analysis~\cite{Abbott:2020niy}.
The template bank covers waveforms with redshifted total masses from \GSTLALBANKMINMASS{} to \GSTLALBANKMAXMASS{}.
The spins are assumed to be parallel (aligned or antialigned) to the orbital angular momentum; the spin magnitudes range from \GSTLALBANKNSSPINLOW{} to \GSTLALBANKNSSPINHIGH{} for components with redshifted masses $< \GSTLALPASTROBNSMASSMAX{} \Msun$, and from \GSTLALBANKBHSPINLOW{} to \GSTLALBANKBHSPINHIGH{} for components with redshifted masses $> \GSTLALPASTROBNSMASSMAX{} \Msun$. 
The template bank is constructed using a stochastic placement method in five different regions of the parameter space that are the same as those defined for the \GWTCTWO{} analysis~\cite{Abbott:2020niy}.
The \SEOBNRROM{} waveform approximant~\cite{Bohe:2016gbl} is used for templates with chirp mass $\geq \GSTLALBANKWFCHIRPMASS$; this waveform is a frequency-domain reduced-order model~\cite{Purrer:2015tud} of the time-domain inspiral--merger--ringdown model \SEOBNRFOUR{} which models quasicircular, nonprecessing \acp{BBH} based upon the \ac{EOB} equations of motion~\cite{Bohe:2016gbl}.
The \TFTWO{} waveform approximant~\cite{Blanchet:1995ez,Poisson:1997ha,Damour:2001bu,Mikoczi:2005dn,Blanchet:2005tk,Arun:2008kb,Buonanno:2009zt,Bohe:2013cla,Bohe:2015ana,Mishra:2016whh} is used for lower-mass systems; this waveform is a frequency-domain, inspiral-only model of aligned-spin \ac{CBC} systems built from closed-form \ac{PN} approximations. 
The template bank is constructed such that any template in the continuous parameter space is certain to match at least one template in the discrete space to greater than a chosen minimum match, where the match used is that given in Eq.~\eqref{eq:overlap}, maximized over the phase and time of coalescence.
The value of the minimum match is chosen to ensure that the \ac{SNR} loss due to the templates not exactly matching the signals is acceptable while keeping the total number of templates small enough to be computationally feasible.
The minimum match is dependent on the region of the parameter space, but is never smaller than \GSTLALTEMPLATEMINMATCH{}~\cite{Abbott:2020niy}.

Triggers are defined by maximizing the matched-filter \ac{SNR} for each template, in each detector, over $\GSTLALWINDOW{}$ time windows~\cite{Messick:2016aqy}.
We use a \ac{SNR} threshold of $\rho > \GSTLALSNRTHRESH{}$ to define triggers.
Triggers from the same template that are time coincident in multiple detectors are grouped together to form events~\cite{Messick:2016aqy}.
The \GSTLAL{} analysis uses single-detector triggers from \ac{HL} coincident time (when either \ac{HL} or \ac{HLV} were operating) to estimate background statistics in bins according to template mass. 
This is due to the low probability of a real signal appearing above threshold in only \ac{LIGO} Hanford or \ac{LIGO} Livingston when both detectors are operating. 
Triggers from single-detector time, or times when only \ac{HV} or \ac{LV} were operating, are excluded from the background estimation to avoid significant contamination by true astrophysical signals.

The likelihood ratio is informed by observables such as the matched-filter \ac{SNR} from each detector, detector sensitivities at the time of coincidence, as well as the output of signal-based-veto tests, and time and phase differences between triggers~\cite{Sachdev:2019vvd}. 
The events are ranked by the likelihood-ratio statistic which compares the probability in the signal hypothesis of finding the given observables to the probability of the same observables in the noise hypothesis.
In addition, the likelihood ratio includes a term from \IDQ{}~\cite{Godwin:2020weu}, a statistical inference framework that identifies short-duration non-Gaussian artifacts in the strain data~\cite{Essick:2020qpo} (described in Appendix~\ref{sec:data-methods}). 
As discussed in the \GWTCTWOFINAL{} paper~\cite{LIGOScientific:2021usb}, \IDQ{} time series were regenerated offline using an acausal binning scheme and a larger set of auxiliary witness channels, making its data products more sensitive in identifying noise artifacts compared to their online counterpart.
An increased sampling rate in the offline configuration also allowed for better resolution of short-duration glitches.
Because of these changes, \IDQ{} had an improved performance in identifying glitches. 
Accordingly, starting in \ac{O3b}, \IDQ{} now has the capability to increase the significance of candidates during times in which no noise artifacts are identified: whereas in the previous procedure used for \GWTCTWO{}~\cite{Abbott:2020niy}, the \IDQ{} term in the \GSTLAL{} likelihood ratio was restricted to be positive, so that it could only decrease the significance of candidates~\cite{Godwin:2020weu}, it may now be either positive or negative.
Additionally, \IDQ{} is now applied to both coincident and single-detector candidates.

Since \ac{O2}, the \GSTLAL{} pipeline has allowed for the possibility of single-detector candidates~\cite{Sachdev:2019vvd}.
This includes two cases: triggers from a time when only one detector was operational, and noncoincident triggers from one detector even when multiple detectors were operational.
Single-detector candidates are required to pass the \ac{SNR} threshold as well as a preliminary likelihood-ratio threshold.
However, single-detector candidates are downweighted with a singles penalty in the likelihood-ratio statistic, depending on the detector in which it was observed and the sensitivities of the detectors which were on at the trigger time~\cite{LIGOScientific:2021usb}.

\subsection{\ac{MBTA}}
\label{sec:mbta-methods}

\ac{MBTA}~\cite{Adams:2015ulm,Aubin:2020goo} uses a template bank covering binaries with redshifted component masses ranging from $\COMPONENTBANKMINMASS$ to $\MBTACOMPONENTBANKMAXMASS$, with the additional constraints that the maximum total mass is $\MBTABANKMAXMASS$, and if the secondary object has a mass lower than $\MBTABANKNSMASS$, then the maximum mass of the primary object is $\MBTABANKMAXNSBHMASS$.
Objects are assumed to have spins parallel to the orbital momentum with maximum dimensionless values $\MBTABANKMAXNSSPIN$ if their masses are below $\MBTABANKNSMASS$, and $\MBTABANKMAXBHSPIN$ otherwise. 
The templates are generated in the time domain, using the \STTFOUR{} waveform approximant~\cite{Buonanno:2002fy,Arun:2008kb,Racine:2008kj,Buonanno:2009zt,Vines:2011ud,Marsat:2013caa,Marsat:2014xea,Bohe:2015ana} if both objects have masses below $\MBTABANKNSMASS$, and the \SEOBNRFOUR{} waveform~\cite{Bohe:2016gbl} otherwise. 
The \STTFOUR{} waveform is an inspiral-only, time-domain model for \ac{CBC} systems based on the \ac{PN} equations of motion, while \SEOBNRFOUR{} is a full inspiral--merger--ringdown waveform appropriate for \acp{BBH}.
The template bank is produced using a stochastic placement method.

The \ac{MBTA} pipeline starts with a preprocessing step, where data are downsampled then gated at (externally or internally) identified times of bad data quality.
To mitigate safety issues in the gating procedure, a subset of the template bank is also analyzed without applying the gating procedure, albeit with higher \ac{SNR} thresholds ($\rho > \MBTASNRTHRESHREGIONFOURH{}$ in Hanford, $\MBTASNRTHRESHREGIONFOURL{}$ in Livingston and $\MBTASNRTHRESHREGIONFOURV{}$ in Virgo).
\ac{MBTA} splits the parameter space into three regions treated as independent searches.
The regions can be considered to cover the \ac{BNS}, \ac{NSBH} and \ac{BBH} source types, although the transition between \ac{NS} and \ac{BH} is conservatively taken to be $\MBTABANKNSMASS$ (to allow for any heavier object to possibly have high spin)~\cite{LIGOScientific:2021usb}.
Single-detector triggers are ranked according to a statistic based on the matched-filter \ac{SNR} modified to take into account the consistency with an astrophysical signal (quantified from the quadratic average of the difference between the \ac{SNR} time series around its maximum and the template autocorrelation) and the local data quality (quantified from the overall pipeline response).
Coincidences are ranked according to a statistic based on the quadratic sum of the single-detector triggers ranking statistics modified to take into account the consistency of some parameters across the various detectors.

\ac{MBTA} initially assigns a \ac{FAR} to events depending on the \emph{coincidence type} (whether \ac{HL}, \ac{HV}, \ac{LV} or \ac{HLV}, depending on which detectors provided a trigger), and the parameter-space region.
The \ac{FAR} is then modified to take into account trials factors from the various coincidence types and regions.
For double coincidences the \ac{FAR} at a given ranking statistic threshold is estimated from the rate of false coincidences (built from single-detector triggers in that region) that are as loud or louder.
Single-detector triggers that are known to be part of foreground coincidences with a combined ranking statistic above $\MBTACRSTHRESHOLDTOREMOVESINGLES$ are excluded from this process, as such statistic values typically correspond to low \acp{FAR} (typically less than $\MBTATYPICALFARATCRSTEN$ for \ac{HL} coincidences) indicating a probable astrophysical origin.
The \ac{FAR} for triple coincidences is derived from that of double coincidences.
Equal trials factors are applied for the three parameter-space regions, whereas for coincidence types, trials factors are applied according to the likelihood of astrophysical sources being detected as coincidences of each type, considering the relative detector sensitivities.

\subsection{\PYCBC{}}
\label{sec:pycbc-methods}

We employ two offline \PYCBC{} configurations in this work~\cite{Davies:2020tsx,Nitz:2017svb,Usman:2015kfa,Canton:2014ena,Allen:2005fk,Allen:2004gu,pycbc-software}.
The first, the \PYCBCHYPERBANK{} analysis, is designed to search for as many different types of signal as possible, and probes a wide range of masses and spins.
Following previous searches~\cite{Nitz:2019hdf,Abbott:2020niy,Nitz:2021uxj,LIGOScientific:2021usb}, we also perform an analysis focusing on the \ac{BBH} region of the parameter space in which we have seen most of our signals so far, making use of a population prior~\cite{Dent:2013cva}. 
This second approach is the \PYCBCBBH{} analysis.

The \PYCBCBBH{} analysis focuses on a region ranging in primary component mass from \PYCBCBBHMASSONEMIN{} to \PYCBCBBHMASSONEMAX{}, with mass ratios from \PYCBCBBHQMIN{} to \PYCBCBBHQMAX{}, and aligned, equal component spins ranging from \PYCBCBBHSPINMIN{} to \PYCBCBBHSPINMAX{}.
The \PYCBCHYPERBANK{} template bank covers a similar parameter space as the \GSTLAL{} template bank, but with a few significant changes.
Both the \PYCBCHYPERBANK{} and \PYCBCBBH{} analyses use the \SEOBNRROM{}~\cite{Bohe:2016gbl} waveform approximant for templates with total mass above $\PYCBCBANKWFMASS$, and \TFTWO{}~\cite{Blanchet:1995ez,Poisson:1997ha,Damour:2001bu,Mikoczi:2005dn,Blanchet:2005tk,Arun:2008kb,Buonanno:2009zt,Bohe:2013cla,Bohe:2015ana,Mishra:2016whh} for lower-mass systems.
The templates within the template bank are placed using a hybrid geometric--random method~\cite{Roy:2017qgg,Roy:2017oul}, and no template is used that has a duration of less than $\PYCBCBANKTMIN$~\cite{DalCanton:2017ala}, meaning there is an upper limit on the mass of the systems. 
If this duration limit is relaxed and additional vetoes on transient data artifacts are applied, higher sensitivity to high-mass systems may be obtained~\cite{Chandra:2021wbw,LIGOScientific:2021tfm}.

Both the \PYCBCHYPERBANK{} and \PYCBCBBH{} analyses use data from all detectors, searching for coincident triggers in two or more detectors.
For each coincident event, we calculate a ranking statistic which is compared to the background to calculate the significance, finally combining the significances from each possible combination of triggers from the available detectors (the coincidence type) into a single result.

The search in the three-detector network is done by performing coincident searches in each coincidence type, and then combining \acp{FAR} depending on the available coincident combinations.
For example, if an event is seen as a \ac{HL} coincidence, the ranking statistic would be calculated, and the \ac{FAR} estimated by counting higher-ranked events in a time-shifted background.
If the Virgo detector is observing, then the \ac{FAR} from the detected event would be added to the \ac{FAR} at that ranking statistic from each of \ac{HV}, \ac{HL} and \ac{HLV} backgrounds.
This method means that we effectively apply a trials factor where it is needed, but not when the coincidence type in which the candidate was found is the only one available such that a trials factor would be inappropriate.

The \PYCBC{} pipelines use a ranking statistic based on the ratio of the expected signal rate and the measured noise rate~\cite{Davies:2020tsx,Nitz:2019hdf}.
This choice of ranking statistic has two consequences.
First, we are able to incorporate more information about the detectors into our assessment of whether an apparent signal is real or not.
For example, we now account for the sensitive volume of the detector network at the time of a candidate and combine the single-detector rates of noise triggers with the time window for coincidences in order to estimate the coincident-trigger rate.
Second, we are able to directly combine \acp{FAR} by summing the \acp{FAR} at the ranking statistic of the event for each coincidence type available at the time; by adding the \acp{FAR} rather than using a trials factor, we take into account the vastly different \acp{FAR} for different coincidence types at the same ranking statistic.

The \PYCBCBBH{} ranking statistic uses a chirp-mass weighting designed to approximately model the \ac{BH} population.
In previous analyses~\cite{Nitz:2019hdf}, this weighting took a different functional form above a template chirp mass of $\PYCBCBBHMAXCHIRPMASSWEIGHTING$, suppressing higher-mass triggers; however, in this analysis, due to an omission in implementation, the same functional form was continued over all templates.
This change resulted in a higher background noise level than expected due to triggers in high-mass templates caused by glitches, potentially reducing the sensitivity of the search.
From the sensitivity studies in Sec.~\ref{sec:vt}, we see that the sensitivity reduction was likely to be small.

Recent alterations to the \PYCBC{} analysis allow the use of \ac{GPU} cores or distributed computing through the Open Science Grid~\cite{Pordes:2007zzb,Sfiligoi:2010zz} in order to perform matched filtering more quickly.

The \PYCBC{} analysis used in this work did not analyze single-detector signals, though recent work allows this feature~\cite{Nitz:2020naa,Davies:2022thw}.
Usually, triggers from significant signals are removed from the background of lower-ranked events within the analysis, in a process called hierarchical removal~\cite{Capano:2016uif}, but as we did not calculate single-detector significance, we have no metric by which to remove these triggers, and so signal triggers can remain in the background.
As a result, these loud triggers from signals can match noise triggers in the time-shifted background and cause an excess of highly ranked background events.
In order to prevent the contamination of the background, \PYCBC{} analyses were performed twice: first with all triggers in place, and then again with the triggers removed from catalog candidates that did not form coincidences in the preliminary analysis.
To ensure that this process matched the usual hierarchical removal procedure, we used the list of candidates from other analyses for this catalog that have a \ac{FAR} of less than \PYCBCSINGLESREMOVALSIGNIFICANCE{}, and compared these to the list of coincident events in the \PYCBC{} analyses.
If no coincident event (of any significance) was found in the \PYCBC pipeline, then a window of \PYCBCSINGLESREMOVALWINDOW{} was removed.
The triggers removed from the background in the \PYCBCHYPERBANK{} pipeline are from around \PYCBCHYPERBANKREMOVEDSINGLES, and from the \PYCBCBBH{} pipeline we remove the triggers from around \PYCBCBBHREMOVEDSINGLES{}.
The sensitive hypervolume estimates of Sec.~\ref{sec:vt} use the analysis with the triggers from single-detector events removed. 
Only a small subset of the analysis chunks are significantly affected by this change, and this effect is particularly muted at the threshold we consider for \VT{} estimates~\cite{Davies:2022thw}.

In addition to the offline analyses described above, we also used \PYCBCLIVE{}~\cite{Nitz:2018rgo,DalCanton:2020vpm} to search for signals in low latency.
The \PYCBCLIVE{} algorithm uses the data and data-quality information that are available in low latency (as described in Appendix~\ref{sec:data-methods}) without human vetting.
\PYCBCLIVE{} uses a more computationally simple ranking statistic than the one used in offline analyses.
This simpler ranking statistic is used in order to maintain speed in a low-latency environment and does not contain all of the information used in the offline statistic.
The reduced $\chi^2$-reweighted \ac{SNR}~\cite{Allen:2005fk} and a sine--Gaussian veto~\cite{Nitz:2017lco,Nitz:2018imz} are used to assess significance of single-detector triggers.
These single-detector triggers are then tested for coincidence, and the coincident ranking statistic is calculated.
The ranking statistic is compared to the time-shifted background from five hours of data to estimate \ac{FAR}.

\subsection{\ac{SPIIR}}
\label{sec:spiir-methods}

The \ac{SPIIR} pipeline~\cite{Luan:2011qx,Chu:2017ovg,Chu:2020pjv} ran as an online low-latency modeled coherent search.
\SPIIR{} is a time-domain equivalent to matched filtering that uses infinite impulse response filters~\cite{Hooper:2011rb,Chu:2020pjv} to approximate waveforms with high accuracy and, in theory, constructs the \ac{SNR} at zero latency.
In \ac{O3} the pipeline operated in two low-latency, parallel modes: one to search using data from the two \ac{LIGO} detectors, and another using data from all three detectors. 
\ac{SPIIR} searches templates with primary component mass ranging from $\SPIIRPRIMARYBANKMINMASS$ to $\SPIIRPRIMARYBANKMAXMASS$, a subset of the \GSTLAL{} template bank~\cite{Chu:2020pjv}. 
For online low-latency analyses, this method is more computationally efficient than traditional Fourier methods, with latency $\SPIIRLATENCY$ in \ac{O3}~\cite{Chu:2020pjv}. 
The filtering process~\cite{Liu:2012vw,Guo:2018tzs,Guo:2018mkw} and coherent candidate selection~\cite{Luan:2011qx} are accelerated using \acp{GPU}.

The pipeline ranks the triggers by a combination of the coherent network \ac{SNR} and a $\chi^2$-distributed signal-consistency statistic from the individual detectors \cite{Chu:2020pjv,Messick:2016aqy}. 
It computes the background of the search by performing $\SPIIRTIMESHIFTS$ time shifts per foreground trigger with \ac{SNR} greater than $\SPIIRSNRTHRESH$. 
The $k$-nearest-neighbors technique was used to estimate the significance for triggers~\cite{Chu:2020pjv}. 
The \ac{FAR} for each trigger is estimated over three timescales (two hours, one day and one week) of collected background triggers for robustness, with the most conservative used for candidates.

\subsection{\ac{CWB}}
\label{sec:cwb-methods}

The \ac{CWB} pipeline detects and reconstructs transient signals with minimal assumptions~\cite{Klimenko:2004qh,Klimenko:2008fu,Necula:2012zz,Klimenko:2015ypf,Salemi:2019uea} by coherently analyzing data from multiple observatories.
The sensitivity of \ac{CWB} approaches that of matched-filter methods for coalescing stellar-mass \acp{BBH} with high chirp masses~\cite{CalderonBustillo:2017skv,LIGOScientific:2021tfm}, such that it can detect high-mass \ac{CBC} sources, and also sources that are not well represented in current template banks such as eccentric systems or large mass asymmetry, precessing \ac{BBH} systems~\cite{Salemi:2019owp}.
It was used in previous \ac{CBC} searches by the \ac{LVK}~\cite{Abbott:2016blz,TheLIGOScientific:2016uux,LIGOScientific:2018mvr,Abbott:2020niy}.

The \ac{CWB} algorithm analyzes whitened data using the Wilson--Daubechies--Meyer wavelet transform~\cite{Klimenko:2004qh,Necula:2012zz} to compute a time--frequency representation.
The algorithm selects excess-energy data in the time--frequency representation and clusters them to define a trigger.
Next, it identifies coherent signal power with the constrained maximum-likelihood method~\cite{Klimenko:2015ypf}, and reconstructs the source sky location and the signal waveforms.

After identifying clusters of coherent data, \ac{CWB} outputs several statistics.
These include the total cluster energy for each detector; the coherent energy $E_\mathrm{c}$ of the reconstructed signal obtained by cross-correlating the normalized signal waveforms reconstructed in different detectors; the residual noise energy $E_\mathrm{n}$ estimated after the reconstructed waveforms are subtracted from the data, and the estimate of the coherent \ac{SNR} in each detector.
The residual noise energy is used to form a chi-squared statistic $\chi^2 = E_\mathrm{n}/N_\mathrm{df}$, where $N_\mathrm{df}$ is the number of independent wavelet amplitudes describing the trigger.
We estimate the signal \acp{SNR} from the reconstructed waveforms. 
Then, by combining the \acp{SNR} of the individual detectors, we calculate the network \ac{SNR}.
The network correlation coefficient $c_\mathrm{c} = E_\mathrm{c}/(E_\mathrm{c}+E_\mathrm{n})$ is another derived statistic that compares coherent and null energies; it approaches $1$ when coherence is high, as expected for real signals.
The \ac{CWB} detection statistic is $\eta_\mathrm{c} \propto [{E_\mathrm{c}/\max(\chi^2,1)}]^{1/2}$, where the $\chi^2$ correction is applied to reduce the contribution of non-Gaussian noise.

For robustness against glitches and to reduce the \ac{FAR} of the pipeline, \ac{CWB} uses signal-independent vetoes, which include Burst \CATTWO{} data-quality flags in the processing step and \CATTHREE{} in the postproduction phase~\cite{Smith:2011an,Aasi:2012wd}. 
To further reduce background, the \ac{CWB} analysis applies cuts based on the network correlation coefficient $c_\mathrm{c}$ and on the $\chi^2$, and employs signal-dependent vetoes based on basic properties of the time--frequency evolution of \ac{CBC} signals~\cite{Tiwari:2015bda,Szczepanczyk:2020osv}.

A generic search for \ac{CBC} systems covers a large parameter space and it is not possible to design a search that is optimized for all such systems because of the wide frequency range in which the signals fall. 
With the setup used for this catalog, \ac{CWB} can reconstruct \ac{GW} signals with durations up to a few seconds in the detectors' frequency range, which makes it better suited to identify \ac{BBH} signals than longer \ac{NSBH} or \ac{BNS} signals. 
A \ac{CBC} signal has a peak frequency inversely proportional to the redshifted total mass, so that less massive binary systems merge at higher frequencies, while more massive systems merge at lower frequencies.
Therefore, just as for the \GWTCTWO{} analysis~\cite{Abbott:2020niy}, the \ac{CWB} analyses in this catalog are performed with two pipeline configurations targeting the detection of high-mass ($f_\mathrm{c}<\CWBCENTRALFREQ$) and low-mass ($f_\mathrm{c}>\CWBCENTRALFREQ$) \ac{BBH} systems.
These configurations use different signal-dependent vetoes defined a priori to alleviate the large variability of nonstationary noise in the detectors' bandwidth.

We estimate the \ac{FAR} of triggers by time shifting the data of one detector with respect to the other in each detector pair, with time lags so large (typically multiples of \CWBTSLIDE) that actual astrophysical signals are excluded, and repeating this for a large number of different time lags over a total time $T_\mathrm{bkg}$ which is of the order of \CWBTBKG.
We count the number of triggers $N_\mathrm{bkg}$ due to background noise having a \ac{SNR} (or another similar ranking statistic) that is at least as large as that of the trigger, and we compute the \ac{FAR} as the $N_\mathrm{bkg}$ divided by $T_\mathrm{bkg}$~\cite{Was:2009vh}.

The detection significance of a trigger identified by either pipeline configuration in a single frequency range is determined by its \ac{FAR} measured by the corresponding \ac{CWB} configuration.
In the end, each configuration reports the selected triggers and their \acp{FAR}.
Whenever the low-mass and high-mass configurations overlap, the trials factor of two (the Bonferroni adjustment for the false alarm probability~\cite{Wright:1992aaa}) is included to determine the final \ac{FAR}~\cite{Abbott:2020tfl}.

The \ac{CWB} algorithm can work with arbitrary detector networks, although the \ac{CWB} analysis presented in this catalog is restricted to the \ac{HL}, \ac{HV} and \ac{LV} pairs. 
The \ac{HLV} network is not included here because it does not improve the significance of the \ac{CWB} candidates for the current sensitivity of the detector network~\cite{LIGOScientific:2021hoh}. 
Thanks to their near alignment, the two \ac{LIGO} detectors select a well-defined \ac{GW} polarization state, and \ac{CWB} can efficiently exploit coherence to mitigate their glitches and make the remaining noise close to Gaussian.
Conversely, the orientation of the Virgo detector differs considerably from that of the \ac{LIGO} detectors so that, at the current sensitivity level, glitches in Virgo data cannot be mitigated as efficiently, and this reduces the discriminating power of current \ac{CWB} \ac{HLV} analyses with respect to \ac{HL} analyses.

\subsection{Search results}
\label{sec:search_results_appendix}

In Sec.~\ref{sec:candidates}, we presented the \pastro{}, \ac{FAR} and network \ac{SNR} of candidates with \ac{CBC} $\pastro{} > \PASTROTHRESHOLD{}$ or \ac{FAR} $<\FARTHRESHYR{}~\mathrm{yr^{-1}}$ in Table~\ref{tab:events} and Table~\ref{tab:marginal_events}, respectively.
Here, we additionally provide the single-detector \acp{SNR} of each candidate in Table~\ref{tab:sngl_ifo_snr}.
The single-detector \acp{SNR} are used as an initial criterion by pipelines to define triggers and determine coincidences, and therefore are an important component in calculating the significance of a detection candidate.
The detectors listed in Table~\ref{tab:events} are those that were operating at the time of each candidate, but whether a candidate was missed or found in a particular detector depends on the matched-filter \ac{SNR} found by each pipeline in the detector's data. 
In particular, each of the single-detector candidates, \FULLNAME{GW200112H}{}, \FULLNAME{GW200302A}{} and \FULLNAME{200105F}{}, were found during times when either \ac{LIGO} Livingston or \ac{LIGO} Hanford were operating simultaneously with the Virgo detector. 
However, Table~\ref{tab:sngl_ifo_snr} shows that these were still classified as single-detector candidates since in each case the \ac{SNR} in Virgo was $< \GSTLALSNRTHRESH{}$. 
Regardless of the number of detectors used for detection, data from all operating detectors is used for inference of the source parameters (described in Appendix~\ref{sec:parameter-estimation-methods}).

Candidates found by multiple analyses typically have comparable \acp{SNR}, but we do not expect the values to be identical because of differences in the template banks and how the pipelines select the most significant template when identifying a candidate. 
The most noticeable difference is in the Livingston \ac{SNR} for \FULLNAME{GW200129D}{}, as discussed in Sec.~\ref{sec:pipeline-comparison}, this is a result of the different analyses' handling of data-quality flags. 

\begin{event_table}
\begin{table*}
% Made by ./scripts/generating_tex_macros/make_single_ifo_snr_table
% DO NOT EDIT THIS FILE DIRECTLY

\begin{ruledtabular}
\begin{tabular}{l cc ccc ccc ccc ccc}
{Name} & \multicolumn{2}{c}{{\CWB{}}} & \multicolumn{3}{c}{{\GSTLAL{}}} & \multicolumn{3}{c}{{\MBTA{}}} & \multicolumn{3}{c}{{\PYCBCHYPERBANK{}}} & \multicolumn{3}{c}{{\PYCBCBBH{}}} \\
\cline{2-3} \cline{4-6} \cline{7-9} \cline{10-12} \cline{13-15} 
& H & L  & H & L & V & H & L & V & H & L & V & H & L & V \\
\hline
\makebox[0pt][l]{\fboxsep0pt\colorbox{lightgray}{\mystrut\hspace*{1.000000\linewidth}}}\!\! \FULLNAME{GW191103A} & 
$ \CWBALLSKYMEETSPASTROTHRESH{GW191103A} \CWBALLSKYHANFORDSNR{GW191103A}$ &
$ \CWBALLSKYMEETSPASTROTHRESH{GW191103A} \CWBALLSKYLIVINGSTONSNR{GW191103A}$ &
$ \GSTLALALLSKYMEETSPASTROTHRESH{GW191103A} \GSTLALALLSKYHANFORDSNR{GW191103A}$ &
$ \GSTLALALLSKYMEETSPASTROTHRESH{GW191103A} \GSTLALALLSKYLIVINGSTONSNR{GW191103A}$ &
$ \GSTLALALLSKYMEETSPASTROTHRESH{GW191103A} \GSTLALALLSKYVIRGOSNR{GW191103A}$ &
$ \MBTAALLSKYMEETSPASTROTHRESH{GW191103A} \MBTAALLSKYHANFORDSNR{GW191103A}$ &
$ \MBTAALLSKYMEETSPASTROTHRESH{GW191103A} \MBTAALLSKYLIVINGSTONSNR{GW191103A}$ &
$ \MBTAALLSKYMEETSPASTROTHRESH{GW191103A} \MBTAALLSKYVIRGOSNR{GW191103A}$ &
$ \PYCBCALLSKYMEETSPASTROTHRESH{GW191103A} \PYCBCALLSKYHANFORDSNR{GW191103A}$ &
$ \PYCBCALLSKYMEETSPASTROTHRESH{GW191103A} \PYCBCALLSKYLIVINGSTONSNR{GW191103A}$ &
$ \PYCBCALLSKYMEETSPASTROTHRESH{GW191103A} \PYCBCALLSKYVIRGOSNR{GW191103A}$ &
$ \PYCBCHIGHMASSMEETSPASTROTHRESH{GW191103A} \PYCBCHIGHMASSHANFORDSNR{GW191103A}$ &
$ \PYCBCHIGHMASSMEETSPASTROTHRESH{GW191103A} \PYCBCHIGHMASSLIVINGSTONSNR{GW191103A}$ &
$ \PYCBCHIGHMASSMEETSPASTROTHRESH{GW191103A} \PYCBCHIGHMASSVIRGOSNR{GW191103A} $ \\
\end{tabular}
\end{ruledtabular}

\caption{
\label{tab:sngl_ifo_snr}
Individual-detector \acp{SNR} for all candidates in Table~\ref{tab:events} and Table~\ref{tab:marginal_events}. 
\ac{LIGO} Hanford, \ac{LIGO} Livingston and Virgo are indicated by H, L and V, respectively.
Numbers in \textit{italics} indicate where a candidate is identified with probability of astrophysical origin $\pastro{} < \PASTROTHRESHOLD{}$. 
Dashes (--) indicate where no significant trigger was identified by a search analysis.
}
\end{table*}
\end{event_table}

\subsection{Search sensitivity and probability of astrophysical origin}
\label{sec:p-astro-methods}

\begin{table*}
\begin{ruledtabular}
\begin{tabular}{c c cccccc }
\multicolumn{2}{c}{ } & Mass & Mass &  Spin & Spin & Redshift & Maximum  \\
\multicolumn{2}{c}{ } & distribution & range (\Msun) & range & orientations & evolution & redshift  \\
\hline
\multirow{6}{*}{{Injections}} & \multirow{2}{*}{\ac{BBH}} &  $\left.p(\massone{}) \propto \massone{}^{\INJBBHMASSONEPOWER{}}\right.$ \rule{0pt}{1.05\normalbaselineskip} & $\INJBBHMASSMIN{} < \massone{} < \INJBBHMASSMAX{}$ & \multirow{2}{*}{$\left|\chi_{1,2}\right| < \INJBBHSPINMAX{}$} & \multirow{2}{*}{Isotropic} & $ \multirow{2}{*}{$\kappa = \INJBBHKAPPAREDSHIFT$}$ & \multirow{2}{*}{$\INJBBHMAXREDSHIFT{}$}  \\
 & & $p(\masstwo{}|\massone{}) \propto \masstwo{}$ & $\INJBBHMASSMIN{} < \masstwo{} < \INJBBHMASSMAX{}$ & & & & \\[0.05\normalbaselineskip]
 & \multirow{2}{*}{\ac{NSBH}} & $\left.p(\massone{}) \propto \massone{}^{\INJNSBHPOWERBBHMASS{}}\right.$ & $\INJNSBHMINBBHMASS{} < \massone{} < \INJNSBHMAXBBHMASS{}$ & $\left|\chi_1\right| < \INJNSBHMAXBBHSPIN{}$ & \multirow{2}{*}{Isotropic} & $ \multirow{2}{*}{$\kappa = \INJNSBHKAPPAREDSHIFT$}$ &  \multirow{2}{*}{$\INJNSBHMAXREDSHIFT{}$} \\
 & & Uniform & $\INJNSBHMINBNSMASS{} < \masstwo{} < \INJNSBHMAXBNSMASS{}$ & $\left|\chi_2\right| < \INJNSBHMAXBNSSPIN{}$ & & &  \\[0.05\normalbaselineskip]
 & \multirow{2}{*}{\ac{BNS}} & \multirow{2}{*}{Uniform} & $\INJBNSMASSMIN{} < \massone{} < \INJBNSMASSMAX{}$ & \multirow{2}{*}{$\left|\chi_{1,2}\right| < \INJBNSSPINMAX{}$} & \multirow{2}{*}{Isotropic} & $ \multirow{2}{*}{$\kappa = \INJBNSKAPPAREDSHIFT$}$ & \multirow{2}{*}{$\INJBNSMAXREDSHIFT{}$}  \\
 & & & $\INJBNSMASSMIN{} < \masstwo{} < \INJBNSMASSMAX{}$ & & & & \\%[0.25\normalbaselineskip]
\hline
{\CWB{} \pastro} &  {\ac{BBH}} & \multicolumn{2}{c}{Same as injections} & & & & \\%[0.25\normalbaselineskip]
\hline
\multirow{6}{*}{\GSTLAL{} \pastro{}} & \multirow{2}{*}{\ac{BBH}} & \multirow{2}{*}{Log-uniform} & $\GSTLALPASTROBBHMASSMIN{} < \massone{} < \GSTLALPASTROBBHMASSMAX{}$ & \multirow{2}{*}{$\left|\chi_{1,2}\right| < \GSTLALPASTROBBHSPINMAX{}$} & \multirow{2}{*}{Aligned} & $ \multirow{2}{*}{$\kappa = \GSTLALPASTROBBHKAPPAREDSHIFT$}$ & \multirow{2}{*}{$\GSTLALPASTROBBHMAXREDSHIFT{}$}  \\ & &  & $\GSTLALPASTROBBHMASSMIN{} < \masstwo{} < \GSTLALPASTROBBHMASSMAX{}$ & & & & \\[0.05\normalbaselineskip]
 & \multirow{2}{*}{{NSBH}} & \multirow{2}{*}{Log-uniform} & $\GSTLALPASTRONSBHMINBBHMASS{} < \massone{} < \GSTLALPASTRONSBHMAXBBHMASS{}$ & $\left|\chi_1\right| < \GSTLALPASTRONSBHMAXBBHSPIN{}$ & \multirow{2}{*}{Aligned} & $ \multirow{2}{*}{$\kappa = \GSTLALPASTRONSBHKAPPAREDSHIFT$}$ &  \multirow{2}{*}{$\GSTLALPASTRONSBHMAXREDSHIFT{}$} \\
 & & & $\GSTLALPASTRONSBHMINBNSMASS{} < \masstwo{} < \GSTLALPASTRONSBHMAXBNSMASS{}$ & $\left|\chi_2\right| < \GSTLALPASTRONSBHMAXBNSSPIN{}$ &  & &  \\[0.05\normalbaselineskip]
 & \multirow{2}{*}{\ac{BNS}} & \multirow{2}{*}{Log-uniform} & $\GSTLALPASTROBNSMASSMIN{} < \massone{} < \GSTLALPASTROBNSMASSMAX{}$ & \multirow{2}{*}{$\left|\chi_{1,2}\right| < \GSTLALPASTROBNSSPINMAX{}$} & \multirow{2}{*}{Aligned} & $ \multirow{2}{*}{$\kappa = \GSTLALPASTROBNSKAPPAREDSHIFT$}$ & \multirow{2}{*}{$\GSTLALPASTROBNSMAXREDSHIFT{}$}  \\
 & & & $\GSTLALPASTROBNSMASSMIN{} < \masstwo{} < \GSTLALPASTROBNSMASSMAX{}$ & & & & \\%[0.25\normalbaselineskip]
\hline
\multirow{7}{*}{\MBTA{} \pastro{}} & \multirow{5}{*}{\ac{BBH}} & \PLPEAK{}~\cite{Abbott:2020gyp} & & \multirow{5}{*}{$\left|\chi_{1,2}\right| < \INJBBHSPINMAX{}$} & \multirow{5}{*}{Isotropic} & $ \multirow{5}{*}{$\kappa = \MBTAPASTROBBHKAPPAREDSHIFT$}$ & \multirow{5}{*}{$\INJBBHMAXREDSHIFT{}$}  \\
 & & with $\alpha=\MBTAPASTROPLPALPHA$, $\beta_q=\MBTAPASTROPLPBETAQ$, & \multirow{2}{*}{$\MBTAPASTROPLPMMIN < m_1 < \MBTAPASTROPLPMMAX$} & & & & \\
 & & $m_{\mathrm{min}}=\MBTAPASTROPLPMMIN\Msun$, $m_{\mathrm{max}}=\MBTAPASTROPLPMMAX\Msun$, & \multirow{2}{*}{$\MBTAPASTROPLPMMIN < m_2 < \MBTAPASTROPLPMMAX$} & & & & \\
 & & $\lambda_{\mathrm{peak}}=\MBTAPASTROPLPLAMBDA$, $\mu_m=\MBTAPASTROPLPMUM\Msun$, & & & & & \\
 & & $\sigma_m=\MBTAPASTROPLPSIGMAM\Msun$, $\delta_m=\MBTAPASTROPLPDELTAM\Msun$ & & & & & \\[0.05\normalbaselineskip]
 & \ac{NSBH} & \multicolumn{2}{c}{Same as injections} & & & & \\[0.05\normalbaselineskip]
 & \ac{BNS} & \multicolumn{2}{c}{Same as injections} & & & & \\%[0.25\normalbaselineskip]
\hline
\multirow{3}{*}{\PYCBCHYPERBANK{} \pastro{}} & \ac{BBH} & & $\Mc{} > \PYCBCPASTROMINMCHIRPBBH{} $ & & & & \\[0.05\normalbaselineskip]
 & \ac{NSBH} & & \hspace{-1.25cm} $ \PYCBCPASTROMINMCHIRPNSBH{} < \Mc{} < \PYCBCPASTROMINMCHIRPBBH{} $ & & & & \\[0.05\normalbaselineskip]

 & \ac{BNS} & & $ \Mc{} < \PYCBCPASTROMINMCHIRPNSBH{} $ & & & & \\%[0.25\normalbaselineskip]
\hline
{\PYCBCBBH{} \pastro{}} & {\ac{BBH}} & & $\Mc{} > \PYCBCPASTROMINMCHIRPBBH{}$ & & & \\
\end{tabular}
\end{ruledtabular}

\caption{
\label{tab:populations}
Parameter distributions used to generate injections and to compute the probability of astrophysical origin \pastro{} for each pipeline. 
The \ac{BNS} injections are generated using the \STTFOUR{} waveform model~\cite{Buonanno:2002fy,Arun:2008kb,Racine:2008kj,Buonanno:2009zt,Vines:2011ud,Marsat:2013caa,Marsat:2014xea,Bohe:2015ana}, while the \ac{BBH} and \ac{NSBH} injections are generated using the \SEOBNRPHM{} model~\cite{Ossokine:2020kjp}, or the \SEOBNRP{} model~\cite{Babak:2016tgq,Bohe:2016gbl,Ossokine:2020kjp} for injections corresponding to binaries with redshifted total mass below $\INJMODELSWITCHMASS{} \Msun$.
We always use the convention that $\massone \geq \masstwo$; this constraint means that the marginalized one-dimensional distributions for the masses will not match the distributions used to define the two-dimensional distributions (as given here) in cases where the $\massone$ and $\masstwo$ distributions overlap. 
Masses are in the source frame, except for the \PYCBC{} rows, where the measured (redshifted) chirp mass is considered. 
The redshift-evolution parameter $\kappa$ controls the injected distribution as described in Eq.~\eqref{eq:redshift evolution}. 
The injection sets are used to estimate sensitive hypervolumes, with weights to match the populations assumed within each \VT{} calculation, including updating the mass, spin, and redshift distributions where appropriate.
}
\end{table*}

To assess search sensitivity, we inject simulated signals into the data, and attempt to identify them with each search analysis. 
The details of the injected populations (which are the same as used for \GWTCTWOFINAL{}~\cite{LIGOScientific:2021usb}) are given in Table~\ref{tab:populations}, and the injected distributions over redshift are defined assuming a flat $\Lambda$--cold dark matter cosmology such that
\begin{equation}\label{eq:redshift evolution}
    p(\redshift{}) \propto \frac{\mathrm{d}\comovingv}{\mathrm{d}\redshift{}} (1+\redshift{})^{\kappa - 1},
\end{equation}
where $\comovingv$ is the comoving volume (see Appendix~\ref{sec:parameter-estimation-methods} for the assumed cosmology~\cite{Ade:2015xua}). 
These injected populations are reweighted to obtain estimates of the sensitive hypervolumes presented in Table~\ref{tab:vt} such that the injected distributions in Table~\ref{tab:populations} do not represent the assumed populations used to estimate search sensitivity. 

The probability of astrophysical origin \pastro{} for a candidate is estimated directly from the ranking statistics \rankstat{} that are used to assess the \ac{FAR}.
By comparing the distributions of ranking statistics under the assumptions of foreground $p(\rankstat | \mathrm{signal})$ or background $p(\rankstat | \mathrm{noise})$, we can calculate a signal-versus-noise Bayes factor for each event. 
This Bayes factor acts as a likelihood in the $\pastro{}$ computation for each event.
The normalization of the astrophysical \rankstat{} distributions depends on merger rates, which are jointly estimated in the calculation, assuming that the triggers are drawn from independent Poisson processes~\cite{Farr:2013yna}. 
For a given \ac{FAR}, $\pastro{}$ will be larger if the true alarm rate is higher.

The construction of the foreground (signal) and background (noise) distributions is specific to individual detection pipelines:
\begin{itemize}
  \item The \PYCBC{} analyses use time-shifted triggers to empirically estimate the rates of background events and their distributions over the search ranking statistic, while foreground distributions are estimated using recovered simulated signals. 
  As in \GWTCTWOFINAL{}~\cite{LIGOScientific:2021usb}, we allow these background and foreground distributions to differ between different combinations of detectors in coincidence, and also allow for a dependence of the foreground distribution and signal rate on which detectors are observing at a given time~\cite{Dent:2021aaa}.   
  In order to model variation of the signal rate over binary masses, the foreground and background estimates are obtained separately over the ranges of template chirp mass given in Table~\ref{tab:populations}; the rate of astrophysical signals is also estimated separately in each range. 

  \item For \GSTLAL{}, the ratio of the foreground-to-background distributions (the signal-to-noise Bayes factor that enters into the \pastro{} calculation) is proportional to the likelihood ratio which is the ranking statistic \rankstat{}.
  Details of the \GSTLAL{} background collection method are given in Appendix~\ref{sec:gstlal-methods}.
  The time--volume sensitivity of the pipeline used in this calculation is estimated based on simulated sources injected into the pipeline and is rescaled to the astrophysical distribution~\cite{Tiwari:2017ndi}.
   We use time--volume ratios to combine triggers from various observation runs and perform the multicomponent analysis yielding \pastro{} and merger rates~\cite{Farr:2013yna,Kapadia:2019uut} inferred from the entire set of available data (from \ac{O1} to \ac{O3b}).

  \item The \ac{MBTA} analysis uses a template bank split into $\MBTAPASTROBINS$ bins in the chirp-mass--mass-ratio parameter space to compute \pastro{} values of events~\cite{Andres:2021vew}.
  The fine binning has the main benefit of allowing the proper tracking over the parameter space of the assumed \acp{CBC} populations used in the foreground distribution.
  It also provides a more tailored estimate of the background rate compared to the \ac{FAR} reported by the analysis, which uses a coarse estimate of the background (integrated over one of the three search regions) that is conservative for signals from high-mass sources.
  It can therefore result in events being assigned a significant \pastro{} in population-rich regions of the parameter space even though they were assigned a high \ac{FAR} value (examples are \FULLNAME{GW200220H}{}, \FULLNAME{GW200306A}{} and \FULLNAME{GW200322G}{}).
  For instance, \FULLNAME{GW200220H}{} is in an \Mc{}--\massratio{} bin that captures $\MBTAEXAMPLEBINSIGNALFRACTION$ of the expected astrophysical signal while it contains only $\MBTAEXAMPLEBINBBHTEMPLATESFRACTION$ of the \ac{BBH} templates.
  For the combined ranking statistics of this candidate, the expected foreground rate density (per unit interval of the ranking statistic squared) is $\MBTAEXAMPLEFOREGROUNDRATEDENSITY$, while the local background rate density is $\MBTAEXAMPLEBACKGROUNDRATEDENSITY$.
  For each of the bins, the background is constructed by making random coincidences of single-detector triggers for each coincidence type (\ac{HL}, \ac{LV}, \ac{HV} or \ac{HLV}) using the templates of the bin considered, but only during \ac{HL} and \ac{HLV} coincidence time to remove single-detector events from the background estimation~\cite{Adams:2015ulm,Aubin:2020goo}.
  This means that the background assigned to an event depends on its coincidence type and on the bin which triggered the associated template.
  The foreground for the \ac{BNS} and \ac{NSBH} categories is estimated using the populations described in Table~\ref{tab:populations}.
  The foreground estimate for the \ac{BBH} uses the \PLPEAK{} population model used to describe the \GWTCTWO{} population~\cite{Talbot:2018cva,Abbott:2020gyp}.

  \item Just as \PYCBC{}, \ac{CWB} also uses time-shifted analysis for significance assessment of background and foreground triggers. 
The distribution of the coherent network \ac{SNR} ranking statistic for the time-shifted triggers is used to estimate the background, and consequently to assign the \ac{FAR}. 
  The foreground is derived from the recovered simulated signals. 
  Since \ac{CWB} is significantly more sensitive to \ac{BBH} systems, only these sources are considered.
\end{itemize}
The precise \pastro{} value depends upon the assumed true population, and hence may be subject to change as we learn more about the astrophysical population of \acp{CBC}. 
The population models used by the various pipelines in their computation of \pastro{} are summarized in Table~\ref{tab:populations}.

When estimating \pastro{} for each candidate, we do so separately for each category of source, as \pastro{} is dependent on the underlying \ac{BNS}, \ac{NSBH} and \ac{BBH} populations. 
We separate the candidates based on their component masses; rather than a rigorous statement of the nature of the component, the \ac{NS} label is used only to identify components whose masses are below \PASTROBBHBOUNDARY{}. 
\ac{BBH}-category candidates are any for which component masses are both above \PASTROBBHBOUNDARY{}, \ac{BNS}-category candidates are the ones for which both component masses fall below this value, and we consider a candidate a part of the \ac{NSBH} category if the primary component mass was above this boundary, and the secondary below it. 
The category chosen for each source is based on the masses of the template found by the search pipeline, rather than a detailed inference of source properties (Sec.~\ref{sec:parameter-estimation}); this may lead to \pastro{} estimates that are biased relative to an ideal calculation using the full information available for the signals.

In Table~\ref{tab:multi_pastro} we give the calculated probabilities that a candidate comes from a system in our \ac{BBH} category \pbbh{}, our \ac{NSBH} category \pnsbh{}, or our \ac{BNS} category \pbns{}.
The probability that a candidate belongs to a specific astrophysical source category (\pbns{}, \pnsbh{} or \pbbh{}) is evaluated from source-class-specific Bayes factors by redistributing the foreground probabilities across astrophysical source classes.
This redistribution makes use of the template-based estimate of the component masses of the candidate, as well as the response of the template bank to an assumed population of \ac{BNS}, \ac{NSBH} and \ac{BBH} signals.
The computation of the probability that a candidate comes from a system in one of the three astrophysical categories requires the choice of a prior on the counts in each category~\cite{Abbott:2016drs}.
\GSTLAL{} used a uniform prior for the \ac{BNS} and \ac{NSBH} categories, and a Poisson--Jeffreys prior for the \ac{BBH} category; \ac{MBTA} used a uniform prior for the \ac{BNS} category, and a Poisson--Jeffreys prior for the \ac{NSBH} and \ac{BBH} categories; \PYCBC{} used a Poisson--Jeffreys prior for all three categories, and \ac{CWB} used a Poisson--Jeffreys prior.
Given the number of candidates, the prior choice does not significantly impact the \ac{BBH} results, but can influence the \ac{BNS} and \ac{NSBH} \pastro{} values (e.g., variations of $\GSTLALALLSKYPASTRODIFFPRIORS{}$ for \FULLNAME{200105F}{}).

In addition to the choice of prior on count, \pbns{}, \pnsbh{} and \pbbh{} also depend upon the assumed foreground and background.
The methods to redistribute the foreground probabilities across astrophysical source classes are specific to individual detection pipelines:
\begin{itemize}
   \item \GSTLAL{} classifies signals into \ac{BNS}, \ac{NSBH}, and \ac{BBH} using a semianalytic template weighting scheme~\cite{Fong:2018elx}, which is needed for a multicomponent \pastro{} calculation~\cite{Kapadia:2019uut}.
   The response of each template to signals from the different categories are computed assuming Gaussian noise~\cite{Fong:2018elx} instead of using simulated signals. 
   For a given trigger, the template identified for this classification is the one which has the highest \ac{SNR} divided by the value of the signal-based-veto test, rather than the one with the highest likelihood ratio.

   \item For \ac{MBTA}, the fraction of recovered simulated signals from each category are used to infer the probabilities~\cite{Andres:2021vew}. 
   Following \GWTCTWOFINAL{}~\cite{LIGOScientific:2021usb}, this analysis assumes an astrophysical population where \acp{BNS} have a maximum component mass of $\MBTAPASTROBNSMASSMAX$, \acp{NSBH} have one component above $\MBTAPASTROBNSMASSMAX$ and one below $\MBTAPASTROBNSMASSMAX$, and \acp{BBH} have both components above $\MBTAPASTROBBHMASSMIN$. 
   While this division between \acp{NS} and \acp{BH} does not match the other analyses, it should preserve our goal of the \ac{BBH} category only including confident \acp{BH} with masses above $\PEBHMassThreshold$, while the \ac{BNS} and \ac{NSBH} categories include any systems that could contain a \ac{NS} (as well as potentially some low-mass \acp{BH}).

   \item For \PYCBC{}, categories are assigned based on the source chirp mass. 
   This is estimated by correcting the redshifted template masses using a luminosity distance derived from the \acp{SNR}~\cite{DalCanton:2020vpm}. 
   As the \PYCBCBBH{} analysis is not sensitive to redshifted chirp masses below $\PYCBCPASTROMINMCHIRPBBH{} \Msun$ (corresponding to an equal-mass binary with components of $\PYCBCPASTROBBHMASSMIN$), we do not calculate \pbns{} for this analysis.

   \item As discussed above, \ac{CWB} is most sensitive to \ac{BBH} signals, and, in this analysis, \acp{BBH} are the only astrophysical source class considered for this pipeline. 
   The assumption that all signals identified by \ac{CWB} correspond to \acp{CBC} is discussed further in Appendix~\ref{sec:cwb-only-events}.
\end{itemize}
Given our current uncertainties on the maximum \ac{NS} mass and minimum \ac{BH} mass, the three categories do not necessarily reflect the true nature of the source, but should serve to highlight candidates of interest if looking for potential \acp{BNS} or \acp{NSBH}, or a clean sample of \acp{BBH}.

The precise values of astrophysical source-class probabilities are generally insensitive to assumptions for candidates confidently identified as noise ($\pastro{} \sim 0$) or signal ($\pastro{} \sim 1$).
However, marginal \pastro{} estimates ($\pastro{} \sim \PASTROTHRESH{}$) tend to fluctuate by $\mathcal{O}(\PASTROUNCERT)$ based on various choices made~\cite{Andres:2021vew}:
\begin{itemize}
   \item The choice of distribution of masses used to estimate the foreground model.
   Since the true distribution of \acp{BNS}, \acp{NSBH} and \acp{BBH} is unknown, the marginal \pastro{} values are subject to this uncertainty.

   \item The choice of injection distributions used to assess the response of the template banks to different astrophysical source classes.
   Given our lack of knowledge of the true distribution of intrinsic parameters for \ac{BNS}, \ac{NSBH} and \ac{BBH} systems, uncertainties germane to this choice are especially pertinent to the \MBTA{} estimations of \pastro{}.
   For \GSTLAL{}, the classification is most sensitive to the choice of upper limit on the \ac{NS} mass distribution, as only triggers falling close to this threshold will have an ambiguous classification.
   For \PYCBC{}, the corresponding uncertainty comes from the choice of threshold on \Mc{} used to assign a candidate to the \ac{BBH} source class.
   Using the response of the template as a means to account for biases in the template-based estimate of intrinsic parameters is itself expected to be suboptimal as compared to a full inference of these parameters, and is therefore itself a source of uncertainty.

   \item The location of the boundary between source classes in mass space.
   The upper limit on the \ac{NS} mass is set at $\PEBHMassThreshold{}$, although the true boundary is unknown.
   Marginal candidates with components close to this boundary could have significantly different \pastro{} depending on which side of the boundary the template estimates of their masses put them.
   For example, a marginal candidate categorized as \ac{BBH} would have a larger \pastro{} than the same candidate categorized as \ac{NSBH}, since \pastro{} depends on the number of foreground candidates pertaining to these source categories; this is the case of \FULLNAME{GW191219E}{}.

   \item Specifically for single-detector candidates, the background distribution must be extrapolated to evaluate the background probability.
   For coincident candidates, the background models are built from random coincidences from data between pairs of detectors time-shifted with respect to each other, which is not possible for single-detector candidates.
\end{itemize}
While the above captures some of the primary factors that affect the values of marginal \pastro{}, the list is not exhaustive. 
Marginal \pastro{} values also depend on other factors which are specific to the analysis methods used by different detection pipelines. 
Additionally, we expect that the estimated values of \pastro{} may change as we learn more about the various astrophysical populations. 
Using the expanded list of candidates including the subthreshold candidates, it is possible to use updated population models to reevaluate \pastro{} and compile revised lists of probable \ac{GW} candidates.

\begin{event_table}
\begin{table*}
\tiny
% Made by ./scripts/generating_tex_macros/make_multicomponent_pastro_table
% DO NOT EDIT THIS FILE DIRECTLY

\begin{ruledtabular}
\begin{tabular}{l c cccc cccc cccc ccc}
 {Name} & {\ac{CWB}} & \multicolumn{4}{c}{\GSTLAL{}} & \multicolumn{4}{c}{\MBTA{}} & \multicolumn{4}{c}{\PYCBCHYPERBANK{}} & \multicolumn{3}{c}{\PYCBCBBH{}} \\
\cline{2-2} \cline{3-6} \cline{7-10} \cline{11-14} \cline{15-17}
 &  $\pastro{}$  & $\pbbh{}$ & $\pnsbh{}$ & $\pbns{}$ & $\pastro{}$  & $\pbbh{}$ & $\pnsbh{}$ & $\pbns{}$ & $\pastro{}$  & $\pbbh{}$ & $\pnsbh{}$ & $\pbns{}$ & $\pastro{}$  & $\pbbh{}$ & $\pnsbh{}$ & $\pastro{}$ \\
\hline
\makebox[0pt][l]{\fboxsep0pt\colorbox{lightgray}{\mystrut\hspace*{1.0\linewidth}}}\!\! \FULLNAME{191118N} &
$ \CWBALLSKYMEETSPASTROTHRESH{191118N} \CWBALLSKYPASTRO{191118N}$ &
$ \GSTLALALLSKYMEETSPASTROTHRESH{191118N} \GSTLALALLSKYPBBH{191118N}$ &
$ \GSTLALALLSKYMEETSPASTROTHRESH{191118N} \GSTLALALLSKYPNSBH{191118N}$ &
$ \GSTLALALLSKYMEETSPASTROTHRESH{191118N} \GSTLALALLSKYPBNS{191118N}$ &
$ \GSTLALALLSKYMEETSPASTROTHRESH{191118N} \GSTLALALLSKYPASTRO{191118N}$ &
$ \MBTAALLSKYMEETSPASTROTHRESH{191118N} \MBTAALLSKYPBBH{191118N}$ &
$ \MBTAALLSKYMEETSPASTROTHRESH{191118N} \MBTAALLSKYPNSBH{191118N}$ &
$ \MBTAALLSKYMEETSPASTROTHRESH{191118N} \MBTAALLSKYPBNS{191118N}$ &
$ \MBTAALLSKYMEETSPASTROTHRESH{191118N} \MBTAALLSKYPASTRO{191118N}$ &
$ \PYCBCALLSKYMEETSPASTROTHRESH{191118N} \PYCBCALLSKYPBBH{191118N}$ &
$ \PYCBCALLSKYMEETSPASTROTHRESH{191118N} \PYCBCALLSKYPNSBH{191118N}$ &
$ \PYCBCALLSKYMEETSPASTROTHRESH{191118N} \PYCBCALLSKYPBNS{191118N}$ &
$ \PYCBCALLSKYMEETSPASTROTHRESH{191118N} \PYCBCALLSKYPASTRO{191118N}$ &
$ \PYCBCHIGHMASSMEETSPASTROTHRESH{191118N} \PYCBCHIGHMASSPBBH{191118N}$ &
$ \PYCBCHIGHMASSMEETSPASTROTHRESH{191118N} \PYCBCHIGHMASSPNSBH{191118N}$ &
$ \PYCBCHIGHMASSMEETSPASTROTHRESH{191118N} \PYCBCHIGHMASSPASTRO{191118N} $ \\
\end{tabular}
\end{ruledtabular}

\caption{
\label{tab:multi_pastro}
Multicomponent \pastro{} values for candidates with $\pastro{} > \PASTROTHRESH{}$ and marginal candidates with \ac{FAR} $<\FARTHRESHYR{}~\mathrm{yr^{-1}}$ where the probability of a \ac{BNS} or \ac{NSBH} category is nonzero in any search analysis.
Since \ac{CWB} does not calculate separate source probabilities, all sources are treated as \acp{BBH} for the purposes of \pastro{} calculation. 
Results in \textit{italics} indicate where an analysis found the candidate with $\pastro{} < \PASTROTHRESHOLD{}$, and a dash (--) indicates that a candidate was not found by an analysis.
Source probability for \ac{BNS} is not given for \PYCBCBBH{}, as the search is not sensitive to redshifted chirp masses below $\PYCBCPASTROMINMCHIRPBBH\Msun$.
This would require extremely high redshifts, to which \ac{LIGO} and Virgo are not sensitive, to correspond to a \ac{BNS} source. 
The \ac{BNS}, \ac{NSBH} and \ac{BBH} categories are defined by the masses associated with the candidate from the search results (as defined in Table~\ref{tab:populations}), and do not necessarily correspond to the true astrophysical population of sources.
}
\end{table*}
\end{event_table}

\subsubsection{Results for all of \ac{O3}}
\label{sec:allo3_results}

Here we present results from all of \ac{O3}, giving sensitivity estimates for the points in parameter space discussed in Sec.~\ref{sec:vt} from injections covering all of \ac{O3}, and the updated \pastro{} for candidates in \ac{O3a} given the updated event-rate information inclusive of \ac{O3b}.

The sensitive hypervolume \VT{} for each search analysis for all of \ac{O3} is presented in Table~\ref{tab:vt_allo3}. 
These results show the same trends as shown in Table~\ref{tab:vt} and Fig.~\ref{fig:vt} for \ac{O3b}. 
However, the values are naturally larger on account of the greater observing time.

\begin{table*}
\footnotesize
% Made by ./scripts/generating_tex_macros/make_allo3_vt_table
% DO NOT EDIT THIS FILE DIRECTLY

\begin{ruledtabular}
\begin{tabular}{ccc ccccc c}
 \multicolumn{3}{c}{ {Binary masses  (\Msun)} }  & \multicolumn{6}{c}{ {Sensitive hypervolume (\Gpcyr)} } \\
\cline{1-3} \cline{4-9}
  \massone{} & \masstwo{} & \Mc{} & {\CWB{}} & {\GSTLAL{}} & {\MBTA{}} & {\PYCBCHYPERBANK{}} & {\PYCBCBBH{}} & {Any} \\
\hline
\makebox[0pt][l]{\fboxsep0pt\colorbox{lightgray}{\mystrut\hspace*{1.000000\linewidth}}}\!\! $ 35.0 $ & $ 35.0 $ & $30.5 $ &
$ \CWBALLSKYALLOTHREEVT{THIRTYFIVETHIRTYFIVE} $ &
$ \GSTLALALLSKYALLOTHREEVT{THIRTYFIVETHIRTYFIVE} $ &
$ \MBTAALLSKYALLOTHREEVT{THIRTYFIVETHIRTYFIVE} $ &
$ \PYCBCALLSKYALLOTHREEVT{THIRTYFIVETHIRTYFIVE} $ &
$ \PYCBCHIGHMASSALLOTHREEVT{THIRTYFIVETHIRTYFIVE} $ &
$ \ANYCBCALLOTHREEVT{THIRTYFIVETHIRTYFIVE} $ \\
\makebox[0pt][l]{\fboxsep0pt{\mystrut\hspace*{1.000000\linewidth}}}\!\! $ 35.0 $ & $ 20.0 $ & $22.9 $ &
$ \CWBALLSKYALLOTHREEVT{THIRTYFIVETWENTY} $ &
$ \GSTLALALLSKYALLOTHREEVT{THIRTYFIVETWENTY} $ &
$ \MBTAALLSKYALLOTHREEVT{THIRTYFIVETWENTY} $ &
$ \PYCBCALLSKYALLOTHREEVT{THIRTYFIVETWENTY} $ &
$ \PYCBCHIGHMASSALLOTHREEVT{THIRTYFIVETWENTY} $ &
$ \ANYCBCALLOTHREEVT{THIRTYFIVETWENTY} $ \\
\makebox[0pt][l]{\fboxsep0pt\colorbox{lightgray}{\mystrut\hspace*{1.000000\linewidth}}}\!\! $ 35.0 $ & $ 1.5 $ & $5.2 $ &
$ \CWBALLSKYALLOTHREEVT{THIRTYFIVEONEPOINTFIVE} $ &
$ \GSTLALALLSKYALLOTHREEVT{THIRTYFIVEONEPOINTFIVE} $ &
$ \MBTAALLSKYALLOTHREEVT{THIRTYFIVEONEPOINTFIVE} $ &
$ \PYCBCALLSKYALLOTHREEVT{THIRTYFIVEONEPOINTFIVE} $ &
$ \PYCBCHIGHMASSALLOTHREEVT{THIRTYFIVEONEPOINTFIVE} $ &
$ \ANYCBCALLOTHREEVT{THIRTYFIVEONEPOINTFIVE} $ \\
\makebox[0pt][l]{\fboxsep0pt{\mystrut\hspace*{1.000000\linewidth}}}\!\! $ 20.0 $ & $ 20.0 $ & $17.4 $ &
$ \CWBALLSKYALLOTHREEVT{TWENTYTWENTY} $ &
$ \GSTLALALLSKYALLOTHREEVT{TWENTYTWENTY} $ &
$ \MBTAALLSKYALLOTHREEVT{TWENTYTWENTY} $ &
$ \PYCBCALLSKYALLOTHREEVT{TWENTYTWENTY} $ &
$ \PYCBCHIGHMASSALLOTHREEVT{TWENTYTWENTY} $ &
$ \ANYCBCALLOTHREEVT{TWENTYTWENTY} $ \\
\makebox[0pt][l]{\fboxsep0pt\colorbox{lightgray}{\mystrut\hspace*{1.000000\linewidth}}}\!\! $ 20.0 $ & $ 10.0 $ & $12.2 $ &
$ \CWBALLSKYALLOTHREEVT{TWENTYTEN} $ &
$ \GSTLALALLSKYALLOTHREEVT{TWENTYTEN} $ &
$ \MBTAALLSKYALLOTHREEVT{TWENTYTEN} $ &
$ \PYCBCALLSKYALLOTHREEVT{TWENTYTEN} $ &
$ \PYCBCHIGHMASSALLOTHREEVT{TWENTYTEN} $ &
$ \ANYCBCALLOTHREEVT{TWENTYTEN} $ \\
\makebox[0pt][l]{\fboxsep0pt{\mystrut\hspace*{1.000000\linewidth}}}\!\! $ 20.0 $ & $ 1.5 $ & $4.2 $ &
$ \CWBALLSKYALLOTHREEVT{TWENTYONEPOINTFIVE} $ &
$ \GSTLALALLSKYALLOTHREEVT{TWENTYONEPOINTFIVE} $ &
$ \MBTAALLSKYALLOTHREEVT{TWENTYONEPOINTFIVE} $ &
$ \PYCBCALLSKYALLOTHREEVT{TWENTYONEPOINTFIVE} $ &
$ \PYCBCHIGHMASSALLOTHREEVT{TWENTYONEPOINTFIVE} $ &
$ \ANYCBCALLOTHREEVT{TWENTYONEPOINTFIVE} $ \\
\makebox[0pt][l]{\fboxsep0pt\colorbox{lightgray}{\mystrut\hspace*{1.000000\linewidth}}}\!\! $ 10.0 $ & $ 10.0 $ & $8.7 $ &
$ \CWBALLSKYALLOTHREEVT{TENTEN} $ &
$ \GSTLALALLSKYALLOTHREEVT{TENTEN} $ &
$ \MBTAALLSKYALLOTHREEVT{TENTEN} $ &
$ \PYCBCALLSKYALLOTHREEVT{TENTEN} $ &
$ \PYCBCHIGHMASSALLOTHREEVT{TENTEN} $ &
$ \ANYCBCALLOTHREEVT{TENTEN} $ \\
\makebox[0pt][l]{\fboxsep0pt{\mystrut\hspace*{1.000000\linewidth}}}\!\! $ 10.0 $ & $ 5.0 $ & $6.1 $ &
$ \CWBALLSKYALLOTHREEVT{TENFIVE} $ &
$ \GSTLALALLSKYALLOTHREEVT{TENFIVE} $ &
$ \MBTAALLSKYALLOTHREEVT{TENFIVE} $ &
$ \PYCBCALLSKYALLOTHREEVT{TENFIVE} $ &
$ \PYCBCHIGHMASSALLOTHREEVT{TENFIVE} $ &
$ \ANYCBCALLOTHREEVT{TENFIVE} $ \\
\makebox[0pt][l]{\fboxsep0pt\colorbox{lightgray}{\mystrut\hspace*{1.000000\linewidth}}}\!\! $ 10.0 $ & $ 1.5 $ & $3.1 $ &
$ \CWBALLSKYALLOTHREEVT{TENONEPOINTFIVE} $ &
$ \GSTLALALLSKYALLOTHREEVT{TENONEPOINTFIVE} $ &
$ \MBTAALLSKYALLOTHREEVT{TENONEPOINTFIVE} $ &
$ \PYCBCALLSKYALLOTHREEVT{TENONEPOINTFIVE} $ &
$ \PYCBCHIGHMASSALLOTHREEVT{TENONEPOINTFIVE} $ &
$ \ANYCBCALLOTHREEVT{TENONEPOINTFIVE} $ \\
\makebox[0pt][l]{\fboxsep0pt{\mystrut\hspace*{1.000000\linewidth}}}\!\! $ 5.0 $ & $ 5.0 $ & $4.4 $ &
$ \CWBALLSKYALLOTHREEVT{FIVEFIVE} $ &
$ \GSTLALALLSKYALLOTHREEVT{FIVEFIVE} $ &
$ \MBTAALLSKYALLOTHREEVT{FIVEFIVE} $ &
$ \PYCBCALLSKYALLOTHREEVT{FIVEFIVE} $ &
$ \PYCBCHIGHMASSALLOTHREEVT{FIVEFIVE} $ &
$ \ANYCBCALLOTHREEVT{FIVEFIVE} $ \\
\makebox[0pt][l]{\fboxsep0pt\colorbox{lightgray}{\mystrut\hspace*{1.000000\linewidth}}}\!\! $ 5.0 $ & $ 1.5 $ & $2.3 $ &
$ \CWBALLSKYALLOTHREEVT{FIVEONEPOINTFIVE} $ &
$ \GSTLALALLSKYALLOTHREEVT{FIVEONEPOINTFIVE} $ &
$ \MBTAALLSKYALLOTHREEVT{FIVEONEPOINTFIVE} $ &
$ \PYCBCALLSKYALLOTHREEVT{FIVEONEPOINTFIVE} $ &
$ \PYCBCHIGHMASSALLOTHREEVT{FIVEONEPOINTFIVE} $ &
$ \ANYCBCALLOTHREEVT{FIVEONEPOINTFIVE} $ \\
\makebox[0pt][l]{\fboxsep0pt{\mystrut\hspace*{1.000000\linewidth}}}\!\! $ 1.5 $ & $ 1.5 $ & $1.3 $ &
$ \CWBALLSKYALLOTHREEVT{ONEPOINTFIVEONEPOINTFIVE} $ &
$ \GSTLALALLSKYALLOTHREEVT{ONEPOINTFIVEONEPOINTFIVE} $ &
$ \MBTAALLSKYALLOTHREEVT{ONEPOINTFIVEONEPOINTFIVE} $ &
$ \PYCBCALLSKYALLOTHREEVT{ONEPOINTFIVEONEPOINTFIVE} $ &
$ \PYCBCHIGHMASSALLOTHREEVT{ONEPOINTFIVEONEPOINTFIVE} $ &
$ \ANYCBCALLOTHREEVT{ONEPOINTFIVEONEPOINTFIVE} $ \\
\end{tabular}
\end{ruledtabular}

\caption{
\label{tab:vt_allo3}
Sensitive hypervolume \VT{} for the various search analyses for all of \ac{O3} at the assessed points in the mass parameter space.
The \emph{Any} results come from calculating the sensitive hypervolume for injections found by at least one search analysis.
The sets of binary masses and distribution of injections found in this bin are the same as given in Table~\ref{tab:vt}.
As in Table~\ref{tab:vt}, where insufficient numbers of injections are recovered such that the sensitive hypervolume cannot be accurately estimated; these cases are indicated by a dash (--).
}
\end{table*}

Finally, in Table~\ref{tab:o3a_pastro} we provide updated calculations of \pastro{} for \ac{O3a} candidates which were published in \GWTCTWOFINAL{}~\cite{LIGOScientific:2021usb} with $\pastro > \PASTROTHRESH{}$ using data from the whole of \ac{O3}. 
For the first time for these candidates, we also report $\pastro{}$ as calculated by the \ac{CWB} pipeline. 
While there are small changes in value compared to the calculation using only \ac{O3a} data, there are no changes in the list of candidates with $\pastro > \PASTROTHRESH{}$.
The change in \pastro{} for \NNAME{GW190425B}, from \GWTCTWOFINALBNSPASTRO{} in \GWTCTWOFINAL{} to $\GSTLALALLSKYPASTRO{GW190425B}$ here, stems from the increased \VT{} with no new confirmed \ac{BNS} detection in \ac{O3b}, and illustrates how medium-range \pastro{} values are subject to vary with our knowledge of source populations. 

\begin{table*}
\tiny
% Made by ./scripts/generating_tex_macros/make_o3a_pastro_table
% DO NOT EDIT THIS FILE DIRECTLY

\begin{ruledtabular}
\begin{tabular}{l ccc ccc ccc ccc ccc}
 {Name} & \multicolumn{3}{c}{\ac{CWB}} & \multicolumn{3}{c}{\GSTLAL{}} & \multicolumn{3}{c}{\MBTA{}} & \multicolumn{3}{c}{\PYCBCHYPERBANK{}} & \multicolumn{3}{c}{\PYCBCBBH{}} \\
\cline{2-4} \cline{5-7} \cline{8-10} \cline{11-13} \cline{14-16}
 & \ac{FAR} & \ac{SNR} & $\pastro{}$  & \ac{FAR} & \ac{SNR} & $\pastro{}$  & \ac{FAR} & \ac{SNR} & $\pastro{}$  & \ac{FAR} & \ac{SNR} & $\pastro{}$  & \ac{FAR} & \ac{SNR} & $\pastro{}$  \\
 & ($\mathrm{yr}^{-1}$) & &  & ($\mathrm{yr}^{-1}$) & &  & ($\mathrm{yr}^{-1}$) & &  & ($\mathrm{yr}^{-1}$) & &  & ($\mathrm{yr}^{-1}$) & &  \\
\hline
\makebox[0pt][l]{\fboxsep0pt\colorbox{lightgray}{\mystrut\hspace*{1.000000\linewidth}}}\!\! {\NNAME{GW190403B}} &
$\CWBALLSKYMEETSPASTROTHRESH{GW190403B} \CWBALLSKYFAR{GW190403B}$ &
$\CWBALLSKYMEETSPASTROTHRESH{GW190403B} \CWBALLSKYSNR{GW190403B}$ &
$\CWBALLSKYMEETSPASTROTHRESH{GW190403B} \CWBALLSKYPASTRO{GW190403B}$ &
$\OTHREEAGSTLALALLSKYMEETSPASTROTHRESH{GW190403B} \GSTLALALLSKYFAR{GW190403B}$ &
$\OTHREEAGSTLALALLSKYMEETSPASTROTHRESH{GW190403B} \GSTLALALLSKYSNR{GW190403B}$ &
$\OTHREEAGSTLALALLSKYMEETSPASTROTHRESH{GW190403B} \GSTLALALLSKYPASTRO{GW190403B}$ &
$\OTHREEAMBTAALLSKYMEETSPASTROTHRESH{GW190403B} \MBTAALLSKYFAR{GW190403B}$ &
$\OTHREEAMBTAALLSKYMEETSPASTROTHRESH{GW190403B} \MBTAALLSKYSNR{GW190403B}$ &
$\OTHREEAMBTAALLSKYMEETSPASTROTHRESH{GW190403B} \MBTAALLSKYPASTRO{GW190403B}$ &
$\OTHREEAPYCBCALLSKYMEETSPASTROTHRESH{GW190403B} \PYCBCALLSKYFAR{GW190403B} $ &
$\OTHREEAPYCBCALLSKYMEETSPASTROTHRESH{GW190403B} \PYCBCALLSKYSNR{GW190403B} $ &
$\OTHREEAPYCBCALLSKYMEETSPASTROTHRESH{GW190403B} \PYCBCALLSKYPASTRO{GW190403B} $ &
$\OTHREEAPYCBCHIGHMASSMEETSPASTROTHRESH{GW190403B} \PYCBCHIGHMASSFAR{GW190403B}$ & 
$\OTHREEAPYCBCHIGHMASSMEETSPASTROTHRESH{GW190403B} \PYCBCHIGHMASSSNR{GW190403B}$ & 
$\OTHREEAPYCBCHIGHMASSMEETSPASTROTHRESH{GW190403B} \PYCBCHIGHMASSPASTRO{GW190403B}$ \\
\end{tabular}
\end{ruledtabular}

\caption{
\label{tab:o3a_pastro}
Updated probability of astrophysical origin \pastro{}, \ac{FAR} and \ac{SNR} values for candidates from \ac{O3a} using data from the whole of \ac{O3}.
We include \pastro{} values for any candidates that were published in \GWTCTWOFINAL{}~\cite{LIGOScientific:2021usb} with $\pastro{} > \PASTROTHRESH{}$. 
Using all of the \ac{O3} data, there are no changes to the list of candidates with $\pastro{} > \PASTROTHRESH{}$. 
As in Table~\ref{tab:events}, results in \textit{italics} indicate where an analysis found the candidate with $\pastro{} < \PASTROTHRESHOLD{}$, and a dash (--) indicates that a candidate was not found by an analysis.
Although \ac{CWB} contributed to the analysis of \NNAME{GW190814H}{}~\cite{Abbott:2020khf}, it is not included in the \ac{CWB} column because it was not detected with \ac{LV} alone with the standard data-quality vetoes, but required a manual override of the \ac{LIGO} Hanford vetoes. 
This table updates Table~I of \GWTCTWOFINAL{}~\cite{LIGOScientific:2021usb}.
}
\end{table*}

\section{Parameter-estimation methods}\label{sec:parameter-estimation-methods}

To determine the astrophysical parameters of each signal's source, we employ statistical inference techniques on the data from the interferometers.
We calculate the posterior probability distribution $\PEposterior$ for the source parameters $\PEparameter$ using Bayes's theorem~\cite{Bayes:1764vd},
\begin{equation}
  \label{eq:pe:bayestheorem}
  \PEposterior \propto \PElikelihood \PEprior,
\end{equation}
where the posterior is proportional to the prior probability distributions on the parameters $\PEprior$, and the likelihood $\PElikelihood$, which is the probability the data $\PEdata$ would be observed given a model with parameters $\PEparameter$. 
Our analysis matches that performed for \GWTCTWOFINAL{}~\cite{LIGOScientific:2021usb}.

Results from a number of analysis pipelines are presented in this work, but the principles used to construct the likelihood are the same for each~\cite{TheLIGOScientific:2016wfe}.
The data from each interferometer are analyzed coherently, making the assumption that the noise can be treated as stationary, Gaussian and independent between each of the interferometers used in the analysis over the duration analyzed for each signal~\cite{LIGOScientific:2019hgc,Berry:2014jja}. 
These assumptions result in a Gaussian likelihood~\cite{Cutler:1994ys} for a single interferometer,
\begin{equation}
  \label{eq:pe:likelihood}
  p(\PEdata^{k}| \PEparameter) \propto
  \exp \left[ - \frac{1}{2} \left\langle \PEdata^{k} - \PEmodelh^{k} \middle|  \PEdata^{k} - \PEmodelh^{k} \right \rangle \right ],
\end{equation}
where $\PEdata^{k}$ is the data and $\PEmodelh^{k}$ the waveform model evaluated at $\PEparameter$ as measured by the interferometer (incorporating the detector response~\cite{Forward:1978zm,Thorne:1987} and adjusted for detector calibration).
The operation $\langle \cdot | \cdot \rangle$ represents the noise-weighted inner product~\cite{Finn:1992wt}, which requires the precalculation of the \ac{PSD} of the noise, and a choice of frequency ranges over which the product should be calculated:  
\begin{itemize}
\item The minimum frequency $\flow$ for the inner product was chosen to be $\PEMinimumFreq$.
\item The maximum frequency was set as $\fhi = \alphaRoll (\fsamp/2)$, where $\fsamp$ is the sampling frequency ($\fsamp/2$ is the Nyquist frequency) and $\alphaRoll$ is included to avoid power loss due to the application of a window function. 
We limit power loss to $\PEPowerLoss$, which for the adopted Butterworth filter~\cite{Veitch:2014wba,Romero-Shaw:2020owr} requires $\alphaRoll = \PENyquistFactor$. 
To limit computational cost, the sampling rate was typically limited to $\fsamp = \PEMaximumSampling$ or $\fsamp = \PEUberSampling$, and a lower rate was used when $\fhi$ was high enough to fully resolve the $(\ell, |m|) = \MaxEllEm$ multipole moments. 
Given current detector sensitivity, we do not expect to gain significant information by using sampling rates above $\fsamp \sim \PEMaximumSampling$.
\item The noise \ac{PSD} for each candidate was estimated using \BAYESWAVE{}~\cite{Littenberg:2014oda,Cornish:2020dwh}. 
The \ac{PSD} was either estimated using the same data used for the likelihood calculation or for an equivalent length of adjacent data. 
We use the median inferred \ac{PSD} value at each frequency~\cite{Chatziioannou:2019zvs,Biscoveanu:2020kat}.
The various \acp{PSD} were precalculated for each candidate and used in each of the parameter-estimation studies for that candidate. 
\end{itemize} 
The duration of the data analyzed for each candidate is chosen such that the evolution of the signal from $\flow$ to merger and ringdown is captured, and that there is $\PEPostMergerTime$ of data postmerger~\cite{LIGOScientific:2021usb}.
The overall likelihood of data from across the detector network is obtained by multiplying together the single-detector likelihoods for the given set of parameters.

As described in Appendix~\ref{sec:prior-sampling}, we marginalize over the uncertainty in the strain calibration. 
The frequency and phase calibration uncertainties are modeled using frequency-dependent splines.
The coefficients of these splines are allowed to vary alongside the signal parameters with prior distributions on each spline node informed by the measured uncertainty at each node~\cite{TheLIGOScientific:2016wfe}. 
Preliminary studies~\cite{Payne:2020myg,Vitale:2020gvb} have shown that, given the \ac{SNR} of the candidates during \ac{O3}, the calibration systematic errors are expected to have negligible impact on the estimation of the astrophysical parameters.

\subsection{Data-quality mitigation}\label{sec:datamitigation}

For candidates affected by transient, non-Gaussian detector noise, as part of the event-validation process described in Sec.~\ref{sec:DQ}, we performed data-quality mitigation prior to performing source-parameter estimation, as summarized in Table~\ref{tab:dq_pe_mitigation}. 
Where possible, noise transients were modeled and subtracted with the \BAYESWAVE{} algorithm~\cite{Cornish:2014kda,Cornish:2020dwh}, or with the \GWSUBTRACT{} algorithm using a witness time series~\cite{Davis:2018yrz,Davis:2022ird}, as described in Appendix~\ref{sec:data-methods}.
Such subtraction was first used to mitigate the effects of a glitch that appeared in data from the \ac{LIGO} Livingston detector overlapping GW170817~\cite{TheLIGOScientific:2017qsa,Pankow:2018qpo}. 

When analyzing Virgo data, the systematic error in calibration around $\EUPOWERGRIDFREQ~\mathrm{Hz}$ described in Sec.~\ref{sec:calibration} was mitigated by setting the \ac{PSD} to a large value ($\PSDLargeValue{}~\mathrm{Hz}^{-1/2}$) for $\VIRGOSUPPRESSLOWERFREQ$--$\VIRGOSUPPRESSHERFREQ~\mathrm{Hz}$, such that the affected data do not influence the results.

\begin{table*}
\begin{ruledtabular}
\begin{tabular}{l l l}
{Candidate} & {Affected detectors} & {Mitigation}\\
\hline
%\FULLNAME{GW191103A}{} & Livingston & Segment length limited to $16~\mathrm{s}$\\%No adjustment needed to PE settings
\FULLNAME{GW191105C}{} & Virgo & BayesWave deglitching\\
\makebox[0pt][l]{\fboxsep0pt\colorbox{lightgray}{\mystrut\hspace*{1\linewidth}}}\!\! \FULLNAME{GW191109A}{} & Hanford, Livingston & BayesWave deglitching\\
\FULLNAME{GW191113B}{} & Hanford & BayesWave deglitching\\
\makebox[0pt][l]{\fboxsep0pt\colorbox{lightgray}{\mystrut\hspace*{1\linewidth}}}\!\! \FULLNAME{GW191127B}{} & Hanford & BayesWave deglitching\\
%\FULLNAME{GW191204G}{} & Hanford, Livingston & Segment length limited to $8~\mathrm{s}$\\%Probably no adjustment needed to PE settings
\FULLNAME{GW191219E}{} & Hanford, Livingston & BayesWave deglitching\\
\makebox[0pt][l]{\fboxsep0pt\colorbox{lightgray}{\mystrut\hspace*{1\linewidth}}}\!\! \FULLNAME{200105F}{} & Livingston & BayesWave deglitching \\
\FULLNAME{GW200115A}{} & Livingston & BayesWave deglitching\\
\makebox[0pt][l]{\fboxsep0pt\colorbox{lightgray}{\mystrut\hspace*{1\linewidth}}}\!\! \FULLNAME{GW200129D}{} & Livingston & Linear subtraction\\
%\FULLNAME{GW200208G}{} & Hanford & Segment length limited to $4~\mathrm{s}$\\%No adjustment needed to PE settings
%\FULLNAME{GW200210B}{} & Livingston & Segment length limited to $16~\mathrm{s}$\\%Probably no adjustment needed to PE settings
%\FULLNAME{GW200216G}{} & Livingston & Segment length limited to $16~\mathrm{s}$\\%No adjustment needed to PE settings
\end{tabular}
\end{ruledtabular}

\caption{
    \label{tab:dq_pe_mitigation}
    List of data used and mitigation methods applied to data surrounding each candidate prior to source-parameter estimation. 
    We list the candidates for which we performed mitigation of instrumental artifacts; there are \MITIGATIONEVENTS{} candidates reported in Table~\ref{tab:events} and the previously reported \FULLNAME{200105F}{}~\cite{LIGOScientific:2021qlt}. 
    For all analyses using Virgo data, calibration error at $\sim\EUPOWERGRIDFREQ~\mathrm{Hz}$ was mitigated by notching out the relevant frequency range.
    The noise-subtraction methods (\BAYESWAVE{}~\cite{Cornish:2014kda,Cornish:2020dwh} glitch modeling and \GWSUBTRACT{} linear subtraction using a witness~\cite{Davis:2018yrz,Davis:2022ird}) used for these candidates are detailed in Appendix~\ref{sec:data-methods}. 
}
\end{table*}

\subsection{Waveforms}\label{sec:waveforms}

The waveform models used to analyze each candidate are selected depending upon the most likely source for the signal.
Each candidate undergoes an initial parameter-estimation analysis shortly after the candidate is first identified.
This is used to roughly infer the component masses (and other properties) of the binary source of the candidate signal, which are used to verify analysis settings. 
A further, more exhaustive set of parameter-estimation analyses are conducted to produce final results. 
To assess potential systematic uncertainties from waveform modeling, we perform analyses with two waveform families~\cite{TheLIGOScientific:2016wfe}.

In cases with component masses in excess of \PEBHMassThreshold, analyses are conducted using the \SEOBNRPHM{}~\cite{Ossokine:2020kjp} and \IMRPhenomXPHM~\cite{Pratten:2020ceb} waveform models. 
The \NRSUR{} \ac{NR} surrogate model~\cite{Varma:2019csw}, previously used in a subset of analyses in \GWTCTWO{}~\cite{Abbott:2020niy}, is restricted in its length to only $\sim \NRSurOrbits$ orbits before the merger, and so not generally applicable for analysis of the candidates in this catalog.
The \SEOBNRPHM{} waveform is part of the \SEOBNR{} waveform family~\cite{Babak:2016tgq,Bohe:2016gbl}.
It is a time-domain model that is constructed by first deriving a time-dependent rotation from the coprecessing to the inertial frame using the \ac{EOB} equations of motion~\cite{Buonanno:1998gg,Buonanno:2000ef} for the spins and orbital angular momentum, and then applying this rotation to the nonprecessing (incorporating only spins parallel to the orbital angular momentum) \SEOBNRHM{} waveform.
The \SEOBNRHM{} model is computed by solving the \ac{EOB} equations, obtained by resumming \ac{PN} corrections, and incorporating information from \ac{NR} simulations and \ac{BH} perturbation theory~\cite{Cotesta:2018fcv}.
To model spin precession, \SEOBNRPHM{} numerically evolves the \ac{EOB} dynamics of the system, including the spins in the time domain~\cite{Ossokine:2020kjp}.
Since \SEOBNRPHM{} inherits its higher-order multipole moment content from \SEOBNRHM{}, it includes the modes $(\ell,|m|)=\{\SEOBNREllEm\}$ in the coprecessing frame.
The \IMRPhenomXPHM{} model is the latest in the \Phenom{} family of phenomenological, frequency-domain \ac{GW} models, and is built upon the higher-order multipole model \IMRPhenomXHM{}~\cite{Garcia-Quiros:2020qpx}.
Each of the available higher-order multipole moments modeled in \IMRPhenomXHM{}, $(\ell,|m|)=\{\IMRPhenomEllEm\}$, has been tuned to \ac{NR} and is rapidly generated through the use of frequency multibanding~\cite{Garcia-Quiros:2020qlt}.
\IMRPhenomXPHM{} includes precession effects by performing a frequency-dependent rotation on the nonprecessing \ac{GW} waveform \IMRPhenomXHM{}~\cite{Schmidt:2010it,Schmidt:2012rh,Hannam:2013oca,Ramos-Buades:2020noq,Garcia-Quiros:2020qpx}. 
The angles used arise from a multiscale expansion of the \ac{PN} equations of motion~\cite{Chatziioannou:2017tdw}. 
Neither waveform models the asymmetry between spherical harmonic modes with positive and negative spherical harmonic index $m$ \cite{Bruegmann:2007bri}, and neither was tuned to \ac{NR} in the precessing sector, but both were validated by comparing to a large set of \ac{BBH} waveforms~\cite{Ossokine:2020kjp,Pratten:2020ceb}.

When the initial parameter estimation provides evidence that the secondary mass is below \PEBHMassThreshold{} then the signal may arise from a \ac{NSBH}. 
In these cases, waveforms that include matter effects can be used to try to identify their imprint on the signal. 
We use the \SEOBNRNSBH~\cite{Matas:2020wab} and \IMRPhenomNSBH~\cite{Thompson:2020nei} waveforms.
Both are nonprecessing, frequency-domain \ac{NSBH} waveforms built upon previous nonprecessing, frequency-domain \ac{BBH} waveform models: \SEOBNRROM~\cite{Bohe:2016gbl} for \SEOBNRNSBH, and a combination of the \IMRPhenomC~\cite{Purrer:2013ojf} amplitude and \IMRPhenomD{}~\cite{Khan:2015jqa} phase for \IMRPhenomNSBH.
These models include corrections to the phase arising from matter effects as in \IMRPhenomPNRTidal{}, but have additional corrections to the amplitude tuned to \ac{NSBH} \ac{NR} waveforms.

For \acp{BBH}, the mass and spin of the final \ac{BH} are calculated from the initial masses and spins using fits to \ac{NR} results~\cite{Abbott:2017vtc,Hofmann:2016yih,Jimenez-Forteza:2016oae,Healy:2016lce,JohnsonMcDaniel:2016aaa}. 
When using \ac{NSBH} waveforms, the mass and spin of the final \ac{BH} are calculated from the initial masses, the initial \ac{BH} spin and the \ac{NS} tidal deformability $\Lambda_2$ using fits to \ac{NR} results~\cite{Zappa:2019ntl}. 
These fits are calibrated to \ac{BBH} fits~\cite{Jimenez-Forteza:2016oae} in order to recover the \ac{BBH} values in the test-mass limit ($\masstwo \to 0$) and in the absence of tides ($\Lambda_2 \to 0$).

None of the waveform models employed for the analyses presented here include the effects of orbital eccentricity, and instead assume that all binaries follow quasicircular orbits.
An eccentric source can be interpreted by a quasicircular analysis to be both higher mass and more equal mass than it truly is~\cite{Martel:1999tm,Lower:2018seu,Lenon:2020oza,OShea:2021ugg,Favata:2021vhw}.
Consequently, if any sources analyzed here have eccentric orbits, their true masses may be lower and their mass ratios more unequal than our inferred values. 
Eccentricity may also influence the inferred spins~\cite{Lenon:2020oza,Romero-Shaw:2020thy,OShea:2021ugg,Favata:2021vhw}. 
Significant eccentricity is not expected for the majority of sources considered here~\cite{Peters:1964zz,Tucker:2021mvo}. 

\subsection{Priors and sampling algorithms}\label{sec:prior-sampling}

To ensure that the parameter space for each candidate is explored adequately, each candidate is analyzed independently, with a choice of prior ranges for parameters that balance the required analysis time with the total volume of parameter space to be sampled.
For all candidates, we choose a uniform prior over spin magnitudes and redshifted component masses, and an isotropic prior over spin orientation, sky location and binary orientation~\cite{LIGOScientific:2018mvr,Abbott:2020niy}. 
The default mass-ratio prior is $\massratio{} \in [\PEMassRatioMin, 1]$ to reflect the range of calibration for our waveform models~\cite{Ossokine:2020kjp,Pratten:2020ceb}. 
However, some candidates show strong support for mass ratios outside of this range (such as \FULLNAME{GW191219E}{}).
In these cases we extend the priors, as biases due to any waveform inaccuracies are likely subdominant to those from truncating the prior, and we consider prior ranges as wide as $\massratio{} \in [\PEMassRatioStretch, 1]$. 
Following \GWTCTWO{}~\cite{Abbott:2020niy}, we reweight posteriors to have a luminosity-distance prior corresponding to a uniform merger rate in the source's comoving frame for a $\Lambda$--cold dark matter cosmology with $\HzeroSymbol = \HzeroValue$ and $\WmSymbol = \WmValue$~\cite{Ade:2015xua}. 

We employed a number of different sampling techniques and their associated parameter-estimation pipelines for the candidate signals presented in this work.
For the majority of candidates, the \BILBY{}~\cite{Ashton:2018jfp,Romero-Shaw:2020owr} and \RIFT{}~\cite{Pankow:2015cra,Lange:2017wki,Wysocki:2019grj} pipelines were used to generate samples from the posterior distributions for each signal.

\BILBY{} provides support for both Markov-chain Monte Carlo samplers and nested sampling techniques~\cite{Ashton:2018jfp}.
We use the \DYNESTY{}~\cite{Speagle:2019ivv} sampler, which uses nested sampling to sample the posterior probability distribution.
Analyses are organized using \BILBYPIPE{} which enables greater automation and reproducibility of analysis pipeline construction~\cite{Romero-Shaw:2020owr}. 
We use \BILBY{} for inferences using the \IMRPhenomXPHM{}~\cite{Pratten:2020ceb} model.

For candidates where more computationally expensive analyses were required, for example, using waveforms that include matter effects, we used the \PBILBY{} code~\cite{Smith:2019ucc}. 
This employs a highly parallel distributed approach to nested sampling that can be run over a large number of processing cores, reducing the wall time of the required computation.

To improve the sampling performance of \BILBY{} and \PBILBY{}, the posterior distribution is analytically marginalized over luminosity distance~\cite{Singer:2015ema} and geocenter time~\cite{Farr:2014aaa,Romero-Shaw:2020owr} prior to sampling. 
We reconstruct posterior distributions for marginalized parameters in postprocessing: for each sample, we interpolate over a one-dimensional likelihood computed at discrete points within the prior of the marginalized parameter, and draw one value from this posterior probability curve~\cite{Thrane:2018qnx,Romero-Shaw:2020owr}. 

For time-domain, computationally expensive waveforms, we use \RIFT{}~\cite{Lange:2018pyp}. 
This algorithm constructs the posterior probability distribution iteratively with two alternating steps. 
First, for a grid of intrinsic-parameter points, a marginalized likelihood is evaluated by integrating over extrinsic parameters (source position, orientation and coalescence time)~\cite{Pankow:2015cra}. 
From this discrete grid of likelihoods, a continuous likelihood distribution is constructed via Gaussian-process regression. 
A new grid is then sampled from the resulting posterior probability distribution; this process is repeated until convergence is reached. 
\RIFT{}'s grid-based approach has been shown to produce results consistent with our stochastic sampling algorithms~\cite{Lange:2018pyp}. 

To marginalize over calibration uncertainty~\cite{Farr:2014aab,TheLIGOScientific:2016wfe}, the calibration coefficients are sampled alongside the source parameters in inferences performed by \BILBY{} and \PBILBY{}~\cite{Romero-Shaw:2020owr}, whereas for \RIFT{}, this marginalization is done using likelihood reweighting (with the same spline calibration model) after the inference of the source parameters~\cite{Payne:2020myg}. 

All sampling algorithms return posterior samples in the same format, and these are postprocessed using \PESUMMARY{}~\cite{Hoy:2020vys} to produce uniform \HDF{} results.
In the preparation of \GWTCTWO{}~\cite{Abbott:2020niy}, we employed some automation to assist with monitoring the parameter-estimation processes as they ran.
For \GWTCTWOFINAL{} and \GWTCTHREE{}, we further developed this automation into the \ASIMOV{}~\cite{Williams:2022pgn} code.
This allowed the creation of analysis pipeline configurations to be fully automated, with the intention of ensuring consistency between analysis settings used for different algorithms.

The settings for the \BILBY{} and \RIFT{} analyses were designed to be as consistent as possible, aside from the differences in waveforms used. 
However, there do exist a number of differences between the analyses, such as the marginalization over time and the tapering applied to time-domain waveforms, that may lead to differences in results. 
Any differences should be negligible for intrinsic parameters such as the masses. 
In cases where the \BILBY{} and \RIFT{} parameters agree, we can be more confident in the robustness of results.

\section{Unconfirmed \ac{CWB}-only candidates }
\label{sec:cwb-only-events}

The minimally modeled \ac{CWB} pipeline (described in Sec.~\ref{sec:cwb}) can identify a range of signal morphologies, including signals unrelated to \ac{CBC} sources~\cite{LIGOScientific:2021hoh}. 
Since \ac{CWB} does not exploit the rich prior information provided by \ac{CBC} waveform templates, its flexibility in identifying many potential signals comes with a reduced sensitivity to \ac{CBC} signals that match such templates as compared to the matched-filter analyses. 
However, for the \ac{O3} analyses, we found that the efficiency of detection of \ac{CWB} becomes comparable to that of matched-filter pipelines for systems with $(1+z)\Mc \gtrsim \CWBMCHIRP$, and it is possible for \ac{CWB} to identify \ac{CBC} signals that would otherwise be omitted from the candidate list. 
In selecting candidates for Table~\ref{tab:events}, we use a criterion that the probability of astrophysical origin \emph{assuming a \ac{CBC} source} is $\pastro{} > \PASTROTHRESH{}$; as explained in Sec.~\ref{sec:candidates}, because we cannot assume that a candidate identified by the \ac{CWB} pipeline is consistent with a \ac{CBC} origin, we require independent support from a template-based search pipeline.

\begin{figure}
\centering
\includegraphics[width=\columnwidth]{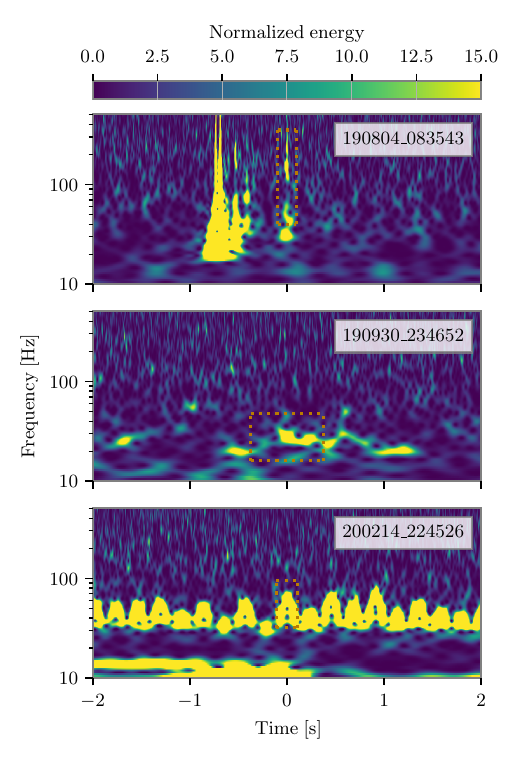} 
\caption{Spectrograms~\cite{Chatterji:2004qg} of data surrounding \FULLNAME{GW190804A}{}, \FULLNAME{GW190930A}{} and \FULLNAME{200214K}{}.  
Time is plotted relative to the central time of each trigger.  
The plotted data are from \ac{LIGO} Hanford for \FULLNAME{GW190930A}{} and from \ac{LIGO} Livingston for \FULLNAME{GW190804A}{} and \FULLNAME{200214K}{}.
The red box represents the bandwidth and duration of the candidate identified by \ac{CWB}.
In all three cases, the data are affected by transient noise at the time of the trigger, and additional excess power is present in the data that is not accounted for as part of the trigger identified by \ac{CWB}.
Although there is power present in the other detectors, the evidence for instrumental origin of the candidate in one detector makes it likely that this is just a chance coincidence.
}
\label{fig:cwb_omega}
\end{figure}
 
\begin{figure}
\centering
\includegraphics[width=0.95\columnwidth]{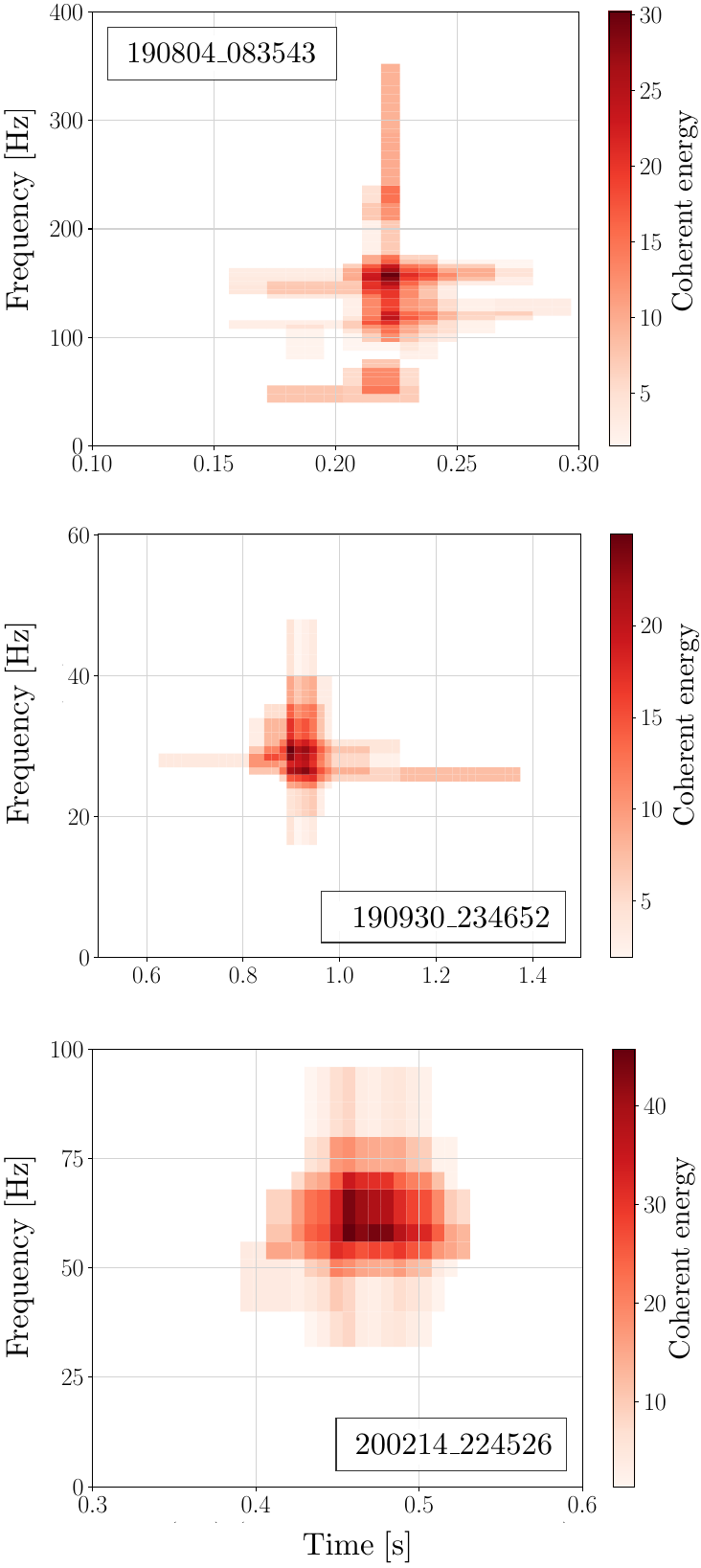} 
\caption{The coherent-energy time--frequency maps of the three candidates identified by only the \ac{CWB} analysis. 
These time--frequency maps are scalograms of the Wilson--Daubechies--Meyer wavelet transform of the candidate signal, where the scale is represented by frequency~\cite{Klimenko:2004qh,Necula:2012zz}, for the coherent energy $E_\mathrm{c}$ (see Appendix~\ref{sec:cwb-methods}).
The normalization of the coherent energy scale is such that the sum of all the pixel values times their area is equal to the power \ac{SNR}.
The time axis corresponds to \ac{GPS} times after adding the appropriate offset.
For \FULLNAME{GW190804A}{}, the offset is $\OFFSETCOA$; for \FULLNAME{GW190930A}{}, the offset is $\OFFSETCOB$, and for \FULLNAME{200214K}{}, the offset is $\OFFSETCOC$.
}
\label{fig:cwb-only}
\end{figure}

Here we discuss three candidates from \ac{CWB} that would have $\pastro{} > \PASTROTHRESH{}$ assuming a \ac{CBC} source, but for which we do not have the counterpart from the matched-filter search pipelines required to corroborate the \ac{CBC} source assumption.
The candidates \FULLNAME{GW190804A}{} and \FULLNAME{GW190930A}{} were found during \ac{O3a}, and \FULLNAME{200214K}{} was found during \ac{O3b}. 
These three candidates have \ac{FAR} $< \FARTHRESHYR{}~\mathrm{yr^{-1}}$, meeting the threshold for marginal candidates. 
The candidate \FULLNAME{GW190804A}{} was also studied in the \ac{O3} minimally modeled search for short-duration transient signals~\cite{LIGOScientific:2021hoh}, and the candidate \FULLNAME{200214K}{} was further studied in the \ac{O3} search for \ac{IMBH} binaries~\cite{LIGOScientific:2021tfm}. 
In each case, we find that the analysis and interpretation of the data is made more difficult by the presence of glitches, as illustrated in Fig.~\ref{fig:cwb_omega}.
The detailed reconstructed signal morphology is shown in Fig.~\ref{fig:cwb-only}, which displays the time--frequency map~\cite{Klimenko:2004qh,Necula:2012zz}.  
For a \ac{CBC} signal, we would typically expect the reconstructed signal to show a chirp from lower to higher frequencies, with higher-mass sources being limited to lower frequencies and shorter durations~\cite{TheLIGOScientific:2016uux,Salemi:2019uea}.
However, we find that the three candidates have a range of signal morphologies. 

The candidate \FULLNAME{GW190804A}{} was identified in low latency by the \ac{CWB} \ac{BBH} search analyzing \ac{HL} network data, and in the offline analysis its \ac{SNR} is $\CWBALLSKYSNR{GW190804A}$ and \ac{FAR} is $\CWBALLSKYFAR{GW190804A}~\mathrm{yr^{-1}}$.
It occurred less than a second after a loud series of glitches in the \ac{LIGO} Livingston detector. 
The time around these glitches was vetoed by a Burst \CATTWO{} flag that measured length sensing and control channels~\cite{Davis:2021ecd}.
Similar sequences of glitches have been observed at other times for both the \ac{LIGO} Livingston and \ac{LIGO} Hanford detectors~\cite{Zevin:2016qwy}. 
In \ac{O3}, it was observed that times around these loud glitches produced a higher rate of background triggers in the \ac{CWB} analysis, and we consider this candidate of likely instrumental origin.

The candidate \FULLNAME{GW190930A}{} was identified in low latency by the \ac{CWB} \ac{BBH} search analyzing \ac{HL} network data, and in the offline analysis its \ac{SNR} is $\CWBALLSKYSNR{GW190930A}$ and \ac{FAR} is $\CWBALLSKYFAR{GW190930A}~\mathrm{yr^{-1}}$. 
\SLOWSCATTER{} glitches~\cite{Soni:2020rbu} are present in the \ac{LIGO} Hanford data at the time of the candidate. 
These glitches correlate with the observed motion of the suspension systems and directly overlap the candidate.
At \ac{LIGO} Livingston, excess motion was measured by accelerometers at the time of the candidate that may also account for the observed signal in that detector's data. 
We consider this candidate of likely instrumental origin. 

The candidate \FULLNAME{200214K}{} was identified in low latency by the \ac{CWB} \ac{BBH} search analyzing \ac{HL} network data, and in the offline analysis its  \ac{SNR} is $\CWBALLSKYSNR{200214K}$ and \ac{FAR} is $\CWBALLSKYFAR{200214K}~\mathrm{yr^{-1}}$.
In \ac{LIGO} Livingston, the candidate was associated with a \FASTSCATTER{} glitch~\cite{Soni:2020rbu}; a sequence of such glitches is observed for multiple seconds before and after the candidate.
As shown in Fig.~\ref{fig:cwb_omega}, the glitch overlaps the candidate in \ac{LIGO} Livingston. 
In \ac{LIGO} Hanford, we find evidence of a weak scattering arch that started $\TMINCOC$ before the trigger and lasted $\TLENCOC$ in the frequency range $\FRANGECOC$. 
The candidate was studied in the search for \ac{IMBH} binaries~\cite{LIGOScientific:2021tfm}, where it was listed as the third-ranked candidate (the first ranked being \NNAME{GW190521B}{}). 
However, it was not corroborated by any matched-filter search analysis, and it was concluded that the trigger was due to noise.

For each of \FULLNAME{GW190804A}{}, \FULLNAME{GW190930A}{} and \FULLNAME{200214K}{} there is plausible evidence that the candidate is of instrumental origin.  
Regardless of the instrumental or astrophysical origin of these candidates, their morphologies (as shown in Fig.~\ref{fig:cwb-only}) do not resemble the \ac{CBC} signals so far detected.
The versatility of \ac{CWB} in identifying potential signals without a template means that a variety of sources could be detected, such that the assumption of a \ac{CBC} source is not assured and must be verified. 
Under the alternative assumption of a non-\ac{CBC} source, the probability of astrophysical origin would be reduced, making any candidates less plausible as \ac{GW} signals. 
Detection of new source types, and inference of their rates, would enable calculation of $\pastro$ for a range of sources in addition to \acp{CBC}.

\clearpage

\bibliography{o3catalog}

\end{document}